\documentclass[twoside]{report} 
\usepackage{epsf,graphicx}
\usepackage{amsmath}
\usepackage{latexsym,amssymb}
\usepackage{setspace,cite} 

\usepackage{fancyhdr}
\usepackage{a4}

\DeclareSymbolFont{AMSb}{U}{msb}{m}{n}
\DeclareMathSymbol{\N}{\mathbin}{AMSb}{"4E}
\DeclareMathSymbol{\Z}{\mathbin}{AMSb}{"5A}
\DeclareMathSymbol{\R}{\mathbin}{AMSb}{"52}
\DeclareMathSymbol{\C}{\mathbin}{AMSb}{"43}

\newcommand{\rp}{\mathbb{R}\mathbb{P}^3}

\begin{document}
\newpage

\pagenumbering{roman}
\setcounter{page}{1} \pagestyle{fancy}
\renewcommand{\chaptermark}[1]{\markboth{\chaptername%
\ \thechapter:\,\ #1}{}}
\renewcommand{\sectionmark}[1]{\markright{\thesection\,\ #1}}


\newpage

\singlespacing  

\newpage


\thispagestyle{empty}

\singlespacing

\vspace*{1cm}
\begin{center}
{\Large\bf Imprints of Spacetime Topology in the Hawking-Unruh Effect\\}
\vspace{2cm}
{\large
by Paul Pierre Langlois, MMath\\
Division of Applied Mathematics\\
\vspace{0.5cm}
School of Mathematical Sciences\\
University of Nottingham}

\vspace{3cm}

{\large
Thesis submitted to the University of Nottingham\\
for the Degree of Doctor of Philosophy, August 2005\\
\vspace{0.3cm}
}
\end{center}

\newpage

\doublespacing

\begin{abstract}


The Unruh and Hawking effects are investigated on certain families of
topologically non-trivial spacetimes using a variety of
techniques. The thesis is split into three main parts.
The first presents the Bogolubov transformation between Rindler
and Minkowski quantizations on two flat spacetimes with topology 
${\R}^3\times{S^1}$ (denoted $M_0$ and $M_-$) for massive 
Dirac spinors. The two inequivalent spin structures on each
spacetime are considered. Results show modifications to the 
Minkowski space Unruh effect. 
This provides a flat space model for the Hawking 
effect on the Kruskal and $\rp$ geon black hole 
spacetimes which is the subject of the rest of this
part.

The second part presents the expectation values of the stress
tensor for massive scalar and spinor fields on $M_0$ and $M_-$, 
and for massive scalar fields on Minkowski space with a single infinite 
plane boundary, in the Minkowski-like vacua. 

The final part investigates particle detector models. The aim 
is two-fold. First we investigate a recent paper 
by Schlicht which highlights the importance of
choosing a suitable regularization when calculating the 
response of monopole detectors in a causal way. We extend
Schlicht's regularization to an arbitrary spacetime dimension, 
to quotient spaces of Minkowski space, to non-linear couplings of the scalar 
field, to a massless Dirac field, and to conformally flat
spacetimes. Secondly we present 
some detector responses. 
Our main new results are the time dependent responses of inertial 
and uniformly accelerated detectors on $M_-$ and $M$ with boundary 
with motion perpendicular to the boundary.
Responses are also considered for static observers in the 
exterior of the $\rp$ geon and comoving observers
in $\rp$ de Sitter space, via those in the associated
global embedding Minkowski spaces (GEMS). The response
on $\rp$ de Sitter space, found both directly and 
in its GEMS, provides support for the validity of the GEMS procedure
applied to detector responses.

\end{abstract}

\newpage

\chapter*{Acknowledgements}
\addcontentsline{toc}{chapter}
         {\protect\numberline{Acknowledgements\hspace{-96pt}}}
         
Firstly I would like to thank my main thesis supervisor Dr Jorma Louko for 
suggesting to me these interesting topics, for many interesting and stimulating
discussions and for reading this manuscript. Thanks also to Professor Edward Armour
as a secondary supervisor and for many useful discussions.

I would like to thank the University of Nottingham for a University of Nottingham
postgraduate studentship
and the States of Guernsey for additional funding during my writing up
period.

Thanks of course go to my parents for all their support in the past and for never
discouraging me from doing anything I choose to do.

Thanks to all my friends in Nottingham and all over the world who have helped make
my life away from my little Island both stimulating and very enjoyable, in particular
thanks to all those associated with BSTN, you know who you are, and to Elaine and
Eadie in Taiwan.

Finally I would like the thank Belinda for all the love and care she has shown towards
me and for making my life complete.
\vspace{11mm}

\large
\hspace{-5mm}\textbf{Examiners}

\hspace{-2mm}Bernard Kay
\vspace{-2mm}

\hspace{-2mm}John Barrett
\normalsize

\singlespacing

\tableofcontents

\doublespacing

\listoffigures
\listoftables

\newpage

\pagenumbering{roman}

\pagenumbering{arabic}

\chapter{Introduction}


In this thesis we consider quantum field theory on certain families of topologically non-trivial
flat and curved spacetimes. We present a collection of new results in three
related areas of research. These include the Unruh
effect on topologically non-trivial flat spacetimes and Hawking 
radiation on black hole spacetimes via Bogolubov transformations, the evaluation
of expectation values of the stress-energy tensor, and the theory of particle 
detector models.
A good first reference to all these subjects is the well known book by Birrell 
and Davies \cite{bd:book}~\footnote{Other noteworthy books on the topic of 
quantum fields in curved spacetimes include those of Fulling \cite{fu:Fulling}
and Wald \cite{wald-qft}.}. 

Quantum field theory in curved spacetime has been an active area of research 
now for many decades, and its results and methods have been of interest to 
researchers in many related fields. The theory was constructed as a step in the
direction of a full, and not yet realised, theory of quantum gravity coupled to 
matter. The theory considers quantized fields propagating on a fixed curved
background manifold, neglecting the dynamical degrees of freedom of the
gravitational field. The gravitational field therefore is left unquantized and 
we are making the assumption that the back reaction of the quantum field on the
spacetime is negligible.
It is clear that quantum field theory in curved spacetime must be a low energy
approximation to a deeper theory whose effects become important as we 
approach the Planck scale.
Despite this limited, and as yet not fully understood, domain in which the 
theory is considered to be an accurate representation of reality, quantum field
theory in curved spacetime has had remarkable successes. Further it is expected
that many of its features and predictions must be included and explained within 
a full theory of quantum gravity.

One of the major successes of the theory is the description of particle 
production by time dependent gravitational fields.
Particle creation effects may occur due to the non 
uniqueness in choosing a time coordinate within classical general
relativity. In quantum field theory in 
Minkowski space we define particles using those solutions to the field 
equations which are positive frequency with respect to the 
timelike Killing vector $\partial_{t}$, in a standard Minkowski 
coordinate system $(t,x,y,z)$. Due to Lorentz invariance a positive frequency 
solution in one inertial frame is positive frequency in all inertial frames. 
Thus the number of particles and in particular the vacuum state
(defined as the state free from positive frequency solutions) is the same 
for all inertial observers. In a general curved spacetime however we no 
longer have global Lorentz invariance and there may be no Killing vectors at all 
with respect to which we can define positive frequency modes. Here two observers 
may not agree on which parts of a solution are positive frequency, even when
both observers follow geodesic worldlines, and as a 
result one observer sees particles in the other's vacuum state. Particles 
have been created. As we shall see this may be true even
in Minkowski space if at least one of the observers is not inertial. The first thorough 
accounts of particle creation 
in a gravitational background were given by Parker~\cite{p:part} 
in the late 60's in relation to cosmology. An expanding universe may
create particles in the process of expansion. Since then a large
amount of literature has been written on various aspects of these effects.

In 1975 Hawking published his much celebrated result~\cite{h:rad} that
a black hole formed by a star collapse radiates with a thermal spectrum 
at temperature 
\begin{equation}
T=\frac{\hbar\kappa}{2\pi{c}{k}}
 \ ,
\end{equation}
as seen by an observer at infinity, where $\kappa$ is the hole's surface gravity.				
This finding influenced a large number of papers in the following years 
as a systematic study of black hole radiation was undertaken.
Further this demonstration of black hole temperature
put the speculations of Bekenstein~\cite{b:bht} and others, on 
an analogy between the laws of classical thermodynamics and certain
relations in black hole mechanics, on a firmer basis. In particular it led 
to the acceptance of one quarter of the area \cite{b:bht} of the horizon 
as the physical entropy of the hole
\footnote{The leading order corrections to this entropy, due to quantum gravity effects
have been investigated
extensively in the recent literature. The leading corrections are logarithmic in
the area (see e.g. \cite{mu:majumdar,mu:majumdar2,s:sdas} and references therein.)}.
It was realised shortly after
the collapsing star analysis that the same temperature and entropy could 
be obtained by considering the Kruskal-Szekeres \cite{k:krusk,S:kj} extension of 
Schwarzschild and on it a quantum state that describes a black hole 
in thermal equilibrium with its environment
\cite{hh-vacuum,is:thermo,u:unruh,GH1,bd:book,kay-wald,wald-qft}.
The
defining characteristics of this Hartle-Hawking state are that it is
regular everywhere on the Kruskal manifold and invariant under the
continuous isometries~\cite{kay-wald,wald-qft}.

In 1976 Unruh~\cite{u:unruh}, following work by Fulling~\cite{f:part},
showed that these thermal effects are not restricted to curved spacetime. He 
found that a uniformly accelerated observer in Minkowski space perceives 
the usual Minkowski vacuum as a thermal bath at temperature
\begin{equation}
\label{eqn:unruhtemp}
T=\frac{\hbar{a}}{2\pi{c}k}
 \ ,
\end{equation}
where $a$ is the observer's proper acceleration. This effect is widely known
as the Unruh effect. Since Unruh's original work the Unruh effect has attracted a lot of interest
building a firm foundation for it, 
including the possibility that the effect could
be experimentally observed. Perhaps the most promising proposal so far, though
far from being realised in practice, is that of using high intensity lasers to 
accelerate electrons at very high magnitudes by Chen and Tajima \cite{ct:chentajima}.

In this thesis we consider aspects of Hawking radiation and the Unruh effect
on certain topologically non-trivial spacetimes. In particular we consider 
Hawking radiation for Dirac spinors on the spacetime
known in the terminology of \cite{fw:topcen} as the $\rp$
geon (for earlier work on the classical properties of this spacetime,
see \cite{mw:geon,Gi:thesis}).
The $\rp$ geon is a $\mathbb{Z}_2$ quotient of Kruskal, it is space and time orientable,
it contains a black and white hole, but it only has one static region,
isometric to standard exterior Schwarzschild.
The geon is an eternal black hole and so as with the Kruskal manifold we 
do not expect to see one in the sky, as an astrophysical object formed in a
star collapse, but
its interest lies in the fact that it is an unconventional black hole
and in particular has non-trivial topology. It provides an interesting arena
in which we can probe our understanding of black hole physics.
In particular here we concentrate on the effect of non-trivial topology on the 
thermal properties.
From the physical point of view, a puzzling feature of the
Hartle-Hawking state on Kruskal spacetime is its reliance on the whole 
manifold. The manifold has two static regions, causally disconnected
from each other and separated by a bifurcate Killing horizon, but the
thermal properties manifest themselves when the state is probed in
only one of the static regions. To explore the significance
of the second exterior region, we may consider a spacetime such as the
$\rp$ geon which is locally isometric to Kruskal but which only contains
one exterior region. In \cite{lm:geon} Louko and Marolf
investigated scalar field quantization on the $\rp$
geon. It was shown that the Hartle-Hawking like quantum state on the geon 
does not appear thermal to all observers in the
exterior region, but it does appear thermal in the standard Hawking
temperature when probed by suitably restricted operators. In
particular, the state appears thermal in the standard temperature to
every operator far from the hole and with support at asymptotically
late (or early) Schwarzschild times.

In chapter \ref{ch:hawkingbog} we extend the scalar field analysis of
\cite{lm:geon} to a massive Dirac field. The main new issue with
fermions is that while exterior Schwarzschild and Kruskal both have
spatial topology ${\mathbb{R}} \times S^2$ and hence a unique spin
structure, the ${\mathbb{RP}}^3$ geon has spatial topology
${\mathbb{RP}}^3 \setminus \{ \textrm{point} \}$ and thus admits two
inequivalent spin structures. The geon thus has \emph{two\/}
Hartle-Hawking like states for fermions, one for each spin
structure. Our first aim is to examine whether these states appear
thermal when probed in the exterior region: We shall find that they
do, in a limited sense similar to what was found for the scalar field
in~\cite{lm:geon}. Our second aim is to examine whether these two
states can be distinguished by observations limited to the exterior
region. We shall find that they can be in principle distinguished by
suitable interference experiments: The states contain Boulware-type
excitations in correlated pairs, and the spin structure affects the
relative phase between the members of each pair. This means that the
restriction of the Hartle-Hawking type state to the geon exterior not
only tells that the classical geometry behind the horizons differs
from Kruskal but also is sensitive to a quantization ambiguity whose
existence cannot be deduced from the exterior geometry. In this sense,
probing the quantum state in the exterior region reveals in principle
both classical and quantum information from behind the horizons. How
this information might be uncovered in practice, for example by
particle detectors will be investigated in chapter \ref{ch:pdchapt}.

As a preparation for this analysis on the $\rp$ geon, we begin in chapter \ref{ch:unruhbog} by 
analysing the Unruh effect for the massive Dirac field on a family of Rindler spaces,
denoted by $M_0$ and $M_-$, whose
topology respectively mimics that of Kruskal and the geon~\cite{lm:geon}. The results
are interesting in their own right as they illustrate a 
particular example of the effect of topology on the thermality of the
Unruh effect. They further serve as a flat space model of the 
Hawking radiation on the geon. While the results are
qualitatively similar to those on the geon, the effects of the spin
structure appear in a much more transparent form, and these Rindler
spaces thus offer a testing ground for localised particle detector
models that aim to resolve the phase factors determined by the spin
structure. 
As a by-product, we also obtain the Bogolubov transformation for massive
Dirac fermions on (ordinary) Rindler space in $(3+1)$ dimensions,
which to the knowledge of the author has not appeared in the
literature.
Chapters \ref{ch:unruhbog} and \ref{ch:hawkingbog} are based
on the paper \cite{pl:langlois}.

One of the most important observables in quantum field theory
in curved spacetime is the stress-energy tensor. Classically
the stress-energy tensor provides all the information on the 
energy and momentum content of the spacetime, and provides the
source of the gravitational field on the right hand side of
Einstein's equations. In the quantum theory the expectation value of the stress-energy
tensor in a given state provides information on the energy and momentum
content of that state, and it may further be used in a semiclassical 
analysis as the source of a semiclassical Einstein equation which 
provides information on the back reaction of the quantum field on the
spacetime geometry. In chapter \ref{ch:emchapter} we calculate the
expectation values of the stress-energy tensor for massive scalar and
spinor fields on the flat spacetimes $M_0$ and $M_-$, introduced 
in chapter \ref{ch:unruhbog}, in the Minkowski-like vacuum states.
We also present the expectation values for massless two-component spinors, showing that
the values are independent of the handedness.
This chapter is mainly based on the paper \cite{pl:langlois2} and, with the 
massless spinor results, also on \cite{pl:langlois}.

Finally in chapter \ref{ch:pdchapt} we investigate particle 
detector models. We begin in section \ref{sec-background3} with a discussion of the 
history and background of particle detector models in 
quantum field theory in curved spacetime. 
We then provide a number of new results.
The focus of the chapter is a recent model proposed by
Schlicht \cite{sc:schlicht,sc:schlicht2}. The model
is that of a monopole detector as first introduced in the
context of quantum field theory in curved spacetime by
DeWitt \cite{d:dewitt}. In \cite{sc:schlicht,sc:schlicht2}
the author considers  monopole detectors following general 
trajectories in Minkowski space which are switched
off at a finite value of the proper time, in contrast to the
asymptotic detections usually considered. Although not
the first to consider such finite time detections Schlicht
seems to be the first to highlight the importance of the
regularization of the Wightman function in such calculations.
In particular if the Wightman function is regularized using the
usual $i\epsilon$ procedure \cite{bd:book} then we are led
to an unphysical result when considering a uniformly accelerated 
detector switched on in the infinite past $\tau_0=-\infty$ and
off at $\tau<\infty$. Schlicht \cite{sc:schlicht,sc:schlicht2}
offers an alternative regularization of the correlation function
by considering a model of a spatially extended detector. The massless scalar field
operator which the detector is coupled to is smeared in the detector's
proper reference frame against a suitable window function which represents
the shape of the detector. The response depends on the detector's size;
however, in the limit as the size goes to $0$ the usual result for
a uniformly accelerated  detector is recovered (that is, a thermal response at temperature
(\ref{eqn:unruhtemp}) \cite{u:unruh,d:dewitt}).
The aim of the chapter is twofold. The first aim is to extend
Schlicht's model in a number of directions. We first make some comments on
Schlicht's regularization and some possible alternatives. We then extend the model to
Minkowski space in arbitrary dimension, to the massive scalar field,
to fields on quotient spaces of Minkowski space, to the massless
non-linearly coupled scalar field and massless Dirac field and to
conformally flat spacetimes.
The second aim is to investigate some specific detector responses.
In particular we consider inertial and uniformly accelerated detectors
on the flat, topologically non-trivial spacetimes $M_0$ and $M_-$ 
introduced in chapter \ref{ch:unruhbog} and on Minkowski space
with an infinite plane boundary. The time dependent responses on $M_-$ and 
on Minkowski space with boundary as the detector approaches the
boundary are presented for the first time.
Further we address the issue, in the case of the Dirac field,
as to whether or not such a detector can distinguish the 
two spin structures on $M_0$ and $M_-$.
Finally we consider static detectors in the exterior of the
$\rp$ geon and comoving observers in $\rp$ de Sitter space 
via their global embedding Minkowski spaces (GEMS) (see
e.g \cite{dl:deser}).
Although until now the GEMS procedure has only been applied
to kinematical arguments we expect that at least in some 
cases the response of detectors in the original curved spaces and
the corresponding ones in their GEMS should be related in some way.
The case of a comoving detector on $\rp$ de Sitter space is
a particularly interesting one as we are able to present the 
calculation both in the curved space and in the embedding space.
It is found that the responses are qualitatively very similar.
This case should therefore be very useful in assessing the
validity of applying the GEMS procedure to cases involving
quotient spaces and time dependent detector responses.
Chapter \ref{ch:pdchapt} is an extended version of \cite{pl:langlois3}.

I end this introduction with a note on the terminology and conventions
used throughout this thesis.
I use the term ``Unruh effect'' when discussing the thermal (or partial 
thermal) properties of vacuum states in all flat (possibly topologically 
non-trivial) spacetimes from the point of view of uniformly accelerated 
observers. The term ``Hawking effect'' is used when discussing thermal
properties of states in black hole spacetimes.
This terminology is in contrast to that used in some of the literature
(for example \cite{wald-qft}), where ``Hawking effect'' is reserved for 
particle production in a collapsing star spacetime and ``Unruh effect''
is used for the existence of a thermal equilibrium state in a spacetime
with a bifurcate Killing horizon.

We work throughout in natural units $\hbar=c=G=1$ and with metric signature 
$(+,-,\ldots,-)$. In $d$-dimensional Minkowski space, the spatial 
$(d-1)$-vectors are denoted by bold face 
characters $\mathbf{x}\in\R^{d-1}$ with $\cdot$ the usual scalar product 
in $\R^{d-1}$, while $d$-vectors 
(used occasionally) are given by an italic script $\mathit{x}$ 
with $\mathit{x}\cdot\mathit{y}=g_{\mu\nu}x^\mu{y^\nu}$, where 
$g_{\mu\nu}$ is the Minkowski metric.

\chapter[Unruh effect for massive spinors]{Unruh effect for massive spinors on flat spacetimes of topology 
$\mathbb{R}^3\times{S^1}$}
\label{ch:unruhbog}

\section{Introduction}

In this chapter we consider the Unruh effect for a
massive Dirac field, 
using a Bogolubov transformation
analysis, on two flat spacetimes with non-trivial
topology. Denoted here as $M_0$ and $M_-$, these two spacetimes are built as
quotients of Minkowski space under the groups generated by the discrete isometries 
$J_0:(t,x,y,z)\mapsto(t,x,y,z+2a)$ and $J_-:(t,x,y,z)\mapsto(t,-x,-y,z+a)$
respectively, where $a$ is a prescribed positive constant.
As a by-product we obtain the Bogolubov transformation for massive Dirac fermions
on (ordinary) Rindler space in $(3+1)$ dimensions, which to the knowledge of the
author has not appeared in the literature.\footnote{Thermality for massive fermions on
$(3+1)$-dimensional Rindler space is demonstrated by other methods in
\cite{tk:takagi,cd:candeld}. The massive $(1+1)$-dimensional case is
considered in~\cite{smg:rin}. The massive $(3+1)$ case is addressed in
\cite{or:oriti,or:oriti2} but the Rindler modes constructed therein are not
suitably orthonormal in the Dirac inner product.}
The underlying 
interest in $M_0$ and $M_-$ arises from their role in modelling, 
in the context of accelerated observers in flat spacetimes, the Hawking (-Unruh)
effect on the Kruskal manifold and the $\rp$ geon respectively \cite{lm:geon}.
The Hawking effect on these black hole spacetimes for massive spinors is the topic
of chapter \ref{ch:hawkingbog}. 

The chapter is constructed as follows.
We begin in section \ref{sec:m0m-intro} by describing the spacetimes $M_0$ and $M_-$
and their properties. In section \ref{sec:m0m-quant} we then quantize the spinor
field in these spacetimes in terms of globally defined Minkowski-like modes and
construct the Minkowski-like vacuum states $|0_0\rangle$ and $|0_-\rangle$.
Uniformly accelerated observers in $M_0$ are introduced in 
section \ref{sec:m0unruh}, and the Rindler particle content of the 
Minkowski like vacuum $|0_0\rangle$ is found from the explicit Bogolubov 
transformation. A similar analysis for uniformly accelerated observers in
$M_-$ is given in section \ref{sec:m-unruh}.

\section{The spacetimes $M_0$ and $M_-$}
\label{sec:m0m-intro}

Let $M$ denote $(3+1)$-dimensional Minkowski space and let $(t,x,y,z)$ 
be a standard set of Minkowski coordinates. The metric reads
\begin{equation}
ds^2=dt^2-dx^2-dy^2-dz^2 \ \ .
\end{equation}
The flat spacetimes $M_0$ and $M_-$ are built as quotients 
of $M$ under the isometry groups generated respectively by the
isometries
\begin{eqnarray} 
M_0:=M/{J_0} \ , & J_0:(t,x,y,z)\mapsto(t,x,y,z+2a) \ , 
\\
M_-:=M/{J_-} \ , & J_-:(t,x,y,z)\mapsto(t,-x,-y,z+a) \ ,
\end{eqnarray}	
where $a>0$ is a prescribed constant. 
$J_0$ and $J_-$ preserve space and time 
orientation and act freely and properly
discontinuously. $M_0$ and $M_-$ are thus
space and time orientable flat Lorentzian manifolds.

As $M$ is the universal covering space of both $M_0$ and $M_-$, 
we may understand $M_0$ and $M_-$ to be coordinatized by $(t,x,y,z)$ with the 
identifications 
\begin{eqnarray} 
M_0:  &  (t,x,y,z)\sim(t,x,y,z+2a) \ ,    
\\
M_-:  &  (t,x,y,z)\sim(t,-x,-y,z+a) \ .
\end{eqnarray}
Note further that $M_0$ is a double cover of $M_-$. 
Field theory on $M_0$ and $M_-$ can be considered via field theory on 
$M$ by the method of images (see e.g \cite{bD:auto,bd:em}).

\begin{figure}[htbp]      
\includegraphics{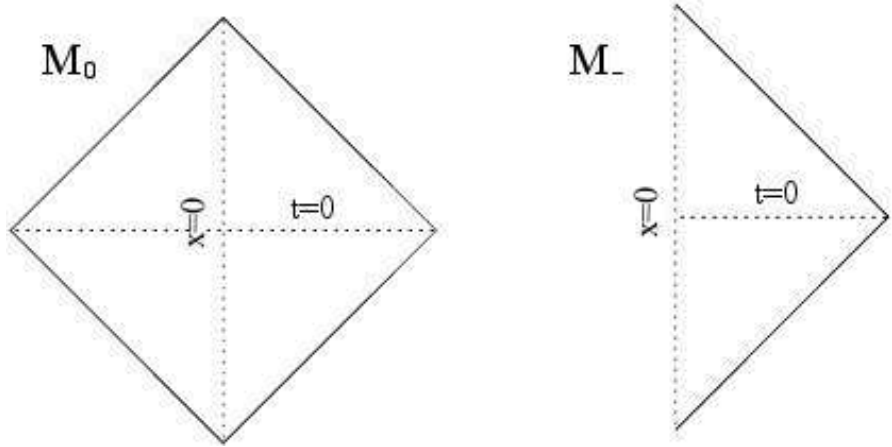}
\centering
\caption[Conformal diagrams of $M_0$ and $M_-$.]{Conformal diagrams of the spacetimes $M_0$ and $M_-$. On $M_0$ each point on the diagram represents a cylinder of circumference $2a$. On $M_-$ the region $x>0$ is identical to that on $M_0$, however at $x=0$ each point represents a suppressed Mobius strip. The infinities of the figures are not faithful to the infinity structures of the spacetimes due to the suppressed dimensions.}
\label{fig:con}
\end{figure}

Like $M$, $M_0$ and $M_-$ are static spacetimes with respect to the global 
timelike Killing vector $\partial_t$. They are globally hyperbolic and 
have spatial topology $\R^2\times{S^1}$. $M_0$ admits seven Killing vectors, 
which generate translations in all $4$ coordinates, boosts in $x$ and
$y$ and rotations in the $(x,y)$-plane. $M_-$ only admits three Killing
vectors, generating translations in $t$ and $z$ and rotations 
in the $(x,y)$-plane.
Conformal diagrams for $M_0$ and $M_-$ in which the local $y$, $z$ 
coordinates have been suppressed are given in figure~\ref{fig:con}.

Due to the $S^1$ factor in the spatial topology, $M_0$ and $M_-$ each admit two 
inequivalent spin structures (see e.g \cite{is:spin}). 
That is there are two inequivalent but equally consistent ways of attaching 
spinors to each. 
Consider first $M_0$. We refer to the vierbein
\begin{eqnarray} 
 V_0=\partial_t &  V_2=\partial_y \nonumber 
\\ 
\label{eqn:minkvier} 
V_1=\partial_x  & V_3=\partial_z \ ,
\end{eqnarray}
as the standard vierbein on $M_0$.
In the standard vierbein the two spin structures amount to imposing respectively 
periodic and antiperiodic boundary conditions as we go around the compact $z$-direction. 
Labelling the spin structures by the index $\eta\in\{1,-1\}$, this means 
\begin{equation} 
\psi(t,x,y,z+2na)=\eta^n\psi(t,x,y,z) \ ,
\end{equation}
where $n\in\Z$ and $\eta=1$ for periodic spinors (untwisted spinors in 
the terminology of \cite{bd:book,is:spin}) and $\eta=-1$ for 
antiperiodic (twisted) spinors. 

Alternatively, another useful vierbein on $M_0$ is one which 
rotates counterclockwise by $2\pi$ in the $(x,y)$-plane as $z$ 
increases by $2a$, that is,
\begin{eqnarray} 
V_0 & = & \partial_t 
\nonumber
\\
V_1 & = & \cos({{\pi}z}/a)\partial_x+\sin({{\pi}z}/a)\partial_y 
\nonumber
\\
V_2 & = & -\sin({{\pi}z}/a)\partial_x+\cos({{\pi}z}/a)\partial_y 
\nonumber
\\
\label{eqn:rotet}
V_3 & = & \partial_z  \ .
\end{eqnarray}
Spinors that are periodic with respect to this rotating 
vierbein are antiperiodic with respect to the 
standard vierbein and vice versa. Both these vierbeins are invariant 
under the action of $J_0$. One could further introduce a vierbein 
that rotates clockwise by $2\pi$ in the $(x,y)$-plane as $z$ increases
by $2a$ (replace $\pi$ with $-\pi$ in (\ref{eqn:rotet})), but
periodic (respectively antiperiodic) boundary conditions in this 
vierbein are equivalent to periodic (antiperiodic) boundary conditions 
in vierbein (\ref{eqn:rotet}). This shows that neither spin stucture on
$M_0$ involves a preferred spatial orientation.

Now consider $M_-$. The standard Minkowski vierbein (\ref{eqn:minkvier})
is not invariant under $J_-$ and therefore does not provide a globally defined
vierbein on $M_-$. However both the counterclockwise rotating vierbein 
(\ref{eqn:rotet}) and its clockwise rotating analogue are invariant under
$J_-$ and hence are well defined on $M_-$. We may therefore specify the
two spin structures on $M_-$ by working in the vierbein (\ref{eqn:rotet})
and imposing respectively periodic and antiperiodic boundary conditions
under $J_-$.
Labelling the spin structures by the index $\rho\in\{1,-1\}$ this means 
\begin{equation} 
\psi(t,(-1)^nx,(-1)^ny,z+na)=\rho^n\psi(t,x,y,z) \ ,
\end{equation}
where $n\in\Z$ and $\rho=1$ for the periodic spinors and $\rho=-1$ for the
antiperiodic spinors with respect to vierbein (\ref{eqn:rotet}).
We could equivalently work in the clockwise rotating vierbein and interchange the
periodic and antiperiodic boundary conditions. 
This shows that the choice of a spin structure on $M_-$ determines a
preferred spatial orientation. For concreteness, we shall throughout 
specify the spin structure with respect to vierbein (\ref{eqn:rotet}).

\section{Spinor field quantization on $M_0$ and $M_-$}
\label{sec:m0m-quant}


In this subsection we consider the quantum theory of a free Dirac field $\psi$ 
with mass $m\geq{0}$ on $M_0$ and $M_-$. We introduce the Minkowski-like vacua
$|0_0\rangle$ and $|0_-\rangle$ for which positive and negative frequencies 
are defined with respect to the global timelike Killing vector $\partial_t$. 
We work in the rotating tetrad (\ref{eqn:rotet}) throughout.

In a general curved spacetime the spinor Lagrangian in the 
vierbein formalism is \cite{bd:book}
\begin{equation} 
\mathcal{L}
=
\mathrm{det}
V\left(\frac{1}{2}i\left[\bar{\psi}\gamma^{\mu}\nabla_{\mu}\psi
-(\nabla_{\mu}\bar{\psi})\gamma^{\mu}\psi\right]-m\bar{\psi}\psi\right) \ ,
\end{equation}
where $V^{\mu}_{\alpha}$ is a set of vierbeins, 
$V_\alpha=V^{\mu}_\alpha\partial_{\mu}$, and 
$\gamma^{\mu}=V^{\mu}_{\alpha}\gamma^{\alpha}$ are the curved space 
counterparts of the Dirac matrices which satisfy
\begin{equation}
\{\gamma^{\mu},\gamma^{\nu}\}=2g^{\mu\nu} \ .
\end{equation}
$\nabla_{\alpha}=V^{\mu}_{\alpha}(\partial_{\mu}+\Gamma_{\mu})$ 
is a spinor covariant derivative with
\begin{equation}
\Gamma_{\mu}
=
\frac{1}{8}
V^{\nu}_{\alpha}V_{\beta\nu;\mu}
[\gamma^{\alpha},\gamma^{\beta}] \ .
\end{equation}
Variation of the action with respect to $\bar{\psi}\equiv{\psi^{\dagger}}\gamma^{0}$ 
yields the covariant Dirac equation
\begin{equation}
\label{eqn:gendir}
i\gamma^{\mu}\nabla_{\mu}\psi-m\psi=0 \ .
\end{equation}

It will be useful to work in the local Minkowski coordinates $(t,x,y,z)$ 
and in the rotating vierbein (\ref{eqn:rotet}) which is well defined on $M_0$
and $M_-$. 
The Dirac equation (\ref{eqn:gendir}) then becomes
\begin{eqnarray} 
& i\{\gamma^0\partial_t+\gamma^1\left(\cos\left(\frac{\pi{z}}{a}\right)\partial_x
+\sin\left(\frac{\pi{z}}{a}\right)\partial_y\right)
+\gamma^2\left(-\sin\left(\frac{\pi{z}}{a}\right)\partial_x
+\cos\left(\frac{\pi{z}}{a}\right)\partial_y\right)
\nonumber
\\
\label{eqn:dir} 
& +\gamma^3\left(\partial_z-\frac{\pi}{4a}\left(\gamma^1\gamma^2-\gamma^2\gamma^1\right)\right)+im\}
\psi=0 \ .
\end{eqnarray}
The inner product on $M$ is
\begin{equation}
\langle{\psi_1,\psi_2}\rangle=\int{dx}\,{dy}\,{dz}\;\psi_1^\dagger{\psi_2} \ ,
\end{equation}
on a constant $t$ hypersurface. 
We denote the inner products on $M_0$ and $M_-$ by 
$\langle{\psi_1,\psi_2}\rangle_0$ and 
$\langle{\psi_1,\psi_2}\rangle_-$ respectively.

Consider first~$M_0$. To construct solutions to (\ref{eqn:dir}) that
are positive and negative frequency with respect to $\partial_t$, we
begin with the Minkowski space positive and negative frequency
solutions in the standard vierbein (\ref{eqn:minkvier}) (see for
example~\cite{bj:plane,it-zub})
and transform to the rotating vierbein
(\ref{eqn:rotet}) by the spinor transformation associated with a
rotation by $\pi$ in the $(x,y)$-plane as $z$ increases
by~$a$. Working here and throughout this section in the standard
representation of the $\gamma$ matrices~\cite{it-zub}, $\gamma^0 =
\bigl(
\begin{smallmatrix} 1 & 0 \\ 0 & -1
\end{smallmatrix} 
\bigr)$ 
and 
$\gamma^i = 
\bigl(
\begin{smallmatrix} 
0 & \sigma_i \\ -\sigma_i & 0 
\end{smallmatrix}
\bigr)$, where $\sigma_i$ are the Pauli matrices,
$\sigma_1 =
\bigl(
\begin{smallmatrix} 0 & 1 \\ 1 & 0
\end{smallmatrix} 
\bigr)$,  
$\sigma_2 = 
\bigl(
\begin{smallmatrix} 
0 & -i \\ i & 0 
\end{smallmatrix}
\bigr)$
and 
$\sigma_3 = 
\bigl(
\begin{smallmatrix} 
1 & 0 \\ 0 & -1
\end{smallmatrix}
\bigr)$, this transformation
reads 
\begin{equation} 
\psi\mapsto{e^{-\frac{\gamma^1\gamma^2\pi{z}}{2a}}\psi}
=
\mathrm{diag} 
\left(
e^{\frac{i\pi{z}}{2a}} , 
e^{\frac{-i\pi{z}}{2a}} , 
e^{\frac{i\pi{z}}{2a}} , 
e^{\frac{-i\pi{z}}{2a}}
\right)
\psi \ .
\end{equation}
Solutions on $M_0$ can be found from the Minkowski space solutions 
via the method of images. The periodic and antiperiodic boundary conditions
will restrict the momentum in the $z$-direction.
We find that a complete set of
normalized positive frequency solutions is $\{U_{j,k_x,k_y,k_z}\}$,
where
\begin{equation}
\label{eqn:minkmodes} 
U_{j,k_x,k_y,k_z}=\frac{1}{4\pi}\sqrt{\frac{(\omega+m)}{a\omega}}
e^{-i\omega{t}+ik_{x}x+ik_{y}y+ik_{z}z}
\, u_{j,k_x,k_y,k_z} \ ,
\end{equation}
with
\begin{equation}
u_1=
\left(
\begin{array}{c}
e^{\frac{i\pi{z}}{2a}}
\\ 
0 
\\ 
e^{\frac{i\pi{z}}{2a}}\frac{k_z}{\omega+m} 
\\
e^{-\frac{i\pi{z}}{2a}}\frac{k_+}{\omega+m} 
\end{array} 
\right)
\ \ \ , \ \ \ 
u_2=
\left(
\begin{array}{c}
0 
\\ 
e^{-\frac{i\pi{z}}{2a}} 
\\ 
e^{\frac{i\pi{z}}{2a}}\frac{k_-}{\omega+m}
\\ 
e^{-\frac{i\pi{z}}{2a}}\frac{-k_z}{\omega+m} 
\end{array} 
\right) \ \ , 
\end{equation}
$k_{\pm}=k_x\pm{i}k_y$ and 
$\omega=\sqrt{m^2+k_x^2+k_y^2+k_z^2}$. 
For spinors that are periodic in the standard
vierbein~(\ref{eqn:minkvier}), 
$k_z = n\pi/a$ with $n\in{\mathbb{Z}}$,
and in the other spin structure 
$k_z= ( n+\tfrac12 )\pi/a$ with $n\in{\mathbb{Z}}$. 
$k_x$~and $k_y$ take all real values. 
The orthonormality relation is 
\begin{equation}
\langle{U_{i,k_x,k_y,k_z},U_{j,k'_x,k'_y,k'_z}}\rangle_0
=\delta_{ij}\delta_{nn'}\delta(k_x-k'_x)\delta(k_y-k'_y) \ .
\end{equation}
Note that the corresponding delta-normalized modes on Minkowski space are
$\sqrt{\frac{a}{\pi}}U_{j,k_x,k_y,k_z}$ with $k_z \in \mathbb{R}$. 

Consider now $M_-$. As $M_0$ is a double cover of $M_-$,
a complete set of modes can be built from (\ref{eqn:minkmodes})
as linear combinations which are invariant under  $J_-$, that is
by superposing the modes (\ref{eqn:minkmodes}) and their images under $J_-$ with
phase factors that lead to the appropriate (anti-)periodicity
properties. We choose the set $\{V_{j,k_x,k_y,k_z}\}$ given by 
\begin{eqnarray} 
&&V_{1,k_x,k_y,k_z}=U_{1,k_x,k_y,k_z}
+\rho{i}{e^{ik_za}}U_{1,-k_x,-k_y,k_z} \ ,
\nonumber 
\\
&&V_{2,k_x,k_y,k_z}=U_{2,k_x,k_y,k_z}
-\rho{i}{e^{ik_za}}U_{2,-k_x,-k_y,k_z} \ ,
\end{eqnarray}
where $k_z= ( n+\tfrac12 )\pi/a$, $n\in{\mathbb{Z}}$ and $\rho=1$
($\rho=-1$) gives spinors that are periodic (antiperiodic) in the
rotating vierbein~(\ref{eqn:rotet}). As with the scalar field
\cite{lm:geon} there is a redundancy in these $V$-modes in that
$V_{j,k_x,k_y,k_z}$ and $V_{j,-k_x,-k_y,k_z}$ are proportional, and we
understand this redundancy to be eliminated by taking for example
$k_y>0$.
The orthonormality condition reads
\begin{equation}
\label{eqn:orthV}
\langle{V_{i,k_x,k_y,k_z},V_{j,k'_x,k'_y,k'_z}}\rangle_-
=
\delta_{ij}\delta_{nn'}\delta(k_x-k'_x)\delta(k_y-k'_y) \ .
\end{equation}
Note that $k_z$ takes in both spin structures the same set of values,
which coincides with the set of modes in the twisted spin structure
on~$M_0$.

Given these complete sets of modes we may proceed to canonically quantize 
the field in the usual way. That is, we expand the field in these sets and 
impose the usual anticommutation relations
on the coefficients, promoted to operators (operator valued distributions). 
Let $|0\rangle$ be the
usual Minkowski vacuum on~$M$, defined by the set
$\{U_{j,k_x,k_y,k_z}\}$. We denote by $|0_0\rangle$ the vacuum on
$M_0$ defined by the set $\{U_{j,k_x,k_y,k_z}\}$ with the suitably
restricted values of $k_z$ and by $|0_-\rangle$ the vacuum on $M_-$
defined by the set $\{V_{j,k_x,k_y,k_z}\}$. $|0_0\rangle$~and
$|0_-\rangle$ both depend on the respective spin structures, but in
what follows we will not need an explicit index to indicate this
dependence.

\section[Bogolubov transformation on $M_0$]{The Unruh construction and Bogolubov transformation on Minkowski
space and $M_0$}
\label{sec:m0unruh}

In this section we consider a uniformly accelerated observer on 
the spacetimes $M$ and $M_0$. We find the Rindler particle content
of the Minkowski vacuum $|0\rangle$ on $M$ and the Minkowski-like vacuum $|0_0\rangle$ 
on $M_0$.


Let $R_0$ be the right-hand-side Rindler wedge of $M_0$, $x>|t|$. We 
introduce in $R_0$ the usual Rindler coordinates $(\tau,\xi,y,z)$ 
by
\begin{eqnarray}  
t & = & \xi\sinh\tau \ ,
\nonumber 
\\ 
\label{eqn:rin} 
x & = & \xi\cosh\tau \ ,
\end{eqnarray}
with $\xi>0$ and $-\infty<\tau<\infty$, understood with the
identification
$(\tau,\xi,y,z)\sim(\tau,\xi,y,z+2a)$. The metric reads 
\begin{equation}
{ds}^2={\xi}^2{d\tau}^2-{d\xi}^2-{dy}^2-{dz}^2
 \ \ . 
\end{equation} 
$R_0$ is a globally hyperbolic spacetime with the complete timelike
Killing vector $\partial_{\tau}=t\partial_x+x\partial_t$, which
generates boosts in the $(t,x)$-plane. The worldines at constant
$\xi$, $y$ and $z$ are those of observers accelerated uniformly in
the $x$-direction with acceleration $\xi^{-1}$ and proper time
$\xi\tau$.
\begin{figure}[htbp] 
\includegraphics{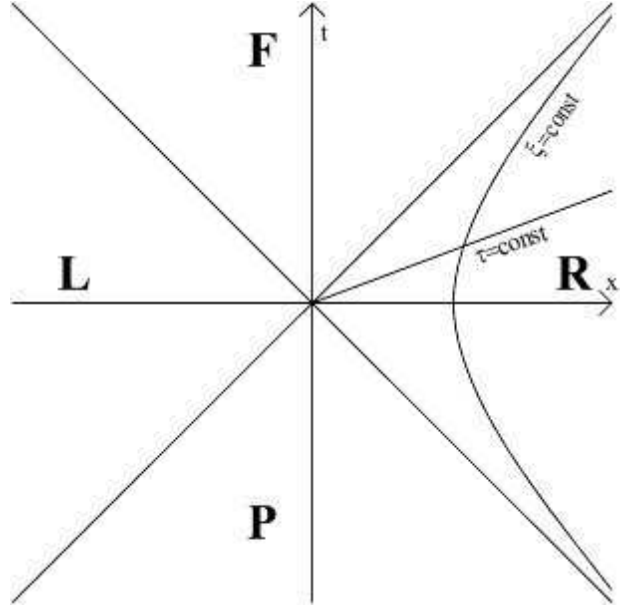}
\centering
\caption[Rindler space.]{Rindler space. The surfaces $x=t$ and $x=-t$ form horizons for 
the Rindler observers. For example an observer in $R$ following the hyperbolic 
path shown can neither send any information to region $L$ nor receive any 
information from there. The two regions are causally disconnected.}
\label{fig:mink}
\end{figure}

We need in $R_0$ a set of orthonormal field modes that are positive
frequency with respect to~$\partial_\tau$. Working in the vierbein aligned
along the Rindler coordinate axes,
\begin{equation}
\label{eqn:rinvier}
V^{\mu}_{a}=\mathrm{diag}({\xi}^{-1},1,1,1)
 \ , 
\end{equation}
the Dirac 
equation (\ref{eqn:gendir}) becomes 
\begin{equation}
\label{eqn:dirrin}
(i\partial_\tau
+i\xi\gamma^0\gamma^1\partial_\xi
+i\xi\gamma^0\gamma^2\partial_y
+i\xi\gamma^0\gamma^3\partial_z
+{i\gamma^0\gamma^1}/2-m\xi\gamma^0)
\psi
=0
\ ,
\end{equation}
where the $\gamma$ matrices are the usual flat space
$\gamma{'s}$. A complete set of mode solutions, positive frequency with respect to 
Rindler Killing time $\partial_\tau$, on $R_0$ can 
then be found by directly separating~(\ref{eqn:dirrin}) by an ansatz of
simultaneous eigenfunctions of $-i\partial_y$, $-i\partial_z$ 
and the Rindler Hamiltonian.  In 
view of comparison with $M_-$ in section~\ref{sec:m-unruh}, we wish 
the solutions to have simple transformation properties 
under~$J_-$. Modes that achieve this are 
\begin{equation}
\label{eqn:rinsol}
\psi^R_{j,\Omega,k_y,k_z}(\tau,\xi,y,z)
=N_j\left(X^R_jK_{i\Omega-\frac{1}{2}}(\kappa\xi)
+Y^R_jK_{i\Omega+\frac{1}{2}}(\kappa\xi)\right)e^{-i\Omega\tau+ik_yy+ik_zz} \ ,
\end{equation}
where
\begin{align} 
X^R_1
= & \left(\begin{array}{ccccccccccccccccccccc}\frac{k_z}{|k_z|}(k_y-im) 
\\ -i(|k_z|-\kappa) 
\\ -i(|k_z|-\kappa) 
\\ \frac{k_z}{|k_z|}(k_y-im) 
\end{array} 
\right) \ ,
 &           
Y^R_1
= & \left(\begin{array}{ccccccccccccccccccccc}\frac{k_z}{|k_z|}(|k_z|-\kappa) 
\\ i(k_y-im) 
\\ -i(k_y-im)
\\ -\frac{k_z}{|k_z|}(|k_z|-\kappa) 
\end{array} 
\right) \ ,
\nonumber 
\\ 
X^R_2
= & \left(\begin{array}{ccccccccccccccccccccc}\frac{k_z}{|k_z|}(|k_z|-\kappa) 
\\ i(k_y+im) 
\\ i(k_y+im)
\\ \frac{k_z}{|k_z|}(|k_z|-\kappa) 
\end{array} 
\right) \ , 
 &    
Y^R_2
= & \left(\begin{array}{ccccccccccccccccccccc}\frac{k_z}{|k_z|}(k_y+im) 
\\ -i(|k_z|-\kappa) 
\\ i(|k_z|-\kappa) 
\\ -\frac{k_z}{|k_z|}(k_y+im) 
\end{array} 
\right) \ , 
\end{align} 
and 
\begin{align} 
N_1=
& \frac{e^{-\frac{i\pi}{4}}\sqrt{\cosh(\pi{\Omega})(\kappa^2-k_z^2)}}
{4\pi(k_y-im)\sqrt{a\pi(\kappa-|k_z|)}} \ , 
\nonumber
\\ 
\label{eqn:norm123} 
N_2=
& \frac{e^{-\frac{i\pi}{4}}\sqrt{\cosh(\pi{\Omega})(\kappa^2-k_z^2)}}
{4\pi(k_y+im)\sqrt{a\pi(\kappa-|k_z|)}} \ , 
\end{align}
$j=1,2$, $\kappa=(m^2+{k_y}^2+{k_z}^2)^{1/2}$, $\Omega>0$ and
$k_y{\in}{\mathbb{R}}$. In the spin structure where the spinors are
periodic (respectively antiperiodic) in the nonrotating
vierbein~(\ref{eqn:rinvier}), the values of $k_z$ are $n\pi/a$
(respectively $(n+\tfrac12)\pi/a$) with
$n\in{\mathbb{Z}}$. $K_{iM+\frac{1}{2}}$ is a modified Bessel
function~\cite{zw:zwillinger}. For $k_z=0$, we understand
the formulas in (\ref{eqn:rinsol})--(\ref{eqn:norm123}) and in what
follows to stand for their limiting values
as $k_z \to {0_+}$, 
$$\psi^R_{j,\Omega,k_y,0}:=\lim_{k_z\rightarrow{0^+}}\psi^R_{j,\Omega,k_y,k_z}.$$ 
The modes are orthonormal as 
\begin{equation}
\langle{\psi^R_{i,\Omega,k_y,k_z},\psi^R_{j,\Omega',k_y',k_z'}}\rangle_{R_0}
=\delta_{ij}\delta_{nn'}\delta(\Omega-\Omega')\delta(k_y-k_y') 
\ , 
\label{eq:psi-ON}
\end{equation}
where the inner product is (see e.g \cite{cd:candeld})
\begin{equation}
\label{eqn:inner}
\langle{\psi_1,\psi_2}\rangle_{R_0} 
=\int{d}\xi\,{dy}\,{dz}\;\psi_1^\dagger{\psi_2} \ ,
\end{equation}
taken on an $\tau=\mathrm{constant}$ hypersurface.

While the above modes would be sufficient for quantizing in $R_0$ in
its own right, they are not suitable for analytic continuation
arguments across the horizons, as the vierbein (\ref{eqn:rinvier})
becomes singular in the limit $x\to |t|$. We therefore express the
modes in the rotating vierbein~(\ref{eqn:rotet}), which is globally defined
on~$M_0$. This choice of vierbein will further make the comparison to $M_-$
transparent in section~\ref{sec:m-unruh}. The Lorentz transformation
between (\ref{eqn:rinvier}) and (\ref{eqn:rotet}) is 
a boost by rapidity $-\tau$ in the
$(\tau,\xi)$-plane followed by a rotation by $\pi$ as $z\mapsto{z+a}$
in the $(x,y)$-plane. The corresponding transformation on the spinors
is
\begin{equation}
\psi
\mapsto 
e^{-\frac{\gamma^1\gamma^2\pi{z}}{2a}}
e^{\frac{\gamma^0\gamma^1\tau}{2}}
\psi  \ ,
\end{equation}
which in our conventions reads
\begin{equation}
\left(\begin{array}{cccc}
e^{\frac{i\pi{z}}{2a}}\cosh{\frac{\tau}{2}} & 0 & 0 & e^{\frac{i\pi{z}}{2a}}\sinh{\frac{\tau}{2}} \\
0 & e^{\frac{-i\pi{z}}{2a}}\cosh{\frac{\tau}{2}}& e^{\frac{-i\pi{z}}{2a}}\sinh{\frac{\tau}{2}} & 0 \\
0 & e^{\frac{i\pi{z}}{2a}}\sinh{\frac{\tau}{2}} & e^{\frac{i\pi{z}}{2a}}\cosh{\frac{\tau}{2}} & 0 \\
e^{\frac{-i\pi{z}}{2a}}\sinh{\frac{\tau}{2}} & 0 & 0 & e^{\frac{-i\pi{z}}{2a}}\cosh{\frac{\tau}{2}} \\
\end{array}\right)\psi \ . 
\end{equation}
In the vierbein~(\ref{eqn:rotet}), our solutions thus become 
\begin{align}
\psi^R_{j,\Omega,k_y,k_z}(t,x,y,z) 
& =
N_j
\left(
X'^R_jK_{i\Omega-\frac{1}{2}}(\kappa\xi)e^{-(i\Omega-\frac{1}{2})\tau}\right. 
\nonumber
\\
\label{eqn:rinmodes} 
&
\hspace{7ex} 
\left.\mbox{}+Y'^R_jK_{i\Omega+\frac{1}{2}}(\kappa\xi)e^{-(i\Omega+\frac{1}{2})\tau}
\right)
e^{ik_yy+ik_zz}
\ , 
\end{align}
where
\begin{align} 
X'^R_1
=& 
\left(
\begin{array}{ccccccccccccccccccccc}
e^{\frac{i\pi{z}}{2a}}\frac{k_z}{|k_z|}(k_y-im) 
\\ 
-ie^{-\frac{i\pi{z}}{2a}}(|k_z|-\kappa) 
\\ 
-ie^{\frac{i\pi{z}}{2a}}(|k_z|-\kappa) 
\\ 
e^{-\frac{i\pi{z}}{2a}}\frac{k_z}{|k_z|}(k_y-im) 
\end{array} 
\right) \ , 
& 
Y'^R_1=&
\left(
\begin{array}{ccccccccccccccccccccc}
e^{\frac{i\pi{z}}{2a}}\frac{k_z}{|k_z|}(|k_z|-\kappa) 
\\ 
ie^{-\frac{i\pi{z}}{2a}}(k_y-im) 
\\ 
-ie^{\frac{i\pi{z}}{2a}}(k_y-im)
\\ 
-e^{-\frac{i\pi{z}}{2a}}\frac{k_z}{|k_z|}(|k_z|-\kappa) 
\end{array} 
\right) \ , 
\nonumber 
\\
X'^R_2=&
\left(
\begin{array}{ccccccccccccccccccccc}
e^{\frac{i\pi{z}}{2a}}\frac{k_z}{|k_z|}(|k_z|-\kappa) 
\\ 
ie^{-\frac{i\pi{z}}{2a}}(k_y+im) 
\\ 
ie^{\frac{i\pi{z}}{2a}}(k_y+im)
\\ 
e^{-\frac{i\pi{z}}{2a}}\frac{k_z}{|k_z|}(|k_z|-\kappa) 
\end{array} 
\right) \ , 
& 
Y'^R_2=&
\left(
\begin{array}{ccccccccccccccccccccc}
e^{\frac{i\pi{z}}{2a}}\frac{k_z}{|k_z|}(k_y+im) 
\\ 
-ie^{-\frac{i\pi{z}}{2a}}(|k_z|-\kappa) 
\\ 
ie^{\frac{i\pi{z}}{2a}}(|k_z|-\kappa) 
\\ 
-e^{-\frac{i\pi{z}}{2a}}\frac{k_z}{|k_z|}(k_y+im) 
\end{array} 
\right) \ . 
\end{align}

In the left hand Rindler wedge $x<-|t|$ of $M_0$, 
denoted $L_0$, we may proceed similarly.
On $L_0$ we define Rindler coordinates $(\tau,\xi,y,z)$ by
\begin{align}
t&=  -\xi\sinh\tau \ ,
\nonumber
\\ 
x&=  -\xi\cosh\tau \ , 
\label{eqn:rinl} 
\end{align}
again with $\xi>0$ and $-\infty<\tau<\infty$ and with the identifications
$(\tau,\xi,y,z)\sim(\tau,\xi,y,z+2a)$ understood. Note that
$\partial_\tau$ is now past-pointing. In the
vierbein~(\ref{eqn:rotet}), a complete orthonormal set of positive
frequency modes with respect to~$\partial_\tau$ is
$\{\psi^L_{j,\Omega,k_y,k_z}\}$ with
\begin{align}
\psi^L_{j,\Omega,k_y,k_z}(t,x,y,z) 
& = 
N_j
\left(
X'^L_jK_{i\Omega-\frac{1}{2}}(\kappa\xi)e^{-(i\Omega-\frac{1}{2})\tau}
\right. 
\nonumber
\\
&
\hspace{7ex} 
\left.
\mbox{}+Y'^L_jK_{i\Omega+\frac{1}{2}}(\kappa\xi)e^{-(i\Omega+\frac{1}{2})\tau}
\right)
e^{ik_yy+ik_zz} \ ,
\label{eqn:rinmodesl}
\end{align}
where
\begin{align}
X'^L_1=&
\left(
\begin{array}{ccccccccccccccccccccc}
e^{\frac{i\pi{z}}{2a}}\frac{k_z}{|k_z|}(k_y-im) 
\\ 
-ie^{-\frac{i\pi{z}}{2a}}(|k_z|-\kappa) 
\\ 
-ie^{\frac{i\pi{z}}{2a}}(|k_z|-\kappa) 
\\ 
e^{-\frac{i\pi{z}}{2a}}\frac{k_z}{|k_z|}(k_y-im) 
\end{array} 
\right) \ , 
& 
Y'^L_1=&
\left(
\begin{array}{ccccccccccccccccccccc}
-e^{\frac{i\pi{z}}{2a}}\frac{k_z}{|k_z|}(|k_z|-\kappa) 
\\ 
-ie^{-\frac{i\pi{z}}{2a}}(k_y-im) 
\\ 
ie^{\frac{i\pi{z}}{2a}}(k_y-im)
\\ 
e^{-\frac{i\pi{z}}{2a}}\frac{k_z}{|k_z|}(|k_z|-\kappa) 
\end{array} 
\right) \ , 
\nonumber 
\\
X'^L_2=&
\left(
\begin{array}{ccccccccccccccccccccc}
-e^{\frac{i\pi{z}}{2a}}\frac{k_z}{|k_z|}(|k_z|-\kappa) 
\\ 
-ie^{-\frac{i\pi{z}}{2a}}(k_y+im) 
\\ 
-ie^{\frac{i\pi{z}}{2a}}(k_y+im)
\\ 
-e^{-\frac{i\pi{z}}{2a}}\frac{k_z}{|k_z|}(|k_z|-\kappa) 
\end{array} 
\right) \ , 
&
Y'^L_2=&
\left(
\begin{array}{ccccccccccccccccccccc}
e^{\frac{i\pi{z}}{2a}}\frac{k_z}{|k_z|}(k_y+im) 
\\ 
-ie^{-\frac{i\pi{z}}{2a}}(|k_z|-\kappa) 
\\ 
ie^{\frac{i\pi{z}}{2a}}(|k_z|-\kappa) 
\\ 
-e^{-\frac{i\pi{z}}{2a}}\frac{k_z}{|k_z|}(k_y+im) 
\end{array} 
\right) \ ,
\end{align}
and the 
ranges of $\Omega$, $k_y$ 
and $k_z$ are as in~$R_0$. When $k_z=0$, we define
$$\psi^L_{j,\Omega,k_y,0}:=\lim_{k_z\rightarrow{0^+}}\psi^L_{j,\Omega,k_y,k_z} \ .$$ 
These solutions satisfy the orthonormality relation
\begin{equation}
\langle{\psi^L_{i,\Omega,k_y,k_z},\psi^L_{j,\Omega',k_y',k_z'}}\rangle_{L_0}
=\delta_{ij}\delta_{nn'}\delta(\Omega-\Omega')\delta(k_y-k_y') \ ,
\end{equation}
where the inner product reads as in (\ref{eqn:inner}) with $R_0$ replaced by
$L_0$.

We may now quantize the field in $R_0$ and $L_0$ in the usual
manner. A~complete set of positive frequency modes with respect to the
future-pointing timelike Killing vector is
$\{\psi^R_{j,\Omega,k_y,k_z}(t,x,y,z)\}$ in $R_0$ and 
$\{\psi^L_{j,-\Omega,k_y,k_z}(t,x,y,z)\}$ in $L_0$, both with $\Omega>0$: The
minus sign in $\psi^L$ arises because 
$\partial_\tau$ is past-pointing in~$L_0$. 
The expansion of the field in these modes and their charge conjugates
reads 
\begin{gather}
\Psi=
\sum_{j}\sum_{n}\int_{0}^{\infty}d\Omega\int_{-\infty}^{\infty} 
\!\!\!\!
dk_y 
\left(
a^R_{j,\Omega,k_y,k_z}\psi^{R}_{j,\Omega,k_y,k_z}
+a^L_{j,\Omega,k_y,k_z}\psi^{L}_{j,-\Omega,k_y,k_z} 
\right.
\nonumber
\\     					  
\label{eqn:rinexp}		        
\left.
\mbox{}+b^{R\dagger}_{j,\Omega,k_y,k_z}\psi^{R,c}_{j,\Omega,k_y,k_z}
+b^{L\dagger}_{j,\Omega,k_y,k_z}\psi^{L,c}_{j,-\Omega,k_y,k_z}
\right) \ . 
\end{gather}
In the standard representation the charge conjugation reads
\begin{equation}\psi^c=i\gamma^2\psi^*=\left(\begin{array}{cccc}
0 & 0 & 0 & 1 \\
0 & 0 & -1 & 0 \\
0 & -1 & 0 & 0 \\
1 & 0 & 0 & 0 \\
\end{array}\right)\psi^* \ ,
\end{equation}
where the superscript ${}^c$ stands for charge conjugation 
and ${}^*$ for normal complex conjugation. It follows that
\begin{align}
\psi^{R,c}_{1,\Omega,k_y,k_z}(t,x,y,z)
= &\ -i\psi^{R}_{1,-\Omega,-k_y,-k_z}(t,x,y,z) \ , 
\nonumber
\\
\label{eq:chargeconjR}
\psi^{R,c}_{2,\Omega,k_y,k_z}(t,x,y,z)
= &\ i\psi^{R}_{2,-\Omega,-k_y,-k_z}(t,x,y,z) \ ,  
\end{align}
with analogous expressions for $\psi^L_1$ and $\psi^L_2$. 
For modes at $k_z=0$ these expressions are understood in the 
limit $\lim_{k_z\rightarrow{0^+}}$. 
We now impose the usual anticommutation relations for the annihilation 
and creation operators
\begin{align}
\left\{a^R_{i,\Omega,k_y,k_z},a^{R\dagger}_{j,\Omega',{k'_y},{k'_z}}\right\}
= &\ \delta_{ij}\delta_{nn'}\delta(\Omega-\Omega')\delta(k_y-{k'_y}) \ , 
\nonumber
\\
\left\{b^R_{i,\Omega,k_y,k_z},b^{R\dagger}_{j,\Omega',{k'_y},{k'_z}}\right\}
= &\ \delta_{ij}\delta_{nn'}\delta(\Omega-\Omega')\delta(k_y-{k'_y}) \ ,
\end{align}
with similar relations holding for $a^L_{i,\Omega,k_y,k_z}$ and $a^{L\dagger}_{j,\Omega',{k'_y},{k'_z}}$ 
and all mixed anticommutators
vanishing. This construction defines the Rindler-like vacuum states
$|0_{R_0}\rangle$, and $|0_{L_0}\rangle$ on $M_0$ as the states 
annihilated by all the appropriate annihilation operators.
\begin{eqnarray}
a^R_{i,\Omega,k_y,k_z}|0_{R_0}\rangle & = & a^L_{i,\Omega,k_y,k_z}|0_{L_0}\rangle=0 \ ,
\nonumber 
\\
b^R_{i,\Omega,k_y,k_z}|0_{R_0}\rangle & = & b^L_{i,\Omega,k_y,k_z}|0_{L_0}\rangle=0 \ ,\qquad \forall{i,\Omega,k_y,k_z} \ .
\end{eqnarray}

Now, we wish to find the Rindler mode content of the Minkowski-like
vacuum $|0_0\rangle$. While we could proceed directly by calculating the 
Bogolubov transformation between the modes~(\ref{eqn:minkmodes})
and~(\ref{eqn:rinmodes}) it is easier to follow Unruh's
analytic continuation method~\cite{u:unruh}. That is, we build from the Rindler 
modes (the $\psi$ modes) a complete set of linear combinations, denoted here as $W$-modes, 
which are purely positive frequency 
with respect to the global timelike Killing vector $\partial_t$. 
The vacua of these $W$-modes is $|0_0\rangle$ and the Bogolubov transformation may be read off easily
from the expansion of the field in them.

As the positive frequency Minkowski modes can be defined by the property that 
they are analytic and bounded in the entire lower half complex $t$ plane, 
we may construct the $W$-modes by analytically continuing the set $\{\psi^R_{j,\Omega,k_y,k_z}\}$ 
across the horizons into the other wedges in the lower half complex $t$-plane. 
The continuation into the $F$ wedge transforms 
our solutions as 
\footnote{More details of this analytic continuation are given in appenix \ref{ch:appendix1}.}
\begin{equation}
\psi^R_{j,\Omega,k_y,n}(t,x,y,z)\mapsto\psi^{F}_{j,\Omega,k_y,n}(t,x,y,z) \ ,
\end{equation}
where
\begin{eqnarray}
\label{eqn:fmode}
\psi^{F}_{1,\Omega,k_y,k_z}(t,x,y,z) & = & 
e^{(i\Omega-\frac{1}{2})\frac{i\pi}{2}}
N_1\left(X'^F_1K_{i\Omega-\frac{1}{2}}({i}{\kappa}\xi)e^{-(i\Omega-\frac{1}{2})\tau} 
\right. 
\nonumber 
\\
&   & \left.\mbox{}+Y'^F_1K_{i\Omega+\frac{1}{2}}({i}{\kappa}\xi)
e^{-(i\Omega+\frac{1}{2})\tau}\right)e^{ik_yy+ik_zz} \ ,
\\
\label{eqn:fmode2}               
\psi^{F}_{2,\Omega,k_y,k_z}(t,x,y,z) & = & 
e^{(i\Omega+\frac{1}{2})\frac{i\pi}{2}}
N_2\left(X'^F_2K_{i\Omega-\frac{1}{2}}({i}{\kappa}\xi)e^{-(i\Omega-\frac{1}{2})\tau} 
\right. 
\nonumber 
\\
&   & \left.\mbox{}+Y'^F_2K_{i\Omega+\frac{1}{2}}({i}{\kappa}\xi)
e^{-(i\Omega+\frac{1}{2})\tau}\right)e^{ik_yy+ik_zz} \ ,
\end{eqnarray}
the modified Bessel function 
$K_{i\Omega\pm\frac{1}{2}}({i}{\kappa}\xi)$ can be given in terms of 
Hankel functions as \cite{zw:zwillinger},
\begin{equation}
K_{i\omega\pm\frac{1}{2}}({i}{\kappa}\xi)=-\frac{i\pi}{2}
e^{-\frac{i\pi}{2}(i\Omega\pm{\frac{1}{2}})}H^{(2)}_{i\Omega\pm\frac{1}{2}}({\kappa}\xi) \ .
\end{equation}
The spinorial factors are given by
\begin{align}
X'^F_1=&
\left(
\begin{array}{ccccccccccccccccccccc}
e^{\frac{i\pi{z}}{2a}}\frac{k_z}{|k_z|}(k_y-im) 
\\ 
-ie^{-\frac{i\pi{z}}{2a}}(|k_z|-\kappa) 
\\ 
-ie^{\frac{i\pi{z}}{2a}}(|k_z|-\kappa) 
\\ 
e^{-\frac{i\pi{z}}{2a}}\frac{k_z}{|k_z|}(k_y-im) 
\end{array} 
\right) \ , 
& 
Y'^F_1=&
\left(
\begin{array}{ccccccccccccccccccccc}
ie^{\frac{i\pi{z}}{2a}}\frac{k_z}{|k_z|}(|k_z|-\kappa) 
\\ 
-e^{-\frac{i\pi{z}}{2a}}(k_y-im) 
\\ 
e^{\frac{i\pi{z}}{2a}}(k_y-im)
\\ 
-ie^{-\frac{i\pi{z}}{2a}}\frac{k_z}{|k_z|}(|k_z|-\kappa) 
\end{array} 
\right) \ , 
\nonumber 
\\
X'^F_2=&
\left(
\begin{array}{ccccccccccccccccccccc}
-ie^{\frac{i\pi{z}}{2a}}\frac{k_z}{|k_z|}(|k_z|-\kappa) 
\\ 
e^{-\frac{i\pi{z}}{2a}}(k_y+im) 
\\ 
e^{\frac{i\pi{z}}{2a}}(k_y+im)
\\ 
-ie^{-\frac{i\pi{z}}{2a}}\frac{k_z}{|k_z|}(|k_z|-\kappa) 
\end{array} 
\right) \ , 
&
Y'^F_2=&
\left(
\begin{array}{ccccccccccccccccccccc}
e^{\frac{i\pi{z}}{2a}}\frac{k_z}{|k_z|}(k_y+im) 
\\ 
-ie^{-\frac{i\pi{z}}{2a}}(|k_z|-\kappa) 
\\ 
ie^{\frac{i\pi{z}}{2a}}(|k_z|-\kappa) 
\\ 
-e^{-\frac{i\pi{z}}{2a}}\frac{k_z}{|k_z|}(k_y+im) 
\end{array} 
\right) \ ,
\end{align}
and the coordinates $(\tau,\xi)$ are in $F$ defined by
\begin{eqnarray} 
t & = & \xi\cosh\tau \ ,
\nonumber 
\\ 
\label{eqn:milne} 
x & = & \xi\sinh\tau \ , 
\end{eqnarray}
with  $\xi>0$, $-\infty<\tau<\infty$.
These modes are by construction positive frequency with respect to $\partial_t$ 
for all $\Omega\in{\R}$. Continuing further into $L_0$,
we find
\begin{eqnarray}
\psi^{F}_{1,\Omega,k_y,k_z} & \mapsto & 
{e^{(i\Omega-\frac{1}{2}){i{\pi}}}}\psi^L_{1,\Omega,k_y,k_z}
 \ ,
\nonumber
\\
\psi^{F}_{2,\Omega,k_y,k_z} & \mapsto & 
{e^{(i\Omega+\frac{1}{2}){i{\pi}}}}\psi^L_{2,\Omega,k_y,k_z}
 \ ,
\label{eqn:modefrigl}
\end{eqnarray}
where $(\tau,\xi)$ are in $L$ given by~(\ref{eqn:rinl}).
Making use of the analogues of (\ref{eq:chargeconjR}) in $L_0$,
we thus see that
a complete set of normalised positive frequency $W$-modes 
$\{W_{j,\Omega,k_y,k_z}\}$ on $M_0$ is given by
\begin{equation}
W_{j,\Omega,k_y,k_z}(t,x,y,z)=
\frac{1}{\sqrt{2\cosh(\pi{\Omega})}}
\left(
e^{\frac{\pi{\Omega}}{2}}\psi^R_{j,\Omega,k_y,k_z}
+{e^{-\frac{\pi{\Omega}}{2}}\psi^{L,c}_{j,-\Omega,-k_y,-k_z}}
\right)
 \ ,
\end{equation}
with $\Omega\in{\R}$. 

It will be useful to split these modes into two sets, as
\begin{align}
W^{(1)}_{j,\Omega,k_y,k_z}(t,x,y,z)= 
&\ 
\frac{1}{\sqrt{2\cosh(\pi{\Omega})}}
\left(
e^{\frac{\pi{\Omega}}{2}}\psi^R_{j,\Omega,k_y,k_z}
+{e^{-\frac{\pi{\Omega}}{2}}\psi^{L,c}_{j,-\Omega,-k_y,-k_z}}
\right) \ ,
\nonumber
\\ 
W^{(2)}_{j,\Omega,k_y,k_z}(t,x,y,z)= 
&\ 
\frac{1}{\sqrt{2\cosh(\pi{\Omega})}}
\left(
-e^{-\frac{\pi{\Omega}}{2}}\psi^{R,c}_{j,\Omega,-k_y,-k_z}
+{e^{\frac{\pi{\Omega}}{2}}\psi^{L}_{j,-\Omega,k_y,k_z}}
\right) \ ,
\label{eqn:wmodesRM0}
\end{align}
where now $\Omega>0$, and we have made further use of (\ref{eq:chargeconjR})
and their analogues in $L_0$. For $k_z=0$ we interpret (\ref{eqn:wmodesRM0})
in the sense $k_z\to{0^+}$.
The normalization is 
\begin{equation}
\langle{W_{i,\Omega,k_y,k_z},W_{j,\Omega',k'_y,k'_z}}\rangle_0
=\delta_{ij}\delta_{nn'}\delta(\Omega-\Omega')\delta(k_y-k'_y) \ .
\end{equation}

We may now expand the field in terms of the $W$-modes (\ref{eqn:wmodesRM0}) as
\begin{gather}
\Psi=\sum_{j}\sum_{n}
\int_{0}^{\infty}d\Omega
\int_{-\infty}^{\infty} \!\!\!\!dk_y 
\left(
c^{(1)}_{j,\Omega,k_y,k_z}W^{(1)}_{j,\Omega,k_y,k_z}
+c^{(2)}_{j,\Omega,k_y,k_z}W^{(2)}_{j,\Omega,k_y,k_z}
\right. 
\nonumber
\\
\label{eqn:Wexp}
\left.
\mbox{}+d^{(1)\dagger}_{j,\Omega,k_y,k_z}W^{(1),c}_{j,\Omega,k_y,k_z}
+d^{(2)\dagger}_{j,\Omega,k_y,k_z}W^{(2),c}_{j,\Omega,k_y,k_z}
\right) 
\end{gather}
and impose the usual anticommutation relations on the 
annihilation and creation operators in~(\ref{eqn:Wexp}). 
The vacuum state for the $W$-modes is by construction $|0_0\rangle$,
with
\begin{eqnarray}
c^{(a)}_{i,\Omega,k_y,k_z}|0_0\rangle  =  d^{(a)}_{i,\Omega,k_y,k_z}|0_0\rangle = 0 \ ,\qquad \forall{a,i,\Omega,k_y,k_z} \ .
\end{eqnarray}
Equating the expressions~(\ref{eqn:Wexp}) and~(\ref{eqn:rinexp}) 
and taking inner products with the $\psi$ modes, we find that the Bogolubov transformation 
between the Rindler modes and $W$-modes reads
\begin{align}
a^{R}_{j,\Omega,k_y,k_z}= 
&\ 
\frac{1}{\sqrt{2\cosh(\pi{\Omega})}}
\left(
e^{\frac{\pi{\Omega}}{2}}c^{(1)}_{j,\Omega,k_y,k_z}
-{e^{-\frac{\pi{\Omega}}{2}}}d^{(2)\dagger}_{j,\Omega,-k_y,-k_z}
\right) \ ,
\nonumber
\\ 
a^{L}_{j,\Omega,k_y,k_z}= 
&\ 
\frac{1}{\sqrt{2\cosh(\pi{\Omega})}}
\left(
e^{\frac{\pi{\Omega}}{2}}c^{(2)}_{j,\Omega,k_y,k_z}
+{e^{-\frac{\pi{\Omega}}{2}}}d^{(1)\dagger}_{j,\Omega,-k_y,-k_z}
\right) \ ,
\nonumber
\\ 
b^{R,\dagger}_{j,\Omega,k_y,k_z}= 
&\ 
\frac{1}{\sqrt{2\cosh(\pi{\Omega})}}
\left(
e^{\frac{\pi{\Omega}}{2}}d^{(1)\dagger}_{j,\Omega,k_y,k_z}
-{e^{-\frac{\pi{\Omega}}{2}}}c^{(2)}_{j,\Omega,-k_y,-k_z}
\right) \ , 
\nonumber
\\
b^{L,\dagger}_{j,\Omega,k_y,k_z}= 
&\ 
\frac{1}{\sqrt{2\cosh(\pi{\Omega})}}
\left(
e^{\frac{\pi{\Omega}}{2}}d^{(2)\dagger}_{j,\Omega,k_y,k_z}
+{e^{-\frac{\pi{\Omega}}{2}}}c^{(1)}_{j,\Omega,-k_y,-k_z}
\right) \ .
\label{eqn:bog}
\end{align}

To find the expectation value of the number 
operator for Rindler modes in the vacuum $|0_0\rangle$, 
we use (\ref{eqn:bog}) to compute
\begin{equation}
\label{eqn:number}
\langle0_0|
a^{R\dagger}_{i,\Omega,k_y,k_z}
a^{R}_{j,{\Omega'},{k'_y},{k'_z}}
|0_0\rangle
=
\frac{1}{(e^{2\pi{\Omega}}+1)}
\delta_{ij}\delta_{n{n'}}\delta(k_y-{k'_y})\delta(\Omega-{\Omega'}) \ .
\end{equation}
From (\ref{eqn:number}) we see immediately that the Minkowski-like vacuum on $M_0$ 
contains Rindler particles. Further, the expected number of Rindler particles in the
Minkowski-like vacuum is given by the Fermi distribution function. 
Recall that an observer with constant $\xi$ is uniformly accelerating with proper time
$\xi\tau$ and acceleration $\xi^{-1}$. The proper energy of a Rindler particle seen
by such an observer is thus not $\Omega$ but $\Omega/\xi$.
The usual interpretation of (\ref{eqn:number}) is then that 
a Rindler observer at constant $\xi$ sees the Minkowski vacuum as a thermal bath at 
temperature $T=\frac{a}{2\pi}$ where $a=1/\xi$ is the observer's acceleration.  We must note of course that setting 
primed and unprimed indices in (\ref{eqn:number}) equal gives a divergent result, due 
to the continuum normalization of our modes. A finite result may 
be obtained by considering wave packets,
as done for the scalar field in~\cite{lm:geon}. 

We can gain more insight into the situation by writing the Minkowski-like 
vacuum $|0_0\rangle$ in terms of the Rindler vacuum $|0_{R_0}\rangle$. Making use
of the inverse Bogolubov transformation to~(\ref{eqn:bog}), we find
\begin{equation}
\label{eqn:sum}
|0_0\rangle
=\prod_{\sigma,j,\Omega,k_y,k_z}
\frac{1}{({e^{-2\pi{\Omega}}+1)^{\frac{1}{2}}}}
\sum_{q=0,1}
(-1)^qe^{-\pi{\Omega}q}
|q\rangle^{R}_{\sigma,j,\Omega,k_y,k_z}
|q\rangle^{L}_{-\sigma,j,\Omega,-k_y,-k_z} 
\ , 
\end{equation}
where we have introduced $\sigma=+1, (-1)$ to label particle (antiparticle) modes respectively and 
the notation on the right hand side is adapted to the tensor
product structure of the Hilbert space considered: 
\begin{eqnarray}
|q\rangle^{R}_{+,j,\Omega,k_y,k_z}
=(a^{R\dagger}_{j,\Omega,k_y,k_z})^q|0_{R_0}\rangle 
&
|q\rangle^{L}_{+,j,\Omega,k_y,k_z}
=(a^{L\dagger}_{j,\Omega,k_y,k_z})^q|0_{L_0}\rangle 
\ ,
\nonumber
\\
|q\rangle^{R}_{-,j,\Omega,k_y,k_z}
=(b^{R\dagger}_{j,\Omega,k_y,k_z})^q|0_{R_0}\rangle 
&
|q\rangle^{L}_{-,j,\Omega,k_y,k_z}
=(b^{L\dagger}_{j,\Omega,k_y,k_z})^q|0_{L_0}\rangle 
\ . 
\end{eqnarray}
The result (\ref{eqn:sum}) is the massive fermion version of the
familiar bosonic result~\cite{bd:book}, indicating an entangled state
in which the right and left Rindler excitations appear in correlated
pairs. 
Suppose we consider an operator $\hat{A}^{(1)}$ 
whose support is in the right hand Rindler wedge only. This operator 
will not couple to any of the 
degrees of freedom of the field in the left hand wedge, and we must trace 
out these degrees of freedom. The operator has expectation
value $\langle0_0|\hat{A}^{(1)}|0_0\rangle=
\mathrm{Tr}(\hat{A}^{(1)}\rho^{(1)})$, where $\rho^{(1)}$ is a
fermionic thermal density matrix in~$R_0$,
\begin{equation}
\rho^{(1)}
=\prod_{\sigma,j,\Omega,k_y,k_z}
\sum_{q=0,1}
\frac{e^{-2q\pi{\Omega}}}{\sum_{m=0,1}e^{-2m\pi{\Omega}}}
|q\rangle^{R}_{\sigma,j,\Omega,k_y,k_z}\;{}^{\phantom{\sigma,j,\Omega,k_y,}R}_{\sigma,j,\Omega,k_y,k_z}
\!\!\langle{q}| \ . 
\end{equation}
We can conclude that such an operator sees $|0_0\rangle$ as a thermal 
bath at the usual Unruh temperature. Similar conclusions can be made 
for any operator whose support is entirely in the left hand wedge.

It is important to note the formal nature of (\ref{eqn:sum}): 
The normalisation factor is actually divergent, owing to the
infinite product over modes and indicating the 
non-existence of a unitary operator linking the two vacua. 
This is not a problem however as the thermality arguments can in fact be made mode 
for mode (see e.g~\cite{tk:takagi}). Further we note that
the usual discussion may follow about the relation of (\ref{eqn:sum})
to Thermofield dynamics (as first discussed by 
Israel in \cite{is:thermo}).

The result (\ref{eqn:sum}) incorporates the two spin structures on
$M_0$, $R_0$ and $L_0$ in the allowed values of~$k_z$, and these
values are the same in all our mode sets. The twisted (respectively
untwisted) $|0_0\rangle$ thus induces twisted (untwisted) thermal
states in both $R_0$ and~$L_0$.

To end this section, we note that the Bogolubov transformation on
Minkowski space can be read off from our formulas on $M_0$ with minor
systematic changes. There is now only one spin structure and $k_z$
takes all real values. The expressions for the various mode functions
include the additional factor $\sqrt{a/\pi}$, sums over $n$ become
integrals over~$k_z$, and in the normalization and anticommutation
relations the discrete delta $\delta_{n n'}$ is replaced by the
delta-function $\delta(k_z - k_z')$. The formulas involve still $a$
because the spinors are expressed in the rotating
vierbein~(\ref{eqn:rotet}). Translation into the standard vierbein
(\ref{eqn:minkvier}) can be accomplished by the appropriate spinor
transformation.
 
\section[Bogolubov transformation on $M_-$]{The Unruh construction and Bogolubov transformation on $M_-$}
\label{sec:m-unruh}

In this section we repeat the analysis of the previous section for $M_-$.

Let $R_-$ denote the Rindler wedge on~$M_-$, given in our local
coordinates by $|x|>|t|$. As $R_-$ is isometric to~$R_0$, we may
introduce in $R_-$ the Rindler-coordinates $(\tau,\xi,y,z)$
by~(\ref{eqn:rin}), again with the identifications
$(\tau,\xi,y,z)\sim(\tau,\xi,y,z+2a)$, and quantize as in~$R_0$,
defining the positive and negative frequencies with respect to the
Killing vector~$\partial_\tau$.
The set $\{\psi^R_{j,\Omega,k_y,k_z}\}$~(\ref{eqn:rinmodes}) 
is therefore a complete set of positive frequency Rindler modes on $R_-$. 
We could work directly with this set, but for convenience of phase factors in
the $W$-modes (\ref{eqn:m-wmode12}) below, we use instead the mode set
$\{\Psi^R_{j,\Omega,k_y,k_z}\}$, defined as in (\ref{eqn:rinmodes}) except
with the normalization factors (\ref{eqn:norm123}) replaced by
\begin{align}
N_1
= 
&\ 
\frac{e^{-\frac{ik_za}{2}}\sqrt{\cosh(\pi{\Omega})(\kappa^2-k_z^2)}}
{4\pi(k_y-im)\sqrt{a\pi(\kappa-|k_z|)}} \ ,
\nonumber
\\
\label{eqn:normm-}
N_2= 
&\ 
\frac{e^{\frac{ik_za}{2}}\sqrt{\cosh(\pi{\Omega})(\kappa^2-k_z^2)}}
{4\pi(k_y+im)\sqrt{a\pi(\kappa-|k_z|)}} \ .
\end{align} 

A key difference from $R_0$ arises now, from the requirement that
the spinors on $R_-$ must be extendible into spinors in one of the two
spin structures on~$M_-$. By the discussion in
section~\ref{sec:m0unruh}, this implies that $k_z$ is restricted to
the values $k_z=(n+\tfrac12)\pi/a$ with $n\in{\mathbb{Z}}$. Both of
the spin structures on $M_-$ thus induce on $R_-$ the same spin
structure, in which the spinors are antiperiodic in the nonrotating
vierbein~(\ref{eqn:minkvier})
\footnote{This is easily seen by noting that a vierbein which rotates by $\pi$ (repectively $-\pi$)
as $z\rightarrow{z+a}$ thus rotates by $2\pi$ ($-2\pi$) as $z\rightarrow{z+2a}$. As we noted
earlier these two vierbeins on $M_0$ are equivalent when used to specify the spin structure.}.
$\{\Psi^R_{j,\Omega,k_y,k_z}\}$ is a complete set of positive frequency solutions 
in $R_-$, written in the globally defined rotating vierbein~(\ref{eqn:rotet}). 
The inner product on $R_-$ is as in~(\ref{eqn:inner}) and the orthonormality relation is
\begin{equation}
\langle{\Psi^R_{i,\Omega,k_y,k_z},\Psi^R_{j,\Omega',k_y',k_z'}}\rangle_{R_-}
=\delta_{ij}\delta_{nn'}\delta(\Omega-\Omega')\delta(k_y-k_y') \ .
\end{equation}

We quantize the field in $R_-$ by expanding $\Psi$ as
\begin{gather}
\Psi
=\sum_{n}\int_{0}^{\infty}d\Omega\int_{-\infty}^{\infty}dk_y
\left(
a_{1,\Omega,k_y,k_z}\Psi^{R}_{1,\Omega,k_y,k_z}
+a_{2,\Omega,k_y,k_z}\Psi^{R}_{2,\Omega,k_y,k_z}
\right.
\nonumber 
\\
\label{eqn:expan-}
\left.
\mbox{}+b^\dagger_{1,\Omega,k_y,k_z}\Psi^{R,c}_{1,\Omega,k_y,k_z}
+b^\dagger_{2,\Omega,k_y,k_z}\Psi^{R,c}_{2,\Omega,k_y,k_z}
\right) 
\end{gather}
and imposing the usual anticommutation relations for the annihilation 
and creation operators. The Rindler-vacuum $|0_{R_-}\rangle$ on
$R_-$ is the state annihilated by all the annihilation operators
in~(\ref{eqn:expan-}),
\begin{eqnarray}
a_{i,\Omega,k_y,k_z}|0_{R_-}\rangle = b_{i,\Omega,k_y,k_z}|0_{R_-}\rangle = 0 \ , \qquad \forall{i,\Omega,k_y,k_z} \ .
\end{eqnarray}

To find the Rindler-mode content of~$|0_-\rangle$, we again 
use the analytic continuation method. Working in the 
local coordinates $(t,x,y,z)$, we continue the modes 
$\{\Psi^R_{j,\Omega,k_y,k_z}\}$ across the horizons in 
the lower half-plane in complexified $t$ and form 
linear combinations that are globally well-defined on~$M_-$.
We begin by continuing the $\{\Psi^R_{j,\Omega,k_y,k_z}\}$ into 
the $F$ region in the lower half complex $t$-plane as on $M_0$.
We obtain the modes $\{\Psi^F_{j,\Omega,k_y,k_z}\}$ identical to the 
$\{\psi^F_{j,\Omega,k_y,k_z}\}$ of~(\ref{eqn:fmode}) and~(\ref{eqn:fmode2})
but with $N_j$ as in~(\ref{eqn:normm-}). These 
modes are a complete set of $W$-modes in the $F$ region of $M_0$ 
(with the suitable restrictions on $k_z$). The analogous modes 
on $M_-$ may be built from these via the method of images. 
As we are working in the globally defined rotating vierbein which 
is invariant under $J_-$, the $W$-modes
on $F_-$ are simply given by (see e.g.~\cite{bd:em})
\begin{equation}
W^{F}_{j,\Omega,k_y,k_z}(t,x,y,z)\propto\Psi^{F}_{j,\Omega,k_y,k_z}(t,x,y,z)
+\rho{\Psi^{F}_{j,\Omega,k_y,k_z}(t,-x,-y,z+a)} \ ,
\end{equation}
where $\Omega\in{\R}$ and we have used the parameter $\rho=+1, (-1)$ 
to label $W$-modes with periodic (antiperiodic) boundary conditions 
respectively. Now it is easy to show
\begin{eqnarray}
\label{eqn:mapj-1}
\Psi^{F}_{1,\Omega,k_y,k_z}(t,-x,-y,z+a) & = & e^{-\pi{\Omega}}\Psi^{F}_{2,-\Omega,-k_y,k_z}(t,x,y,z) \ ,
\\
\label{eqn:mapj-2}
\Psi^{F}_{2,\Omega,k_y,k_z}(t,-x,-y,z+a) & = & e^{-\pi{\Omega}}\Psi^{F}_{1,-\Omega,-k_y,k_z}(t,x,y,z) \ ,
\end{eqnarray}
so that
\begin{eqnarray}
W^{F}_{1,\Omega,k_y,k_z}(t,x,y,z) & \propto & \Psi^{F}_{1,\Omega,k_y,k_z}(t,x,y,z)
+\rho{e^{-\pi{\Omega}}\Psi^{F}_{2,-\Omega,-k_y,k_z}(t,x,y,z)} \ ,
\label{eqn:wmodeM-f1}
\\
W^{F}_{2,\Omega,k_y,k_z}(t,x,y,z) & \propto & \Psi^{F}_{2,\Omega,k_y,k_z}(t,x,y,z)
+\rho{e^{-\pi{\Omega}}\Psi^{F}_{1,-\Omega,-k_y,k_z}(t,x,y,z)} \ ,
\label{eqn:wmodeM-f2}
\end{eqnarray}
with $\Omega\in{\R}$. There is a redundency in the $W$-modes (\ref{eqn:wmodeM-f1})
and (\ref{eqn:wmodeM-f2}) in that
\begin{equation}
\label{eqn:redund}
W^{F}_{1,\Omega,k_y,k_z}\propto{W^{F}_{2,-\Omega,-k_y,k_z}} \ ,
\end{equation}
and we eliminate this redundancy by taking $\Omega>0$ and $-\infty<k_y<\infty$.
Continuing back to $R_-$ in the lower half complex $t$ plane, 
$\Psi^{F}_{j,\Omega,k_y,k_z}\mapsto\Psi^{R}_{j,\Omega,k_y,k_z}$, and making use of the relations
\begin{eqnarray}
\Psi^{R,c}_{1,\Omega,k_y,k_z}(t,x,y,z) & = & -\Psi^{R}_{1,-\Omega,-k_y,-k_z}(t,x,y,z) \ ,
\nonumber 
\\
\Psi^{R,c}_{2,\Omega,k_y,k_z}(t,x,y,z) & = & \Psi^{R}_{2,-\Omega,-k_y,-k_z}(t,x,y,z)  \ ,
\end{eqnarray}
we find a complete set $\{W^R_{j,\Omega,k_y,k_z}\}$ of $W$-modes on $R_-$ to be  
\begin{align}
&{W^{R}_{1,\Omega,k_y,k_z}(t,x,y,z)}=\frac{1}{\sqrt{2\cosh{\pi{\Omega}}}}
\left(
{e^{\frac{\pi{\Omega}}{2}}}\Psi^{R}_{1,\Omega,k_y,k_z}
+\rho{e^{-\frac{\pi{\Omega}}{2}}}\Psi^{R,c}_{2,\Omega,k_y,-k_z}
\right) \ ,
\nonumber
\\
\label{eqn:m-wmode12}
&{W^{R}_{2,\Omega,k_y,k_z}(t,x,y,z)}=\frac{1}{\sqrt{2\cosh{\pi{\Omega}}}}
\left(
{e^{\frac{\pi{\Omega}}{2}}}\Psi^{R}_{2,\Omega,k_y,k_z}
-\rho{e^{-\frac{\pi{\Omega}}{2}}}\Psi^{R,c}_{1,\Omega,k_y,-k_z}
\right) \ ,
\end{align}
with $\Omega>0$. The orthonormality condition is 
\begin{equation}
\label{eqn:orthW}
\langle{W^R_{i,\Omega,k_y,k_z},W^R_{j,\Omega',k'_y,k'_z}}\rangle_{R_-}
=\delta_{ij}\delta_{nn'}\delta(\Omega-\Omega')\delta(k_y-k'_y)
 \ .
\end{equation}

The expansion of the field in $W$-modes reads 
\begin{gather}
\Psi
=\sum_{n}\int_{0}^{\infty}d\Omega\int_{-\infty}^{\infty}dk_y 
\left(
c_{1,\Omega,k_y,k_z}W^R_{1,\Omega,k_y,k_z}
+c_{2,\Omega,k_y,k_z}W^R_{2,\Omega,k_y,k_z}
\right.
\nonumber 
\\
\label{eqn:Wexp2}                                                   
\left.
\mbox{}+d^\dagger_{1,\Omega,k_y,k_z}W^{R,c}_{1,\Omega,k_y,k_z}
+d^\dagger_{2,\Omega,k_y,k_z}W^{R,c}_{2,\Omega,k_y,k_z}
\right) \ ,
\end{gather}
where $k_z=\frac{(2n+1)\pi}{2a}$.
The annihilation and creation operators in (\ref{eqn:Wexp2}) satisfy
the usual anticommutation relations, and $|0_-\rangle$ is the state
annihilated by the annihilation operators.
It follows that the
Bogolubov transformation between the Rindler-modes and the $W$-modes
reads
\begin{align}
a_{1,\Omega,k_y,k_z}= 
&\ 
\frac{1}{\sqrt{2\cosh(\pi{\Omega})}}
\left(
e^{\frac{\pi{\Omega}}{2}}c_{1,\Omega,k_y,k_z}
-\rho{e^{-\frac{\pi{\Omega}}{2}}}d^{\dagger}_{2,\Omega,k_y,-k_z}
\right) \ ,
\nonumber
\\
a_{2,\Omega,k_y,k_z}= 
&\ 
\frac{1}{\sqrt{2\cosh(\pi{\Omega})}}
\left(
e^{\frac{\pi{\Omega}}{2}}c_{2,\Omega,k_y,k_z}
+\rho{e^{-\frac{\pi{\Omega}}{2}}}d^{\dagger}_{1,\Omega,k_y,-k_z}
\right) \ ,
\nonumber
\\
b^\dagger_{1,\Omega,k_y,k_z}= 
&\ 
\frac{1}{\sqrt{2\cosh(\pi{\Omega})}}
\left(
e^{\frac{\pi{\Omega}}{2}}d^\dagger_{1,\Omega,k_y,k_z}
-\rho{e^{-\frac{\pi{\Omega}}{2}}}c_{2,\Omega,k_y,-k_z}
\right) \ , 
\nonumber
\\
\label{eqn:bog3}
b^\dagger_{2,\Omega,k_y,k_z}= 
&\ 
\frac{1}{\sqrt{2\cosh(\pi{\Omega})}}
\left(
e^{\frac{\pi{\Omega}}{2}}d^\dagger_{2,\Omega,k_y,k_z}
+\rho{e^{-\frac{\pi{\Omega}}{2}}}c_{1,\Omega,k_y,-k_z}
\right) \ . 
\end{align}

Now consider the expectation values of the Rindler
mode number operators in the vacuum $|0_-\rangle$. 
Using (\ref{eqn:bog3}), we find
\begin{equation}
\langle0_-|a^\dagger_{i,\Omega,k_y,k_z}a_{j,{\Omega'},{k_y}',{k_z}'}|0_-\rangle
=\frac{1}{(e^{2\pi{\Omega}}+1)}\delta_{ij}\delta_{n{n'}}\delta(k_y-{k_y}')\delta(\Omega-{\Omega'})
 \ ,
\end{equation}
which is identical to the $M_0$ result (\ref{eqn:number}) and hence compatible with
a bath of Rindler particles at the usual Unruh temperature. 
Note further that the expectation values are the same in both spin structures. 
More information may however be gained by expressing
$|0_-\rangle$ in terms of $|0_{R_-}\rangle$. 
The expression analogous to~(\ref{eqn:sum}) is 
\begin{equation}
\label{eqn:geonitmink}
|0_-\rangle
=\prod_{\sigma,\Omega,k_y,k_z}\frac{1}{({e^{-2\pi{\Omega}}+1)^{\frac{1}{2}}}}
\sum_{q=0,1}(-\rho)^qe^{-\pi{\Omega}q}
|q\rangle_{\sigma,1,\Omega,k_y,k_z}
|q\rangle_{-\sigma,2,\Omega,k_y,-k_z} \ ,
\end{equation}
where $\sigma=+1 (-1)$ and
\begin{eqnarray}
|q\rangle_{+,j,\Omega,k_y,k_z}=(a^{\dagger}_{j,\Omega,k_y,k_z})^q|0_{R_-}\rangle 
\ ,
\nonumber
\\
|q\rangle_{-,j,\Omega,k_y,k_z}=(b^{\dagger}_{j,\Omega,k_y,k_z})^q|0_{R_-}\rangle 
\ .
\end{eqnarray}
From (\ref{eqn:geonitmink}) it is seen that $|0_-\rangle$ is in fact an
entangled state of Rindler-excitations, and the correlations are
between a particle and an antiparticle with opposite eigenvalues
of~$k_z$ both located in $R_-$.
This is in some sense what we would expect as the involution giving $M_-$
from $M_0$ identifies the two Rindler wedges.
As all the excitations are in the unique Rindler
wedge~$R_-$, the expectation values of generic operators in $R_-$ are
not thermal.
However, for any operator that couples to one member of 
each correlated pair in $R_-$, the expectation values
will be indistinguishable from those in the corresponding state $|0_0\rangle$
in~$R_0$, indicating thermality in the standard Unruh
temperature.
Consider for example an operator $\hat{A}$, concentrated on an accelerated worldine in $R_-$,
such that $\hat{A}$ only couples to one of the Rindler modes, say $\psi_{1,\Omega,k_y,k_z}$. 
The expectation value of $\hat{A}$ is then
\begin{eqnarray}
\label{eqn:operatorM-}
\langle0_-|\hat{A}|0_-\rangle
& = & 
\prod_{\sigma,\Omega,k_y,k_z}\frac{1}{(1+e^{-2\pi{\Omega}})}\sum_{q=0,1}
e^{-2q\pi{\Omega}}{}_{\sigma,1,\Omega,k_y,k_z}\!\langle{q}|\hat{A}|q\rangle_{\sigma,1,\Omega,k_y,k_z} \ ,
\nonumber
\\
& = & 
\mathrm{Tr}\,(\hat{A}\rho^{(1)}) \ ,
\end{eqnarray}
where $\rho^{(1)}$ has the form of a thermal density matrix. We can conclude that $\hat{A}$
sees $|0_-\rangle$ as a thermal bath at the usual temperature.
This is also the case for example for any operator that only
couples to excitations with a definite sign of~$k_z$. In particular,
we have seen that the number operator expectation values are indistinguishable
from~(\ref{eqn:number}). Further it can be argued from the isometries as
in
\cite{lm:geon} 
that the experiences of a Rindler-observer become asymptotically
thermal at early or late times.

A key result of our analysis is that while both spin structures on
$M_-$ induce a state in the same spin structure in~$R_-$, the explicit
appearance of $\rho$ in (\ref{eqn:geonitmink}) shows that the two
states differ. A~Rindler-observer in $R_-$ can therefore in principle
detect the spin structure on $M_-$ from the non-thermal
correlations. How these correlations could be detected in practice,
for example by particle detectors with a local coupling to the field,
is an interesting question that we will touch upon in 
chapter \ref{ch:pdchapt}. As the restriction
of $|0_-\rangle$ to $R_-$ is not invariant under the Killing
vector~$\partial_{\tau}$, and as the isometry arguments show that the
correlations disappear in the limit of large~$|\tau|$, investigating
this question requires a particle detector formalism that can
accommodate time dependent situations~\cite{sc:schlicht}.

\chapter[Hawking effect for massive spinors]{Hawking (-Unruh) effect for massive spinors on the Kruskal manifold
and the $\rp$ geon}
\label{ch:hawkingbog}

\section{Introduction}

In this chapter we analyze the Hawking effect on the $\rp$ geon black hole spacetime 
for a massive Dirac field. Section~\ref{sec:geonintro} begins by reviewing some 
relevant properties of the Kruskal manifold and the $\rp$ geon. 
Here we also discuss the spin structures that these spacetimes admit. 
In section~\ref{sec:boulware} we consider spinor field theory on 
the geon and we construct the Boulware vacuum $|0_B\rangle$ in one 
exterior region. This construction is analogous to that of the
Rindler vacuum $|0_R\rangle$ in section~\ref{sec:m0m-quant}. In 
section~\ref{sec:geonbog} we construct the Hartle-Hawking-like vacuum 
$|0_G\rangle$ on the geon and compute the Bogolubov 
transformation between $|0_B\rangle$ and $|0_G\rangle$. 
We also give an expression for $|0_G\rangle$ in terms of 
$|0_B\rangle$, analogous to equation~(\ref{eqn:geonitmink}).

\section{Kruskal spacetime and the $\rp$ geon}
\label{sec:geonintro}

\begin{figure}[htbp]
\includegraphics{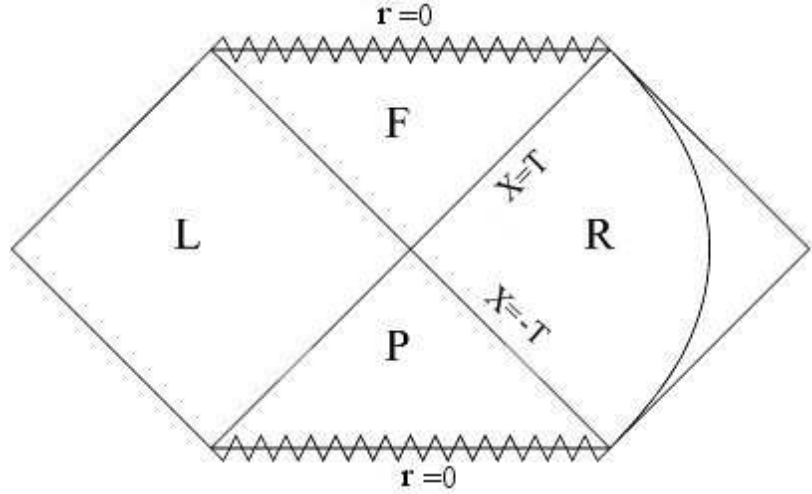}
\centering
\caption[Conformal diagram of Kruskal spacetime.]{Conformal diagram for the Kruskal Spacetime. 
Each point on the diagram represents a suppressed 2-sphere.}
\label{fig:krusk}
\end{figure}

In the notation of~\cite{lm:geon}, the Kruskal 
metric in the Kruskal coordinates 
$(T,X,\theta,\phi)$ reads 
\begin{equation}
{ds}^2=
\frac{32M^3}{r}e^{-r/(2M)}
({dT}^2-{dX}^2)
-r^2({d\theta}^2+{\sin^2{\theta}}{d\phi}^2) \ , 
\end{equation}
where $M>0$, $T^2 - X^2 < 1$ and $r$ is determined as a function of
$T$ and $X$ by $T^2 - X^2 = 1 - r/(2M)$. The manifold consists of the
right and left exteriors, denoted respectively by $R$ and $L$, and the
black hole (future) and white hole (past) interiors, denoted respectively by $F$ and $P$,
separated from each other by the bifurcate Killing horizon at $|T| =
|X|$ (see the conformal diagram, figure \ref{fig:krusk}).
The four regions may be individually covered by the Schwarzschild
coordinates $(t,r,\theta,\phi)$, in which the metric reads
\begin{equation}
{ds}^2=
\left(1-\frac{2M}{r}\right){dt}^2
-\frac{{dr}^2}{\left(1-\frac{2M}{r}\right)}
-r^2({d\theta}^2+{\sin^2{\theta}}{d\phi}^2) \ , 
\label{eq:schw}
\end{equation}
where $2M < r < \infty$ and $t$ ($r$) is timelike (spacelike) in the exteriors 
while $0 < r < 2M$ and $t$ ($r$) is spacelike (timelike) in the
interiors. The transformation between Kruskal and Schwarzschild coordinates
reads
\begin{align}
T= &\ \pm\left(\frac{r}{2M}-1\right)^{\frac{1}{2}}e^{r/(4M)}
\sinh{\left(\frac{t}{4M}\right)} \ ,
\nonumber
\\
X= &\ \pm\left(\frac{r}{2M}-1\right)^{\frac{1}{2}}e^{r/(4M)}
\cosh{\left(\frac{t}{4M}\right)} \ , 
\label{eq:schw-kruskal-transf}
\end{align}
in $R$ $(+)$ and $L$ $(-)$, and 
\begin{align}
T= &\ \pm\left(-\frac{r}{2M}+1\right)^{\frac{1}{2}}e^{r/(4M)}
\cosh{\left(\frac{t}{4M}\right)} \ ,
\nonumber
\\
X= &\ \pm\left(-\frac{r}{2M}+1\right)^{\frac{1}{2}}e^{r/(4M)}
\sinh{\left(\frac{t}{4M}\right)} \ , 
\end{align}
in $F$ $(+)$ and $P$ $(-)$.
The exteriors are static, with the timelike Killing vector~$\partial_t$. 
The Kruskal manifold is a globally hyperbolic, inextendible black hole spacetime. 
The analogy between the accelerated observers in Minkowski space 
with constant $\xi$, $y$, $z$ and the static observers in the Schwarzschild 
region $R$ with constant $r$, $\theta$, $\phi$ is clear. 
The vector $\partial_t$ in Kruskal is analogous to $\partial_\tau$ which generated 
boosts in the flat spaces, while $\partial_T$ is analogous to $\partial_t$ in 
Minkowski space (however $\partial_T$ is not a Killing vector 
on the Kruskal manifold).
The Kruskal manifold has two static regions, each isometric to the
Schwarzschild spacetime, causally disconnected
from each other and separated by a bifurcate Killing horizon.
No event in $R$ can causally influence any event in $L$, and vice versa, as the two regions 
are connected only by spacelike curves. 

\begin{figure}[htbp]
\includegraphics{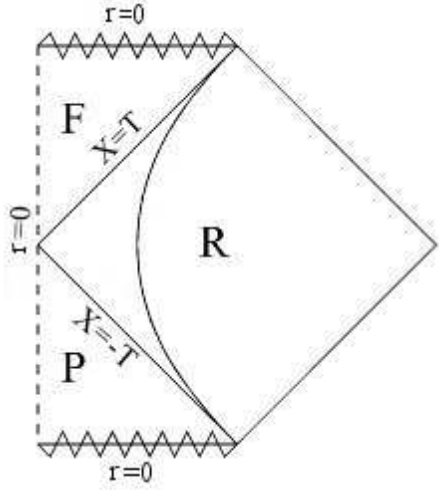}
\centering
\caption[Conformal diagram of the $\rp$ geon.]{Conformal diagram of the $\rp$ geon spacetime. 
The regions $X>0$ of Kruskal and the $\rp$ geon are isometric. Points $X>0$ represent a suppressed $2$-sphere
while points at $X=0$ represent a copy of $\mathbb{RP}^2$.}
\label{fig:geon}
\end{figure}

The ${\mathbb{RP}}^3$ geon is the quotient of the Kruskal manifold 
under the $\mathbb{Z}_2$ isometry group generated by the map 
\begin{equation}
\label{eqn:involJ}
J:(T,X,\theta,\phi)\mapsto(T,-X,\pi-\theta,\phi+\pi) \ . 
\end{equation}
The construction is analogous to that of $M_-$ from $M_0$ in
section~\ref{sec:m0m-intro}. 
Taking the quotient of the Kruskal manifold under the map
(\ref{eqn:involJ}) identifies the two exterior regions with 
a reflection in $X$ accompanied by an antipodal map on the 2-sphere. 
(\ref{eqn:involJ}) preserves space and time orientation and acts freely 
and properly discontinuously. The $\rp$ geon is thus a globally hyperbolic 
inextendible black hole spacetime which is locally isometric to Kruskal
but which contains only the one exterior region $R$. The regions $X>0$ of 
the Kruskal manifold and the $\rp$ geon are isometric. Figure
\ref{fig:geon} is a conformal diagram of the geon.

As the Kruskal manifold has spatial topology
${\mathbb{R}}\times{S^2}$, it is simply connected and has a unique
spin structure. The quotient construction implies \cite{hatcher} that
the geon has fundamental group~${\mathbb{Z}}_2$ and admits two spin
structures. As in section~\ref{sec:m0m-intro}, we describe these spin
structures in terms of periodic and antiperiodic boundary conditions
for spinors in a specified vierbein. On Kruskal, a standard reference
vierbein is
\begin{eqnarray}
\label{eqn:kruskvier} 
V_0=\sqrt{\frac{r}{32M^3}}e^{\frac{r}{4M}}\partial_T & V_2=\frac{1}{r}\partial_{\theta}
\nonumber 
\\ 
V_1=\sqrt{\frac{r}{32M^3}}e^{\frac{r}{4M}}\partial_X  & V_3=\frac{1}{r\sin\theta}\partial_{\phi} \ ,  
\end{eqnarray}
which is aligned along Kruskal coordinate axes.
This vierbein is not invariant under $J$. 
A second useful vierbein which is invariant under~$J$ is 
\begin{eqnarray} 
V_0 & = & \sqrt{\frac{r}{32M^3}}e^{\frac{r}{4M}}\partial_T 
\nonumber 
\\
V_1 & = & \cos{\phi}\sqrt{\frac{r}{32M^3}}e^{\frac{r}{4M}}\partial_X+\frac{\sin{\phi}}{r}\partial_{\theta} 
\nonumber 
\\
V_2 & = & -\sin{\phi}\sqrt{\frac{r}{32M^3}}e^{\frac{r}{4M}}\partial_X+\frac{\cos{\phi}}{r}\partial_{\theta} 
\nonumber 
\\
V_3 & = & \frac{1}{r\sin\theta}\partial_{\phi} \ ,
\label{eq:geon-bein} 
\end{eqnarray}
which rotates by $\pi$ in the $(X,\theta)$ tangent plane as $\phi$ increases by $\pi$.
The vierbein
(\ref{eq:geon-bein}) is well defined on the geon, and when the spinors
are written with respect to it, the two spin structures correspond to respectively
periodic and antiperiodic boundary conditions as
$\phi\mapsto{\phi+\pi}$. 
Alternatively we could consider an invariant vierbein which rotates 
by $-\pi$ as $\phi\mapsto{\phi+\pi}$ again imposing suitable boundary 
conditions. Periodic spinors with respect to this vierbein would 
be antiperiodic with respect to (\ref{eq:geon-bein}) and vice versa.
Both (\ref{eqn:kruskvier}) and
(\ref{eq:geon-bein}) are singular at $\theta=0$ and $\theta=\pi$, but
these coordinate singularities on the sphere 
can be handled by usual methods
and will not affect our discussion.  

In practice, we will work with the 
standard Kruskal vierbein~(\ref{eqn:kruskvier}). Although this vierbein is not 
invariant under $J$ it is the simplest one to work with on the Kruskal manifold. 
The boundary conditions appropriate
for the two geon spin structures will be found by the method-of-images
technique of the Appendix of~\cite{bd:em}.

\section[Boulware vacuum]{Spinor field theory on Schwarzschild spacetime and the Boulware vacuum}
\label{sec:boulware}

In this section we review the construction of the Boulware 
vacuum $|0_B\rangle$ for a massive Dirac field in one exterior region of either Kruskal 
or the $\rp$ geon~\cite{bo:boulware}. The Boulware vacuum is that 
empty of particles when positive frequency is defined with 
respect to Schwarzschild Killing time $\partial_t$. While this 
vacuum as such is well known, we will need to decompose the 
field in a novel basis in order to make contact with the geon 
in subsection~\ref{sec:geonbog}.

We consider the quantization of the spinor field $\psi$ in one 
exterior region $R$ in Schwarzschild coordinates. Here there are 
two natural choices of vierbein to work with. The first is
aligned along Schwarzschild coordinate axes, the second aligned along a 
set of axes oriented relative to fixed orthogonal directions \cite{bw:brilwheel}.
We choose to work with a vierbein aligned along the Schwarzschild coordinate axes, given
by 
\begin{equation}
\label{eqn:schwarzvier}
V^{\mu}_{a}
=\mathrm{diag}
\left(
\frac{1}{\left(1-\frac{2M}{r}\right)^{\frac{1}{2}}},
\left(1-\frac{2M}{r}\right)^{\frac{1}{2}},
\frac{1}{r},
\frac{1}{r\sin{\theta}}
\right) \ . 
\end{equation} 
This choice has the disadvantage of a more complex angular 
component analysis but the advantage of being analogous to the vierbein~(\ref{eqn:rinvier}) used in our
Rindler calculations. 

The Dirac equation~(\ref{eqn:gendir}) then reads
\begin{eqnarray}
\left[
m
+\frac{\gamma^2}
{ir\sin^{\frac{1}{2}}\theta}
\partial_{\theta}\sin^{\frac{1}{2}}\theta 
+\frac{\gamma^3}
{ir\sin{\theta}}
\partial_{\phi}
+\frac{\gamma^0}
{i\left(1-\frac{2M}{r}\right)^{\frac{1}{2}}}
\partial_t 
\right.
\nonumber 
\\
\label{eqn:dirinschwarz}
\left.
\mbox{}
+\frac{\gamma^1}
{ir}\left(1-\frac{2M}{r}\right)^{\frac{1}{4}}
\partial_r\left(1-\frac{2M}{r}\right)^{\frac{1}{4}}r
\right]
\psi=0 \ ,
\end{eqnarray}
where the $\gamma$
matrices are flat space $\gamma$'s.

Let us at this point 
choose a particular representation of the $\gamma$ matrices. 
We take
$$\begin{array}{cc}\gamma^0=\left(\begin{array}{cccc}
0 & 0 & 1 & 0 \\
0 & 0 & 0 & 1 \\
1 & 0 & 0 & 0 \\
0 & 1 & 0 & 0 \\
\end{array} \right) \ , &
\gamma^1=\left(\begin{array}{cccc}
0 & 0 & -1 & 0 \\
0 & 0 & 0 & -1 \\
1 & 0 & 0 & 0 \\
0 & 1 & 0 & 0 \\
\end{array}\right) \ , \end{array} $$
\begin{equation}\label{eqn:gammarep}
\begin{array}{cc}\gamma^2=\left(\begin{array}{cccc}
-i & 0 & 0 & 0 \\
0 & i & 0 & 0 \\
0 & 0 & i & 0 \\
0 & 0 & 0 & -i \\
\end{array}\right) \ , &
\gamma^3=\left(\begin{array}{cccc}
0 & -i & 0 & 0 \\
-i& 0 & 0 & 0 \\
0 & 0 & 0 & i \\
0 & 0& i & 0 \\
\end{array}\right) \ . 
\end{array} 
\end{equation}
This particular choice  is made in order to simplify the 
angular analysis and the radial equations. 
It has the advantage that charge conjugation takes the simple form 
\begin{equation}
\psi^c=
\left(
\begin{array}{cccc}
1 & 0 & 0 & 0 \\
0 & 1 & 0 & 0 \\
0 & 0 & -1 & 0 \\
0 & 0 & 0 & -1 \\
\end{array}
\right)
\psi^* \ ,
\label{eq:chargeconjS}
\end{equation}
where ${}^*$ stands for complex conjugation as in chapter \ref{ch:unruhbog}.

We seek a complete set of orthonormal solutions to (\ref{eqn:dirinschwarz}), positive 
frequency with respect to Schwarzschild Killing time $\partial_t$. 
We use the separation ansatz 
\begin{equation}
\label{eqn:dirsolution}
\psi_{\omega,k',m'}(t,r,\theta,\phi)
=N\frac{e^{-i\omega{t}}}{r\left(1-\frac{2M}{r}\right)^{\frac{1}{4}}}
\left(
\begin{array}{c}
F(r)Y^{k'}_{m'}(\theta,\phi) \\
G(r)Y^{k'}_{m'}(\theta,\phi) \\
\end{array}
\right)_{\omega{k'}} \ ,
\end{equation}
where $\omega>0$ for positive frequency modes. 
The angular eigenfunctions $Y^{k'}_{m'}(\theta,\phi)$ are an orthonormal 
set of eigenfunctions of $J_3$, $J^2$ and $K$ where
\begin{eqnarray}
J_3 & = & -i\partial_{\phi}
 \ ,
\nonumber
\\
J_{\pm} & = & e^{\pm{i}\phi}
\left(\pm\partial_{\theta}-\cot{\theta}\frac{1}{i}\partial_{\phi}
+\frac{1}{2}\frac{i\gamma^2\gamma^3}{\sin{\theta}}\right)
 \ ,
\nonumber
\\
K & = & i\beta\gamma^1
\left(
\gamma^2\frac{1}{i\sin^{\frac{1}{2}}\theta}\partial_{\theta}\sin^{\frac{1}{2}}\theta
+\gamma^3\frac{1}{i\sin{\theta}}\partial_{\phi}
\right)
 \ ,
\end{eqnarray}
with the properties
\begin{eqnarray}
J_3{Y^{k'}_{m'}(\theta,\phi)} & = & m'{Y^{k'}_{m'}(\theta,\phi)}
 \ ,
\nonumber
\\
J_{\pm}{Y^{k'}_{m'}(\theta,\phi)} & = & 
[j(j+1)-m'(m'\pm{1})]^{\frac{1}{2}}{Y^{k'}_{m'\pm{1}}(\theta,\phi)} 
 \ ,
\nonumber
\\
K{Y^{k'}_{m'}(\theta,\phi)} & = & k'{Y^{k'}_{m'}(\theta,\phi)}
 \ ,
\label{eqn:angularoper}
\end{eqnarray}
where $j=\frac{1}{2},\frac{3}{2},\frac{5}{2},\ldots$, $m'=-j,-j+1,\ldots,j$ and  
$k'=\pm(j+\frac{1}{2})$. These operators are 
obtained from the usual angular operators $J_3$, $J_\pm$, $K$  used in Minkowski 
space quantum field theory (i.e relative to a Cartesian set of axes) 
using the relevant spinor transformation which corresponds to the transformation 
between a Cartesian vierbein oriented relative to fixed orthogonal directions 
and the vierbein (\ref{eqn:schwarzvier}). 
This transformation is discussed in Boulware~\cite{bo:boulware} (see also \cite{bw:brilwheel} for a discussion
of these two simple choices of vierbein in the Schwarzschild geometry).
$J_3$, for example, is the total (orbital plus spin) angular momentum about the $x^3$-axis and so $m'$ takes 
half odd integer values. 
This is clear when written in the Cartesian vierbein oriented relative to fixed orthogonal directions 
but is not clear from the form in
(\ref{eqn:angularoper}) as (\ref{eqn:angularoper}) is written with respect to vierbein
(\ref{eqn:schwarzvier}).
Following Boulware (see Appendix A of \cite{bo:boulware}) we may now 
construct explicit expressions for the $Y^{k'}_{m'}(\theta,\phi)$ eigenfunctions. 
We find
\begin{equation}
Y^{k'}_{m'}(\theta,\phi)=\frac{e^{im'\phi}}{(2\pi)^{\frac{1}{2}}}
\left[\frac{(j+m')!}{(j-m')!}\right]^{\frac{1}{2}}\frac{(\tan\frac{\theta}{2})^{\sigma_2/2}}{\sin^{m'}\theta}
\left(\frac{\partial}{\partial\cos\theta}\right)^{j-m'}\frac{\sin^{2j}\theta(\tan\frac{\theta}{2})^{-\sigma_2}}
{2^j(j-\frac{1}{2})!}\psi_{k'} \ ,
\end{equation}
where 
$\psi_{k'}=\frac{1}{\sqrt{2}|k'|}\left(\begin{array}{c}|k'| \\
k' \\
\end{array}\right)$. Further we find the following 
relations which will be useful later
\begin{eqnarray}
\label{eqn:angrel1}
Y^{k'}_{m'}(\theta,\phi)^* & = & -i(-1)^{m'-1/2}\frac{k'}{|k'|}Y^{-k'}_{-m'}(\theta,\phi) \ ,
\\
\label{eqn:angrel2}
Y^{k'}_{m'}(\theta,\phi) & = & -i\frac{k'}{|k'|}\sigma_2Y^{-k'}_{m'}(\theta,\phi) \ ,
\\
\label{eqn:angrel3}
Y^{k'}_{m'}(\pi-\theta,\phi+\pi) & = & i(-1)^{j-1/2}\frac{k'}{|k'|}\sigma_1Y^{k'}_{m'}(\theta,\phi) \ ,
\end{eqnarray}
where $\sigma_j$ are Pauli matrices.
These spinor harmonics are normalized such that
\begin{equation}
\int_{\mathrm{angles}}\sin\theta\,{d\theta}\,{d\phi}\;
Y^{k''}_{m''}(\theta,\phi)^{\dagger}
Y^{k'}_{m'}(\theta,\phi)=\delta^{k''k'}\delta_{m''m'} \ .
\end{equation}

For a complete set of orthonormal mode solutions we must 
now find a basis of solutions for the radial functions 
$F(r)$ and $G(r)$ which are suitably orthonormal.
The radial functions satisfy
\begin{subequations}
\label{eqn:rad12}
\begin{align}
\label{eqn:rad1}
\left(
1-\frac{2M}{r}
\right)
\partial_rF-i\omega{F}= 
&\ 
\left(
1-\frac{2M}{r}
\right)
^{\frac{1}{2}}
\left(
\frac{k'}{r}-im
\right)
G \ , 
\\
\label{eqn:rad2}
\left(
1-\frac{2M}{r}
\right)
\partial_rG+i\omega{G}= 
&\ 
\left(
1-\frac{2M}{r}
\right)
^{\frac{1}{2}}
\left(
\frac{k'}{r}+im
\right)
F \ . 
\end{align}
\end{subequations}
Following Chandrasekhar~\cite{ch:chbook}, we reduce the radial
equations (\ref{eqn:rad12}) to a pair of Schr\"odinger-like
equations. Writing
\begin{equation}
\left(
\begin{array}{c} 
F(r) 
\\
G(r) 
\end{array}
\right)
=\left(
\begin{array}{c} 
\frac{1}{2}(Z_++Z_-)e^{(-\frac{i}{2}\tan^{-1}(\frac{mr}{k'}))} 
\\ 
\frac{1}{2}(Z_+-Z_-)e^{(\frac{i}{2}\tan^{-1}(\frac{mr}{k'}))}
\end{array}
\right) \ ,
\label{eq:FGvZ}
\end{equation}
we find that  $Z_{\pm}$ satisfy 
\begin{equation}
\label{eqn:zplusminus} 
\left(
\frac{d}{d\hat{r}_*}{\mp}W\right)Z_{\pm}
=i\omega{Z_{\mp}} \ , 
\end{equation}
where $\hat{r}^*=r^*+\frac{1}{2\omega}\tan^{-1}(\frac{mr}{k'})$, 
$r^*=r+2M\ln(|r-2M|/2M)$ and
\begin{equation}
W=
\frac{(r^2-2Mr)^{\frac{1}{2}}(k'^2+m^2r^2)^{\frac{3}{2}}}{r^2(k'^2+m^2r^2)
+\frac{k'm}{2\omega}(r^2-2Mr)} \ .
\end{equation}
$Z_{\pm}$ hence satisfy the one-dimensional time-independent 
Schr\"odinger equations
\begin{equation}
\label{eqn:scat}
\left(
\frac{d^2}{d\hat{r}^2_*}
+\omega^2\right)Z_{\pm}=V_{\pm}Z_{\pm} \ ,
\end{equation}
where 
\begin{equation}
V_{\pm}=W^2\pm\frac{dW}{d\hat{r}_*} 
\ . 
\end{equation}
From the usual one-dimensional scattering theory
the eigenvalue spectrum for $\omega$ is seen to be continuous
and consists of the entire positive real line \cite{me:messiah}.

Suppose first $\omega^2>m^2$. The eigenvalue 
spectrum is then degenerate of order 2. One way we can break this degeneracy and
obtain a complete set of orthonormal solutions would be to choose the 
solutions that have the scattering theory asymptotic form,
\begin{subequations}
\label{eqn:zleftright}
\begin{equation}
\label{eqn:zleft}
\stackrel{\leftarrow}{Z_\pm}
=
\left\{
\begin{array}{cc}
\stackrel{\leftarrow}{B_\pm}e^{-i\omega{\hat{r_*}}} 
&  
\hat{r_*}\rightarrow{-\infty} 
\\ 
e^{-i(p\hat{r_*}+\frac{Mm^2}{p}\ln(\frac{\hat{r_*}}{2M}))}
+\stackrel{\leftarrow}{A_\pm}e^{i(p\hat{r_*}
+\frac{Mm^2}{p}\ln(\frac{\hat{r_*}}{2M}))} 
& 
\hat{r_*}\rightarrow{\infty} \ \ ,
\end{array}
\right. 
\end{equation}
\begin{equation}
\label{eqn:zright}
\stackrel{\rightarrow}{Z_\pm}
=
\left\{
\begin{array}{cc}e^{i\omega\hat{r_*}}
+\stackrel{\rightarrow}{A_\pm}e^{-i\omega\hat{r_*}} 
&  
\hat{r_*}\rightarrow{-\infty} 
\\ 
\stackrel{\rightarrow}{B_\pm}e^{i(p\hat{r_*}
+\frac{Mm^2}{p}\ln(\frac{\hat{r_*}}{2M}))} 
& 
\hat{r_*}\rightarrow{\infty} \ \ , 
\end{array}
\right. 
\end{equation}
\end{subequations}
where $p=\sqrt{(\omega^2-m^2)}$. $\stackrel{\leftarrow}{Z_\pm}$ is
purely ingoing at the horizon and $\stackrel{\rightarrow}{Z_\pm}$ is
purely outgoing at infinity. Considering the Wronskian
between these solutions in each asymptotic region 
yields relations between the transmission and reflection
coefficients
\begin{eqnarray}
(1-|\stackrel{\leftarrow}{A_\pm}|^2) & = & \frac{\omega}{p}|\stackrel{\leftarrow}{B_\pm}|^2 \ ,
\nonumber
\\
(1-|\stackrel{\rightarrow}{A_\pm}|^2) & = & \frac{p}{\omega}|\stackrel{\rightarrow}{B_\pm}|^2 \ ,
\nonumber
\\
\omega\stackrel{\leftarrow}{B_\pm} & = & p\stackrel{\rightarrow}{B_\pm} \ ,
\nonumber
\\
-{\stackrel{\rightarrow}{B_\pm}}^*\stackrel{\leftarrow}{A_\pm} & = & \omega{\stackrel{\rightarrow}{A_\pm}}^*\stackrel{\leftarrow}{B_\pm} \ ,
\nonumber
\\
{\stackrel{\leftarrow}{A_\pm}}^* & = & -\frac{\stackrel{\rightarrow}{B_\pm}^*\stackrel{\rightarrow}{A_\pm}}{\stackrel{\rightarrow}{B_\pm}} \ ,
\nonumber
\\
|\stackrel{\rightarrow}{A_\pm}|^2 & = & |\stackrel{\leftarrow}{A_\pm}|^2 \ .
\label{eqn:wronskiancond}
\end{eqnarray}
From (\ref{eqn:zplusminus}) it further follows that
$\stackrel{\leftarrow}{B_+}=-\stackrel{\leftarrow}{B_-}$ and
$\stackrel{\rightarrow}{A_+}=-\stackrel{\rightarrow}{A_-}$.
We may therefore take for our radial modes 
the following basis,
\begin{eqnarray}
\stackrel{\longleftarrow}{\left(\begin{array}{c} F(r) \\G(r) \end{array}\right)} & = & \sqrt{\frac{\omega}{2\pi{p}}}\left\{\begin{array}{cc}\left(\begin{array}{c} 0 \\ \stackrel{\leftarrow}{B_+} \end{array}\right)e^{-i\omega{r_*}} & r_*\rightarrow{-\infty} \\ \left(\begin{array}{c}1 \\ 0 \end{array}\right) e^{-iq_{\hat{r}_*}}+\left(\begin{array}{c} 0 \\ \stackrel{\leftarrow}{A_+}\end{array}\right)e^{iq_{\hat{r}_*}} & \hat{r_*}\rightarrow{\infty}\end{array} \ ,\right. 
\nonumber
\\
\stackrel{\longrightarrow}{\left(\begin{array}{c} F(r) \\G(r) \end{array}\right)} & = & \sqrt{\frac{1}{2\pi}}\left\{\begin{array}{cc}\left(\begin{array}{c} 1 \\ 0 \end{array}\right)e^{i\omega{r_*}}+\left(\begin{array}{c} 0 \\ \stackrel{\rightarrow}{A_+} \end{array}\right)e^{-i\omega{r_*}} & r_*\rightarrow{-\infty} \\  \left(\begin{array}{c} 0 \\ \stackrel{\rightarrow}{B_+}\end{array}\right)e^{iq_{\hat{r}_*}} & \hat{r_*}\rightarrow{\infty}\end{array} \ ,\right. 
\end{eqnarray}
where $q_{\hat{r}_*}={(\frac{1}{2}\tan^{-1}(\frac{mr}{k'}))}+(p\hat{r_*}+\frac{Mm^2}{p}\ln(\frac{\hat{r_*}}{2M}))$. 
Here $\stackrel{\longleftarrow}{\left(\begin{array}{c} F(r) \\G(r) \end{array}\right)}$ is a 
purely ingoing radial mode at the horizon, while 
$\stackrel{\longrightarrow}{\left(\begin{array}{c} F(r) \\G(r) \end{array}\right)}$ 
is purely outgoing at infinity.
However, to be able to handle the geon in
subsection~\ref{sec:geonbog}, we will need modes that transform simply
under charge conjugation~(\ref{eq:chargeconjS}) and under $J$ (\ref{eqn:involJ}) when 
continued analytically into the $F$ region. Using the
properties of the reflection and transmission coefficients 
(\ref{eqn:wronskiancond}) and the properties of the spinor spherical
harmonics~(\ref{eqn:angrel1})-(\ref{eqn:angrel3}), 
we find after considerable effort that a
convenient set of positive frequency Boulware modes 
is $\{\Psi^{\pm}_{\omega,k',m'}\}$, given by
\begin{subequations}
\label{eqn:posmode12}
\begin{align}
\label{eqn:posmode1}
\Psi^+_{\omega,m',k'}
&=\frac{
e^{-\frac{i\pi}{2}(j+m'+(\frac{k'}{|k'|}-1)/2)}
e^{-i\omega{t}}
}
{r(1-\frac{2M}{r})^{\frac{1}{4}}}
{{
\left(
\begin{array}{c}
u(r)Y^{k'}_{m'}(\theta,\phi) 
\\
v(r)Y^{k'}_{m'}(\theta,\phi) 
\\
\end{array}
\right)
}}^+_{\omega{k'}} \ ,
\\
\label{eqn:posmode2}
\Psi^-_{\omega,m',k'}
&=\frac{
e^{-\frac{i\pi}{2}(j-m'+(\frac{k'}{|k'|}-1)/2)}
e^{-i\omega{t}}
}
{r(1-\frac{2M}{r})^{\frac{1}{4}}}
{{
\left(
\begin{array}{c}
u(r)Y^{k'}_{m'}(\theta,\phi) 
\\
v(r)Y^{k'}_{m'}(\theta,\phi) 
\\
\end{array}
\right)
}}^-_{\omega{k'}} \ ,
\end{align}
\end{subequations}
where the radial functions with superscript ${}^+$ are
specified by the horizon asymptotic behaviour
\begin{subequations}
\label{eq:deg-nearhor}
\begin{align}
\left(
\begin{array}{c}u(r) 
\\
v(r) 
\\
\end{array}
\right)^+_{\omega{k'}}
&=
\frac{1}{\sqrt{4\pi}}
\left\{
\sqrt{1+\sqrt{1-|\stackrel{\rightarrow}{A}|^2}}
\left(
\begin{array}{c}
1 
\\
0 
\\
\end{array}
\right)
e^{i\omega{r_*}}
\vphantom{
\frac{\stackrel{\rightarrow}{A_+}}
{\sqrt{1+\sqrt{1-|\stackrel{\rightarrow}{A}|^2}}}
}
\right.
\nonumber
\\
\label{eqn:+mode}
&
\hspace{5ex}
\left.
+\frac{\stackrel{\rightarrow}{A_+}}
{\sqrt{1+\sqrt{1-|\stackrel{\rightarrow}{A}|^2}}}
\left(
\begin{array}{c}
0 
\\
1 
\\
\end{array}
\right)
e^{-i\omega{r_*}}
\right\},
\ \ \ \ \ \hat{r}^*\to{-\infty} 
\ , 
\end{align}
and
\begin{equation}
\label{eqn:-mode}
\left(
\begin{array}{c}
u(r) 
\\
v(r) 
\\
\end{array}
\right)^-_{\omega{k'}}
=
\left(
\begin{array}{cc} 
0 & 1
\\ 
1 & 0 
\end{array}
\right)
\left(
\begin{array}{c}
u(r) 
\\
v(r) 
\\
\end{array}
\right)^{+*}_{\omega{k'}} \ . 
\end{equation}
\end{subequations}
The key property for charge conjugation is~(\ref{eqn:-mode}). 
The modes $\{\Psi^{\pm}_{\omega,k',m'}\}$ are complete for $\omega^2 > m^2$ 
and delta-orthonormal in the Dirac inner product 
\begin{equation}
\label{eqn:innerprod}
\langle\psi_1,\psi_2\rangle
=\int_{\mathrm{angles}}\sin\theta\,
{d\theta}\,
{d\phi}\,
\int^{\infty}_{2M}dr\,\frac{r^2}{\left(1-\frac{2M}{r}\right)^{\frac{1}{2}}}
\psi_1^{\dagger}\psi_2 \ ,
\end{equation}
taken on a constant $t$ hypersurface. 

Suppose then $0<\omega^2<m^2$ in~(\ref{eqn:scat}). There is now only
one linearly independent delta-normalizable solution for
each~$\omega$. This solution vanishes at infinity and has at the
horizon the behaviour
\begin{equation}
Z_{\pm}=a_{\pm}\cos(\omega\hat{r}^*+\delta_{\pm})  \ ,
\ \ \ 
\hat{r}^*\to{-\infty} 
\ , 
\end{equation}
where $a_{\pm}$, and $\delta_{\pm}$ are real constants. Physically
these solutions correspond to particles that do not reach
infinity. Proceeding as above, we find that a convenient set of
positive frequency Boulware modes, complete for $0<\omega^2<m^2$ and
delta-orthonormal in the Dirac inner product~(\ref{eqn:innerprod}), is
\begin{equation}
\label{eqn:posmode3}
\psi_{\omega,k',m'}(t,r,\theta,\phi)
=e^{-\frac{i\pi}{2}(j+|m'|+(1-\frac{k'}{|k'|})/2)}
\frac{e^{-i\omega{t}}}{r(1-\frac{2M}{r})^{\frac{1}{4}}}
\left(
\begin{array}{c}
F(r)Y^{k'}_{m'}(\theta,\phi) \\
G(r)Y^{k'}_{m'}(\theta,\phi) \\
\end{array}
\right)_{\omega{k'}} \ , 
\end{equation}
where $0<\omega < m$ and the radial functions are specified by the
horizon asymptotic behaviour
\begin{equation}
\label{eqn:wmodewm}
\left(
\begin{array}{c}
F(r) 
\\
G(r) 
\\
\end{array}
\right)_{\omega{k'}}
=
\frac{1}{\sqrt{2\pi}}
\left\{
\left(
\begin{array}{c} 
e^{i\delta_+} \\
0 \\
\end{array}
\right)
e^{i\omega{r_*}}+
\left(
\begin{array}{c}
0 \\
e^{-i\delta_+} \\
\end{array}
\right)
e^{-i\omega{r_*}}
\right\} \ ,
\ \ \ 
\hat{r}^*\mapsto{-\infty} \ . 
\end{equation}

Up to this point we have used the Schwarzschild
vierbein~(\ref{eqn:schwarzvier}). To make contact with the geon in
subsection~\ref{sec:geonbog}, we need to express the modes in a vierbein
that is regular at the horizons. We therefore now transform our modes
to the Kruskal
vierbein (\ref{eqn:kruskvier}) by the spinor transformation
$\psi\mapsto{e^{\frac{t}{8M}\gamma^0\gamma^1}\psi}$.
We suppress the
explicit transformed expressions and 
continue to use the same symbols for the mode
functions.

We are now ready to quantize. The field is expanded in our orthonormal
modes and their charge conjugates as
\begin{eqnarray}       
\Psi                   
& = & 
\sum_{k',m'}
\int_{0}^{m}d\omega
\left(
a_{\omega,k',m'}\psi_{\omega,k',m'}
+b^{\dagger}_{\omega,k',m'}\psi^c_{\omega,k',m'} 
\right) 
\nonumber 
\\ 
&   & 
\mbox{}
+\sum_{k',m'}
\int_{m}^{\infty}d\omega
\left(
a_{+,\omega,k',m'}\Psi^+_{\omega,k',m'}+
a_{-,\omega,k',m'}\Psi^-_{\omega,k',m'}
\right. 
\nonumber 
\\
\label{eqn:boulexp}
&   & 
\hspace{17ex}
\left.
\mbox{}
+b^\dagger_{+,\omega,k',m'}\Psi^{+,c}_{\omega,k',m'}
+b^\dagger_{-,\omega,k',m'}\Psi^{-,c}_{\omega,k',m'}
\right) \ ,  
\end{eqnarray}
where the annihilation and creation operators satisfy the usual
anticommutation relations. The vacuum annihilated by the annihilation
operators is the Boulware vacuum~$|0_B\rangle$
\begin{equation}
a_{\omega,k',m'}|0_B\rangle=
b_{\omega,k',m'}|0_B\rangle=
a_{\pm,\omega,k',m'}|0_B\rangle=
b_{\pm,\omega,k',m'}|0_B\rangle=0 \ , \quad \forall{\omega,k',m'} \ .
\end{equation}
The Boulware vacuum here is analogous to the Rindler 
vacuum $|0_{R_-}\rangle$ in section~\ref{sec:m-unruh}. 
$|0_B\rangle$ is by
construction the state void of particles defined with respect to the
Schwarzschild Killing time.
It is the natural vacuum for static observers in the
exterior region $R$.


\section[Bogolubov transformation on the geon]{The Hartle-Hawking-like vacuum and Bogolubov transformation on the geon}
\label{sec:geonbog}

In this section we decompose the geon Hartle-Hawking-like vacuum
into Boulware excitations. We use the analytic continuation method,
following closely subsection~\ref{sec:m-unruh}.

The Hartle-Hawking vacuum on the Kruskal manifold is defined by mode functions that
are purely positive frequency with respect to the horizon generators
and hence analytic in the lower half-plane in the complexified Kruskal
time~$T$. It follows that on Kruskal we can construct $W$-modes whose
vacuum is the Hartle-Hawking vacuum by analytically continuing the
Boulware-modes across the horizons in the lower half-plane in~$T$. The
quotient from Kruskal to the geon defines in each spin structure on
the geon the Hartle-Hawking-like vacuum $|0_G\rangle$, by restriction
to the Kruskal $W$-modes that are invariant under the
map $J$~(\ref{eqn:involJ}). Our task is to find these modes. 

We begin by taking our positive 
frequency ($\omega>0$) modes~(\ref{eqn:posmode1}), (\ref{eqn:posmode2})
and~(\ref{eqn:posmode3}) written in the Kruskal vierbein (\ref{eqn:kruskvier}) 
and continuing them across the future horizon into the interior black hole region 
$F$ in the lower half complex $T$-plane.
Let us consider a general positive frequency mode solution to 
equation~(\ref{eqn:dirinschwarz}) near the horizon $r=2M$. 
In the Kruskal vierbein this can take the form
\begin{equation}
\label{eqn:genmode}
\psi^R(t,r,\theta,\phi)=
N\frac{e^{-i\omega{t}}}{r(1-\frac{2M}{r})^{\frac{1}{4}}}\left\{\left(\begin{array}{c} A \\
0 \\
\end{array}\right)e^{i\omega{r_*}+\frac{t}{8M}}+\left(\begin{array}{c}0 \\
B \\
\end{array}\right)e^{-i\omega{r_*}-\frac{t}{8M}}\right\}Y^{k'}_{m'}(\theta,\phi) \ ,
\end{equation}
where $A$ and $B$ are arbitrary constants, $N$ is a normalization factor and $\omega>0$. 
Writing this in Kruskal coordinates, it becomes
\begin{eqnarray}
\psi^R(T,X,\theta,\phi)=N\left[\left(\begin{array}{c}A \\
0 \\
\end{array}\right)_{\omega{k'}}(X-T)^{4M\omega{i}}\left(\frac{X+T}{X-T}\right)^{\frac{1}{4}} 
\right.
\nonumber
\\
\left.
+\left(\begin{array}{c}0 \\
B \\
\end{array}\right)_{\omega{k'}}(X+T)^{-4M\omega{i}}
\left(\frac{X-T}{X+T}\right)^{\frac{1}{4}}\right]
\frac{1}{r}\frac{(r/2M)^{1/2}e^{r/{8M}}}{(X^2-T^2)^{\frac{1}{4}}}Y^{k'}_{m'}(\theta,\phi) 
 \ .
\end{eqnarray}
Continuation to $F$ in the lower half complex $T$-plane is then implemented 
by the substitutions $(X-T)^{1/4}\mapsto(T-X)^{1/4}e^{i\pi/4}$ and
$\ln(X-T)\mapsto\ln(T-X)+i\pi$, under which $\psi^R\mapsto\psi^F$ where
\begin{eqnarray}
\psi^F(T,X,\theta,\phi)=N\left[\left(\begin{array}{c}A \\
0 \\
\end{array}\right)_{\omega{k'}}(T-X)^{4M\omega{i}}
e^{-4M\omega{\pi}-\frac{i\pi}{2}}\left(\frac{X+T}{T-X}\right)^{\frac{1}{4}} 
\right.
\nonumber
\\
\label{eqn:psiF}
\left.
+\left(\begin{array}{c}0 \\
B \\
\end{array}\right)_{\omega{k'}}(X+T)^{-4M\omega{i}}\left(\frac{T-X}{X+T}\right)^{\frac{1}{4}}\right]
\frac{1}{r}\frac{(r/2M)^{1/2}e^{r/{8M}}}{(T^2-X^2)^{\frac{1}{4}}}Y^{k'}_{m'}(\theta,\phi) 
 \ .
\end{eqnarray}

We need to build from (\ref{eqn:psiF}) modes that are well-defined in the $F$ region of the
geon by the method of images.
If we were working in the rotating Kruskal vierbein which is invariant under 
the involution $J$ the method of images could be applied directly and 
we would have
\begin{equation}
W^F(T,X,\theta,\phi)=
\psi^F(T,X,\theta,\phi)+\epsilon\psi^F(T,-X,\pi-\theta,\phi+\pi)
 \ ,
\end{equation}
where $\epsilon=1,(-1)$ labels fields with periodic (antiperiodic) boundary 
conditions respectively. However we are instead working in the standard Kruskal 
vierbein which is not invariant under $J$. The spinor transformation associated 
with the transformation between the rotating Kruskal vierbein and the standard one 
(a rotation by $-\pi$ in the $(X,\theta)$-plane as $\phi\mapsto{\phi+\pi}$) is given by 
\begin{equation} 
W^F(T,X,\theta,\phi)\mapsto
{e^{\frac{\phi\gamma^1\gamma^2}{2}}}
W^F(T,X,\theta,\phi)=W^F_s(T,X,\theta,\phi)
 \ .
\end{equation}
Now it is easy to show that in terms of this standard Kruskal vierbein
\begin{equation}
\label{eqn:methofim}
W^F_s(T,X,\theta,\phi)=\psi^F_s(T,X,\theta,\phi)
+\epsilon{e^{-\frac{\pi\gamma^1\gamma^2}{2}}}\psi^F_s(T,-X,\pi-\theta,\phi+\pi)
 \ .
\end{equation}
This is the expression we require our $W$-modes on the geon 
to satisfy when working in this non-invariant vierbein.
Now applying~(\ref{eqn:psiF}) to~(\ref{eqn:methofim}) we find 
positive frequency $W$-modes on the geon in $F$ near $r=2M$ given by
\begin{eqnarray}
W^F(T,X,\theta,\phi) & = & \psi^F(T,X,\theta,\phi)
\nonumber
\\
& & +
{\epsilon}i(-1)^jN\left[\left(\begin{array}{c}0 \\
A \\
\end{array}\right)_{\omega{k'}}(T+X)^{4M\omega{i}}
e^{-4M\omega{\pi}-\frac{i\pi}{2}}\left(\frac{T-X}{T+X}\right)^{\frac{1}{4}} 
\right.
\nonumber
\\
& & 
+\left.\left(\begin{array}{c}B \\
0 \\
\end{array}\right)_{\omega{k'}}(T-X)^{-4M\omega{i}}\left(\frac{T+X}{T-X}\right)^{\frac{1}{4}}\right]
\frac{1}{r}\frac{(r/2M)^{1/2}e^{r/{8M}}}{(T^2-X^2)^{\frac{1}{4}}}Y^{-k'}_{m'}(\theta,\phi) \ .
\nonumber
\\
\end{eqnarray}
On continuation back to the right hand exterior in the lower half 
complex $T$-plane, these modes become
\begin{eqnarray}
W^R(T,X,\theta,\phi) & = & \psi^R(T,X,\theta,\phi)
\nonumber
\\
& & +  
{\epsilon}(-1)^{j+1}
e^{-4M\omega{\pi}}N\left[\left(\begin{array}{c}0 \\
-A \\
\end{array}\right)_{\omega{k'}}(T+X)^{4M\omega{i}}\left(\frac{X-T}{T+X}\right)^{\frac{1}{4}} 
\right.
\nonumber
\\
&  & +
\left.\left(\begin{array}{c}B \\
0 \\
\end{array}\right)_{\omega{k'}}(X-T)^{-4M\omega{i}}\left(\frac{T+X}{X-T}\right)^{\frac{1}{4}}\right]
\frac{1}{r}\frac{(r/2M)^{1/2}e^{r/{8M}}}{(X^2-T^2)^{\frac{1}{4}}}Y^{-k'}_{m'}(\theta,\phi) \ .
\nonumber
\\
\label{eqn:genWinR}
\end{eqnarray}

We are now ready to specialise to the Boulware modes found in the previous section. 
For $\omega>m$ comparing~(\ref{eqn:genmode}) 
and~(\ref{eqn:+mode},\ref{eqn:-mode}), and making use of the relations 
(\ref{eqn:angrel1})-(\ref{eqn:angrel3}) we find in the $R$ region a complete set of $W$-modes 
given by
\begin{align}
W^+_{\omega,k',m'}(t,r,\theta,\phi)
&=\frac{1}{\sqrt{2\cosh(4\pi{M}\omega)}}
\left(
e^{2\pi{M}\omega}\Psi^+_{\omega,k',m'}
+\epsilon{e^{-2\pi{M}\omega}}\Psi^{-,c}_{\omega,k',-m'}
\right) \ ,
\nonumber
\\
W^-_{\omega,k',m'}(t,r,\theta,\phi)
&=\frac{1}{\sqrt{2\cosh(4\pi{M}\omega)}}
\left(
e^{2\pi{M}\omega}\Psi^-_{\omega,k',m'}
-\epsilon{e^{-2\pi{M}\omega}}\Psi^{+,c}_{\omega,k',-m'}
\right) \ .
\label{eqn:wmodesgeon}
\end{align}
(Note the similarity between (\ref{eqn:wmodesgeon}) and~(\ref{eqn:m-wmode12})). 
These modes are delta-orthonormal with respect to the inner product (\ref{eqn:innerprod}).
For $0<\omega<m$, comparing~(\ref{eqn:genmode}) and~(\ref{eqn:wmodewm}) 
we find $W$-modes given by
\begin{equation}
W_{\omega,k',m'}(t,r,\theta,\phi)
=\frac{1}{\sqrt{2\cosh(4\pi{M}\omega)}}
\left(
e^{2\pi{M}\omega}\psi_{\omega,k',m'}
+\epsilon\frac{m'}{|m'|}{e^{-2\pi{M}\omega}}\psi^{c}_{\omega,k',-m'}
\right) \ . 
\label{eq:W-geon-deg}
\end{equation}
again orthonormal with respect to the Kruskal inner product (\ref{eqn:innerprod}). 
It can be verified that the
factors $m'/|m'|$ appearing in (\ref{eq:W-geon-deg}) cannot be
absorbed into the phase factors of the Boulware modes. 

On the geon, the expansion of the field in the $W$-modes reads 
\begin{eqnarray}
\Psi 
& = & 
\sum_{k',m'}
\int_{0}^{m}d\omega
\left(
c_{\omega,k',m'}W_{\omega,k',m'}
+d^{\dagger}_{\omega,k',m'}W^c_{\omega,k',m'} 
\right)
\nonumber 
\\ 
&   &
\mbox{}
+\sum_{k',m'}
\int_{m}^{\infty}d\omega
\left(
c_{+,\omega,k',m'}W^+_{\omega,k',m'}
+c_{-,\omega,k',m'}W^-_{\omega,k',m'} 
\right. 
\nonumber
\\
\label{eqn:boulexp2}
&   &
\hspace{17ex}
\left.
\mbox{}
+d^\dagger_{+,\omega,k',m'}W^{+,c}_{\omega,k',m'}
+d^\dagger_{-,\omega,k',m'}W^{-,c}_{\omega,k',m'}
\right) \ ,
\end{eqnarray}
with the usual anticommutation relations imposed on the creation and annihilation operators. 
We denote by $|0_G\rangle$ the state annihilated by all the annihilation 
operators,
\begin{equation} 
c_{\omega,k',m'}|0_G\rangle=d_{\omega,k',m'}|0_G\rangle
=c_{\pm,\omega,k',m'}|0_G\rangle=d_{\pm,\omega,k',m'}|0_G\rangle
=0
 \ , \quad \forall{\omega,k',m'} \ .
\end{equation}
$|0_G\rangle$ is the Kruskal-like vacuum on the geon, induced
by the usual Hartle-Hawking vacuum on Kruskal.
Comparison of
(\ref{eqn:boulexp}) and (\ref{eqn:boulexp2}) gives the Bogolubov 
transformation 
\begin{align}
a_{\omega,k',m'}= 
&\ 
\frac{1}{\sqrt{2\cosh(4\pi{M}\omega)}}
\left(
e^{{2\pi{M}\omega}}c_{\omega,k',m'}
+\epsilon\frac{m'}{|m'|}{e^{-{2\pi{M}}\omega}}d^{\dagger}_{\omega,k',-m'}
\right) \ ,
\nonumber 
\\
b^{\dagger}_{\omega,k',m'}= 
&\ 
\frac{1}{\sqrt{2\cosh(4\pi{M}\omega)}}
\left(
e^{{2\pi{M}\omega}}d^{\dagger}_{\omega,k',m'}
+\epsilon\frac{m'}{|m'|}{e^{-{2\pi{M}}\omega}}c_{\omega,k',-m'}
\right) \ ,
\nonumber
\\
a_{+,\omega,k',m'}= 
&\ 
\frac{1}{\sqrt{2\cosh(4\pi{M}\omega)}}
\left(
e^{2\pi{M}\omega}c_{+,\omega,k',m'}
-\epsilon{e^{-2\pi{M}\omega}}d^{\dagger}_{-,\omega,k',-m'}
\right) \ , 
\nonumber
\\
a_{-,\omega,k',m'}= 
&\ 
\frac{1}{\sqrt{2\cosh(4\pi{M}\omega)}}
\left(
e^{2\pi{M}\omega}c_{-,\omega,k',m'}
+\epsilon{e^{-2\pi{M}\omega}}d^{\dagger}_{+,\omega,k',-m'}
\right) \ ,
\nonumber
\\
b^\dagger_{+,\omega,k',m'}= 
&\ 
\frac{1}{\sqrt{2\cosh(4\pi{M}\omega)}}
\left(
e^{2\pi{M}\omega}d^\dagger_{+,\omega,k',m'}
-\epsilon{e^{-2\pi{M}\omega}}c_{-,\omega,k',-m'}
\right) \ ,
\nonumber
\\
b^\dagger_{-,\omega,k',m'}= 
&\ 
\frac{1}{\sqrt{2\cosh(4\pi{M}\omega)}}
\left(
e^{2\pi{M}\omega}d^\dagger_{-,\omega,k',m'}
+\epsilon{e^{-2\pi{M}\omega}}c_{+,\omega,k',-m'}
\right) \ . 
\label{eqn:bogtrangeon}
\end{align}

For the expectation values of 
Boulware mode number operators, (\ref{eqn:bogtrangeon}) gives e.g.
\begin{equation}
\langle0_G|a^\dagger_{+,\omega,k',m'}a_{+,\omega',k'',m''}|0_G\rangle
=\frac{1}{(e^{8\pi{M}\omega}+1)}\delta(\omega-\omega')\delta_{k'k''}\delta_{m'm''}
 \ ,
\end{equation}
and a discussion similar to that given on $M_-$ follows. 
Note in particular that the geon spin structure does not show up in the
expectation value. 
The number operator expectation value does not distinguish 
between the state $|0_G\rangle$ on the geon and the usual 
Hartle-Hawking state on Kruskal, it is representative of
a thermal bath at temperature $T=(8\pi{M})^{-1}$.

Expressing $|0_G\rangle$ in terms of $|0_B\rangle$, we find
\begin{align}
|0_G\rangle 
& = 
\prod_{\substack{0<\omega < m \\ k',m'}}
\frac{1}{({e^{-8\pi{M}\omega}+1)^{\frac{1}{2}}}}
\sum_{q=0,1}
{\left(
\frac{\epsilon{m'}}{|m'|} 
\right)}^q
e^{-4\pi{M}\omega{q}}
(a^{\dagger}_{\omega,k',m'}b^{\dagger}_{\omega,k',-m'})^q
|0_B\rangle 
\nonumber
\\
& 
\hspace{3ex}
\times 
\prod_{\substack{\omega > m \\ \sigma,k',m'}}
\frac{1}{({e^{-8\pi{M}\omega}+1)^{\frac{1}{2}}}}\!\!
\sum_{q=0,1}(-\epsilon)^qe^{-4\pi{M}\omega{q}}
|q\rangle_{\sigma,+,\omega,k',m'}
|q\rangle_{-\sigma,-,\omega,k',-m'} 
\ ,
\label{eqn:geonvac}
\end{align}
where $\sigma=+1 (-1)$ and
\begin{eqnarray}
|q\rangle_{+,+,\omega,k',m'}= 
(a^{\dagger}_{+,\omega,k',m'})^q|0_B\rangle 
&
|q\rangle_{+,-,\omega,k',m'}= 
(a^{\dagger}_{-,\omega,k',m'})^q|0_B\rangle 
\ ,
\nonumber
\\
|q\rangle_{-,+,\omega,k',m'}= 
(b^{\dagger}_{+,\omega,k',m'})^q|0_B\rangle 
&
|q\rangle_{-,-,\omega,k',m'}= 
(b^{\dagger}_{-,\omega,k',m'})^q|0_B\rangle 
\ .
\end{eqnarray}
A comparison of (\ref{eqn:geonvac}) and (\ref{eqn:geonitmink}) shows
that $|0_G\rangle$ is closely similar to the state $|0_-\rangle$
on~$M_-$.
The Boulware modes are correlated in pairs in the Hartle-Hawking-like vacuum state. 
The correlations now are between a particle in the right hand exterior region and an 
antiparticle in the same exterior with opposite $m'$ eigenvalue. 
$|0_G\rangle$~does not appear thermal to generic
static observers in~$R$, but it appears thermal in the standard
Hawking temperature ${(8\pi M)}^{-1}$ near the infinity when probed
by operators that only couple to one member of each correlated pair
in~(\ref{eqn:geonvac}), such as operators that only couple to a
definite sign of the angular momentum quantum number~$m'$. In
particular, number operator expectation values are thermal, as seen above, and the
isometry arguments of \cite{lm:geon} show that the experiences of any
static observer become asymptotically thermal in the large $|t|$
limit. 

The explicit appearance of $\epsilon$ in (\ref{eqn:geonvac}) shows
that the non-thermal correlations in $|0_G\rangle$ reveal the geon spin
structure to an observer in~$R$. This is a phenomenon that could not
have been anticipated just from the geometry of~$R$, which in its own
right has only one spin structure.


\chapter[Stress tensor]{Stress tensor on $M_0$, $M_-$ and Minkowski space with plane boundary}
\label{ch:emchapter}

\section{Introduction}

In this chapter the expectation values of the energy-momentum tensor in the Minkowski-like vacua
on $M_0$, $M_-$ and Minkowski space with an infinite plane boundary are presented.
In section \ref{sec:masscalarem} we calculate $\langle{0}|T_{\mu\nu}|0\rangle$
for the massive scalar field on all three spacetimes. 
The values on $M_0$ agree with the literature \cite{th:scalar} while to
the knowledge of the author those on $M_-$ are new. 
In the massless limit these values agree with those in \cite{lm:geon,bd:em,dw:dewitthi}. 
Various limits are discussed and qualitative differences
between the massive and massless values noted.
In section \ref{sec:masslessem} we calculate $\langle{0}|T_{\mu\nu}|0\rangle$
for left and right handed $2$-component Weyl spinors on the spacetimes $M_0$ and $M_-$. 
It is shown that the values are independent of handedness but are
dependent on the spin structure. Limits are discussed and the differences between
$\langle{0}|T_{\mu\nu}|0\rangle$ on the two spacetimes noted.
Finally in section \ref{sec:masspinorem} we calculate $\langle{0}|T_{\mu\nu}|0\rangle$
for massive $4$-component spinors on $M_0$ and $M_-$. The results on $M_0$ agree with those
in \cite{an:spinor}~\footnote{
Except for an extra factor of $2$ here as in \cite{an:spinor} Majorana
spinors are considered.}~\label{emfootnote}. 
The results on $M_-$ are new.
Again various limits are discussed. In the massless limit the results
agree with those in section \ref{sec:masslessem} (see also \cite{pl:langlois,bD:auto}).

\section{Massive scalar field}
\label{sec:masscalarem}

The stress tensor for  massive scalar fields in flat spacetimes with non-trivial
topology was considered by Tanaka and Hiscock in \cite{th:scalar}. 
In particular they consider $\langle{0}|T_{\mu\nu}|0\rangle$ on flat spacetimes 
with topology and product space geometry $\R^3\times{S^1}$ (which is denoted $M_0$ here), 
$\R^2\times{T^2}$ and $\R^1\times{T^3}$.
It is seen in \cite{th:scalar} that the magnitude of the energy density decreases with an increasing 
field mass.
Here we reproduce the result on $M_0$ and present the expectation values on $M_-$
and on Minkowski space with a single infinite plane boundary at $x=0$.

The energy-momentum tensor for the massive scalar field in a general $4$-dimensional
curved spacetime $\tilde{M}$ in our conventions is \cite{bd:book}
\begin{eqnarray}
T_{\mu\nu} & = &
(1-2\xi)\phi_{;\mu}\phi_{;\nu}
+
\left(2\xi-\frac{1}{2}\right)g_{\mu\nu}g^{\rho\sigma}\phi_{;\rho}\phi_{;\sigma}
-
2\xi\phi_{;\mu\nu}\phi
\nonumber
\\
&  & +
\frac{1}{2}\xi{g_{\mu\nu}}\phi\Box\phi
-\xi
\left[
R_{\mu\nu}
-\frac{1}{2}Rg_{\mu\nu}+\frac{3}{2}\xi{R}g_{\mu\nu}\right]
\phi^2
\nonumber
\\
& & + \frac{1}{2}\left[1-3\xi\right]m^2g_{\mu\nu}\phi^2
 \ ,
\end{eqnarray}
which, applying the field equation $[\Box+m^2+\xi{R}]\phi=0$, may be
written on a flat spacetime ($R=0$) as
\begin{eqnarray}
T_{\mu\nu} & = &
(1-2\xi)\phi_{;\mu}\phi_{;\nu}
+
\left(2\xi-\frac{1}{2}\right)g_{\mu\nu}g^{\rho\sigma}\phi_{;\rho}\phi_{;\sigma}
-
2\xi\phi_{;\mu\nu}\phi
\nonumber
\\
&  & +
2\xi{g_{\mu\nu}}\phi\Box\phi
+\frac{1}{2}m^2g_{\mu\nu}\phi^2
 \ ,
\end{eqnarray}
where $\xi$ gives the coupling to the gravitational field ($\xi=0$ for minimal
coupling and $\xi=1/6$ for conformal coupling). By the point splitting 
technique \cite{ch:pointsplit}, 
where we split the points in the quadratic expressions above and take the coincidence
limit at the end, the renormalized expectation value of $T_{\mu\nu}$ in  
vacuum state $|0\rangle$ may be written as 
\begin{eqnarray}
\langle{0}|T_{\mu\nu}|0\rangle & = &
\frac{1}{2}
\lim_{x'\rightarrow{x}}
\left[
(1-2\xi)\nabla_\mu\nabla_{\nu'}
+
\left(2\xi-\frac{1}{2}\right)g_{\mu\nu}\nabla_\alpha\nabla^{\alpha'}
-
2\xi\nabla_\mu\nabla_{\nu}
\right.
\nonumber
\\
\label{eqn:emtensor}
&  & +
\left.
2\xi{g_{\mu\nu}}\nabla_\alpha\nabla^{\alpha}
+\frac{1}{2}m^2g_{\mu\nu}
\right]
\left(G^{(1)}_{\tilde{M}}(x,x')-G^{(1)}(x,x')\right)
 \ ,
\end{eqnarray}
where $G^{(1)}_{\tilde{M}}(x,x')=\langle{0}|\{\phi(x),\phi(x')\}|0\rangle=G^+_{\tilde{M}}(x,x')+G^-_{\tilde{M}}(x,x')$ 
is the scalar Hadamard function in ${\tilde{M}}$ and $G^{(1)}(x,x')$ is the
Hadamard function in Minkowski space.

On Minkowski space $G^{(1)}(x,x')$ may be 
calculated by considering the expansion of the field in a complete set of mode solutions
to the Klein Gordon equation
\begin{equation}
\label{eqn:modeexpantion}
\phi(t,\mathbf{x})=
\frac{1}{(2\pi)^{3/2}}
\int
\frac{d^{3}k}{(2\omega)^{1/2}}
\left(
a(\mathbf{k})e^{-i(\omega{t}-\mathbf{k}\cdot\mathbf{x})}
+a^{\dagger}(\mathbf{k})e^{i(\omega{t}-\mathbf{k}\cdot\mathbf{x})}
\right) \ ,
\end{equation}
where $\omega=\left(\mathbf{k}^2+m^2\right)^{1/2}$.
Then
\begin{eqnarray}
\langle{0}|\{\phi(x),\phi(x')\}|0\rangle
& = &
\frac{1}{(2\pi)^3}
\int
\frac{d^{3}k}{2\omega}
e^{-i(\omega{(t-t')}-\mathbf{k}\cdot(\mathbf{x}-\mathbf{x'}))}
\nonumber
\\
&   & 
+
\frac{1}{(2\pi)^3}
\int
\frac{d^{3}k}{2\omega}
e^{i(\omega{(t-t')}-\mathbf{k}\cdot(\mathbf{x}-\mathbf{x'}))}
 \ ,
\end{eqnarray}
where we have made use of the anticommutation relations for the annihilation and
creation operators. The integrals may now be performed by transforming to
spherical coordinates in $k$-space. After doing the angular integral we find
\begin{eqnarray}
\langle{0}|\{\phi(x),\phi(x')\}|0\rangle
 & = &  
\frac{1}{(2\pi)^2R}
\int_0^\infty
dk\,
\frac{k}{\left(k^2+m^2\right)^{1/2}}
\sin(kR)
\left[e^{-i\left(k^2+m^2\right)^{1/2}{(t-t')}}
\right.
\nonumber
\\
&   & \left.+e^{i\left(k^2+m^2\right)^{1/2}{(t-t')}}\right] \ ,
\label{eqn:hadinter}
\end{eqnarray}
where $R=|\mathbf{x}-\mathbf{x'}|$. (\ref{eqn:hadinter}) may be written
as
\begin{eqnarray}
\langle{0}|\{\phi(x),\phi(x')\}|0\rangle
 & = &  
\frac{-1}{8\pi^2R}
\partial_R
\bigg[
\int_{-\infty}^\infty
dk\,
\frac{1}{\left(k^2+m^2\right)^{1/2}}
\left(e^{-i\left(\left(k^2+m^2\right)^{1/2}{(t-t')}-kR\right)}
\right.
\nonumber
\\
&   &
\left.
+e^{i\left(\left(k^2+m^2\right)^{1/2}{(t-t')}+kR\right)}\right)
\bigg]
 \ ,
\end{eqnarray}
and the above integrals may be performed by changing coordinates to
$k=m\sinh\theta$, $\omega=m\cosh\theta$ and making use of the well known
integral representations for the Bessel functions (see e.g \cite{gr:inttables}).
The result is 
(see appendix C of \cite{bj:relfields}) 
\begin{eqnarray}
G^{(1)}(x,x') & = & \frac{1}{4\pi{R}}\partial_RY_0\left(m\sqrt{(t-t')^2-R^2}\right)\Theta\left((t-t')^2-R^2\right)
\nonumber
\\
& & -\frac{1}{2\pi^2R}\partial_RK_0\left(m\sqrt{R^2-(t-t')^2}\right)\Theta\left(R^2-(t-t')^2\right) \ .
\end{eqnarray}
Now we note that $\partial_z{F_0(z)}=-F_1(z)$ for
$F=K, Y$ and so~\footnote{Note 
the sign difference in the second term from that in Fullings book \cite{fu:Fulling} p85
which is the reference used by Tanaka and Hiscock \cite{th:scalar}. Further 
note the typographical error in \cite{th:scalar} where the $Y$ Bessel has
been replaced by the $I$ Bessel. This does not affect the results in \cite{th:scalar}.}
\begin{eqnarray}
G^{(1)}(x,x') 
& = & 
\frac{m}{4\pi{\left((t-t')^2-R^2\right)^{1/2}}}Y_1\left(m\sqrt{(t-t')^2-R^2}\right)\Theta\left((t-t')^2-R^2\right)
\nonumber
\\
&   & 
+\frac{m}{2\pi^2\left(R^2-(t-t')^2\right)^{1/2}}K_1\left(m\sqrt{R^2\!-\!(t-t')^2}\right)\Theta\left(R^2\!-\!(t-t')^2\right) \ ,
\nonumber
\\
\label{eqn:minkhad}
\end{eqnarray}
where $Y,K$ are Bessel functions. 

The Hadamard function on the quotient spaces $M_0$ and $M_-$ may be found by the method
of images \cite{bD:auto,bd:em},\footnote{It can be shown that (\ref{eqn:m0had}) and (\ref{eqn:m-had})
may be obtained by expanding the fields in modes, obtained from
an image sum, proving that they are true Green's functions, that
is, they are expectation values of the operators in a quantum state.}
\begin{eqnarray}
\label{eqn:m0had}
G^{(1)}_{M_0}(x,x') & = & \langle{0_0}|\{\phi(x),\phi(x')\}|0_0\rangle=\sum_{n\in{\Z}}\eta^nG^{(1)}\left(x,J_0^nx'\right) \ ,
\\
\label{eqn:m-had}
G^{(1)}_{M_-}(x,x') & = & \langle{0_-}|\{\phi(x),\phi(x')\}|0_-\rangle=\sum_{n\in{\Z}}\rho^nG^{(1)}\left(x,J_-^nx'\right) \ ,
\end{eqnarray}
where $J_0:(t,x,y,z)\mapsto(t,x,y,z+2a)$, $J_-:(t,x,y,z)\mapsto(t,-x,-y,z+a)$ and
$\eta=1, (-1)$, $\rho=1, (-1)$ label untwisted and (twisted) fields respectively.
The calculation of $\langle{0}|T_{\mu\nu}|0\rangle$ is now reduced to that
of finding derivatives of these Hadamard functions and applying (\ref{eqn:emtensor}) to (\ref{eqn:m0had})
and (\ref{eqn:m-had}). Next we present the results.

\subsection{$M_0$}
\label{sec-scalaremm0}

For $M_0$, in the coincidence limit $G^{(1)}_{M_0}(x,x')$ becomes
a function only of $(2na)^2$, which is positive. This means that only the
$K$ Bessel term in (\ref{eqn:minkhad}) contributes. Further we note
that $\lim_{x'\rightarrow{x}}\partial_{x^\mu}G^{(1)}_{M_0}(x,x')
=-\lim_{x'\rightarrow{x}}\partial_{{x'}^\mu}G^{(1)}_{M_0}(x,x')$.
These observations simplify the calculations somewhat. The result is
\begin{eqnarray}
\langle{0_0}|T_{tt}|0_0\rangle
=-\langle{0_0}|T_{xx}|0_0\rangle
=-\langle{0_0}|T_{yy}|0_0\rangle
=
-\sum_{n=1}^\infty
\eta^n
\frac{m^2}{2\pi^2(2na)^2}
K_2(2mna)
 \ ,
\nonumber
\\
\label{eqn:scalarM0}
\langle{0_0}|T_{zz}|0_0\rangle
=
\sum_{n=1}^\infty
\eta^n
\left[
\frac{m^2}{2\pi^2(2na)^2}
K_2(2mna)
-
\frac{m^3}{2\pi^2(2na)}
K_3(2mna)
\right] \ ,
\end{eqnarray}
with all other terms vanishing. The results agree with \cite{th:scalar}.
The massless limit may be easily checked by noting that near $z=0$
\begin{equation}
K_\nu(z)
=
\frac{\Gamma(\nu)}
{2}
\left(\frac{z}{2}\right)^{-\nu}
 \ ,
\end{equation}
when the real part of $\nu$ is positive \cite{gr:inttables}. These agree with
\cite{lm:geon,dw:dewitthi} and also with
\cite{bd:em}, the extra factor of $4$ in \cite{bd:em} arising because the
authors consider a multiplet of $2$ complex massless scalar fields.
The leading corrections for small mass are of the order $O\left(m^2\right)$. 
$\langle{0_0}|T_{\mu\nu}|0_0\rangle\rightarrow{0}$ exponentially in the
large mass and large $a$ limits, in contrast to the massless case 
where $\langle{0_0}|T_{\mu\nu}|0_0\rangle$ vanishes as $O\left(a^{-4}\right)$.

\subsection{$M_-$}
\label{eqn:m-emmassivescal}

For $M_-$ in the coincidence limit $G^{(1)}_{M_-}(x,x')$ becomes
a function only of $\left(x-(-1)^nx\right)^2+\left(y-(-1)^ny\right)^2+(na)^2$, which is positive. Again
only the $K$ Bessel term in (\ref{eqn:minkhad}) contributes.
Further we note the following
\begin{eqnarray}
\lim_{x'\rightarrow{x}}\partial_{\mu}G^{(1)}_{M_-}(x,x')
& = & -\lim_{x'\rightarrow{x}}\partial_{{\mu'}}G^{(1)}_{M_-}(x,x')
\;\; \textrm{for}\; \mu\in\{t,z\} \ ,
\nonumber
\\
\lim_{x'\rightarrow{x}}\partial_{\mu}G^{(1)}_{M_-}(x,x')
& = & -(+)\lim_{x'\rightarrow{x}}\partial_{{\mu'}}G^{(1)}_{M_-}(x,x')
\;\;
\textrm{when $n$ even (odd) and}\; \mu\in\{x,y\} \ .
\nonumber
\\
\label{eqn:m-relation}
&   & 
\end{eqnarray}
As $G^{(1)}_{M_-}(x,x')=\sum_{n\in{\Z}}\rho^nG^{(1)}\left(x,J_-^nx'\right)$, where 
$\rho=+1, (-1)$ labels untwisted (twisted) fields,
the result here may be split into two parts where the part coming from the
even terms in the sum above leads to the same expectation values
as on $M_0$ for untwisted fields. Therefore we write
$\langle{0_-}|T_{\mu\nu}|0_-\rangle=\langle{0_0}|T_{\mu\nu}|0_0\rangle_{(\eta=1)}+\rho
\langle{0_-}|T_{\mu\nu}|0_-\rangle_{\mathrm{odd}}$ where we find
\begin{eqnarray}
\langle{0_-}|T_{tt}|0_-\rangle_{\mathrm{odd}}
& = & 
\sum_{n\in{\Z}}
\left[
-(4\xi-1)\frac{m^3}{4\pi^2\sigma_n}
K_3(m\sigma_n)
\left[
1-\frac{(2na+a)^2}{\sigma_n^2}
\right]
\right.
\nonumber
\\
&   &\;\;\;\;\;\;\;+\left.
\left(
2\xi-\frac{3}{4}
\right)
\frac{m^2}{\pi^2\sigma_n^2}
K_2(m\sigma_n)\right] \ ,
\nonumber
\\
\langle{0_-}|T_{xx}|0_-\rangle_{\mathrm{odd}}
& = & 
\sum_{n\in{\Z}}
\left[
(4\xi-1)\frac{m^3y^2}{\pi^2\sigma_n^3}
K_3(m\sigma_n)
-
\left(
2\xi-\frac{1}{2}
\right)
\frac{m^2}{2\pi^2\sigma_n^2}
K_2(m\sigma_n)\right] \ ,
\nonumber
\\
\langle{0_-}|T_{yy}|0_-\rangle_{\mathrm{odd}}
& = & 
\sum_{n\in{\Z}}
\left[
(4\xi-1)\frac{m^3x^2}{\pi^2\sigma_n^3}
K_3(m\sigma_n)
-
\left(
2\xi-\frac{1}{2}
\right)
\frac{m^2}{2\pi^2\sigma_n^2}
K_2(m\sigma_n)\right] \ ,
\nonumber
\\
\langle{0_-}|T_{zz}|0_-\rangle_{\mathrm{odd}}
& = & 
\sum_{n\in{\Z}}
\left[
\frac{m^3}{4\pi^2\sigma_n}
K_3(m\sigma_n)
\left[
(4\xi-1)-\frac{4\xi(2na+a)^2}{\sigma_n^2}
\right]
\right.
\nonumber
\\
&   &\;\;\;\;\;\;\;-\left.
\left(
2\xi-\frac{3}{4}
\right)
\frac{m^2}{\pi^2\sigma_n^2}
K_2(m\sigma_n)\right] \ ,
\nonumber
\\
\langle{0_-}|T_{xy}|0_-\rangle_{\mathrm{odd}}
& = & 
\sum_{n\in{\Z}}
(1-4\xi)\frac{m^3xy}{\pi^2\sigma_n^3}
K_3(m\sigma_n) \ ,
\end{eqnarray}
with $\sigma_n=\left((2x)^2+(2y)^2+(2na+a)^2\right)^{1/2}$, and the
sum is over all $n$ including $n=0$. Other components vanish. Again
it is a simple matter to check the massless limit here.
The results agree with those of \cite{lm:geon} and \cite{bd:em} in this limit.
The leading correction for small mass is $O\left(m^2\right)$, and $\langle{0_-}|T_{\mu\nu}|0_-\rangle$
vanishes exponentially in the large mass and, for non zero mass, the
large $a$ limits. The difference between $\langle{0_-}|T_{\mu\nu}|0_-\rangle$ 
on $M_-$ and $\langle{0_0}|T_{\mu\nu}|0_0\rangle$ on $M_0$ (with $\eta=1$)
vanishes exponentially as $r^2:=x^2+y^2\rightarrow\infty$.
This behaviour is qualitatively different to the
massless case where the difference vanishes as $O\left(r^{-3}\right)$
\cite{lm:geon}.
The result in the massive case is to our knowledge new.

\subsection{Minkowski space with infinite plane boundary}
\label{sec:emonmwithbound}

In this subsection we consider $\langle{0_B}|T_{\mu\nu}|0_B\rangle$ 
in $4$-dimensional Minkowski space with an infinite plane boundary
at $x=0$. While the stress-energy for a massless field is well known 
(see e.g. \cite{bd:book,d:dewitt}), the massive results are to 
our knowledge new. 
There is some similarity with
$M_-$ as both spaces may be considered as quotients of Minkowski space
with the quotient group including a reflection in $x$ about $x=0$.

Again here the scalar Hadamard function is given by the method of 
images,
\begin{equation}
G^{(1)}_B(x,x')=\langle{0_B}|\{\phi(x)\phi(x')\}|0_B\rangle=G^{(1)}(x,x')+\eta{G^{(1)}\left(x,J_Bx'\right)}
 \ ,
\end{equation}
where $J_B:(t,x,y,z)\mapsto(t,-x,y,z)$ and $\eta=+1, (-1)$ labels
Neumann (Dirichlet) boundary conditions on the plate. 
From (\ref{eqn:emtensor}) the result is
\begin{eqnarray}
\langle{0_B}|T_{tt}|0_B\rangle & = & -\langle{0_B}|T_{yy}|0_B\rangle
=-\langle{0_B}|T_{zz}|0_B\rangle
\nonumber
\\
& = & \eta\left[
\frac{(1-4\xi)m^3}{4\pi^2|2x|}K_3(m|2x|)
+
\frac{(2\xi-1)m^2}{8\pi^2x^2}K_2(m|2x|)
\right] \ ,
\nonumber
\\
\langle{0_B}|T_{xx}|0_B\rangle & = & 0 \ .
\end{eqnarray}
In the massless limit these agree with the literature \cite{bd:book}.
For the massive field $\langle{0_B}|T_{\mu\nu}|0_B\rangle$ is non-zero
and has nonvanishing trace for both conformal and minimal coupling. The mass
breaks the conformal invariance of the field.
$\langle{0_B}|T_{\mu\nu}|0_B\rangle$ vanishes exponentially as the mass and as $x$ 
go to infinity,
in contrast to the massless case which behaves as $O\left(x^{-4}\right)$ for
the minimally coupled field and is identically $0$ for conformal coupling.
It is interesting to note that for conformal coupling the leading order
correction for small mass is $O\left(m^2\right)$, while for minimal coupling it is 
$O\left(m^4\right)$. This is a qualitative difference from both $M_0$ and $M_-$.

\section{Massless spinor field}
\label{sec:masslessem}

In this section we present the expectation value of the the energy-momentum tensor 
for free left and right handed 2-component massless Dirac fields in the Minkowski-like 
vacua $|0_0\rangle$ and $|0_-\rangle$ on $M_0$ and $M_-$ respectively. The corresponding
values for the massless scalar field were given in \cite{lm:geon} (and can be obtained
from the results in section \ref{sec:masscalarem} in the massless limit). We find that 
the values depend on the spin structure but are independent of the handedness of the spinor. 
We begin by describing the reduction of the 4-component massive spinor formalism of 
chapter \ref{ch:unruhbog} to 2-component massless spinors.

Given the formalism set out in chapter \ref{ch:unruhbog} we set $m=0$ and introduce 
a Weyl representation for the $\gamma$ matrices
\begin{equation}
\begin{array}{cc}
\gamma^0=\left(\begin{array}{cc} 
0 & -I 
\\ 
-I & 0 
\end{array}\right)
 \ , 
& 
\gamma^i=\left(\begin{array}{cc} 
0 & \sigma_i 
\\ 
-\sigma_i & 0 \end{array}\right) 
\end{array}
 \ ,
\end{equation} 
where the $\sigma_i$ are the usual Pauli matrices. 
With this representation the 4-component spinor $\psi$ decouples into two 2-component 
spinors $\psi_L$ and $\psi_R$ labelled as left and right handed respectively.
$\psi=\left(\begin{array}{c} \psi_R \\ \psi_L \end{array}\right)$. 
The 4-component Lagrangian then reduces to
\begin{equation}
\mathcal{L}_L=\mathrm{det} 
V\frac{1}{2}iV^{\mu}_{a}\left[{\psi}^\dagger_L\sigma^{a}_L\nabla_{\mu}\psi_L
-\left(\nabla_{\mu}{\psi}^\dagger_L\right)\sigma^{a}_L\psi_L\right]
 \ ,
\end{equation}
and
\begin{equation}
\mathcal{L}_R=\mathrm{det} 
V\frac{1}{2}iV^{\mu}_{a}\left[{\psi}^\dagger_R\sigma^{a}_R\nabla_{\mu}\psi_R
-\left(\nabla_{\mu}{\psi}^\dagger_R\right)\sigma^{a}_R\psi_R\right]
 \ ,
\end{equation}
where
\begin{eqnarray}
\sigma^{a}_L & = & (I,-\sigma_x,-\sigma_y,-\sigma_z) \ ,
\nonumber
\\
\label{eqn:sigma}
\sigma^{a}_R & = & (I,\sigma_x,\sigma_y,\sigma_z) \ .
\end{eqnarray}
Further the energy-momentum tensor for the $4$-component field
\cite{bd:book},
\begin{equation}
\label{em:spinorfield}
T_{\mu\nu}
=
\frac{i}{2}
\left[
\bar{\psi}\gamma_{(\mu}\nabla_{\nu)}\psi-(\nabla_{(\mu}\bar{\psi})\gamma_{\nu)}\psi
\right] \ ,
\end{equation}
where $A_{(\mu}B_{\nu)}=1/2(A_\mu{B_\nu}+A_\nu{B_\mu})$, becomes the sum
of left and right handed parts,
\begin{eqnarray}
\label{eqn:spinorem1}
T^L_{\mu\nu}
& = & 
\frac{i}{2}
\left[\psi^\dagger_L\sigma^L_{(\mu}\nabla_{\nu)}\psi_L-\left(\nabla_{(\mu}\psi_L^\dagger\right)\sigma^L_{\nu)}\psi_L
\right] \ ,
\\
T^R_{\mu\nu}
& = & 
\frac{i}{2}
\left[\psi^\dagger_R\sigma^R_{(\mu}\nabla_{\nu)}\psi_R-\left(\nabla_{(\mu}\psi_R^\dagger\right)\sigma^R_{\nu)}\psi_R
\right] \ ,
\label{eqn:spinorem23}
\end{eqnarray}
where $\sigma^L_{\mu}=V_\mu^\alpha\sigma^L_\alpha$ and $\sigma^L_\alpha=\eta_{\alpha\beta}\sigma_L^\beta$.
Writing (\ref{eqn:spinorem1}) and (\ref{eqn:spinorem23}) in the form
\begin{equation}
\label{eqn:spinortmunu1}
T_{\mu\nu}
=
\frac{i}{4}
\mathrm{Tr}
\left(
\sigma_{(\mu}\left[\nabla_{\nu)}\psi,\psi^\dagger\right]-\sigma_{(\mu}\left[\psi,\nabla_{\nu)}\psi^\dagger\right]
\right) \ ,
\end{equation}
where the $L,R$ superscripts and subscripts have been dropped, and further point
splitting \cite{ch:pointsplit} the quadratic terms in (\ref{eqn:spinortmunu1}), the 
renormalized expectation value 
of $T_{\mu\nu}$, on the flat spacetimes of interest here,
in the vacuum state $|0\rangle$ may now be expressed in terms of the spinor 
Hadamard function 
$S^{(1)}_{\alpha\beta{\tilde{M}}}(x,x')=\langle{0}|\left[\psi_\alpha(x),{\psi^\dagger}_\beta(x')\right]|0\rangle$
on $\tilde{M}$ which in Minkowski space is a solution to $i\sigma^\alpha\partial_\alpha{S^{(1)}(x,x')}=0$,
\begin{equation}
\label{eqn:spinorem2}
\langle{0}|T_{\mu\nu}|0\rangle
= 
\frac{i}{4}
\lim_{x'\rightarrow{x}}
\mathrm{Tr}
\,
\sigma_{(\mu}
\left(\nabla_{\nu)}
-\nabla_{\nu')}\right)\left(S^{(1)}_{\tilde{M}}(x,x')-S^{(1)}(x,x')\right) \ .
\end{equation}
We may further express the spinor Hadamard function
in terms of the scalar Hadamard function
\begin{eqnarray}
S^{(1)}_L(x,x')=-iL^\alpha\nabla_\alpha{G^{(1)}(x,x')} \ ,
\\
S^{(1)}_R(x,x')=-iR^\alpha\nabla_\alpha{G^{(1)}(x,x')} \ ,
\end{eqnarray}
where $L^\alpha=(\mathbb{I},\sigma_x,\sigma_y,\sigma_z)$
and $R^\alpha=(\mathbb{I},-\sigma_x,-\sigma_y,-\sigma_z)$,
and so 
\begin{eqnarray}
\langle{0}|T^L_{\mu\nu}|0\rangle
& = & 
\frac{1}{8}
\lim_{x'\rightarrow{x}}
\mathrm{Tr}
\left[\left(\sigma^L_{\mu}\nabla_{\nu}+\sigma^L_{\nu}\nabla_{\mu}\right)
-
\left(\sigma^L_{\mu}\nabla_{\nu'}+\sigma^L_{\nu}\nabla_{\mu'}\right)\right]
\nonumber
\\
&  & \times
L^\rho\nabla_\rho
\left(G^{(1)}_{\tilde{M}}(x,x')-G^{(1)}(x,x')\right) \ ,
\nonumber
\\
\label{eqn:emL}
\\
\langle{0}|T^R_{\mu\nu}|0\rangle
& = & 
\frac{1}{8}
\lim_{x'\rightarrow{x}}
\mathrm{Tr}
\left[\left(\sigma^R_{\mu}\nabla_{\nu}+\sigma^R_{\nu}\nabla_{\mu}\right)
-
\left(\sigma^R_{\mu}\nabla_{\nu'}+\sigma^R_{\nu}\nabla_{\mu'}\right)\right]
\nonumber
\\
&  & \times
R^\rho\nabla_\rho
\left(G^{(1)}_{\tilde{M}}(x,x')-G^{(1)}(x,x')\right) \ .
\nonumber
\\
\label{eqn:emR}
\end{eqnarray}
$G^{(1)}(x,x')$ may be obtained as the massless limit of
(\ref{eqn:minkhad}), 
\begin{equation}
\label{eqn:minkhadmasl}
G^{(1)}(x,x')=-\frac{1}{\left[2\pi^2\left((t-t')^2-(x-x')^2-(y-y')^2-(z-z')^2\right)\right]} \ ,
\end{equation}
which satisfies $\partial^\alpha\partial_\alpha{G^{(1)}(x,x')}=0$.
We find therefore that the 2-component spinor Hadamard function in Minkowski space is
\begin{eqnarray}S^{(1)}(x,x') 
&    =   & -\frac{i}{{\pi}^2[(t-t')^2-(r-r')^2]^2} 
\nonumber 
\\
& \times & \left[\begin{array}{cc} (t-t')-\lambda(z-z') & \lambda(-(x-x')+i(y-y')) 
\\ 
\lambda(-(x-x')-i(y-y')) & (t-t')+\lambda(z-z') \end{array} \right] \ ,
\end{eqnarray}
where we have introduced the parameter $\lambda=1 (-1)$ for left-handed (right-handed) 
spinors respectively.
The Hadamard functions on $M_0$ and $M_-$ are given by the method of images,
but care must be taken with the local frames with respect to which the
spinors are expressed.
We shall work throughout in a vierbein aligned along the usual Minkowski 
coordinate axes. This has the advantage of making covariant and 
partial derivatives coincide. As this vierbein is invariant under $J_0$
the calculation on $M_0$ is reduced to a straightforward calculation of 
derivatives of the Hadamard function. The Minkowski vierbein however is not invariant 
under $J_-$ and more care must be taken on $M_-$.

\subsection{$M_0$}

On $M_0$, the vierbein aligned along Minkowski coordinate axes is invariant 
under $J_0$ and the Hadamard function is simply given by the method of images.
We therefore have
\begin{equation}
\label{eqn:m0had2}
S^{(1)}_{M_0}(x,x')
=\sum_{n\in\Z}\eta^nS^{(1)}\left(x,J^n_0(x')\right) \ ,
\end{equation}
where again $\eta=1, (-1)$ for spinors with periodic (antiperiodic) boundary 
conditions with respect to this vierbein. In this expression $G^{(1)}(x,x')$ is the 
Minkowski Hadamard function (\ref{eqn:minkhadmasl}), and $(x,x')$ on the left of the equality labels points in 
$M_0$ while on the right it labels points in Minkowski space. To find 
$\langle{0_0}|T_{\mu\nu}|0_0\rangle$ we apply (\ref{eqn:emL}) and (\ref{eqn:emR}) to
(\ref{eqn:m0had2}). The result is
\begin{equation}
\langle{0_0}|T^L_{tt}|0_0\rangle=-\langle{0_0}|T^L_{xx}|0_0\rangle=-\langle{0_0}|T^L_{yy}|0_0\rangle
=\frac{1}{3}\langle{0_0}|T^L_{zz}|0_0\rangle=\frac{1}{8{\pi}^2}\sum_{n=1}^\infty\frac{\eta^{{n}}}{{n}^4a^4}
 \ ,
\label{eqn:masslessemm0}
\end{equation}
with all other components vanishing.
The values for $\langle{0_0}|T^R_{\mu\nu}|0_0\rangle$ are the same.
The remaining
summation 
may be calculated explicitly using the result \cite{zw:zwillinger}
\begin{equation} 
\sum_{l'=1}^{\infty}\frac{\eta^{l'}}{l'^4}
=\left(-\frac{7}{8}\right)^{\frac{1-\eta}{2}}\frac{\pi^4}{90} \ ,
\end{equation}
and so
\begin{equation}
\langle{0_0}|T^L_{tt}|0_0\rangle=-\langle{0_0}|T^L_{xx}|0_0\rangle=-\langle{0_0}|T^L_{yy}|0_0\rangle
=\frac{1}{3}\langle{0_0}|T^L_{zz}|0_0\rangle=\frac{\pi^2}{720a^4}\left(-\frac{7}{8}\right)^{\frac{1-\eta}{2}}
 \ .
\end{equation}
Expectation values are non-zero in contrast to Minkowski space, and are
$-2$ times those for the massless scalar field (see e.g \cite{lm:geon}
or the massless limits of the results in section \ref{sec-scalaremm0}).
The stress tensor is diagonal and invariant under all isometries and it 
is independent of the handedness of the spinor but dependent on the
spin structure. The $\eta=-1$ spin structure is energetically 
preferred.

\subsection{$M_-$}
\label{sec-masslessspinm-}

On $M_-$, the standard Minkowski vierbein is not invariant under
$J_-$. Care must therefore be taken in applying the method of images
to expressions written with respect to this vierbein. 
In an invariant vierbein the spinor Hadamard function on $M_-$ would be 
given directly by the method of images, that is
\begin{equation}
S^{(1)}_{M_-}(x,x')
=
\sum_{n\in{\Z}}
\rho^n
S^{(1)}\left(x,J_-^nx'\right) \ ,
\end{equation}
with $\rho=+1, (-1)$,
however as we choose to work in the Minkowski vierbein the image
expression is different. Consider vierbein (\ref{eqn:rotet}) which is invariant under $J_-$. 
(\ref{eqn:rotet}) rotates by $\pi$ 
in the $(x,y)$-plane as $z\rightarrow{z+a}$. The transformation from this vierbein to
the standard Minkowski one is clearly the corresponding rotation by $-\pi$. The
associated transformation on the $2$-component spinor Hadamard function is
\begin{equation}
S^{(1)}_{SM}(x,x')
=
e^{-\frac{\pi\sigma_x\sigma_y}{2a}z}
S^{(1)}_{RM}(x,x')
e^{\frac{\pi\sigma_x\sigma_y}{2a}z'} \ ,
\end{equation}
where the $R (S)$ subscript denotes the rotating (standard) vierbein respectively.
Therefore on $M_-$ in the standard vierbein
\begin{equation}
S^{(1)}_{SM_-}(x,x')
=
\sum_{n\in{\Z}}
\rho^n
S^{(1)}_{SM_-}\left(x,J_-^nx'\right)
e^{-\frac{n\pi\sigma_x\sigma_y}{2}} \ .
\end{equation}
In terms of the scalar Hadamard function this translates
to using the following expression in (\ref{eqn:emL}) and (\ref{eqn:emR})
\begin{equation}
\label{eqn:hadexprmassless}
G^{(1)}_{M_-}(x,x')
=
\sum_{n\in{\Z}}
\rho^n
G^{(1)}\left(x,J_-^nx'\right)
e^{-\frac{n\pi\sigma_x\sigma_y}{2}} \ .
\end{equation}
Here $\rho=+1, (-1)$ labels periodic (antiperiodic)
spinors with respect to the vierbein which rotates by $\pi$ as $z\rightarrow{z+a}$.
$\rho$ thus labels the two inequivalent spin structures on $M_-$ \cite{pl:langlois}.
Applying (\ref{eqn:emL}) and (\ref{eqn:emR}) to
(\ref{eqn:hadexprmassless}), the result is
$\langle{0_-}|{T^L_{\mu\nu}}|0_-\rangle=\langle{0_0}|{T^L_{\mu\nu}}|0_0\rangle_{(\eta=-1)}
+\rho\langle{0_-}|{T^L_{\mu\nu}}|0_-\rangle_{\mathrm{odd}}$, where
\begin{eqnarray}
\langle{0_-}|{T^L_{xz}}|0_-\rangle_{\mathrm{odd}}=-\frac{4}{{\pi}^2}\sum_{n\in\Z}\frac{(-1)^na(2n+1)y}{\{(2x)^2+(2y)^2+(2n+1)^2a^2\}^3}
 \ ,
\nonumber
\\
\label{eqn:tmunu}\langle{0_-}|{T^L_{yz}}|0_-\rangle_{\mathrm{odd}}=\frac{4}{{\pi}^2}\sum_{n\in{\Z}}\frac{(-1)^na(2n+1)x}{\{(2x)^2+(2y)^2+(2n+1)^2a^2\}^3}
 \ ,
\end{eqnarray}
with all other components vanishing. The values for right handed spinors
are the same. 

The stress tensor is invariant under all continuous isometries of $M_-$,
as it must be by construction, but now there are non-zero shear components,
$\langle{0_-}|{T_{xz}}|0_-\rangle$ and $\langle{0_-}|{T_{yz}}|0_-\rangle$, 
which change sign under isometries that reverse the spatial 
orientation. 
These components do not depend on the handedness but they have the
opposite signs in the two spin structures.
The spatial orientation determined by the spin 
structure can thus be detected from the shear part of the stress tensor.

As on $M_0$, the summations may be evaluated in terms of elementary functions. 
The diagonal terms are given by the values on $M_0$ in the
$\eta=-1$ spin structure there,
while the non-diagonal terms may be evaluated using the calculus of residues. 
The results are best expressed in terms of the orthonormal frame 
$\{dt,dr,\omega^{\hat{\phi}},dz\}$ defined by
\begin{eqnarray}
x & = & r\cos{\phi} \ ,
\nonumber
\\
y & = & r\sin{\phi} \ ,
\end{eqnarray}
with $\omega^{\hat{\phi}}=rd\phi$. 
We find
\begin{eqnarray}
\langle0_-|T_{tt}|0_-\rangle
=-\langle0_-|T_{rr}|0_-\rangle
& = & -\langle0_-|T_{\hat{\phi}\hat{\phi}}|0_-\rangle
=\frac{1}{3}\langle0_-|T_{zz}|0_-\rangle
=-\left(\frac{7}{8}\right)\frac{\pi^2}{720a^4}
 \ ,
\nonumber
\\
\langle0_-|T_{\hat{\phi}z}|0_-\rangle
& = & -\frac{\rho\pi^2}{64a^4}\frac{d}{dq}\left(\frac{\sinh{q}}{q\cosh^2{r}}\right) \ ,
\end{eqnarray}
where $q=\frac{\pi}{a}r$. For $q>0$ we find 
$\frac{d}{dq}\left(\frac{\sinh{q}}{q\cosh^2{r}}\right)<0$ and so 
$\langle0_-|T_{\hat{\phi}z}|0_-\rangle$ picks up the sign of $\rho$.
The shear part in either spin structure vanishes at
$r=0$ and tends to $0$ exponentially as $r\rightarrow\infty$.
$\langle0_-|T_{\mu\nu}|0_-\rangle$ thus
reduces to $\langle0_0|T_{\mu\nu}|0_0\rangle$ in the $\eta=-1$ spin structure 
at large $r$. Numerical evidence however suggests that there is
a range for $r$ where the shear part is in fact the dominant part of the
stress-energy. 

We summarise the results in table~\ref{table:T-components}.
\begin{table}[t!]
\begin{center}
\begin{tabular}{|c|c|c|c|}
\hline
& 
$M_0$, $\eta=1$ 
& 
$M_0$, $\eta=-1$ 
& 
$M_-$ 
\\
\hline
$\langle{T_{00}}\rangle$ 
& 
$\displaystyle \frac{\pi^2}{720 a^4}$ 
&
$\displaystyle -\frac{7 \pi^2}{5760 a^4}$ 
&
$\displaystyle {-\frac{7 \pi^2}{5760 a^4}}^{\vphantom{A^A}}_{\vphantom{A_A}}$ 
\\
$\langle{T_{rr}}\rangle=\langle{T_{\hat{\phi}\hat{\phi}}}\rangle$ 
& 
$\displaystyle - \frac{\pi^2}{720 a^4}$ 
& 
$\displaystyle \frac{7 \pi^2}{5760 a^4}$ 
&
$\displaystyle {\frac{7 \pi^2}{5760 a^4}}^{\vphantom{A^A}}_{\vphantom{A_A}}$ 
\\
$\langle{T_{zz}}\rangle$ 
& 
$\displaystyle \frac{\pi^2}{240 a^4}$ 
&
$\displaystyle -\frac{7 \pi^2}{1920 a^4}$ 
&
$\displaystyle {-\frac{7 \pi^2}{1920
a^4}}^{\vphantom{A^A}}_{\vphantom{A_A}}$ 
\\
$\langle{T_{\hat{\phi}z}}\rangle$ 
& 
0 
& 
0 
&
$\displaystyle {-\frac{\rho\pi^2}{64a^4}\frac{d}{dq}
\!
\left(\frac{\sinh{q}}{q\cosh^2 \! {q}}\right)
}^{\vphantom{A^A}}_{\vphantom{A_{\displaystyle A}}}$
\\
\hline
\end{tabular}
\end{center}
\caption[$\langle{T_{\mu\nu}}\rangle$ for a left-handed two-component spinor in
the Minkowski-like vacua on $M_0$ and $M_-$ in the orthonormal frame
$\{dt,dr,\omega^{\hat{\phi}},dz\}$.]{The nonvanishing components of 
$\langle{T_{\mu\nu}}\rangle$ for a left-handed two-component spinor in
the Minkowski-like vacua on $M_0$ and $M_-$ in the orthonormal frame
$\{dt,dr,\omega^{\hat{\phi}},dz\}$. The values for a right-handed spinor are
identical.}
\label{table:T-components}
\end{table}
Note the difference in the results between $M_0$ and $M_-$. Thus we see that 
quantum fields can probe the topology. Any observer who is sensitive to the 
energy-momentum tensor can tell the difference between the two vacuum states.

\section{Massive spinor field}
\label{sec:masspinorem}

In this section we repeat the calculations above for a massive
Dirac field. The result on $M_0$ was given in a recent paper \cite{an:spinor}
where it is shown that the magnitude of energy density 
decreases with increasing field mass, as for the scalar field.
We present the results for $M_0$ and $M_-$.

The energy-momentum tensor for the massive Dirac field is given in (\ref{em:spinorfield}).
By a calculation analogous to that leading from (\ref{eqn:spinorem1}) to
(\ref{eqn:spinorem2}) the renormalized expectation value of the energy-momentum tensor 
in the state $|0\rangle$ may be written
\begin{equation}
\langle{0}|T_{\mu\nu}|0\rangle
= 
\frac{i}{4}
\lim_{x'\rightarrow{x}}
\mathrm{Tr}
\,
\gamma_{(\mu}
\left(\nabla_{\nu)}
-\nabla_{\nu')}\right)\left(S^{(1)}_{\tilde{M}}(x,x')-S^{(1)}(x,x')\right) \ ,
\end{equation}
where $S^{(1)}_{\alpha\beta}(x,x')=\langle{0}|\left[\psi_\alpha(x),\bar{\psi}_\beta(x')\right]|0\rangle$
is the massive Dirac field Hadamard function. $S^{(1)}(x,x')$ satisfies the 
homogeneous Dirac equation and may be expressed in terms of the
scalar Hadamard function by $S^{(1)}(x,x')=-(i\gamma^\rho\nabla_\rho+m)G^{(1)}(x,x')$.
Hence
\begin{eqnarray}
\langle{0}|T_{\mu\nu}|0\rangle
& = &  
\frac{1}{8}
\lim_{x'\rightarrow{x}}
\mathrm{Tr}
[(\gamma_{\mu}\nabla_{\nu}+\gamma_{\nu}\nabla_{\mu})
-
(\gamma_{\mu}\nabla_{\nu'}+\gamma_{\nu}\nabla_{\mu'})]
\nonumber
\\
&   & 
\times
\gamma^\rho\nabla_\rho
\left(G^{(1)}_{\tilde{M}}(x,x')-G^{(1)}(x,x')\right) \ .
\label{eqn:spinexpect}
\end{eqnarray}

Again the Hadamard function on $M_0$ and $M_-$ may be found 
by the method of images, with the same care taken with the local Lorentz
frames as discussed for the massless field in the
previous section. Again we shall work throughout 
with a vierbein aligned along the usual Minkowski coordinate axes. 

\subsection{$M_0$}

Recall that on $M_0$
\begin{equation}
G^{(1)}_{M_0}(x,x')
=
\sum_{n\in{\Z}}
\eta^n
G^{(1)}\left(x,J_0^nx'\right) \ .
\end{equation}
Here again we note that in the coincidence limit $G^{(1)}_{M_0}(x,x')$ becomes
a function only of $(2na)^2$, which is positive, and so only the
$K$ Bessel term in (\ref{eqn:minkhad}) contributes. Note also that
$\lim_{x'\rightarrow{x}}\partial_{x^\mu}G^{(1)}_{M_0}(x,x')
=-\lim_{x'\rightarrow{x}}\partial_{{x'}^\mu}G^{(1)}_{M_0}(x,x')$.
With these observations (\ref{eqn:spinexpect}) reduces to
\begin{equation}
\langle{0_0}|T_{\mu\nu}|0_0\rangle
=
\lim_{x'\rightarrow{x}}
\nabla_\mu\nabla_\nu
\left(G^{(1)}_{M_0}(x,x')-G^{(1)}(x,x')\right) \ ,
\end{equation}
and we find for the non-zero expectation values
\begin{eqnarray}
\langle{0_0}|T_{tt}|0_0\rangle
=-\langle{0_0}|T_{xx}|0_0\rangle
=-\langle{0_0}|T_{yy}|0_0\rangle
=
\sum_{n=1}^\infty
\eta^n
\frac{2m^2}{\pi^2(2na)^2}
K_2(2mna)
 \ ,
\nonumber
\\
\langle{0_0}|T_{zz}|0_0\rangle
=
\sum_{n=1}^\infty
\eta^n
\left[
-\frac{2m^2}{\pi^2(2na)^2}
K_2(2mna)
+
\frac{2m^3}{\pi^2(2na)}
K_3(2mna)
\right] \ ,
\end{eqnarray}
where $\eta=+1, (-1)$ labels periodic (antiperiodic) spinors with respect to the standard 
Minkowski vierbein.

The results are $-4$ times 
those of the massive scalar field (\ref{eqn:scalarM0}). The factor of $-1$ is 
due to the different statistics while the factor of $4$ is due to degrees 
of freedom. 
Thus, as with the scalar field, the leading corrections for small mass are of the order $O\left(m^2\right)$,
and $\langle{0_0}|T_{\mu\nu}|0_0\rangle\rightarrow{0}$ exponentially in the
large mass and large $a$ limits, in contrast to the massless case 
where $\langle{0_0}|T_{\mu\nu}|0_0\rangle$ vanishes as $O\left(a^{-4}\right)$.
The massless limit agrees  with twice the massless $2$-component expectation values 
(\ref{eqn:masslessemm0}) (see also \cite{pl:langlois,bd:em,dw:dewitthi}) as expected.

\subsection{$M_-$}
\label{sec:massivediracemm-}

Again we choose to work here in the Minkowski vierbein which is
not invariant under $J_-$. A similar discussion therefore follows
to that in section \ref{sec-masslessspinm-} when applying the method
of images to the Hadamard functions.
For $4$-component spinors the transformation on the Hadamard function
from the rotating vierbein
(\ref{eqn:rotet}) to the standard Minkowski vierbein (a rotation by
$-\pi$ in the $(x,y)$-plane) is 
\begin{equation}
S^{(1)}_{SM}(x,x')
=
e^{\frac{\pi\gamma^1\gamma^2}{2a}z}
S^{(1)}_{RM}(x,x')
e^{-\frac{\pi\gamma^1\gamma^2}{2a}z'} \ ,
\end{equation}
where the $R$, $(S)$ subscript denotes the rotating (standard) vierbein respectively.
Therefore, on $M_-$ in the standard vierbein we obtain
\begin{equation}
S^{(1)}_{SM_-}(x,x')
=
\sum_{n\in{\Z}}
\rho^n
S^{(1)}_{SM_-}\left(x,J_-^nx'\right)
e^{\frac{n\pi\gamma^1\gamma^2}{2}} \ ,
\end{equation}
where $\rho=+1, (-1)$ labels periodic (antiperiodic)
spinors with respect to the vierbein which rotates by $\pi$ as $z\rightarrow{z+a}$.
In terms of the scalar Hadamard function this translates
to using in (\ref{eqn:spinexpect}) the expression 
\begin{equation}
\label{eqn:hadexpr}
G^{(1)}_{M_-}(x,x')
=
\sum_{n\in{\Z}}
\rho^n
G^{(1)}\left(x,J_-^nx'\right)
e^{\frac{n\pi\gamma^1\gamma^2}{2}} \ .
\end{equation}
In the coincidence limit $G^{(1)}_{M_-}(x,x')$ becomes
a function only of $\left(x-(-1)^nx\right)^2+\left(y-(-1)^ny\right)^2+(na)^2$, which is positive, so that again
only the $K$ Bessel term in (\ref{eqn:minkhad}) contributes. Further we note again
the relations (\ref{eqn:m-relation}). We now therefore just apply 
(\ref{eqn:spinexpect}) to (\ref{eqn:hadexpr}).
The calculation is made easier by splitting the sum in (\ref{eqn:hadexpr})
into odd and even terms. The even terms lead to the expectation values on $M_0$ in the
twisted spin structure there and $\langle{0_-}|T_{\mu\nu}|0_-\rangle=
\langle{0_0}|T_{\mu\nu}|0_0\rangle_{(\eta=-1)}+\rho\langle{0_-}|T_{\mu\nu}|0_-\rangle_{\mathrm{odd}}$
with
\begin{eqnarray}
\langle{0_-}|T_{xz}|0_-\rangle_{\mathrm{odd}}
& = & 
-\sum_{n\in{\Z}}
(-1)^n\frac{m^3y(2na+a)}{\pi^2\sigma_n^3}
K_3(m\sigma_n) \ ,
\nonumber
\\
\langle{0_-}|T_{yz}|0_-\rangle_{\mathrm{odd}}
& = & 
\sum_{n\in{\Z}}
(-1)^n\frac{m^3x(2na+a)}{\pi^2\sigma_n^3}
K_3(m\sigma_n) \ ,
\end{eqnarray}
where $\sigma_n=\left((2x)^2+(2y)^2+(2na+a)^2\right)^{1/2}$.

We see that the spinor expectation values on $M_-$ are not 
$-4$ times those of the scalar field. In particular it is interesting to note that
for the scalar field the only non-zero off-diagonal term is $\langle{0_-}|T_{xy}|0_-\rangle$,
while for the spinor field this term is $0$ but $\langle{0_-}|T_{xz}|0_-\rangle$
and $\langle{0_-}|T_{yz}|0_-\rangle$ are non-zero. Note also that these
two terms for the spinor field change sign under a change of spin structure.
In the massless limit the results agree with twice those of the
massless $2$-component spinor field found in section \ref{sec:masslessem}.
The leading order corrections for small mass are $O\left(m^2\right)$.
The non-diagonal terms, and therefore the difference from the values on $M_0$
(with $\eta=-1$), are vanishing at $x=y=0$, and they tend exponentially
to zero in the limits of large mass, large $a$ and large $x^2+y^2$.

\chapter{Particle detector models}
\label{ch:pdchapt}

\section{Introduction}

In this chapter I present a number of new results and extensions on the
theory of particle detector models within quantum field theory in curved spacetimes 
\cite{u:unruh,d:dewitt}. I shall begin in 
section~\ref{sec-background3} with a review of the history and current literature
on such models. The quantity of literature on this subject is quite vast and as such 
my review will not aim to be complete, but I will aim to cover all papers relevant
for the new results presented in the subsequent chapters. Our main focus here
will be with a recent paper by Schlicht \cite{sc:schlicht}. Schlicht 
presents a model which is able to handle time-dependent responses in Minkowski space, that is 
detectors following trajectories which are not stationary in the sense of Letaw
\cite{le:letaw}. This is true also of the model of DeWitt \cite{d:dewitt}; however, to 
the knowledge of the author, Schlicht is the first to point out the 
importance of the regularization of the correlation function in the calculations. 
In particular, as will be discussed in more
detail below, Schlicht shows that the usual $i\epsilon$-prescription is inadequate in
this context. He then introduces an alternative regularization by considering the monopole detector 
of \cite{d:dewitt} as the zero size limit of a detector with spatial extent,
and reproduces the usual thermal 
response \cite{u:unruh,d:dewitt} on a uniformly accelerated trajectory. 

We have two main aims.
The first is to extend the model of Schlicht in a number of directions. 
In particular we extend it
to Minkowski space in arbitrary dimension, to spacetimes built as quotients of Minkowski space 
under discrete isometry groups, and to the non-linearly coupled scalar field
and the massless Dirac field. We also discuss briefly the detector coupled to the
massive scalar field.
The second aim is to consider the response of such detectors in various specific
spacetimes and along various specific trajectories of interest. In particular we consider the responses
of inertial and uniformly accelerated detectors in $d$-dimensional Minkowski space with a single
infinite plane boundary, on a particular conical spacetime~\cite{do:dowker}
(and a higher dimensional generalisation)
and on the quotient spaces $M_0$ and $M_-$ introduced in chapter
\ref{ch:unruhbog}. 
We also consider static detectors on the $\rp$ geon and comoving detectors 
on $\rp$ de Sitter space via those in the associated global embedding Minkowski
spaces (GEMS).

\section{Background}
\label{sec-background3}

Particle detector models in the context of quantum field 
theory in curved spacetime began to be discussed in the 
mid 1970's. At that time one of the most important questions 
within the theory was, and remains to be, ``what is a suitable 
definition of a particle in a general curved spacetime''. In 
Minkowski space quantum field theory particles are defined with 
great input from the invariance group, the Poincare group. 
In particular a particle mode solution to a field equation in
Minkowski space 
is one which is positive frequency with respect to the usual 
Minkowski time coordinate $t$. Under Poincare transformations
positive frequency solutions transform to positive frequency solutions
and so the concept of particle is the same for every 
inertial observer, in the sense that all inertial observers agree
on the number of particles present.
Further the Minkowski vacuum state, defined as the state with no particles present,
is invariant under the Poincare group.
The problem in a general curved spacetime however becomes clear. 
General relativity, in particular general covariance, teaches us 
that there is no notion of a preferred time coordinate and so no 
preferred notion of positive frequency. Further the Poincare group 
is not a symmetry group in a general curved spacetime, and so the 
definition of particle in Minkowski space cannot be extended there. 
As a possible resolution to the problem it was suggested by DeWitt 
and others that a possible operational definition of a particle could 
be that a particle is something that a particle detector detects
(although note the circularity in the definitions if we define a 
particle detector to be something that detects a particle).

The original detector considered by Unruh \cite{u:unruh} is that of
a particle in a box coupled to a quantum field in a curved 
background manifold. The detection of a particle is indicated 
by an excitation of the detector by the field, and it is shown that 
this may correspond to both absorption or emission of a field quantum.
An inertial detector in Minkowski space is not excited in the
Minkowski vacuum but may be de-excited by the possibility of
spontaneous emission.
Unruh \cite{u:unruh} was the first to show that a uniformly accelerated detector in
Minkowski space responds as if immersed in a thermal bath at temperature
$T=a/(2\pi)$ (In natural units) with the associated Planckian spectrum, which is now well known
as the Unruh effect. What an accelerated 
detector sees as an absorption of a field quantum an inertial observer 
regards as the emission of one by the detector.
Further Unruh also showed in \cite{u:unruh} that the response of
a static detector close to a Schwarzschild black hole is closely related to the case of
the accelerated detector in Minkowski space and is thermal, while a geodesic detector
near the horizon will not see this thermal flux of particles.

DeWitt's model \cite{d:dewitt} \footnote{Note that the first published paper where
such a detector is discussed seems to be that of Candelas and Sciama \cite{sc:sciamacand}.} 
is a simplification of
Unruh's. He considers a quantum mechanical particle with many
energy levels linearly coupled to a massless scalar field via a monopole moment operator.
This is the model that appears most in the literature, and the one we will
use throughout this chapter. Again the
response of a uniformly accelerated detector in Minkowski space, where
the detector is switched on in the infinite past and off in the infinite future
is seen to be thermal and Planckian.

A good early review article which discusses particle detector
models is that of Sciama, Candelas and Deutsch \cite{sc:sciamacanddeu}.
This is quite a thorough article on the relation
between quantum field theory, horizons and 
thermodynamics. They discuss the uniformly accelerated detector 
from the point of view of the balance between the thermal 
fluctuations of the field along the trajectory and those of the
detector itself. They find that
``when the detector accelerates, the correlations in the zero-point fluctuations
of the field \textit{along the detector's world-line} no longer precisely balance
its own zero-point fluctuations. 
The detector is consequently excited.''
They also discuss the reason why the effect is exactly thermal,
due to the Gaussian nature of the processes.
Further they consider the response in the asymptotic regions
($r\rightarrow{2M,\infty}$) in the Kruskal manifold, in the
Boulware, Unruh and Hartle-Hawking vacuum states,
and discuss the thermal response of the inertial 
detector in $4$-dimensional de Sitter space.
All for detections carried out over the entire wordline of the
detector.

Since these original papers, a large number of papers have been written on many 
different aspects of these models. Takagi~\cite{tk:takagi1,tk:takagi3,tk:takagi} 
considers the response of a linearly coupled monopole detector
for a massive scalar field in $d$-dimensional Minkowski space. For the 
uniformly accelerated detector he notes for the first time the 
curious phenomena that the
response for a scalar field in even dimensional spacetime has a 
spectrum which includes a Planckian factor whereas for odd dimensions a Fermi 
factor replaces the Planck one. This is often called the
``apparent statistics inversion'' highlighting the fact that the statistics
inversion is just in the power spectrum of the detector and 
does not point to any change of statistics for the quantum field itself.
As Unruh shows~\cite{u:unruh2}, the Fermi factor in the case of odd 
dimension results from the integral over all modes the detector
is sensitive to, with each mode
having a Planckian population as expected for the scalar field.
In~\cite{tk:takagi3,tk:takagi}. Takagi also shows that the 
response of a uniformly accelerated detector is dependent on the mass
of the field. He does not find an explicit expression
for the response function but does prove that the response as the mass
tends to infinity tends to $0$ in contrast with the thermal massless result.
Further Takagi~\cite{tk:takagi4,tk:takagi} extends the model of
DeWitt to a detector coupled to the massless Dirac field via the
scalar quantity $\bar{\psi}\psi$
and also to a detector coupled to the electromagnetic field.
In all cases he shows that the response for the uniformly accelerated detector
satisfies the KMS condition \cite{kms:kubo,kms:martinschwinger}, and 
is thus thermal in this sense. Takagi's paper
\cite{tk:takagi} offers a good review of the
Unruh effect up until 1986, along with a good discussion of the KMS condition
in the context of detector responses.

In a nice paper by Letaw~\cite{le:letaw} the stationary worldlines in 
Minkowski space are defined by the criterion that
the particle detector responses along them are time independent. 
That is the Wightman function along them is invariant under translations
in proper time $\tau$. Further these worldlines are shown to be
the orbits of the timelike Killing vector fields.
The
stationary worldlines are classified into six types (only three of
which, inertial, uniform linear acceleration and uniform circular motion
being well known)
and responses on a representative trajectory from each class are 
considered. In all cases except the inertial one, the response
is non-zero. An interesting case is that of uniform circular motion. The
response is non-zero 
even though, as pointed out in an earlier paper by Letaw and Pfautsch 
\cite{le:letaw2}, the Bogolubov transformation between the two relevant 
coordinate systems is trivial. Further as the Minkowski vacuum and that
adapted to the circular motion are shown to coincide, the response
of a detector in circular motion moving through the vacuum state adapted to it
is also non-zero \cite{ddm:daviesdrm}. Davies et al \cite{ddm:daviesdrm} clarify the
situation by considering a detector in uniform circular motion in Minkowski space
with an infinite cylindrical boundary and also in a compactified Minkowski 
space. It is seen that the response of the detector vanishes
if the boundary is at a radius less than the speed-of-light surface
(that is, no corotating points within the spacetime have a velocity 
greater than $c$).
It is only in this case that the rotating vacuum is unambiguously defined.
The same conclusions were found earlier in $2+1$-dimensions for
a detector in a cylindrical cavity in~\cite{lpa:levinpera}.
See also~\cite{ni:nico} for a corresponding discussion in a general
axisymmetric static curved spacetime and a comment on the electromagnetic
and Dirac cases.
The related issue of defining a Hartle-Hawking vacuum (i.e. invariant under the
isometries and regular everywhere) on the Kerr spacetime has also been discussed. 
A theorem by Kay and Wald \cite{kay-wald} proves the
non-existence of any stationary Hadamard state (and therefore any Hartle-Hawking
state) on Kerr. 
However Ottewill and Winstanley, for example, \cite{wo:winstanott} discuss various attempts to define
a thermal state with most (but not all) of the properties of the Hartle-Hawking state.
In a very recent 
paper Duffy and Ottewill \cite{ot:ottduff} have shown that a state with the defining features
of the Hartle-Hawking vacuum can be constructed when the region outside the
speed-of-light surface is removed by introducing a mirror.

These results show that one
must be cautious in how one interprets the particle detector as 
detecting particles, as discussed in \cite{le:letaw2} and for
example by Sriramkumar and Padmanabhan in~\cite{sp:srirpad2}.
The response of a uniformly accelerated detector in the Minkowski vacuum
and the number of Rindler particles there, found by canonically 
quantizing the field, agree, as of course does the response of 
an inertial detector and the number of Minkowski particles in the
Minkowski vacuum. However the response of a circular motion detector and the
number of ``rotating particles'', found by canonically quantizing the
field with respect to a rotating coordinate system do not agree, the
first being non-zero while the latter being $0$.
An alternative viewpoint is expressed by De Lorenci, De Paola and
Svaiter~\cite{slp:svaitlorenpaol,sl:svaitloren,sp:svaitlpaola}
who propose an alternative canonical quantization and rotating vacuum state.
We shall not discuss circular motion detectors further in this chapter.

A number of early papers discussed detectors coupled to the scalar
field with more complicated couplings than the linear coupling considered
by DeWitt. For example Hinton~\cite{h:hinton2,h:hinton3}
considers detectors non-linearly coupled to the field and those coupled
to derivatives of the field. He shows, as expected with different interaction
Lagrangians, that the responses for these different couplings differ.
He then makes an attempt to introduce a classification scheme for such detectors.
More recently Sriramkumar has shown~\cite{sr:sriramk,sr:sriramk2}
that, for a uniformly accelerated detector non-linearly coupled to the
scalar field in a $d$-dimensional Minkowski space, when $d$ is even the
response includes the Planckian factor for all couplings, while for
$d$ odd there is a Fermi factor when the power of the coupled field
is odd and a Planckian factor when this power is even.
Detectors which couple to components of the energy momentum tensor have
been considered by Padmanabahn and Singh~\cite{ps:padsingh}.
Ford and Roman \cite{fr:fordrom} further answer a challenge set by
Padmanabhan and Singh by constructing a detector which is able to measure the
stress tensor of a quantized scalar field by non-gravitational means. The prescription
requires a detector coupled to the derivative of the field and switched on and off
rapidly.

In all the above papers the detector is that of a monopole detector.
That is, the detector is coupled to the field at a point along the
detector's trajectory. A step closer to a more realistic model is
to consider, as Unruh does in his original paper~\cite{u:unruh},
a detector extended in space. Of the two models considered by
Unruh, one is not fully relativistic while
the other has infinite extent. In~\cite{go:groveott} the authors
define the concept of rigid detectors, where rigidity is given in the
detector's rest frame. Their detector model has a truly finite 
spatial extent. The main example they use is that of the 
detector in uniform circular motion in Minkowski space. The analysis is complicated here
by the effects of radiation of the walls of the detector itself.
Once these are properly taken into account, the discrepancies
found in \cite{le:letaw2,le:letaw} between the detector response and 
a Bogolubov transformation analysis are reconciled.
In this chapter we shall not consider such finite extent detectors,
although we will make some more comments on them in section \ref{sec:altreg}.
The model we consider will instead be that of a monopole detector
but considered as a limit of an infinitely extended one 
(see \cite{sc:schlicht,sc:schlicht2}).

In most of the previous papers (with the exception of \cite{le:letaw,h:hinton3}),
only responses taken over the entire detector worldline are considered.
Another step towards a more realistic detection process is to consider detectors
switched on only for a finite time interval. The first paper to consider such
a response for an inertial and uniformly accelerated detector in Minkowski 
space appears to be that of Svaiter and Svaiter~\cite{ss:svaiter}. Here the
switching of the detector is performed instantaneously. Even in the inertial case
the detector is excited due to the switching. In the limit of large time interval the
usual results are recovered. For both the inertial and uniformly accelerated detectors
the probability of detection is seen to have a logarithmic divergence, but the
transition rate (the proper time derivative of the detection probability) at any
time after the switching is finite, although unbounded close to the switching. It 
is worth noting that an earlier paper by Grove~\cite{gr:grove}
also considers a finite time detector. His main interest is in the detection of the 
negative energy flux outside a moving mirror and the complicated situation necessarily 
hides the conclusions mentioned here, and half way through his calculation he moves to 
a detector switched on at $-\infty$.
In~\cite{hm:higuchi} the authors prove that the
logarithmic divergences in the probability seen by Svaiter and Svaiter are due to the
instantaneous switching. They consider instead a continuous switching window function.
The probability is then seen to be finite, and the usual results are recovered when the
time of detection goes to infinity. For a more extensive discussion on this and the
consideration of different continuous switching functions see~\cite{sp:srirpad}.
All these finite time calculations are done for the massless scalar field.
It is worth noting in a discussion of finite time detectors
the papers \cite{rm:rovellimartin,ma:matinetti} where
the Unruh effect for finite time observers is considered from the point of 
view of the ``thermal time hypothesis'' (this has also recently been extended to observers
in de Sitter space \cite{yu:tian}).
Here the authors are able to associate a temperature with such observers,
and even to inertial observers with a finite lifetime. It would be 
interesting to relate their results to the responses of finite
time particle detectors. The relation is not so straightforward however 
as the two approaches have quite different theoretical foundations. 
Further the temperature introduced in \cite{rm:rovellimartin} depends
on the lifetime of the observer as well as on the proper time along the trajectory. 
The approach is therefore not causal in the sense that
the temperature experienced at a given time depends on how long the 
observer will live in his future. The particle detector calculations
on the other hand are causal in that the response only depends on the
past motion of the detector and not on what will happen in his future.
We shall not pursue the relation between these approaches further here.

Our main reference in this chapter will be Schlicht~\cite{sc:schlicht,sc:schlicht2},
who is concerned with the time dependent response of a monopole detector
linearly coupled to the massless scalar field travelling along non-stationary 
trajectories (those which are not the orbits of any timelike Killing vector field) in Minkowski 
space. 
The focus of this paper is the regularization
of the correlation function that appears in the expression for the detector transition rate.
In particular Schlicht shows that for a detector uniformly accelerated
in Minkowski space the usual $i\epsilon$ regularization of the Wightman function
leads to a time dependent, and hence presumably physically inappropriate,
result. He then argues for an alternative regularization by 
considering the monopole detector as the limit of a spatially extended detector.
The usual result is reproduced in the case of the uniformly accelerated detector.
Further a numerical calculation of the response of a detector following a 
non-stationary worldline that interpolates between inertial and uniformly
accelerated motion is presented. The response interpolates between zero and the
Plankian response in a way that appears physically reasonable.
It is worth noting here that some papers mention
Schlicht's paper~\cite{sc:schlicht} as a reference
for finite time detectors (in particular~\cite{ma:matinetti,ja:Jacobson,dc:coule}). Although~\cite{sc:schlicht}
can easily be adapted to this case it is no more a paper on it,
than the original ones e.g.~\cite{d:dewitt,le:letaw} because all calculations are done
for detectors which are switched on at $\tau_0=-\infty$.

The layout of the rest of this chapter is as follows.
In section \ref{sec-unruh} we first briefly review the formalism for the monopole
detector and extend Schlicht's regularization to a Minkowski space of arbitrary
dimension. We reproduce the expected results in the case of inertial
and uniformly accelerated detectors. In contrast to Schlicht we shall consider
both detectors switched on in the infinite past and detectors that are switched on 
at a finite time. In the finite time case we consider an instantaneous
switching of the detector, no logarithmic divergences are encountered as we shall
work throughout with the transition rate rather than the response function. 
Further we comment on the case of a detector coupled to
a massive scalar field and present the analogue of Schlicht's 
correlation function there.
The section ends with a reformulation of Schlicht's regularization
in terms of a point-like detector and a momentum cut-off. The 
advantage of this view is that
it makes the extension to curved spacetimes, such as Friedman-Robertson-Walker
cosmologies more straightforward. Further it is shown that
a similar view may be taken of a regularization obtained from 
that of Schlicht by modifying the detector shape profile.

In section \ref{sec:automorphic} we consider a Schlicht-type detector on some spacetimes 
built as quotient spaces of
Minkowski space under certain discrete isometry groups. We use the formalism
of automorphic field theory on Minkowski space as discussed by
Banach and Dowker~\cite{bD:auto,bd:em}. That is, we consider
quantum field theory on these spaces as quantum field theory on Minkowski
space under certain restrictions imposed by the quotient group. 
We are in particular interested in responses on the spacetimes $M_0$ and $M_-$ 
introduced in section \ref{sec:m0m-intro}. We will present an analytic 
expression for the response of the inertial detector on $M_0$ and $M_-$ and for the 
uniformly accelerated detector on $M_0$, in all cases with motion in the $x$-direction. 
The time dependent response a uniformly accelerated detector 
on $M_-$ is also considered, giving some analytic arguments for its general behaviour
as well as discussing its numerical evaluation. This response is related to that of
a static detector in the single exterior region of the $\rp$ geon.
We also consider $4$-dimensional Minkowski space with an infinite plane 
boundary and inertial and uniformly accelerated detectors
with motion parallel and perpendicular to the boundary.
This situation has relevance to the investigations of quantum inequality bounds 
and the detection of negative energy
\cite{fo:ford,fo:ford2,fo:phenford,fo:phenford2,fo:phenfordrom,f:fewster,f:fewster2,f:fewster3}. 
Ford, Svaiter and Lyra \cite{fsv:fordsvaitlyra} and more recently 
Davies and Ottewill \cite{do:daviesott} have considered
an inertial detector travelling parallel to the boundary, with Dirichlet
boundary conditions,
and switched on at a finite time, with a number of different 
switching functions. It is shown that the presence of the boundary,
presumably for reasons related to the negative vacuum energy, reduces the
excitations that result from the finite time switching.\footnote{This conclusion was also reached earlier by Grove \cite{gr:grove}
with a more complicated analysis.}
We find that for the motions not parallel to the boundary the picture
is considerably more subtle.
Finally we consider detectors on a $4$-dimensional conical spacetime 
and on some higher dimensional generalizations.
The $4$-dimensional spacetime may be considered as the spacetime
outside a particular idealized cosmic string, of infinite length and
zero radius (see e.g. \cite{vi:vilenkin}).
The higher dimensional generalizations have relevance to the responses 
of a static detector in the exterior of the
$\rp$ geon and comoving detectors in $\rp$ de Sitter space via their 
global embedding Minkowski spaces, as will be discussed further in sections
\ref{sec:Geon} and \ref{sec-desitrp3}. 

In section \ref{sec:nonlin}
we extend Schlicht's formalism to a non-linear coupling of the scalar field, first in Minkowski
space (section \ref{sec:nonlinmink}) and then in the quotient spaces (section \ref{sec:nonlinauto}). 
We comment briefly on responses in various specific situations.

In section \ref{sec:dirac} we extend the formalism further to the more
realistic case of a detector coupled to the massless Dirac field in Minkowski space 
(section \ref{sec:diracmink}). In $4$-dimensional 
Minkowski space
we recover the usual result for the transition rate and the power spectrum of the 
Dirac noise as defined by Takagi in (e.g. \cite{tk:takagi}) for a uniformly accelerated 
detector.
That is a power spectrum, with Fermi factor, and response function
characteristic of a thermal response at the Unruh temperature (\ref{eqn:unruhtemp}).
The Dirac detector is futher extended to the quotient spaces in 
section \ref{sec:diracauto} and the
transition rate and power spectrum are discussed for a uniformly 
accelerated detector on $M_0$ and $M_-$.
On $M_-$ our main aim is to address the question
as to whether a uniformly accelerated detector can distinguish the two 
different spin structures~\cite{pl:langlois}. We will find that
at least in the case of any motion in the $(x,y)$-plane no such 
distinction arises.

Section \ref{sec:Geon} 
begins with an introduction to the literature on relating the Hawking 
effect in black hole spacetimes to the Unruh effect in global
embedding Minkowski spaces (or GEMS). It then
presents an embedding of the $\rp$ geon spacetime
in a $7$-dimensional Minkowski space with identifications, as a 
quotient space of the 
embedding of the Kruskal manifold in $7$-dimensional Minkowski space 
discussed by Chamblin and Gibbons~\cite{gb:embed}. The response
of a uniformly accelerated detector in the embedding space, relevant
to that of a static detector on the $\rp$ geon, is then discussed.

Section \ref{sec:conformal} considers a detector coupled linearly to
a conformally
coupled massless scalar field in a conformally flat spacetime.
A regularization of the correlation function in the conformal
vacuum is introduced which is different from that obtained by
applying a naive $i\epsilon$ prescription. The expression obtained
is conformally related to the correlation function found in Minkowski space in
section \ref{sec-linmink}.

Finally section \ref{sec-desitrp3} considers detectors in de Sitter 
and $\rp$ de Sitter space \cite{jk:desit,bm:mcinnes}.
In the case of a comoving observer we show that the regularization 
given in section \ref{sec:conformal} may also
arise from the consideration of a detector with spatial extent.
The usual thermal result for comoving and uniformly accelerated 
detectors in de Sitter space is recovered.
We then discuss a comoving
detector, where motion is orthogonal to the distinguished foliation
in $\rp$ de Sitter space~\cite{jk:desit}. We first consider the 
response from a direct calculation in the $\rp$ de Sitter space
and then from a relevant calculation for the associated worldline
in its embedding space, motivated by the GEMS scheme. 
The embedding space is a $5$-dimensional Minkowski 
space with identifications, related to the usual $5$-dimensional embedding of
$4$-dimensional de Sitter space. These last calculations
should be useful in assessing the validity and determining the relation
between particle detector calculations in the global embedding Minkowski
spaces and those in the original spacetimes of interest.
The results provide support to the view that the GEMS approach captures, to 
a considerable extent, properties of quantum fields.

Appendices \ref{ch:appendixA},\ref{ch:appendixB},\ref{ch:appendixC},\ref{ch:appendixe} contain details of some of the integrations
required in sections \ref{sec-unruh}-\ref{sec-desitrp3}. Appendix \ref{ch:appendixA2} provides
details of the correlation function and responses in $2$-dimensional Minkowski space. Appendix
\ref{ch:appendixd} provides more details for the automorphic field detector.

\section{Causal detector with linear coupling}
\label{sec-unruh}

\subsection{Massless scalar field in $d$-dimensional Minkowski space}
\label{sec-linmink}

In this section we briefly review the model point-like particle 
detectors of DeWitt \cite{d:dewitt}. We discuss the use of spatial sampling functions
to regularize the correlation function 
and extend the regularization in \cite{sc:schlicht}, using a Lorentzian
sampling function, to $d$-dimensional Minkowski space $M$, $d\geq{2}$. 

\subsubsection{DeWitt's monopole detector}

The model is that of a monopole detector, moving along a prescribed classical 
trajectory in $M$ and coupled to a massless scalar field $\phi$. The field has 
Hamiltonian $H_\phi$, and satisfies the massless Klein-Gordon equation. 
The free field operator is expanded in terms of a standard complete set of orthonormal solutions 
to the field equation as
\begin{equation}
\label{eqn:modeexpan}
\phi(t,\mathbf{x})=
\frac{1}{(2\pi)^{(d-1)/2}}
\int
\frac{d^{d-1}k}{(2\omega)^{1/2}}
\left(
a(\mathbf{k})e^{-i(\omega{t}-\mathbf{k}\cdot\mathbf{x})}
+a^{\dagger}(\mathbf{k})e^{i(\omega{t}-\mathbf{k}\cdot\mathbf{x})}
\right) \ ,
\end{equation}
where $(t,x_1,x_2,\ldots,x_{d-1})$ are usual Minkowski coordinates and, 
in the massless case, $\omega=|\mathbf{k}|$. The field is quantized by 
imposing, for the creation and annihilation operators, the usual commutation relations
\begin{equation}
\left[a(\mathbf{k}),a^{\dagger}(\mathbf{k})\right]
=
\delta^{d-1}(\mathbf{k}-\mathbf{k'}) \ . 
\end{equation}
The Minkowski vacuum $|0\rangle$ is the state annihilated by all 
the annihilation operators.

The detector is a quantum mechanical system with a set of energy
eigenstates $\{|0_D\rangle,|E_i\rangle\}$. It moves along 
a prescribed classical trajectory
$t=t(\tau)$, $\mathbf{x}=\mathbf{x}(\tau)$, where 
$\tau$ is the detector's proper time, and it couples to the scalar field via the 
interaction Hamiltonian
\begin{equation}
\label{eq:interhamil}
H_{\mathrm{int}}=
c
m(\tau)
\phi(\tau) \ ,
\end{equation}
where $c$ is a (small) coupling constant and $m(\tau)$ is the detector's monopole moment operator 
\cite{d:dewitt}. The evolution of $m(\tau)$ is given by
\begin{equation}
m(\tau)=
e^{iH_D\tau}m(0)e^{-iH_D\tau} \ . 
\end{equation}
As the interaction takes place at a point along the trajectory at any given 
time, the detector is known as a point-like detector.

Suppose that at time $\tau_0$ the detector and field are in the product state 
$|0,E_0\rangle=|0\rangle|E_0\rangle$, where $|E_0\rangle$ is a detector state with 
energy $E_0$. We want to know the probability that at a later 
time $\tau_1>\tau_0$ the detector is found in state $|E_1\rangle$,
regardless of the final state of the field. We are interested in both excitations,
$E_1>E_0$ and de-excitations, $E_1<E_0$. We find the required expression in the
interaction picture (see e.g. \cite{Gr:greiner}) where all field operators satisfy 
the free field equations and the 
time evolution is shared between operators and states. In this picture the time 
evolution of the product states is governed by the interaction Hamiltonian
\begin{equation}
i\frac{d}{d\tau}
|\psi(\tau)\rangle
=
cm(\tau)\phi(\tau)
|\psi(\tau)\rangle \ . 
\end{equation}
The amplitude for the transition from state $|0,E_0\rangle$ at $\tau=\tau_0$ 
to $|\psi,E_1\rangle$ at $\tau=\tau_1$, by the usual interaction picture theory
is
\begin{equation}
\langle\psi,E_1|0,E_0\rangle
=
\langle\psi,E_1|
T
\exp\left[-i\int^{\tau_1}_{\tau_0}d\tau\,{H_{\mathrm{int}}}(\tau)\right]
|0,E_0\rangle \ ,
\end{equation}
where $T$ is the time ordering operator.
To first order in perturbation theory the expression
reads
\begin{eqnarray}
\label{eqn:prob1stpert}
\langle\psi,E_1|0,E_0\rangle
& = & 
ic
\langle\psi,E_1|
\int^{\tau_1}_{\tau_0}d\tau\,{m(\tau)}\phi(\tau)
|0,E_0\rangle 
\nonumber
\\
& = & 
ic
\langle{E_1}|m(0)|E_0\rangle
\int^{\tau_1}_{\tau_0}{d\tau}\,e^{i(E_1-E_0)\tau}\langle\psi|\phi(\tau)|0\rangle
 \ . 
\end{eqnarray}
The transition probability to all possible states of the field is given by
squaring (\ref{eqn:prob1stpert}) and summing over the complete set $\{|\psi\rangle\}$
of final field states, with the result
\begin{equation}
\sum_{\psi}
|\langle\psi,E_1|0,E_0\rangle|^2=
c^2
|\langle{E_1}|m(0)|E_0\rangle|^2
\int^{\tau_1}_{\tau_0}d\tau
\int^{{\tau}_1}_{{\tau}_0}d\tau'
e^{-i(E_1-E_0)(\tau-\tau')}\langle{0}|\phi(\tau)\phi(\tau')|0\rangle
 \ . 
\end{equation}
This expression has two parts. The sensitivity $c^2|\langle{E_1}|m(0)|E_0\rangle|^2$
depends only on the internal details of the detector and is not considered hereafter.
The ``response function''
\begin{equation}
F_{\tau_0,\tau_1}(\omega)=
\int^{\tau_1}_{\tau_0}d\tau
\int^{{\tau}_1}_{{\tau}_0}d\tau'
e^{-i\omega(\tau-\tau')}\langle{0}|\phi(\tau)\phi(\tau')|0\rangle
 \ ,
\end{equation}
where $\omega=E_1-E_0$ ($\omega>0$ for excitations and $\omega<0$ for de-excitations), 
does not depend on the internal details of the detector and so is
common for all such detectors. 

We now follow Schlicht \cite{sc:schlicht} and
change coordinates to $u=\tau$, $s=\tau-\tau'$ for $\tau'<\tau$
and $u=\tau'$, $s=\tau'-\tau$ for $\tau'>\tau$. Then
\begin{equation}
F_{\tau_0,\tau_1}(\omega)=
2\int^{\tau_1}_{\tau_0}du
\int^{u-\tau_0}_{0}ds
\,
Re\left(
e^{-i\omega{s}}\langle{0}|\phi(u)\phi(u-s)|0\rangle
\right)
 \ ,
\end{equation}
because 
$\langle{0}|\phi(\tau')\phi(\tau)|0\rangle=\langle{0}|\phi(\tau)\phi(\tau')|0\rangle^*$ (a 
consequence of $\phi$ being a self 
adjoint operator and the axioms of inner products).
To obtain an expression
for the ``transition rate'', we differentiate with respect to $\tau_1$ and obtain
\begin{equation}
\label{eq:transition}
\dot{F}_{\tau_0,\tau}(\omega)=
2\int^{\tau-\tau_0}_{0}ds\,
Re
\left(
e^{-i\omega{s}}\langle{0}|\phi(\tau)\phi(\tau-s)|0\rangle
\right)
 \ ,
\end{equation}
where we have written $\tau_1=\tau$.
The transition rate is clearly causal in the sense that it does not depend on the state
of motion of the detector after time $\tau$ but only on times $\tau_0<\tau'<\tau$ which
label the past motion of the detector.
If the correlation function $\langle{0}|\phi(\tau)\phi(\tau-s)|0\rangle$ is
invariant under translations in $\tau$ this expression may be simplified to
\begin{equation}
\label{eq:transitiontindep}
\dot{F}_{\tau_0,\tau}(\omega)=
\int^{\tau-\tau_0}_{-\tau+\tau_0}ds\,
\,e^{-i\omega{s}}\langle{0}|\phi(s)\phi(0)|0\rangle
 \ .
\end{equation}

The transition rate, which is proportional to the derivative of the
probability of the transition with respect to the switch off time $\tau$,
may be written as 
\begin{equation}
\dot{F}_{\tau_0,\tau}(\omega)=
\lim_{\delta\tau\rightarrow{0}}
\frac{F_{\tau+\delta\tau,\tau_0}-F_{\tau,\tau_0}}{\delta\tau}
 \ .
\label{eqn:transderivecp}
\end{equation}
From (\ref{eqn:transderivecp}) it is clear that the transition rate is not related to the response of a detector turned on at $\tau$ and
off at $\tau+\delta\tau$ in the limit as $\delta\tau$ goes to
zero. Rather the transition rate compares detector responses
in two different ensembles of detectors, one switched off at
$\tau$ and the other at $\tau+\delta\tau$ in this limit.
$F_{\tau_0,\tau}$, being proportional to a probability
is strictly non-negative. $\dot{F}_{\tau_0,\tau}$, however,
may a priori be negative, and we shall see examples where it indeed is.

The above formalism for a monopole detector is standard in the 
literature (see e.g \cite{bd:book}) and in particular is the model
first introduced by DeWitt \cite{d:dewitt}.
The correlation function $\langle{0}|\phi(x)\phi(x')|0\rangle$
in (\ref{eq:transition}) is the positive frequency Wightman function, which 
can be obtained from the expansion (\ref{eqn:modeexpan}),
\begin{equation}
\label{eqn:integralwight}
\langle{0}|\phi(x)\phi(x')|0\rangle
=
\frac{1}{(2\pi)^{d-1}}
\int
\frac{d^{d-1}k}{2\omega}
e^{-i\omega(t-t')+i\mathbf{k}\cdot(\mathbf{x}-\mathbf{x}')}
\ .
\end{equation}
The integrals for $d>2$ may be performed by moving to hyperspherical coordinates
(polar coordinates and spherical coordinates for $d=3$ and $d=4$ respectively). 
The $|\mathbf{k}|$ integral requires regularization due to ultraviolet 
divergences.
The usual regularization \cite{bd:book}, often 
known in the literature as the $i\epsilon$-prescription, 
is to introduce the exponential cut-off $e^{-\epsilon|\mathbf{k}|}$ in the high
frequency Minkowski modes.
The resulting expression is (see appendix \ref{ch:appendixA})
\begin{equation}
\label{eq:correusu}
\langle{0}|\phi(\tau)\phi(\tau')|0\rangle
=\frac{\Gamma{[d/2-1]}}{4\pi^{d/2}}\frac{1}
{\left[i^2(t(\tau)-t(\tau')-i\epsilon)^2+|\mathbf{x}(\tau)-\mathbf{x}(\tau')|^2\right]^{d/2-1}}
 \ ,
\end{equation}
for $d>2$. 

For $d=2$ it is well known that the massless scalar field theory 
suffers from infrared divergences \cite{wh::whightman}:
The formal expression for the correlation function is
(see appendix \ref{ch:appendixA2})
\begin{equation}
\label{eq:correusu2dim}
\langle{0}|\phi(\tau)\phi(\tau')|0\rangle
=-\frac{1}{4\pi}\ln((t(\tau)-t(\tau')-i\epsilon)^2-(x(\tau)-x(\tau'))^2)+C
 \ ,
\end{equation}
where $C$ is an ill-defined constant.
We may still consider the particle detector in $2$-dimensions dropping the
constant divergent term in the correlation function. Provided we include a temporal window
function which vanishes at infinity and consider an asymptotic (infinite time)
detection the constant terms in the
correlation function do not contribute to the response (see appendix \ref{ch:appendixA2}),
this point was made by Takagi in \cite{tk:takagi}. 

The fundamental observation of reference \cite{sc:schlicht} is that if we now use (\ref{eq:correusu})
in (\ref{eq:transition}) on a uniformly accelerated worldline with acceleration $1/\alpha$ and proper time $\tau$,
\begin{eqnarray}
t & = & \alpha\sinh(\tau/\alpha) \ ,
\nonumber
\\
\label{eq:accel}
x & = & \alpha\cosh(\tau/\alpha) \ ,
\end{eqnarray}
with the detector switched on in the infinite
past, $\tau_0=-\infty$, we obtain a time dependent and 
apparently unphysical result,
instead of the expected time independent thermal result 
(see e.g. \cite{d:dewitt,bd:book}).
Schlicht shows this in $4$ dimensions by specific numerical and analytic 
calculations, and the general $d$ case follows similarly. His
solution is to propose a different regularization of the
correlation function, by considering the monopole detector 
as the limit of a spatially extended one. 
We next discuss Schlicht's regularization and related alternatives.

\subsubsection{Alternative regularizations}
\label{sec:altreg}

In \cite{sc:schlicht} Schlicht introduces a regularization
of the Wightman function by considering the monopole
detector as the limit of a detector extended in space\footnote{See also
\cite{go:groveott} for a discussion of detector models with spatial 
extent.}. 
An argument for the physical reasonableness of the regularization is that
a spatially extended model should provide a more realistic
model of a real particle detector. 
A related attempt 
to introduce a regularization via
a spatially extended detector was made and 
discussed in some depth by Takagi \cite{tk:takagi}. 
Takagi however employs formal limiting arguments in a way that leads to the
usual $i\epsilon$ prescription. The justification of these arguments has been criticised in \cite{sc:schlicht}.

Intuitively the most physically reasonable
model is one which remains rigid in the detector's proper
reference frame. In Minkowski space a reasonable choice for this
reference frame is that built from Fermi coordinates (see e.g. \cite{MTW:grav}
for a thorough discussion on the construction of Fermi coordinates). 
This is the choice adopted by Schlicht.
Having chosen to work in a Fermi coordinate frame there are many ways in 
which we could model the extended detector. 
For one such model of a detector which cannot
resolve distances shorter than $\epsilon$
we can replace the field value at a point, $\phi(\tau)$ 
in the interaction Hamiltonian (\ref{eq:interhamil}), 
by a field average
over a patch of the spatial hypersurface, of characteristic size $\epsilon$
and centered on the worldline. For example we could consider
\begin{equation}
\label{eqn:boxsmear}
\phi(\tau)=\int_{\epsilon\times\ldots\times\epsilon}
d^{d-1}\xi\,
\phi(x(\tau,\mathbf{\xi})) \ ,
\end{equation}
where $(\tau,\mathbf{\xi})$ are the Fermi coodinates, the
integral is over all $-\epsilon/2<\mathbf{\xi}_i<\epsilon/2$ for
all $i$
and $x(\tau,\mathbf{\xi})$ describes the transformation between the Minkowski
and Fermi coordinates.
This model detector has a truly finite spatial extent.
The effect of the spatial averaging is to introduce
a cut-off at small distances. The unregularized integral expression (\ref{eqn:integralwight}) 
for the Wightman function 
is recovered from $\langle{0}|\phi(\tau)\phi(\tau')|0\rangle$
with $\phi(\tau)$ given by (\ref{eqn:boxsmear}) in the
limit as $\epsilon$ goes to $0$.

Of course we are not limited to considering the spatial
averaging (\ref{eqn:boxsmear}).
Other possibilities arise, for example, if we smear the field operator with a suitable
window function of characteristic length $\epsilon$,
\begin{equation}
\label{eq:smear}
\phi(\tau)=
\int
d^{d-1}\xi
\,
W_\epsilon(\mathbf{\xi})
\phi(x(\tau,\mathbf{\xi})) \ ,
\end{equation}
which is the procedure considered in \cite{sc:schlicht}.
The box averaging in (\ref{eqn:boxsmear}) 
is equivalent to a window function which is the
product of step functions.
In effect this again introduces a cut-off at short distances
of the order of $\epsilon$ in size.
We may choose a window function with infinite support, such
as that used in \cite{sc:schlicht}, if the window function
decreases sufficiently rapidly at large distances. Or
we may consider a detector of truly finite extent with the use
of a window function with compact support.
We require the window function to be normalised as 
\begin{equation}
\int
d^{d-1}\xi
\,
W_\epsilon(\mathbf{\xi})
=1 \ ,
\end{equation}
so that smearing a constant function
will 
return that constant value.
We also require the window function to approximate the
$(d-1)$-dimensional Dirac $\delta$ function, so that
in the limit as $\epsilon$ tends to $0$, (\ref{eq:smear})
formally gives the field value $\phi(x(\tau))$.
The window function hence staisfies $W_\epsilon(\mathbf{\xi})\approx{0}$
for $|\mathbf{\xi}|>>\epsilon$ and $W_\epsilon(\mathbf{\xi})\propto{\epsilon^{-(d-1)}}$
for $|\mathbf{\xi}|<<\epsilon$.

\subsubsection{Schlicht's correlation function in $d$ dimensions}

We now extend Schlicht's regularization of the Wightman function
\cite{sc:schlicht} to $d$-dimensional
Minkowski space. Although a large number
of different window functions could be considered, the one chosen
in \cite{sc:schlicht} seems to be the easiest for obtaining a
closed expression for the Wightman function on an arbitrary
trajectory. Further, as we shall see in section \ref{sec-ultra}, the regularization 
introduced by this window function is similar to that of the
usual $i\epsilon$ regularization except for the important 
difference that the 
cut-off of high frequency modes is as seen in the reference frame 
of the detector, whereas the usual $i\epsilon$ procedure introduces
a cut-off of the high frequency modes as seen by inertial observers.

The $d$-dimensional analogue of the window function considered 
in \cite{sc:schlicht} is
\begin{equation}
\label{eq:shape}
W_{\epsilon}(\mathbf{\xi})
=
\frac{\Gamma[d/2]}{\pi^{d/2}}
\frac{\epsilon}{(\mathbf{\xi}^2+\epsilon^2)^{d/2}} \ .
\end{equation}
(\ref{eq:shape}) is sometimes referred to 
as a Lorentzian window or sampling function.
It approximates a $(d-1)$-dimensional Dirac $\delta$ function as it scales as 
\begin{equation}
W_{\epsilon}(\mathbf{\xi})
=
\frac{1}{\epsilon^{d-1}}
W_1(\mathbf{\xi}/\epsilon) \ ,
\end{equation}
and
\begin{equation}
\int
d^{d-1}\xi\,
W_{\epsilon}(\mathbf{\xi})
=1 \ .
\end{equation}
The advantage of this window function is that in the
calculation of the correlation function
$\langle{0}|\phi(\tau)\phi(\tau')|0\rangle$ all integrals
may be done explicitly.

Now using (\ref{eq:smear}) and (\ref{eq:shape}),
we find
\begin{eqnarray}
\langle{0}|\phi(\tau)\phi(\tau')|0\rangle
& = &
\frac{1}{(2\pi)^{d-1}}
\int
\frac{d^{d-1}k}{2\omega}
\int
d^{d-1}\xi\,
W_{\epsilon}(\mathbf{\xi})
e^{-i({\omega}t(\tau,\mathbf{\xi})-\mathbf{k}\cdot\mathbf{x}(\tau,\mathbf{\xi}))}
\nonumber
\\
\label{eq:expan}
& & \times\int
d^{d-1}\xi'\,
W_{\epsilon}(\mathbf{\xi'})
e^{i({\omega}t(\tau',\mathbf{\xi'})-\mathbf{k}\cdot\mathbf{x}(\tau',\mathbf{\xi'}))} \ .
\end{eqnarray}
Consider then the integral over $\mathbf{\xi}$ in (\ref{eq:expan}),
\begin{equation}
\label{eqn:gfun}
g_\epsilon(\mathbf{k};\tau)
=
\int
d^{d-1}\xi\,
W_\epsilon(\mathbf{\xi})
e^{-i(\mathit{k}\cdot\mathit{x}(\tau,\mathbf{\xi}))} \ ,
\end{equation}
where we have written $(t,\mathbf{x})$, $(\omega,\mathbf{k})$,as a 
$d$-vectors $\mathit{x}$ and $\mathit{k}$
respectively. The
transformation to Fermi coordinates is
\begin{equation}
\mathit{x}(\tau,\mathbf{\xi})
=
\mathit{x}(\tau)
+\xi_1\mathit{e_{\xi_1}}(\tau)
+
\cdots
+\xi_{d-1}\mathit{e_{\xi_{d-1}}}(\tau) \ ,
\end{equation}
where $\dot{\mathit{x}}(\tau),\mathit{e_{\xi_1}}(\tau),\ldots,\mathit{e_{\xi_{d-1}}}(\tau)$
form an orthonormal basis which is Fermi transported along the worldline. Now 
\begin{eqnarray}
g_\epsilon(\mathbf{k};\tau)
& = & 
\int
d^{d-1}\xi\,
W_\epsilon(\mathbf{\xi})
e^{-i(\mathit{k}\cdot\mathit{x}(\tau,\mathbf{\xi}))} \nonumber
\\
& = &
e^{-i(\mathit{k}\cdot\mathit{x}(\tau))}
\int
d^{d-1}\xi\,
W_\epsilon(\mathbf{\xi})
e^{-i(\xi_1\mathit{k}\cdot\mathit{e_{\xi_1}}(\tau)+\cdots+\xi_{d-1}\mathit{k}\cdot\mathit{e_{\xi_{d-1}}}(\tau))}
 \ .
\end{eqnarray}
Writing
$\bar{\mathbf{k}}:=(\mathit{k}\cdot\mathit{e_{\xi_1}}(\tau),\ldots,\mathit{k}\cdot\mathit{e_{\xi_{d-1}}}(\tau))$, we have 
\begin{equation}
\label{eqn:gequation}
g_\epsilon(\mathbf{k};\tau)
=
\frac{\epsilon\Gamma(d/2)}
{\pi^{d/2}}
e^{-i(\mathit{k}\cdot\mathit{x}(\tau))}
\int
d^{d-1}\xi\,
\frac{1}
{(\mathbf{\xi}^2+\epsilon^2)^{d/2}}
e^{-i\bar{\mathbf{k}}\cdot\mathbf{\xi}} \ .
\end{equation}
The integration for $d>2$ may be performed by moving to hyperspherical coordinates in $\mathbf{\xi}$ 
space (see appendix \ref{ch:appendixA}), while that for
$d=2$ may be done directly (see appendix \ref{ch:appendixA2}). The result is
\begin{equation}
g_\epsilon(\mathbf{k};\tau)
=
e^{-i(\mathit{k}\cdot\mathit{x}(\tau))}
e^{-\epsilon|\bar{\mathbf{k}}|}
=e^{-i(\mathit{k}\cdot\mathit{x}(\tau))}
e^{-\epsilon\left[(\mathit{k}\cdot\mathit{e_{\xi_1}}(\tau))^2+\cdots+(\mathit{k}\cdot\mathit{e_{\xi_{d-1}}}(\tau))^2\right]^{(1/2)}}
 \ .
\end{equation}
Now we use the same arguments as Schlicht. We multiply the expression
\begin{equation}
\mathit{k}
=
(\mathit{k}\cdot{\dot{\mathit{x}}})\dot{\mathit{x}}
-
(\mathit{k}\cdot{e_{\xi_1}})e_{\xi_1}
-
\cdots
-
(\mathit{k}\cdot{e_{\xi_{d-1}}})e_{\xi_{d-1}}
\end{equation}
by $\mathit{k}$, giving
\begin{equation}
(\mathit{k}\cdot{\dot{\mathit{x}}})^2
=
(\mathit{k}\cdot{e_{\xi_1}})^2
+
\cdots
+
(\mathit{k}\cdot{e_{\xi_{d-1}}})^2
=
(\omega\dot{t}-\mathbf{k}\cdot\dot{\mathbf{x}})^2 \ .
\end{equation}
As $\mathit{k}\cdot{\dot{\mathit{x}}}$
is the product of a lightlike and a timelike vector, both future-pointing, it is postive. Hence
\begin{equation}
\label{eqn:gepktau}
g_\epsilon(\mathbf{k};\tau)
=e^{-i(\mathit{k}\cdot\mathit{x}(\tau))}
e^{-\epsilon\mathit{k}\cdot{\dot{\mathit{x}}}}
 \ ,
\end{equation}
and so
\begin{equation}
\label{eq:correlation}
\langle{0}|\phi(\tau){\phi}(\tau')|0\rangle=
\frac{1}{(2\pi)^{d-1}}
\int
\frac{d^{d-1}k}{2\omega}
e^{-i\omega(t-t'-i\epsilon(\dot{t}+\dot{t}'))
+i\mathbf{k}\cdot(\mathbf{x}-\mathbf{x'}-i\epsilon(\dot{\mathbf{x}}+\dot{\mathbf{x}}'))} \ .
\end{equation}
For $d>2$ the integral in (\ref{eq:correlation}) may be performed again 
by moving to hyperspherical coordinates (as shown in
appendix \ref{ch:appendixA}). The final result is
\begin{eqnarray}
& \langle{0}|\phi(\tau)\phi(\tau')|0\rangle=
\frac{\Gamma[d/2-1]}{4\pi^{d/2}}
\frac{1}{A^{d/2-1}} \ , \nonumber
\\
\label{eq:corre}
& A=\left[i^2(t(\tau)-t(\tau')-i\epsilon(\dot{t}(\tau)+\dot{t}(\tau')))^2
+(\mathbf{x}(\tau)-\mathbf{x}(\tau')-i\epsilon(\dot{\mathbf{x}}(\tau)+\dot{\mathbf{x}}(\tau')))^2\right] \ .
\end{eqnarray}

For $d=2$,
\begin{equation}
\langle{0}|\phi(\tau){\phi}(\tau')|0\rangle
=
\frac{1}{4\pi}
\int
\frac{dk}{|k|}
e^{-i|k|(t-t'-i\epsilon(\dot{t}+\dot{t}'))
+ik(x-x'-i\epsilon(\dot{x}+\dot{x}'))} \ .
\end{equation}
The integral is performed as that leading to (\ref{eq:correusu2dim}).
The correlation function is thus (see appendix \ref{ch:appendixA2})
\begin{eqnarray}
\langle{0}|\phi(\tau)\phi(\tau')|0\rangle
& = & -\frac{1}{4\pi}\ln[(t(\tau)-t(\tau')-i\epsilon(\dot{t}(\tau)+\dot{t}(\tau'))^2
\nonumber
\\
&   & -(x(\tau)-x(\tau')-i\epsilon(\dot{x}(\tau)+\dot{x}(\tau'))^2]+C \ ,
\label{eqn:2dimcorrfun}
\end{eqnarray}
where $C$ is an ill-defined constant.

We note here that following a similar calculation to that given above
the usual $i\epsilon$ regularization of the correlation function
(i.e. (\ref{eq:correusu}) and (\ref{eq:correusu2dim})) may also be obtained
by considering a spatially smeared field. The difference from the above
regularization being that the smearing is done in the Minkowski 
reference frame, that is always with respect to inertial observers, and
not in the Fermi frame. This model detector is thus not rigid in its 
rest frame.

\subsubsection{Inertial detector in $d$ dimensions}

First we consider the transition rate (\ref{eq:transitiontindep}) with the correlation 
function (\ref{eq:corre})
for a detector following an inertial trajectory. On Minkowski space
the existence of boost Killing vectors implies that the response of
the detector will be independent of velocity $v$. We consider therefore the 
trajectory $t=\tau$, $\mathbf{x}=\mathrm{constant}$,  
where $-\infty<\tau<\infty$.
If the detector is switched on in the infinite past, $\tau_0=-\infty$,
the transition rate for $d>2$ is 
\begin{equation}
\dot{F}_\tau(\omega)=
\frac{\Gamma[d/2-1]}{4\pi^{d/2}}
\int^{\infty}_{-\infty}ds\,
\frac{e^{-i\omega{s}}}
{[i^2(s-2i\epsilon)^2]^{d/2-1}} \ .
\end{equation}
Note that we obtain the same expression in this case if we use instead the
usual correlation function (\ref{eq:correusu}) i.e. with
the $i\epsilon$ regularization. The integral may be done by residues 
and the result as $\epsilon\rightarrow{0}$ is 
\begin{equation}
\label{eqn:inertrespons}
\dot{F}_\tau(\omega)=
\frac{\Gamma[d/2-1](-\omega)^{d-3}}{2\pi^{d/2-1}(d-3)!}
\Theta(-\omega) \ ,
\end{equation}
where $\Theta(-\omega)$ is the Heaviside step function. 
When $d=2$ there 
are added complications due to the constant terms, one of which is infinite,
in the correlation function (\ref{eqn:2dimcorrfun}), but these terms do not contribute to the 
transition rate 
provided we include a temporal window function into the integral
(see appendix \ref{ch:appendixA2}). We find
\begin{equation}
\label{eqn:2diminertraj}
\dot{F}_\tau(\omega)=
-\frac{1}{\omega}
\Theta(-\omega) \ .
\end{equation}
The results given by (\ref{eqn:inertrespons}) and (\ref{eqn:2diminertraj}) are as expected
on Minkowski space. The transition rate vanishes for $\omega>0$ indicating that an inertial detector
is not excited by the Minkowski vacuum. For $\omega<0$ the transition rate is non-zero due to
the possibility of spontaneous emission by the detector.

Following Takagi's analysis of the uniformly accelerated 
detector \cite{tk:takagi} we may relate the transition rate given by (\ref{eqn:inertrespons}) and (\ref{eqn:2diminertraj}) 
to a local density of states. On Minkowski
space or a quotient of Minkowski space with non-compact spatial sections and mode solutions 
\begin{equation}
u_{\mathbf{k}}(t,\mathbf{x})
=
\frac{e^{-i|\mathbf{k}|t}}{(2|\mathbf{k}|)^{1/2}}
h_{\mathbf{k}}(\mathbf{x}) \ ,
\end{equation}
the local density of states is defined to be \cite{to:takagiotte}
\begin{equation}
\label{eqn:localdensity}
\rho(\omega,\mathbf{x})
=
\int
d^{d-1}k\,
|h_{\mathbf{k}}(\mathbf{x})|^2
\delta(|\mathbf{k}|-\omega) \ .
\end{equation}
Due to the normalization of the modes 
\begin{equation}
\rho(\omega)
=
\int
d^{d-1}x\,
\rho(\omega,\mathbf{x})
=
\int
d^{d-1}k\,
\delta(|\mathbf{k}|-\omega) \ ,
\end{equation}
and so if the space is homogeneous then $\rho(\omega,\mathbf{x})$
is independent of $\mathbf{x}$ and is then the density of states per unit volume.
The usual Minkowski density of states is obtained by taking 
$h_{\mathbf{k}}(\mathbf{x})=\frac{e^{i\mathbf{k}\cdot\mathbf{x}}}{(2\pi)^{(d-1)/2}}$,
and then
\begin{equation}
\rho_M(\omega,\mathbf{x})
=
\int
\frac
{d^{d-1}k}
{(2\pi)^{d-1}}\,
\delta(|\mathbf{k}|-\omega) \ .
\end{equation}
For $d>2$ the integral may be done by transforming to hyperspherical coordinates. We find
\begin{equation}
\rho_M(\omega,\mathbf{x})
=
\frac{\omega^{d-2}}{2^{d-2}\pi^{(d-1)/2}\Gamma((d-1)/2)}\Theta(\omega) \ .
\end{equation}
By making use of the properties of Gamma functions it is easy to show that
\begin{equation}
\dot{F}_\tau(\omega)=
-\frac{\pi}{\omega}
\Theta(-\omega)
\rho_M(|\omega|,\mathbf{x}) \ .
\end{equation}
This relation also holds for $d=2$ as shown in appendix \ref{ch:appendixA2}.

For a detector that is switched on at a finite proper time $\tau_0$,
the transition rate, for $d>2$, is given by (\ref{eqn:inertrespons}) plus the 
correction term
\begin{equation}
\label{eqn:finitsvaitsvait}
C_{\Delta\tau}(\omega)
=
-\frac{\Gamma[d/2-1]}{2\pi^{d/2}}
\int^{\infty}_{\Delta\tau}
ds
\,
\frac{\cos(\omega{s})}
{(i^2s^2)^{d/2-1}}
 \ ,
\end{equation}
where $\Delta\tau=\tau-\tau_0$. We note that $C_{\Delta\tau}(\omega)$ is
manifestly finite for all $\Delta\tau>0$. The integration may be performed by successive 
integrations by parts. In the case $d=4$ we recover the result of Svaiter and 
Svaiter~\cite{ss:svaiter},
\begin{equation}
\label{eqn:inertcorrec}
C_{\Delta\tau}(\omega)
=
\frac{\cos(\omega\Delta\tau)}{2\pi^2\Delta\tau}
+\frac{|\omega|}{2\pi^2}\mathrm{si}(|\omega|\Delta\tau) \ ,
\end{equation}
where $\mathrm{si}$ is a shifted sine integral \cite{as:abromowitz},
\begin{equation}
\mathrm{si}(z)=
\int_0^z
dt\,
\frac{\sin{t}}{t}
-\frac{\pi}{2} \ .
\end{equation}
There 
are regions of the $|\omega|\Delta\tau$ parameter space where 
$C_{\Delta\tau}(\omega)$ and therefore $\dot{F}_\tau(\omega)$
is negative (recall that for $\omega>0$ the asymptotic response is $0$
and so $C_{\Delta\tau}(\omega)$ is the total response there). 
A plot of (\ref{eqn:inertcorrec}) can be found in \cite{ss:svaiter}.

Finally we consider the limit of the response as $\Delta\tau\rightarrow{0}$.
If we take this limit directly, (\ref{eqn:inertcorrec}) shows there is
a divergence. It is interesting however to see whether or not keeping $\epsilon\neq{0}$
regularizes this divergence. That is, is the spatial smearing enough
to regularize this ultraviolet divergence? We restrict ourselves here to
$4$ dimensions.
The transition rate for this finite time detection is
\begin{eqnarray}
\dot{F}_\tau(\omega,\epsilon)
& = &
-\frac{1}{4\pi^{2}}
\int^{\Delta\tau}_{-\Delta\tau}
ds
\,
\frac{e^{-i\omega{s}}}
{(s-2i\epsilon)^{2}} \ ,
\nonumber
\\
& = & 
-\frac{1}{4\pi^{2}}
\left[
\frac{e^{i\omega\Delta\tau}}{(-\Delta\tau-2i\epsilon)}
-
\frac{e^{-i\omega\Delta\tau}}{(\Delta\tau-2i\epsilon)}
\right]
+
\frac{i\omega}{4\pi^2}
\int^{\Delta\tau}_{-\Delta\tau}
ds\,
\frac{e^{-i\omega{s}}}{(s-2i\epsilon)} \ ,
\nonumber
\\
& = & 
-\frac{1}{4\pi^{2}}
\left[
\frac{e^{i\omega\Delta\tau}}{(-\Delta\tau-2i\epsilon)}
-
\frac{e^{-i\omega\Delta\tau}}{(\Delta\tau-2i\epsilon)}
\right]
+
\frac{i\omega}{4\pi^2}
e^{2\epsilon\omega}
\left[
\mathrm{E}_\mathrm{i}(-i\omega\Delta\tau-2\epsilon\omega)
\right.
\nonumber
\\
&  & \left.
-\mathrm{E}_\mathrm{i}(i\omega\Delta\tau-2\epsilon\omega)
\right] \ ,
\label{eqn:finiteinertiallim2}
\\
& = & 
-\frac{1}{4\pi^{2}}
\left[
\frac{e^{i\omega\Delta\tau}}{(-\Delta\tau-2i\epsilon)}
-
\frac{e^{-i\omega\Delta\tau}}{(\Delta\tau-2i\epsilon)}
\right]
+
\frac{i\omega}{4\pi^2}
e^{2\epsilon\omega}
\left[
\mathrm{Ci}(\omega\Delta\tau-2i\epsilon\omega)
\right.
\nonumber
\\
\label{eqn:finiteinertiallim}
& & \left.
-i\mathrm{si}(\omega\Delta\tau-2i\epsilon\omega)
-
\mathrm{Ci}(\omega\Delta\tau+2i\epsilon\omega)-i\mathrm{si}(\omega\Delta\tau+2i\epsilon\omega)
\right] \ ,
\end{eqnarray}
where we have now indicated the explicit dependence on $\epsilon$. The
second line follows by integration by parts, the third from \cite{gr:inttables},
$\mathrm{E}_\mathrm{i}$ is the exponential integral and $\mathrm{Ci}$ and $\mathrm{si}$ are cosine and
shifted sine integrals respectively \cite{as:abromowitz}.
Taking the $\epsilon\rightarrow{0}$ limit of (\ref{eqn:finiteinertiallim})
we recover (\ref{eqn:inertcorrec})$+|\omega|\Theta(-\omega)/(2\pi)$ as 
expected.
Suppose however we keep $\epsilon\neq{0}$ and take the $\Delta\tau\rightarrow{0}$
limit of (\ref{eqn:finiteinertiallim}). We see clearly from (\ref{eqn:finiteinertiallim2})
that in this limit the transition rate vanishes.
Thus the spatial smearing is enough to regularize the transition 
function in this limit.
Similar conclusions hold in the $2$-dimensional case, though we
shall not spell out the calculations here.
Similar investigations (though not the exact case considered here) 
of the $\Delta\tau$ and $\epsilon$ limits of the response function 
on inertial and uniformly accelerated worldlines were made in \cite{sp:srirpad}.

\subsubsection{Uniformly accelerated detector in $d$ dimensions}

Next we consider the transition rate for a spatially extended detector
whose centre follows the uniformly accelerated worldline (\ref{eq:accel}).
Inserting the worldline (\ref{eq:accel}) into (\ref{eq:corre})
gives
\begin{equation}
\label{cronaccel}
\langle{0}|\phi(\tau)\phi(\tau')|0\rangle=
\frac{\Gamma[d/2-1]}{i^{d-2}(4\pi)^{d/2}}
\frac{1}{\left(
\alpha\sinh\left(\frac{\tau-\tau'}{2\alpha}\right)-i\epsilon\cosh\left(\frac{\tau-\tau'}{2\alpha}\right)
\right)^{d-2}} \ .
\end{equation}
As the correlation function is invariant under translations in $\tau$
the transition rate is given by
\begin{equation}
\label{transaccel}
\dot{F}_\tau(\omega)=
\frac{\Gamma[d/2-1]}{i^{d-2}(4\pi)^{d/2}}
\int^{\Delta\tau}_{-\Delta\tau}
ds\,
\frac{e^{-i\omega{s}}}{\left(
\alpha\sinh\left(\frac{s}{2\alpha}\right)-i\epsilon\cosh\left(\frac{s}{2\alpha}\right)
\right)^{d-2}} \ .
\end{equation}
In the case of a detector switched on in the infinite past, $\tau_0=-\infty$,
the integral here may be done by residues (see Appendix \ref{ch:appendixB}).
The transition rate is independent of $\tau$ along the trajectory
and is given by
\begin{eqnarray}
\dot{F}_\tau(\omega)= 
&  &
\frac{\pi}{2^{d-2}\pi^{(d-1)/2}\alpha^{d-3}\Gamma((d-1)/2)}
\nonumber
\\
\label{thermal}
& \times &
\left\{
\begin{array}{ll}
\frac{\alpha\omega}{\left(e^{2\pi\omega\alpha}-1\right)}
\prod_{k=1}^{(d-4)/2}
\left(
\left(\frac{d-2}{2}-k\right)^2
+\alpha^2\omega^2
\right)
&
d\;\; \textrm{even}
\\
\frac{1}{\left(e^{2\pi\omega\alpha}+1\right)}
\prod_{k=1}^{(d-3)/2}
\left(
\left(\frac{d-2}{2}-k\right)^2
+\alpha^2\omega^2
\right)
&
d\;\; \textrm{odd}
\end{array}
\right. \ ,
\end{eqnarray}
where for $d=3$ and $d=4$ the products $\prod_{k=1}^{(d-3)/2}\left(
\left(\frac{d-2}{2}-k\right)^2
+\alpha^2\omega^2
\right)$ and \\
$\prod_{k=1}^{(d-4)/2}\left(
\left(\frac{d-2}{2}-k\right)^2
+\alpha^2\omega^2
\right)$ in (\ref{thermal}) are both $1$.
The corresponding result for $d=2$ is (see appendix \ref{ch:appendixA2})
\begin{equation}
\label{eqn:accelresp2}
\dot{F}_{\tau}(\omega)
=
-\frac{1}{\omega(e^{2\pi\alpha\omega}-1)} \ .
\end{equation}
The transition rate (\ref{thermal}) and (\ref{eqn:accelresp2}) is 
as in the literature \cite{sr:sriramk}.
It is thermal with characteristic temperature $T=1/(2\pi\alpha)$
in the sense that is satisfies the KMS condition
\begin{equation}
\label{eqn:kmscondition}
\dot{F}_\tau(\omega)=
e^{-\omega/T}
\dot{F}_\tau(-\omega) \ ,
\end{equation} 
at that temperature (see e.g \cite{tk:takagi}).
Further (\ref{thermal}) contains the expected ``apparent'' statistics inversion 
as we go from odd to even dimensions.

Takagi has shown \cite{tk:takagi} that the transition rate is related to
the local density of Rindler states where we consider Rindler mode
solutions in the definition (\ref{eqn:localdensity}). The relation reads explicitly
\begin{equation}
\dot{F}_\tau(\omega)=
\frac{\pi}{\omega}
\frac{\rho_R(\omega)}{\left(e^{2\pi\alpha\omega}-1\right)} \ .
\end{equation} 
More details can be found in \cite{tk:takagi}.

The correction term for finite time detections is 
\begin{equation}
C_{\Delta\tau}(\omega)
=
-\frac{\Gamma[d/2-1]}{i^{d-2}(4\pi)^{d/2}\alpha^{d-2}}
\int^{\infty}_{\Delta\tau}
ds
\,
\frac{\cos(\omega{s})}
{\sinh^{d-2}\left(\frac{s}{2\alpha}\right)} \ .
\end{equation}
The correction term is even in $\omega$ and so breaks the KMS
condition (\ref{eqn:kmscondition}). The transition rate for
a detector accelerated for only a finite time is thus not thermal
in this sense.
The effect of using a
smoother switching for finite time detections for inertial
and uniformly accelerated motions was investigated in 
\cite{hm:higuchi,sp:srirpad}.

\subsection{Massive scalar field}
\label{sec-massive}

In this section we compute the correlation function for the smeared field operator (\ref{eq:smear}) for
a massive scalar field. 

The detector model is as in section \ref{sec-linmink}.
The field is expanded in modes as in (\ref{eqn:modeexpan}) but now with $\omega=\left(\mathbf{k}^2+m^2\right)^{1/2}$,
where $m$ is the field mass. The transition rate for the detector is given by (\ref{eq:transition}),
where 
$\phi(\tau)$ is the smeared field operator (\ref{eq:smear}). The correlation 
function is given by the expression (\ref{eq:expan}) and we 
are again led to consider $g_\epsilon(\mathbf{k};\tau)$ as in (\ref{eqn:gfun}). The calculation 
of $g_\epsilon(\mathbf{k};\tau)$
is identical to that leading to (\ref{eqn:gepktau}) and so we find 
$g_\epsilon(\mathbf{k};\tau)=e^{-i(\mathit{k}\cdot\mathit{x}(\tau))}
e^{-\epsilon\mathit{k}\cdot{\dot{\mathit{x}}}}$ and
\begin{equation}
\label{eq:correlationmass}
\langle{0}|\phi(\tau){\phi}(\tau')|0\rangle=
\frac{1}{(2\pi)^{d-1}}
\int
\frac{d^{d-1}k}{2\omega}\,
e^{-i\omega(t-t'-i\epsilon(\dot{t}+\dot{t}'))
+i\mathbf{k}\cdot(\mathbf{x}-\mathbf{x'}-i\epsilon(\dot{\mathbf{x}}+\dot{\mathbf{x}}'))} \ ,
\end{equation}
now with $\omega=\left(\mathbf{k}^2+m^2\right)^{1/2}$.

The integrals here
may again be done by moving to hyperspherical coordinates in $k$-space.
We restrict ourselves now to the case $d=4$ (although the arbitrary $d$ case
follows similarly). After performing the
angular integrals, we find
\begin{equation}
\langle{0}|\phi(\tau)\phi(\tau')|0\rangle
=  
\frac{1}{(2\pi)^2R}
\int_0^\infty
dk\,
\frac{k}{\left(k^2+m^2\right)^{1/2}}
\sin(kR)
e^{-i\left(k^2+m^2\right)^{1/2}{(t-t'-i\epsilon(\dot{t}+\dot{t}'))}} \ ,
\end{equation}
where $R=\sqrt{(\mathbf{x}-\mathbf{x'}-i\epsilon(\dot{\mathbf{x}}+\dot{\mathbf{x}}'))^2}$. This may be written
as
\begin{equation}
\label{eqn:massiinteqn}
\langle{0}|\phi(\tau)\phi(\tau')|0\rangle
=  
\frac{-1}{8\pi^2R}
\partial_R
\int_{-\infty}^\infty
dk\,
\frac{1}{\left(k^2+m^2\right)^{1/2}}
e^{-i(\left(k^2+m^2\right)^{1/2}{(t-t'-i\epsilon(\dot{t}+\dot{t}'))}-kR)}
 \ .
\end{equation}
We now change variables by $k=m\sinh{\theta}$, so that $\omega=m\cosh{\theta}$ and
\begin{equation}
\langle{0}|\phi(\tau)\phi(\tau')|0\rangle
=  
\frac{-1}{8\pi^2R}
\partial_R
\int_{-\infty}^\infty
d\theta\,
e^{im(R\sinh{\theta}-(t-t'-i\epsilon(\dot{t}+\dot{t}'))\cosh{\theta})}
 \ .
\end{equation}

For the detector trajectory we consider only timelike worldlines,
$(t-t')>(\mathbf{x}-\mathbf{x}')$. We distinguish two cases.
Firstly, for $(t-t')>0$, we make the substitution $(t-t'-i\epsilon(\dot{t}+\dot{t}'))=
\sqrt{\lambda}\cosh{\theta_0}$, $R=\sqrt{\lambda}\sinh{\theta_0}$, with
$\lambda=(t-t'-i\epsilon(\dot{t}+\dot{t}'))^2-R^2$. We find
\begin{equation}
\langle{0}|\phi(\tau)\phi(\tau')|0\rangle
=  
\frac{-1}{8\pi^2R}
\partial_R
\int_{-\infty}^\infty
d\theta\,
e^{-im\sqrt{\lambda}\cosh(\theta_0-\theta)}
 \ .
\end{equation} 
Now we note
\begin{equation}
\label{eqn:K0}
K_0(z)=\frac{1}{2}\int_{-\infty}^{\infty}dt\,
e^{-z\cosh{t}} \ ,
\end{equation}
valid for $Re(z)>0$ \cite{gr:inttables}, where $K_0$ is a modified Bessel 
function, and we may show for a timelike worldline that $Im(\sqrt{\lambda})<0$.
Hence
\begin{eqnarray}
\langle{0}|\phi(\tau)\phi(\tau')|0\rangle
& = &   
\frac{-1}{4\pi^2R}
\;\partial_R\left[
K_0\left(im\sqrt{\lambda}\right)\right]
\nonumber
\\
\label{eqn:masscorr}
& = &
-\frac{im}{4\pi^2\sqrt{\lambda}}
\;K_1\left(im\sqrt{\lambda}\right) \ ,
\end{eqnarray} 
as $\partial_z{K_0(z)}=K_1(z)$.
The case $(t-t')<0$ is similar except we make the
change of variables $(t-t'-i\epsilon(\dot{t}+\dot{t}'))=
-\sqrt{\lambda}\cosh{\theta_0}$, $R=\sqrt{\lambda}\sinh{\theta_0}$
again we may use the integral representation of $K_0$ and we obtain the
same result (\ref{eqn:masscorr}).

The massless limit is easily checked. We have near $z=0$, $K_1(z)=1/z$
\cite{gr:inttables} and so the correlation function (\ref{eqn:masscorr})
agrees with that of Schlicht in this limit.
Had we used the field operator without smearing we would have obtained (\ref{eqn:massiinteqn}) with
$\epsilon=0$. The usual regularization procedure, as with the massless
case, would then be to introduce a cut-off in the high frequency modes by $t\rightarrow{t-i\eta}$, where
$\eta$ is small. The result would be (\ref{eqn:masscorr}) but with
$\lambda=(t-t'-i\eta)^2-|\mathbf{x}-\mathbf{x}'|^2$.

\subsection{The regularization as an ultraviolet cut-off}
\label{sec-ultra}

In this section we show that the regularization discussed in sections \ref{sec-linmink} and \ref{sec-massive}
may be viewed as the introduction of an ultraviolet cut-off into the integral expression for the
correlation function.

Suppose we begin with a normal monopole detector model like that of DeWitt. The
transition rate is given by (\ref{eq:transition}), and the correlation function is
\begin{equation}
\label{eq:correlationmon}
\langle{0}|\phi(\tau){\phi}(\tau')|0\rangle=
\frac{1}{(2\pi)^{d-1}}
\int
\frac{d^{d-1}k}{2\omega}
\,
e^{-i\omega(t(\tau)-t(\tau'))
+i\mathbf{k}\cdot(\mathbf{x}(\tau)-\mathbf{x}(\tau'))} \ .
\end{equation}
This expression requires regularization. The usual $i\epsilon$ procedure
is to introduce an ultraviolet cut-off in the modes by including
in the integrand the factor $e^{-\epsilon\omega}$. This means that 
high frequency modes as seen by an inertial observer are 
exponentially suppressed.
As discussed in section \ref{sec-linmink} this leads to an unphysical time-dependent result in the
case where the uniformly accelerated detector is switched on in the infinite past.
Let us instead consider an alternative ultraviolet cut-off where it is
the high frequency modes as seen by the detector which are suppressed.
Motivated by the fact that, as Unruh puts it \cite{u:unruh} ``A particle
detector will react to states which have positive frequency with respect to
the detectors proper time, not with respect to any universal time.''.
Here the frequency is the time component of the $d$-momentum. We want
therefore this frequency component in the rest frame of the detector,
that is the frame which is Fermi-Walker transported along the trajectory
with basis vectors $\{u(\tau),\mathbf{e}_{\xi_1}(\tau),\ldots,\mathbf{e}_{\xi_{d-1}}(\tau)\}$
given the $d$-momentum $(\omega,k_{x_1},\ldots,k_{x_{d-1}})$ in the usual Minkwoski
frame. The two frames are of course linked by a Lorentz transformation, and a simple
calculation reveals that the frequency component in the Fermi-Walker frame at time $\tau$
is $\omega'(\tau)=\omega\dot{t}(\tau)-\mathbf{k}\cdot\dot{\mathbf{x}}(\tau)$.
Therefore if we introduce an ultraviolet cut-off in these frequencies at $\tau$ and
$\tau'$ into (\ref{eq:correlationmon}), that is we add a factor of
$e^{-\epsilon(\omega(\dot{t}(\tau)+\dot{t}(\tau'))-\mathbf{k}\cdot(\dot{\mathbf{x}}(\tau)+\dot{\mathbf{x}}(\tau')))}$
and compare with (\ref{eq:correlation}) we see that the resulting correlation function is
that of Schlicht.

Another way to introduce this cut-off is to define operationally
the frequency as seen by the moving detector by $i\frac{d\phi}{d\tau}=\omega'\phi$,
where $\tau$ is the detector's proper time.
A mode $\phi\propto{e^{-i\omega{t}+i\mathbf{k}\cdot\mathbf{x}}}$ with frequency
$\omega$ in the standard Minkowski frame thus has frequency 
$\omega'=\omega\dot{t}(\tau)-\mathbf{k}\cdot\dot{\mathbf{x}}(\tau)$ as seen by the detector.

We have obtained an appropriate regularization of the 
correlation function without 
introducing a detector with infinite spatial extent, that is without smearing the 
field operator in the interaction Hamiltonian. We will be using this reasoning 
as the definition of a regularization in de Sitter space in section \ref{sec-desitrp3}.

\subsection{An alternative detector form}
\label{sec-detectform}

In this section we discuss the effect of an alternative shape for the
detector. In particular we discuss 
a Gaussian detector profile, instead of (\ref{eq:shape}). We show that again
the regularization may be viewed as a cut-off in the high frequency modes as
seen by the detector. 

The model is that of the previous sections with the smeared field operator in the
interaction Hamiltonian given by (\ref{eq:smear}). However now we consider
a Gaussian detector shape
\begin{equation}
\label{eq:shapegua}
W_{\epsilon}(\mathbf{\xi})
=
\frac{1}{\epsilon^{d-1}(2\pi)^{(d-1)/2}}
e^{-\frac{\mathbf{\xi}^2}{2\epsilon^2}} \ .
\end{equation}
This shape function also approximates a $(d-1)$-dimensional $\delta$ function, since it scales as 
\begin{equation}
W_{\epsilon}(\mathbf{\xi})
=
\frac{1}{\epsilon^{d-1}}
W_1(\mathbf{\xi}/\epsilon) \ ,
\end{equation}
and satisfies
\begin{equation}
\int
d^{d-1}\xi\,
W_{\epsilon}(\mathbf{\xi})
=1 \ .
\end{equation}
The associated correlation function again takes the form (\ref{eq:expan}). Following
the arguments after (\ref{eq:expan}), we arrive at
\begin{equation}
\label{eqn:ggaussian}
g_\epsilon(\mathbf{k};\tau)
=
\frac{1}{\epsilon^{d-1}(2\pi)^{(d-1)/2}}
e^{-i(\mathit{k}\cdot\mathit{x}(\tau))}
\int
d^{d-1}\xi\,
e^{-\frac{\mathbf{\xi}^2}{2\epsilon^2}}
e^{-i\bar{\mathbf{k}}\cdot\mathbf{\xi}} \ .
\end{equation}
The integral may be done by transforming to hyperspherical 
coordinates as described in Appendix \ref{ch:appendixA}.
We find
\begin{eqnarray}
g_\epsilon(\mathbf{k};\tau)
& = &
e^{-i(\mathit{k}\cdot\mathit{x}(\tau))}
e^{-\frac{\epsilon^2|\bar{\mathbf{k}}|^2}{2}} \ ,
\nonumber
\\
& = & 
e^{-i(\mathit{k}\cdot\mathit{x}(\tau))}
e^{-\frac{\epsilon^2|\mathit{k}\cdot\dot{\mathit{x}}(\tau)|^2}{2}} \ ,
\end{eqnarray} 
and the correlation function becomes
\begin{eqnarray}
\langle{0}|\phi(\tau){\phi}(\tau')|0\rangle
& = &
\frac{1}{(2\pi)^{d-1}}
\int
\frac{d^{d-1}k}{2\omega}\,
e^{-i\omega(t-t')
+i\mathbf{k}\cdot(\mathbf{x}-\mathbf{x'})}
\nonumber
\\
&   &\times
e^{-\frac{\epsilon^2|\omega\dot{t}(\tau)-\mathbf{k}\cdot\dot{\mathit{x}}(\tau)|^2}{2}}
e^{-\frac{\epsilon^2|\omega\dot{t}(\tau')-\mathbf{k}\cdot\dot{\mathit{x}}(\tau')|^2}{2}}
 \ .
 \label{eqn:gauscorrfun}
\end{eqnarray}

Comparing (\ref{eqn:gauscorrfun}) with the discussion of section \ref{sec-ultra}, we see that the
regularization 
is again equivalent to introducing an ultraviolet cut-off in the modes as seen by the
moving detector: only now the cut-off at time $\tau$ takes the form
$e^{-\frac{\epsilon^2\omega'^2}{2}}$ where $\omega'$
is the frequency of the Minkowski modes as seen from the Fermi-Walker 
transported frame.
This result is expected to be useful in any thorough investigation of
the dependence (or otherwise) of the response on the shape of the detector.
It is clear that for a detector with extent $\epsilon$ the response will
be dependent on the shape, and my intuition is that this dependence will
disappear in the $\epsilon\rightarrow{0}$ limit, that is in this limit
all shape functions will lead to the same response. This however is
still an open question.

\section[Quotients of Minkowski space]{Linearly coupled scalar field detector on quotients of Minkowski space}
\label{sec:automorphic}

In this section we adapt the detector model of section \ref{sec-linmink} to 
spacetimes built as quotients $M/\Gamma$ of Minkowski
space under certain discrete symmetry groups $\Gamma$.
In particular we calculate responses
on $M_0$, $M_-$, Minkowski space with an infinite plane boundary 
and certain conical spacetimes.

We notice immediately that there is a problem in directly applying
the methods of section \ref{sec-linmink} to these spacetimes. As these
quotient spaces do not have infinite spatial sections in all directions,
it does not directly make sense to consider 
a detector with infinite spatial extent as used in (\ref{eq:smear}).
We shall argue however that we may introduce a detector similar to that of 
section \ref{sec-linmink} by working with automorphic fields on $M$~\cite{bD:auto,bd:em}.
That is, we consider fields on the quotient spaces as fields on $M$ satisfying
certain transformation properties under the action of the discrete group.

\subsection{The automorphic field detector}
\label{sec:autodetector}

Consider Minkowski space $M$ in $d$ dimensions, and consider the quotient 
space $M/\Gamma$ where $\Gamma$ is some discrete isometry group. $|\Gamma|$
may be infinite (as indeed is the case on $M_0$ and $M_-$) which will
mean that some of the following expressions remain formal in those cases.
We will find however that these formalities do not interfere as in any calculations
done the infinities and the formally vanishing normalization factors will cancel to
give finite results.

The automorphic 
field $\hat{\phi}$ is constructed from the ordinary field ${\phi}$ as the sum
\begin{equation}
\label{eqn:autofield}
\hat{\phi}(x)
:=
\frac{1}{\left(\sum_{\gamma\in\Gamma}p(\gamma)^2\right)^{1/2}}
\sum_{\gamma\in\Gamma}p(\gamma)\phi(\gamma^{-1}x) \ ,
\end{equation}
where $p(\gamma)$ is a representation of $\Gamma$ in $SL(\R)\simeq\{1,-1\}$.
The normalization in (\ref{eqn:autofield}) has been chosen so that, at equal times
(see appendix \ref{ch:appendixd})
\begin{equation}
\left[\hat{\phi}(x),\dot{\hat{\phi}}(x')\right]
=
i\delta^{(d-1)}(x-x')
+\textrm{image terms} \ .
\end{equation}
The two point function for the automorphic field is then given by the 
method of images as
\begin{equation}
\label{eqn:automorcorre}
\langle{0}|\hat{\phi}(x)\hat{\phi}(x')|0\rangle
=
\sum_{\gamma\in\Gamma}p(\gamma)\langle{0}|\phi(x)\phi(\gamma^{-1}x')|0\rangle \ ,
\end{equation}
where $\langle{0}|\phi(x)\phi(x')|0\rangle$ is the usual correlation 
function on Minkowski space.

As a model of a particle detector on $M/\Gamma$, we introduce on $M$ a
detector linearly coupled to the automorphic field by
\begin{equation}
\label{eqn:hamiauto}
H_{\mathrm{int}}=
c
m(\tau)
\hat{\phi}(\tau) \ ,
\end{equation}
with\footnote{If we considered here $\hat{\phi}(\tau)
=\hat{\phi}(x(\tau))$ with the usual $i\epsilon$ regularization,
again we would find as Schlicht does in Minkowski space an unphysical result
for the response of the uniformly accelerated detector on these spacetimes. 
This is most easily seen by considering that
the $\gamma=I$ term (where $I$ is the identity element) in (\ref{eqn:automorcorre}) 
is that found in Minkowski space.}
\begin{equation}
\label{eqn:autointham}
\hat{\phi}(\tau)
=
\int
d^{d-1}\xi
\,
W_{\epsilon}(\mathbf{\xi})
\hat{\phi}(x(\tau,\mathbf{\xi})) \ .
\end{equation}
One might ask why we have not included in (\ref{eqn:hamiauto}) image terms under
$\Gamma$ (that is one term for each image of the detector). There are two obvious ways
in which such terms could be included. Firstly we could consider each image
term in the sum to be weighted by the representation $p(\gamma)$. In certain situations, such as
on Minkowski space with an infinite plane boundary with Dirichlet boundary conditions
this would however lead to the detector and its image terms cancelling each other
to give a vanishing interaction Hamiltonian. 
Alternatively we could consider image terms without the representation weights.
Then each image term would be equal to that in (\ref{eqn:hamiauto}) and so we 
would obtain the same results with an overall (possibly infinite) normalisation factor, which can 
be absorbed in the coefficient $c$. We therefore work with (\ref{eqn:hamiauto}).

From (\ref{eqn:hamiauto}), a discussion analogous to that in section \ref{sec-linmink}
leads to the transition rate 
(to first order in perturbation theory)
\begin{equation}
\label{eqn:autotrans}
\dot{F}_{\tau}(\omega)=
2\int^{\infty}_{0}ds
\,
Re
\left(
e^{-i\omega{s}}\langle{0}|\hat{\phi}(\tau)\hat{\phi}(\tau-s)|0\rangle
\right) \ ,
\end{equation}
where the correlation function for
the automorphic field in (\ref{eqn:autotrans}), $\langle{0}|\hat{\phi}(\tau)\hat{\phi}(\tau-s)|0\rangle$,
is given by the method of images applied to (\ref{eq:corre})
(see appendix \ref{ch:appendixd}).
Explicitly we have
\begin{equation}
\langle{0}|\hat{\phi}(\tau)\hat{\phi}(\tau')|0\rangle
=
\sum_{\gamma\in\Gamma}p(\gamma)\langle{0}|\phi(\tau)\phi(\gamma^{-1}\tau')|0\rangle \ .
\end{equation}

\subsection{Scalar detector on $M_0$}

Recall that $M_0$ is a quotient of $4$-dimensional Minkowski space where the quotienting group
$\Gamma$ is that generated by the isometry $J_0:(t,x,y,z)\mapsto(t,x,y,z+2a)$.
The transition rate for our detector is given by
(\ref{eqn:autotrans}) with 
\begin{equation}
\label{eqn:corronm0}
\langle{0}|\hat{\phi}(\tau)\hat{\phi}(\tau')|0\rangle=\sum_{n\in{\Z}}\eta^n\langle{0}|\phi(\tau)\phi(J_0^n\tau')|0\rangle
 \ ,
\end{equation}
where $\eta=+1,(-1)$ are the representations of $\Gamma$, labelling untwisted (twisted)
fields respectively. 

\subsubsection{Inertial detector on $M_0$}

Consider first a detector following the inertial trajectory
\begin{eqnarray}
t=\tau(1-v^2)^{-1/2} \ , & z=\tau{v}(1-v^2)^{-1/2}  \ ,\nonumber \\
x=x_0  \ , & y=y_0 \ ,  
\end{eqnarray}
where velocity $-1<v<0$, $-\infty<\tau<\infty$ is the detector's proper time and 
$x_0$, $y_0$ are constants.
Substituting the trajectory into (\ref{eqn:corronm0}) and 
then (\ref{eqn:autotrans}), we find that
the transition rate for a detector switched on in the 
infinite past 
reads
\begin{equation}
\dot{F}_\tau(\omega)=
-\frac{1}{4\pi^2}
\sum_{n=-\infty}^{\infty}
\eta^n
\int^\infty_{-\infty}
ds
\,
\frac{e^{-i\omega{s}}}{(s-2i\epsilon)^2+4nav(s-2i\epsilon)(1-v^2)^{-1/2}-(2na)^2}
\ .
\end{equation}
The integral may be done by residues. The result is
\begin{equation}
\dot{F}_\tau(\omega)=
-\frac{(1-v^2)^{1/2}}{4\pi{a}}
\sum_{n=-\infty}^{\infty}
\eta^n
\frac{\sin(2\omega{n}a(1-v^2)^{-1/2})}{n}
e^{\frac{2\omega{n}{a}vi}{(1-v^2)^{1/2}}}\Theta(-\omega) \ .
\end{equation}
As on Minkowski space the transition rate vanishes for 
$\omega>0$ while it is non-zero for $\omega<0$ although the
rate of spontaneous emission is altered from the Minkowski rate
due to the non-trivial topology.
The transition rate depends on velocity $v$ due to the absence of
a boost Killing vector in the $z$ direction on $M_0$.

\begin{figure}[htbp]
\includegraphics[angle=0, width={3in}]{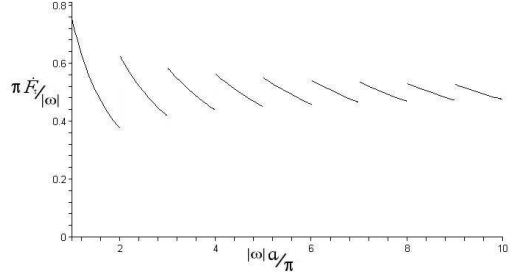}
\centering
\caption[Transition rate of inertial detector on $M_0$ for an untwisted field and $\omega<0$.]{Transition rate of an inertial detector 
on $M_0$ for an untwisted scalar field.
We have taken $\omega<0$ and plotted
$\pi\dot{F}_{\tau}(\omega)/|\omega|$ against $|\omega|a/\pi$.}
\label{fig:M0untwisted}
\end{figure}

Consider now the limit as $v\rightarrow{0}$. By the
isometries of $M_0$ this gives the response of an inertial 
detector that may have arbitrary velocity in the $x$ or $y$ directions.
Then
\begin{equation}
\label{eqn:transm0scalarinert}
\dot{F}_\tau(\omega)=
\left(
-\frac{\omega}{2\pi}
+\frac
{1}
{2\pi{a}}
\sum_{n=1}^{\infty}
\eta^n
\frac{1}
{n}
\sin(-2n\omega{a})
\right)
\Theta(-\omega) \ .
\end{equation}
In the case 
of an untwisted field, $\eta=1$, the summation in (\ref{eqn:transm0scalarinert}) is recognized as 
the Fourier series of the $2\pi$-periodic
function that on the
interval $(0,2\pi)$ takes the form $f(-2\omega{a})=\frac{1}{2}(\pi+2\omega{a})$ (see e.g. \cite{zw:zwillinger}).
We hence find
\begin{equation}
\label{eqn:inerm0}
\dot{F}_\tau(\omega)=
\frac
{\left(\left[\frac{-\omega{a}}{\pi}\right]+\frac{1}{2}\right)}
{2a}
\Theta(-\omega) \ ,
\end{equation}
where $[x]$ denotes the integer part of $x$. 
We plot $\pi\dot{F}_\tau/|\omega|$ against $|\omega|{a}/\pi$ 
for $\omega<0$ in figure \ref{fig:M0untwisted}.
For a twisted field, $\eta=-1$, we note
$(-1)^n\sin(nx)=\sin(n(x+\pi))$ and find
\begin{equation}
\label{eqn:inerm2}
\dot{F}_\tau(\omega)=
\frac
{\left[\frac{-\omega{a}}{\pi}+\frac{1}{2}\right]}
{2a}
\Theta(-\omega) \ .
\end{equation}

\subsubsection{Uniformly accelerated detector}

If we consider a detector following the worldline of uniform 
acceleration (\ref{eq:accel}) we obtain, again for the detector
switched on at $\tau_0=-\infty$,
\begin{equation}
\label{M0response}
\dot{F}_\tau(\omega)=\frac{\omega}{2\pi(e^{2\pi\alpha\omega}-1)}
\left(
1+
\sum_{n=1}^{\infty}
\eta^n
\frac{\sin{[2\alpha\omega{\mathrm{arc}}\sinh{(na/\alpha)}]}}{na\omega\sqrt{1+n^2a^2/\alpha^2}}
\right) \ ,
\end{equation}
where the integral in the transition rate has again been done by residues.
The result agrees with that obtained by Louko and Marolf in \cite{lm:geon}.
The response is independent of $\tau$. Moreover it is thermal in the
sense that it satisfies the KMS condition (\ref{eqn:kmscondition}).
There is however a break
from the purely Planckian form found on Minkowski space due to the
image sum.

\subsection{Minkowski space with an infinite plane boundary}
\label{sec:minkbound}

In this subsection we consider a detector on $d$-dimensional Minkowski space\footnote{Throughout
this section we consider $d>2$.}, with 
Minkowski coordinates $(t,x_1,\ldots,x_{d-1})$,
with an infinite boundary at $x_1=0$. 
Detectors on this spacetime (in particular in the case 
of $4$-dimensions)
have been considered by a number of authors (see e.g. \cite{dlo:detecbound,sz:suzuki})\footnote{The case of 
Dirichlet boundary conditions for $d=4$ has been the focus
of some study recently on the response of detectors to negative energy~\cite{do:daviesott}.
The authors of \cite{do:daviesott} considered the response of a finite time detector travelling inertially
parallel to the boundary. They found that the negative energy outside the boundary (\ref{eqn:tmunu})
has the effect of decreasing the excitations which are present even in Minkowski space 
due to the switching of the detector.}.
However there has not been any presentations (as far as the author is aware)
of the time dependent response for an inertial or uniformly accelerated observer
who approaches the boundary from infinity.
The results in this section are interesting also as a preliminary calculation 
before the response of detectors on $M_-$ is considered. Many of the features
seen here will also be observed there, but in a simpler context as there is no compact
direction and therefore the image sum is finite.
Further we will see in section \ref{sec:rp3} that the response of a uniformly accelerated
detector on $4$-dimensional Minkwoski space with an infinite plane boundary 
is very closely related to the response of an 
inertial detector coupled to a conformal scalar field
in $\rp$ de Sitter spacetime~\cite{jk:desit}.

Again the discussion of 
section \ref{sec:autodetector} follows through and the 
transition rate is given by (\ref{eqn:autotrans})
where now the automorphic correlation function is
\begin{equation}
\label{eqn:correonbound}
\langle{0}|\hat{\phi}(\tau)\hat{\phi}(\tau')|0\rangle
=
\sum_{n=0,1}\beta^n\langle{0}|\phi(\tau)\phi(J_b^n\tau')|0\rangle \ ,
\end{equation}
where
$J_b:(t,x_1,x_2,\ldots,x_{d-1})\mapsto(t,-x_1,x_2,\ldots,x_{d-1})$ and
$\beta=+1,(-1)$, which label Neumann and (Dirichlet) boundary conditions
respectively.
We note here that on $4$-dimensional Minkowski space with boundary at $x=0$ 
the renormalized expectation values $\langle{0}|T_{\mu\nu}|0\rangle$
of the energy-momentum tensor in the vacuum state induced by the Minkowski vacuum
were given in section \ref{sec:emonmwithbound}. For the minimally coupled massless 
scalar field they reduce
to
\begin{equation}
\label{eqn:tmunu2}
\langle0|T_{tt}|0\rangle
=
-\langle0|T_{yy}|0\rangle
=
-\langle0|T_{zz}|0\rangle
=\beta\frac{1}{16{\pi^2}x^4}
 \ ,\;\;
\langle0|T_{xx}|0\rangle
=0 \ .
\end{equation}

\subsubsection{Inertial detector}

\begin{figure}[htbp]
\includegraphics[angle=0, width={3in}]{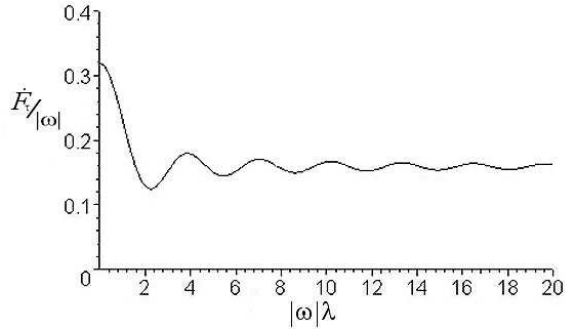}
\centering
\caption[Transition rate of inertial detector on $M$ with boundary, motion parallel to boundary, $\omega<0$.]{Transition rate for inertial detector moving parallel to the boundary, as
a function of distance $\lambda$ from the boundary. We have taken $\omega<0$ and plotted
$\dot{F}_{\tau}(\omega)/|\omega|$ against $|\omega|\lambda$ for
Neumann boundary conditions.}
\label{fig:inertialparal}
\end{figure}

Firstly we consider an inertial detector with motion
parallel to the boundary. Due to the isometries of the
spacetime we may consider, without loss of generality, the trajectory
$t=\tau$, $x_1=\lambda$, $x_i=0$ for $1<i\leq{d-1}$
where $\lambda$, the distance from the boundary, is constant.
The transition rate contains two terms. The first one comes 
from the first term in the boundary space correlation
function (\ref{eqn:correonbound}) and equals the corresponding 
transition rate on Minkowski space. For a detector
switched on in the infinite past this is given by (\ref{eqn:inertrespons}).
The second term, comes from the image term in (\ref{eqn:correonbound}),
and is given by
\begin{equation}
\dot{F}_{B\tau}(\omega)=
\frac{\beta\Gamma(d/2-1)}
{4\pi^{d/2}}
\int_{-\infty}^{\infty}
ds\,
\frac{e^{-i\omega{s}}}
{[i^2(s-2i\epsilon)^2+(2\lambda)^2]^{d/2-1}} \ .
\label{eq:boundparttrans}
\end{equation}
The integral in (\ref{eq:boundparttrans}) can be evaluated by residues.
Specialising to $4$ dimensions, the total transition rate including the Minkowski 
part is found to be
\begin{equation}
\label{eqn:inertialparallel}
\dot{F}_{\tau}(\omega)=
\left(-\frac{\omega}{2\pi}
-\frac{\beta}
{4\pi\lambda}
\sin(2\omega\lambda)
\right)
\Theta(-\omega) \ .
\end{equation}
In figure \ref{fig:inertialparal} we plot $\dot{F}_{\tau}(\omega)/|\omega|$ against
$|\omega|\lambda$ for Neumann boundary conditions.
On the boundary the rate
is twice that in Minkowski space, while far from the boundary the rate becomes that
on Minkowski space.
For Dirichlet boundary conditions the transition rate vanishes on the
boundary as expected.
Our results agree with those in~\cite{dlo:detecbound}.

As on Minkowski space the response can be related to the
local density of states and the difference from the Minkowski result can be
attributed to the difference in the density of states. For example in the
case of $4$-dimensional Minkowski space with Dirichlet boundary
conditions on the boundary we have mode solutions to the Klein
Gordon equation of the form $u_{\mathbf{k}}(\mathbf{x})=\frac{e^{-i\omega{t}}}{(4\pi^3\omega)^{1/2}}
{e^{i(k_yy+k_zz)}}\sin(k_xx)$. From (\ref{eqn:localdensity}) the density of states is thus
\begin{equation}
\rho_B(\omega,\mathbf{x})
=
\int
\frac{d^3k}{2\pi^3}\,\sin^2(k_xx)\delta(|\mathbf{k}|-\omega) \ .
\end{equation}
The integral may be done by moving to spherical coordinates,
with the result 
\begin{equation}
\rho_B(\omega,\mathbf{x})
=\frac{\omega^2}{2\pi^2}
\left(
1-
\frac{\sin(2\omega{x})}{2\omega{x}}
\right)\Theta(\omega) \ .
\end{equation}
Hence $\dot{F}_{\tau}(\omega)=-\frac{\pi}{\omega}\Theta(-\omega)\rho_B(|\omega|,\mathbf{x})|_{x=\lambda}$.

Inertial detectors with motion parallel to the boundary and
switched on at a finite time are considered in \cite{fsv:fordsvaitlyra,do:daviesott}.

Next, restricting ourselves to $d=4$ again, we consider an 
inertial detector approaching the boundary from 
infinity following the worldline 
\begin{eqnarray}
\label{eqn:inertraj}
t=\tau(1-v^2)^{-1/2} \ , & x=\tau{v}(1-v^2)^{-1/2}  \ ,\nonumber \\
y=y_0  \ , & z=z_0 \ ,  
\end{eqnarray}
with $-1<v<0$ and $-\infty<\tau<0$.
We expect the response in this case
to be dependent on proper time $\tau$, as the boundary breaks the
translation invariance of Minkowski space in the $x$-direction.
 
Again the transition rate is in two parts. The 
Minkowski part (the $n=0$ term) will lead again to the Minkowski space rate
(\ref{eqn:inertrespons}). For a detector switched on the the infinite past
this part reads
\begin{equation}
\dot{F}_{M\tau}(\omega)=
-\frac{\omega}
{2\pi}
\Theta(-\omega) \ ,
\end{equation}
where $\Theta(-\omega)$ is a step function, and $M$ denotes
that this is the Minkowski term. The Minkowski term of course 
is independent of $\tau$.
It is the image term in the correlation function which leads to a $\tau$-dependent 
result. The image part of the transition rate is
\begin{equation}
\label{eqn:boundpartmboundinet}
\dot{F}_{B\tau}(\omega)=
-\frac{\beta}{2\pi^2}
(1-v^2)
\lim_{\epsilon\rightarrow{0}}
\int^{\infty}_{0}ds\,
Re
\left(
\frac
{e^{-i\omega{s}}}
{(s-2i\epsilon)^2-v^2({s-2\tau})^2}
\right) \ .
\end{equation}
The integral may be evaluated, using some contour arguments, in terms
of sine and cosine integrals (see 
appendix \ref{ch:appendixC}). We find
\begin{eqnarray}
\dot{F}_\tau(\omega) & = & -\frac{\beta}{2\pi^2(b+c)}
(
-\mathrm{Ci}(b|\omega|)\cos(b|\omega|)
-\mathrm{si}(b|\omega|)\sin(b|\omega|)
\nonumber
\\
\label{eqn:inertonbound}
&  & +\mathrm{Ci}(c|\omega|)\cos(c|\omega|)
+\mathrm{si}(c|\omega|)\sin(c|\omega|)
+2\pi\sin(b\omega)\Theta(-\omega)
) \ ,
\end{eqnarray}
where $\mathrm{Ci}$, $\mathrm{si}$ are the cosine and shifted sine integrals 
\cite{as:abromowitz}, $b=2v\tau/(1+v)$ and $c=2v\tau/(1-v)$.

\begin{figure}[htbp]
\includegraphics[angle=0, width={3in}]{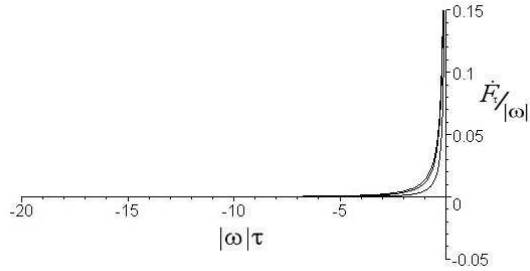}
\centering
\caption[Transition rate of inertial detector on $M$ with boundary, motion perpendicular to boundary, $\omega>0$.]{Transition rate for inertial detector approaching boundary with Neumann boundary conditions
and $\omega>0$.
$\dot{F}_\tau(\omega)/|\omega|$ is plotted against $|\omega|\tau$ 
for $v=-1/2$ (lower curve near the axis), $v=-1/3$ and $v=-1/4$ (upper curve).}
\label{fig:inertialwg0}
\end{figure}

\begin{figure}[htbp]
\includegraphics[angle=0, width={3in}]{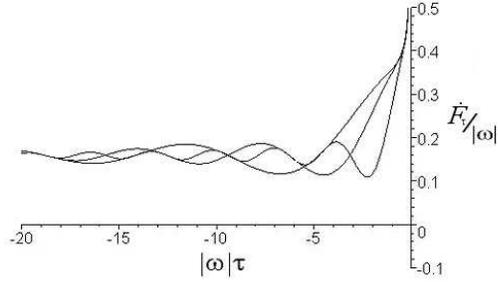}
\centering
\caption[Transition rate of inertial detector on $M$ with boundary, motion perpendicular to boundary, $\omega<0$.]{Transition rate for inertial detector approaching boundary with Neumann boundary conditions
and $\omega<0$.
$\dot{F}_\tau(\omega)/|\omega|$ is plotted against $|\omega|\tau$ for
$v=-1/2$ (lower curve), $v=-1/3$ and $v=-1/4$ (upper curve).}
\label{fig:inertialwl0}
\end{figure}

In figure \ref{fig:inertialwg0} we plot 
$\dot{F}_\tau(\omega)/|\omega|$ against $|\omega|\tau$ for $\omega>0$, Neumann boundary conditions 
and for various values of $v$.
We note that for $\omega>0$ the transition rate is non-zero, in contrast 
to the response of an inertial detector travelling parallel to the boundary,
and diverges as the boundary is reached.
For $\omega<0$ recall that the Minkowski part of the transition rate is non zero.
We plot the total rate for $\omega<0$ in figure \ref{fig:inertialwl0}.
In both cases the response depends on the velocity, as expected since
there is no boost isometry in the $x$-direction.
Further in both cases we may show that the divergence at $x=0$ goes as 
$1/\tau$ and so is weaker than that in the energy expectation values (\ref{eqn:tmunu2}).
It can also be verified that the transition rate dies off at $\tau=-\infty$ as $O(1/\tau^3)$.
Further numerical evidence suggests that the divergences persist
for a detector that is switched on at a finite time.

\subsubsection{Uniformly accelerated detectors}

Consider now a uniformly accelerated detector with acceleration parallel to the boundary and
switched on in the infinite past. We may consider
without loss of generality the worldline
\begin{eqnarray}
t & = & \alpha\sinh(\tau/\alpha) \ ,
\nonumber
\\
x_1 & = & \lambda \ ,
\nonumber
\\
x_2 & = & \alpha\cosh(\tau/\alpha) \ ,
\nonumber
\\
x_i & = & 0 \ , \;\;\;\;\;\;\;\;\;\;\;\;\;\;\;\;2<i\leq{d-1} \ .
\label{eqn:accelboundparr}
\end{eqnarray}

The response again is in two parts.
The first term in (\ref{eqn:correonbound}) leads to the
thermal transition rate an accelerated detector on Minkowski space
(\ref{thermal}). 
Restricting to four dimensions, the boundary part of the transition rate is
\begin{equation}
\label{eqn:boundaryaccelparra}
\dot{F}_{B\tau}(\omega)=
-\frac{\beta}
{4\pi^2}
\int^{\infty}_{-\infty}ds\,
\frac{e^{-i\omega{s}}}
{\left(
4\left(\alpha\sinh(\frac{s}{2\alpha})-i\epsilon\cosh(\frac{s}{2\alpha})\right)^2
-(2\lambda)^2
\right)} \ .
\end{equation}
The integral can be done by residues. 
The result is
\begin{equation}
\label{eqn:boundaryaccelparra2}
\dot{F}_{B\tau}(\omega)=
\frac{\beta}
{4\pi}
\frac{\alpha}{\lambda(\alpha^2+\lambda^2)^{1/2}}
\frac{1}{(e^{2\pi\omega\alpha}-1)}
\sin(2\omega\alpha\mathrm{arcsinh}(\lambda/\alpha)) \ 
\end{equation}
which agrees with~\cite{dlo:detecbound}.
The response is thermal in the sense that it satisfies the KMS
condition at temperature $T=(2\pi\alpha)^{-1}$.
In \cite{dlo:detecbound} the authors further show that the response is
related to the local density of Rindler states here.

Now let us consider the uniformly accelerated worldline (\ref{eq:accel}).
The acceleration is now perpendicular to the boundary.
We begin by considering the detector switched on in the
infinite past. 
The Minkowski part of the correlation function again leads
to the thermal response (\ref{thermal}). The image term
on worldline (\ref{eq:accel}) gives
\begin{equation}
\label{eqn:imtermperpacc}
\langle{0}|\phi(\tau)\phi(J_B\tau')|0\rangle
=
\frac{\beta\Gamma[d/2-1]}
{4\pi^{d/2}\left((4\alpha^2+4\epsilon^2)\cosh^2\left(\frac{\tau+\tau'}{2\alpha}\right)\right)^{d/2-1}} \ ,
\end{equation}
where again $\beta=1,(-1)$ labels Neumann (Dirichlet) boundary conditions. 
It may be argued by the
dominated convergence theorem that the $\epsilon$ can be dropped when calculating the 
transition rate. The geometrical reason is
that the worldline and its image under $J_B$ are totally spacelike
separated, and so the correlation function required in the
transition rate contains no divergences in the integration region.

The image term of the transition rate is thus
\begin{equation}
\label{eqn:boundaryaccel}
\dot{F}_{B\tau}(\omega)=
\frac{\beta\Gamma[d/2-1]}
{2\pi^{d/2}}
\int^{\infty}_{0}ds\,
\frac{\cos(\omega{s})}
{\left(2\alpha\cosh\left(\frac{2\tau-s}{2\alpha}\right)\right)^{d-2}} \ .
\end{equation}
We see immediately that this part of the transition rate is even in $\omega$ and hence
the boundary term breaks the KMS condition (\ref{eqn:kmscondition}). In this sense
the response is non-thermal and non-Planckian.

\begin{figure}[htbp]
\includegraphics[angle=0, width={3in}]{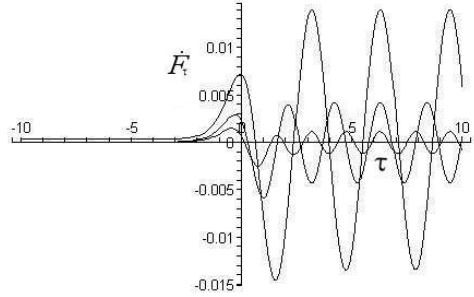}
\centering
\caption[Transition rate of uniformly accelerated detector switched on at $\tau_0=-\infty$ on $M$ with boundary, motion perpendicular to boundary.]{Transition rate for uniformly accelerated detector approaching boundary
with Neumann boundary conditions, for $\alpha=1$, $\omega=1$ (upper curve), $\omega=1.5$ and $\omega=2$ (lower curve).}
\label{fig:accel}
\end{figure}

Consider now the four-dimensional case.
When $\tau=0$ we can do the integral in (\ref{eqn:boundaryaccel})
analytically, with the result
\begin{equation}
\dot{F}_{B0}(\omega)=
\frac{\beta\omega}{4\pi\sinh(\omega\pi\alpha)} \ .
\end{equation}
For general $\tau$ we may compute the integral numerically
for different values of $\alpha$, $\tau$ and $\omega$. We have
done this on Maple. For $\alpha=1$ we find the total transition rate 
(including the thermal part) displayed 
in figure \ref{fig:accel}.
Note that at many times $\tau$ the image part dominates the Minkowski part.
Further we can prove analytically that the image part of the transition rate is given by
\begin{equation}
\dot{F}_{B\tau}(\omega)=
\frac{\beta\omega\cos(2\tau\omega)}{2\pi\sinh(\omega\pi\alpha)}+B_\tau(\omega) \ ,
\end{equation}
where the function $B_\tau(\omega)$ is bounded in absolute value by $\frac{\beta}{2\pi^2\alpha}e^{-\frac{2\tau}{\alpha}}$
(see appendix \ref{ch:appendixe}).
Therefore for
large but finite $\tau$ the image part of the transition rate is found not to tend to
$0$ but instead is periodic in $\tau$ with period $\pi/\omega$.\footnote{In appendix \ref{ch:appendixe} 
we prove that for arbitrary dimension 
the image part of the transition rate consists of a term periodic in $\tau$ with period $\pi/\omega$ plus a term 
bounded in absolute value by $(\mathrm{constant})e^{-\frac{(d-2)\tau}{2\alpha}}$.}
This is a property only of the transition rate of a detector which is
turned on in the infinite past. 

\begin{figure}[htbp]
\includegraphics[angle=0, width={3in}]{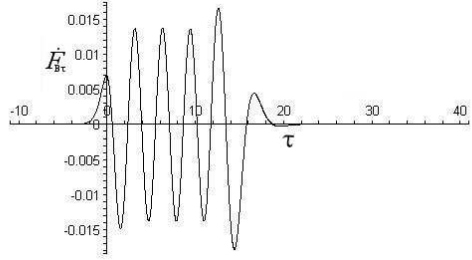}
\centering
\caption[Image term of the transition rate of a uniformly accelerated detector 
switched on at a finite time on $M$ with boundary, motion perpendicular to boundary.]
{Image term of the transition rate of a uniformly accelerated detector approaching the boundary
with Neumann boundary conditions for a detector switched on at $\tau_0=-15$, for $\alpha=1$, $\omega=1$.}
\label{fig:finitetime}
\end{figure}

Considering now a detector switched on at finite time $\tau_0$
(which recall is the more realistic situation). The
image part of the transition rate is given by (\ref{eqn:boundaryaccel})
with the upper limit of the integral replaced by $\tau-\tau_0$.
In figure \ref{fig:finitetime}
we plot this image part of the transition rate only, when the switching of the detector 
is instantaneous for, $\tau_0=-15$, $\alpha=1$ and $\omega=1$.
We see that when the detector is switched on only for a
finite time the transition rate is periodic for some time however 
falls off to the usual thermal response at late $\tau$.
In appendix \ref{ch:appendixe} we present an analytic proof of this, i.e. that 
the boundary part of the response vanishes in the limit as $\tau\rightarrow\infty$
when the detector is switched on at a finite time.

\begin{figure}[htbp]
\includegraphics[angle=0, width={3in}]{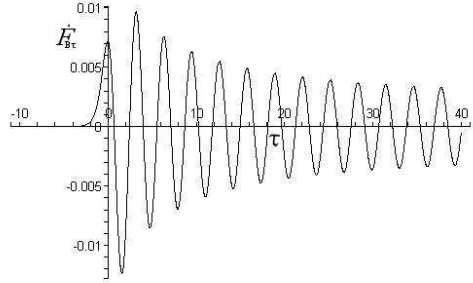}
\centering
\caption[Image term of the transition rate of a uniformly accelerated detector switched 
on at a finite time on $M$ with boundary with exponential switching function.]{Image term of the transition 
rate of a uniformly accelerated detector approaching boundary
with Neumann boundary conditions for a detector switched on at $\tau_0=-15$ with
an exponential switching function, for 
$\alpha=1$, $\omega=1$.}
\label{fig:finitetimewindow}
\end{figure}

\begin{figure}[htbp]
\includegraphics[angle=0, width={3in}]{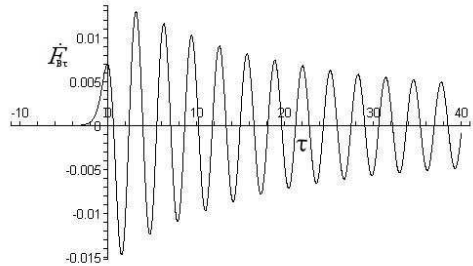}
\centering
\caption[Image term of the transition rate of a uniformly accelerated detector switched 
on at a finite time on $M$ with boundary with a Gaussian switching function.]{Image term of the transition 
rate of a uniformly accelerated detector approaching boundary
with Neumann boundary conditions for a detector switched on at $\tau_0=-15$ with
a Gaussian switching function, for 
$\alpha=1$, $\omega=1$.}
\label{fig:finitetimewindow2}
\end{figure}

Recall that in the case of instantaneous switching it was found that
even for the inertial detector in Minkowski space the response function
includes a logarithmic divergence \cite{ss:svaiter}. The transition rate
however, although altered from the infinite time case, is finite for
all non-zero finite time detections.
Further in \cite{hm:higuchi} it was shown that the divergence in the 
response rate is due to the instantaneous switching: if the detector is switched on
smoothly, no divergence occurs.
It is interesting then to briefly investigate the effect of smooth switching
on the results obtained above.
We introduce therefore a smooth window function in time $\tau$ into 
the transition rate (\ref{eq:transition}), that is we consider the
rate
\begin{equation}
\label{eqn:finitetimetrans}
\dot{F}_{\tau,\tau_0}(\omega)=
2\int^{\infty}_{0}ds\,
W(s,\tau-\tau_0)
Re
\left(
e^{-i\omega{s}}\langle{0}|\phi(\tau)\phi(\tau-s)|0\rangle
\right) \ ,
\end{equation}
where $W(s,\tau-\tau_0)$ is a smooth window function
with characteristic length $\tau-\tau_0$, ie
$W(s,\tau-\tau_0)\approx{1}$ for $s<<\tau-\tau_0$
and $W(s,\tau-\tau_0)\approx{0}$ for $s>>\tau-\tau_0$.
In particular we consider exponential
and Gaussian switching functions
\begin{eqnarray}
\label{eqn:finitewind1}
W_1(s,\tau-\tau_0) & = & e^{-\frac{|s|}{\tau-\tau_0}} \ ,
\\
\label{eqn:finitewind2}
W_2(s,\tau-\tau_0) & = & e^{-\frac{s^2}{2(\tau-\tau_0)^2}} \ .
\end{eqnarray}
The effect of these window functions on the response
of a uniformly accelerated detector in Minkowski space
was investigated in \cite{sp:srirpad}. Here we will
only consider the effect on the image part of the
transition rate on Minkowski space with boundary.
Substituting the image term (\ref{eqn:imtermperpacc}) and one of the
window functions (\ref{eqn:finitewind1}), (\ref{eqn:finitewind2}) into the transition 
rate (\ref{eqn:finitetimetrans}), we may calculate the integral 
numerically. In figures \ref{fig:finitetimewindow} and \ref{fig:finitetimewindow2} we plot the
transition rate in $4$-dimensions for a sample of the parameters
and with exponential and Gaussian switching respectively.
The numerical results suggest that in all cases of finite time detection the
image part of the transition rate tends to $0$ as the detection time tends to 
infinity. That is, the transition rate tends to that on Minkowski space
in this limit, as expected as in this limit the detector recedes infinitely far
from the boundary. It is an interesting result that this was not the case
for a detector switched on in the infinite past.

\subsection{Conical singularity and generalisations}
\label{sec:quotient}

In this section we consider the response of a uniformly accelerated
detector following trajectory (\ref{eq:accel}) on the quotient space of Minkowski space
under the group generated by the involution $J_{c_2}:(t,x,y,z)\mapsto(t,-x,-y,z)$.
Further we consider the higher dimensional generalisation of this 
spacetime constructed as the quotient of $d$-dimensional 
Minkowski space under the involution 
$J_{c_k}:(t,x_1,x_2,\ldots\,x_{d-1})\mapsto(t,-x_1,-x_2,-\ldots\,-x_k,x_{k+1},\ldots,x_{d-1})$
where $1<k<d$.
For reasons discussed later the response on these higher dimensional spaces will be
relevant when we consider the response of static detectors
on the $\rp$ geon in section \ref{sec:Geon} and inertial detectors on $\rp$ 
de Sitter space in section \ref{sec-desitrp3}. 

These spacetimes are conifolds \cite{sw:schleichwitt,sw:schleichwitt2}.
As quotients of Minkowski space under an involution with fixed points they are flat 
away from these $(d-k)$-dimensional hypersurfaces of fixed points but 
may be considered to have a distributional curvature on them
(see e.g \cite{gt:gerochtra,ga:distrbutioncurve}).
The spacetime $M/J_{c_2}$ is sometimes referred to as a conical spacetime as it 
has a conical singularity at $x_1=x_2=0$.
Transforming to cylindrical coordinates by $x_1=r\cos\phi$, $x_2=r\sin\phi$,
the isometry takes the form $J_{c_2}:(t,r,\phi,x_3,\ldots\,x_{d-1})\mapsto(t,r,\phi+\pi,x_3,\ldots,x_{d-1})$
and the metric reads
\begin{equation}
\label{cs:cosmic}
ds^2=
dt^2-dr^2-r^2d\phi^2-(dx_3)^2-\cdots-(dx_{d-1})^2 \ ,
\end{equation}
where $dr^2+r^2d\phi^2$ with the identification $(r,\phi)\sim(r,\phi+\pi)$
is the metric on a cone with deficit angle $\pi$.
In $4$ dimensions $M/J_{c_2}$  may be considered as the spacetime outside
an idealized, cosmic string with ``gravitational mass per unit length''
$\mu=1/8$ (see \cite{an:anderson}).

First we note that for an inertial or uniformly 
accelerated detector whose motion is in any direction $x_i$ with $k<i<d$, the 
response on these spacetimes will be the same as that
of a detector at rest or accelerating parallel to the
boundary on Minkowski space with boundary (where $\lambda$ in
(\ref{eqn:inertialparallel}) and (\ref{eqn:boundaryaccelparra2}) 
is now given by the shortest distance of the detector to the hypersurface of
fixed points $\lambda=\left((x_1)^2+(x_2)^2+\cdots+(x_k)^2\right)^{1/2}$).
This can be clearly seen by directly comparing the correlation
functions in both cases.

Consider now a particle detector uniformly accelerated with 
trajectory (\ref{eq:accel}) in the spacetime $M/J_{c_k}$.
Again as $J_{c_k}$ is an involution the correlation function
consists of two terms, 
\begin{equation}
\langle{0}|\hat{\phi}(\tau)\hat{\phi}(\tau')|0\rangle=\langle{0}|\phi(\tau)\phi(\tau')|0\rangle
+\beta\langle{0}|\phi(\tau)\phi(J_{c_k}\tau')|0\rangle \ ,
\end{equation}
where $\beta=+1, (-1)$ label the two possibilities for the
representations of the group in the automorphic field expansion 
(\ref{eqn:autofield}).
Consider first a detector switched on at $\tau_0=-\infty$.
The first term, when substituted into the transition rate (\ref{eqn:autotrans}) 
on the worldline (\ref{eq:accel}), leads to the thermal response in 
Minkowski space (\ref{thermal}).
The image term is
\begin{equation}
\beta\langle{0}|\phi(\tau)\phi(J_{c_k}\tau')|0\rangle
=
\frac{\beta\Gamma[d/2-1]}
{4\pi^{d/2}\left(4\alpha^2\cosh^2\left(\frac{2\tau-s}{2\alpha}\right)+C_k\right)^{d/2-1}} \ ,
\end{equation}
where $C_k=\sum_{m=2}^{k}(2x_m)^2$. As the trajectory and
its image are totally spacelike separated we have dropped the
regularization in the above expression.
The transition rate for this part is then
\begin{equation}
\label{eqn:boundaryaccel2}
\dot{F}_{I\tau(\omega)}=
\frac{\beta\Gamma[d/2-1]}
{2\pi^{d/2}}
\int^{\infty}_{0}ds\,
\frac{\cos(\omega{s})}
{\left(4\alpha^2\cosh^2\left(\frac{2\tau-s}{2\alpha}\right)+C_k\right)^{d/2-1}} \ .
\end{equation}

(\ref{eqn:boundaryaccel2}) is even in $\omega$ and so the
image term breaks the KMS condition and the response
is non-thermal and non-Planckian.
Further with a similar calculation to that of
appendix \ref{ch:appendixe} we may prove that (\ref{eqn:boundaryaccel2})
consists of a term periodic in $\tau$ with period $\pi/\omega$
and a term which decays
exponentially as $\tau\rightarrow\infty$. The qualitative
behaviour is therefore similar to that of the
uniformly accelerated detector on $M$ with boundary,
investigated in section \ref{sec:minkbound}.
As in section \ref{sec:minkbound} it can be shown also here that for a detector
switched on at $\tau_0>-\infty$ the image part
of the transition rate tends to $0$ as 
$\tau\rightarrow\infty$.

For a detector that accelerates towards the surface
of fixed points of the involution, that is in the
direction $r=((x_1)^2+(x_2)^2+\cdots+(x_k)^2)^{1/2}$,
the response is identical to that on Minkowski space with boundary 
with acceleration perpendicular to the boundary.
In sections \ref{sec:Geon} and \ref{sec-desitrp3}
we shall plot (\ref{eqn:boundaryaccel2}) for some 
specific values of $d$ and $k$ numerically.

We end this section with a comment on more general cosmic string spacetimes.
The methods used above could be easily applied to a larger class of idealized
cosmic string spacetimes for which the metric is (\ref{cs:cosmic}), with
the identification $(r,\phi)\sim(r,\phi+\pi/n)$ where $n\in{\Z}$
(and thus a deficit angle of $\pi(2-1/n)$), as in these
cases the correlation function may be given by a mode sum.
Detectors with motion parallel to such cosmic strings have been considered by
\cite{ss:cosmic,ds:cosdet}. Their main conclusions, which agree with ours
here where they overlap, are that the detector does respond to the presence of
the string in a manner which depends on its distance from the string.
Our results above and in the previous subsection add to the discussion, as we have been 
able to show the behaviour of detectors when motion is perpendicular to the string for
the specific case of $n=1$.
Numerical evaluations of the transition rate for any $n$ could be done in a similar 
way, but we shall not pursue this further here.
It is important to note however this class of cosmic strings does not include
realistic cosmic strings of the GUT scale where the deficit angle is $\approx{10^{-5}}$.

\subsection{Scalar detector on $M_-$}

Finally let us consider $M_-$. Recall that $M_-$ is 
a quotient of Minkowski space (or of $M_0$, it being a double cover of $M_-$) under the map
$J_-:(t,x,y,z)\mapsto(t,-x,-y,z+a)$. Again we may use the method of images
to find the correlation function for the automorphic field, with the result
\begin{equation}
\label{eqn:m-imagesum}
\langle{0}|\hat{\phi}(\tau)\hat{\phi}(\tau')|0\rangle=\sum_{n\in{\Z}}\langle{0}|\phi(\tau)\phi(J_-\tau')|0\rangle \ .
\end{equation}
The transition rate is given by (\ref{eqn:autotrans}).

\subsubsection{Inertial detector}

Considering a detector following the inertial trajectory
(\ref{eqn:inertraj}). The transition rate may again be split into two parts. The first comes from 
the $M_0$ part in the image sum (the even $n$ terms in (\ref{eqn:m-imagesum})) and will lead to
the same response as on $M_0$ (\ref{eqn:inerm0}). The other part (due to odd $n$
terms in (\ref{eqn:m-imagesum})) is similar
to the boundary part of such a detector on $M$ with boundary,
giving in the transition rate the contribution
\begin{eqnarray}
\dot{F}_{I\tau(\omega)} & = & 
-\frac{1}{2\pi^2}
(1-v^2)
\sum_{n=-\infty}^{\infty}
\frac{1}
{C_n+B_n}
(
-\mathrm{Ci}(B_n|\omega|)\cos(B_n|\omega|)
\nonumber
\\
&  & -\mathrm{si}(B_n|\omega|)\sin(B_n|\omega|)
+\mathrm{Ci}(C_n|\omega|)\cos(C_n|\omega|)
+\mathrm{si}(C_n|\omega|)\sin(C_n|\omega|)
\nonumber
\\
&  & +2\pi\sin(B_n\omega)\Theta(-\omega)
) \ ,
\end{eqnarray}
where
\begin{eqnarray}
B_n=\frac{-4\tau{v}^2+(16\tau^2{v^2}+4(1-v^2)^2((2y_0)^2+(2na+a)^2))^{1/2}}
{2(1-v^2)} \ ,
\\
C_n=\frac{+4\tau{v^2}+(16\tau^2{v^2}+4(1-v^2)^2((2y_0)^2+(2na+a)^2))^{1/2}}
{2(1-v^2)} \ .
\end{eqnarray}

In contrast to the analogous result on $M$ with boundary, there is no divergence
here at $x=0$ as there is no obstruction there and the inertial detector on
$M_-$ carries through $x=0$ smoothly. Note that on $M_-$ the energy-momentum
tensor expectation 
values are finite over the whole spacetime (see e.g. \cite{lm:geon} or the massless limits
of the results found in section \ref{eqn:m-emmassivescal}), while on $M$ with
boundary they diverge at $x=0$. For $\omega>0$ the $M_0$ part of the transition rate
vanishes while the image part is odd in $\tau$.
Further, numerical evaluations of the sum indicate that it is non-zero.
Here therefore we have an example of a spacetime and trajectory with no
pathologies at all where the total transition rate is negative for some values
of proper time $\tau$.
The response depends on the velocity as $t\partial_x+x\partial_t$ is not 
a Killing vector on $M_-$.

\subsubsection{Uniformly accelerated detector}

Consider a detector following the uniformly accelerated
worldline (\ref{eq:accel}).
Again the correlation function is in two parts. The part coming from
the $M_0$ part of the image sum  is given by the corresponding response on $M_0$
((\ref{M0response}) with $\eta=-1$). This part satisfies the KMS condition
and so in this sense is thermal. The ``image'' part 
is then somewhat similar to the boundary part found in the case
of $M$ with boundary. The transition rate for this term is 
\begin{equation}
\label{eqn:m-accelbound}
\dot{F}_{I\tau}(\omega)=
\frac{1}{2\pi^2}
\sum_{n=-\infty}^\infty
\int^{\infty}_{0}ds\,
\frac{\cos(\omega{s})}
{4\alpha^2\cosh^2\left({\frac{2\tau-s}{2\alpha}}\right)+4y_0^2+a^2(2n-1)^2} \ ,
\end{equation}
where $y_0$ is the $y$ coordinate of the detector. 
As on Minkowski space with boundary this image part of the transition rate
is even in $\omega$ and so breaks the KMS condition. The transition rate
is thus non-thermal and non-Planckian.
With a similar calculation to that done in appendix \ref{ch:appendixe} we find
\begin{equation}
\dot{F}_{I\tau}(\omega)=
\sum_{n=-\infty}^\infty
\left(
\frac{\alpha\cos(2\tau\omega)\sin\left(\alpha\omega{\mathrm{arccosh}\left(\frac{c_n}{2\alpha^2}\right)}\right)}
{\pi(c_n^2-4\alpha^4)^{1/2}\sinh(\omega\pi\alpha)}
+B_{n,\tau}(\omega)
\right) \ ,
\end{equation}
where $c_n=2\alpha^2+4y_0^2+(2na-a)^2$, and each $B_{n,\tau}(\omega)$ is bounded
by $1/{(2\pi^2\alpha)}e^{-2\tau/\alpha}$. Further we show in appendix \ref{ch:appendixe}
that the sum of $B_{n,\tau}(\omega)$ over $n$ is bounded by a function which exponentially
decays as $\tau\rightarrow\infty$.
We see as with the detector on $M$ with
boundary for large $\tau$ the response becomes periodic in $\tau$ with period $\pi/\omega$.

We may investigate the
general case here numerically. An analytic result is easy to find in the
case when $\tau=0$, with the result
\begin{equation}
\dot{F}_{I0}(\omega)=
\frac{\alpha}{2\pi}
\sum_{n=-\infty}^\infty
\frac{\sin\left(\alpha\omega{\mathrm{arccosh}\left(\frac{c_n}{2\alpha^2}\right)}\right)}
{(c_n^2-4\alpha^4)^{1/2}\sinh(\omega\pi\alpha)}  \ .
\end{equation}

For a detector switched on instantaneously at a finite time $\tau_0>-\infty$ 
an analytic calculation (see appendix \ref{ch:appendixe}) shows that the difference
between the response on $M_-$ and that on $M_0$ dies off as $\tau\rightarrow\infty$,
as would be expected far away from $x=0$. It is an interesting point that this is 
not the case for the detector switched on the infinite past.

This clarifies and adds to the
discussion on particle detectors given in \cite{lm:geon}.

\section{Causal detector with non-linear coupling}
\label{sec:nonlin}

In this section we extend the detector model 
to non-linear couplings. The aim is to see whether 
the formalism adapts easily to this case. The non-linear case 
is also interesting in that it is a step towards the Dirac detector considered 
in section \ref{sec:dirac}.

\subsection{Minkowski space}
\label{sec:nonlinmink}

The detector model is similar to that section \ref{sec-unruh} except in that
the interaction Hamiltonian is taken to be
\begin{equation}
\label{eq:interh2}
H_{\mathrm{int}}=
c
m(\tau)
\phi^n(\tau) \ ,
\end{equation}
where $n$ is an integer greater than $1$.
Again we assume the field is in its ground state and detector in a state with energy
$E_0$ at
$\tau=\tau_0$, and we seek the probability that the detector
is found in state $|E_1\rangle$ at later time $\tau_1>\tau_0$.
This probability is found to be, to first order in perturbation theory
\begin{equation}
\sum_{\psi}
|\langle\psi,E_1|0,E_0\rangle|^2=
c^2
|\langle{E_1}|m(0)|E_0\rangle|^2
\int^{\tau_1}_{\tau_0}d\tau
\int^{{\tau}_1}_{{\tau}_0}d\tau'
e^{-i\omega(\tau-\tau')}\langle{0}|\phi^n(\tau)\phi^n(\tau')|0\rangle
 \ ,
\end{equation}
with $\omega=E_1-E_0$.
The response function is
\begin{equation}
{F}_\tau(\omega)=
\int^{\tau_1}_{\tau_0}d\tau
\int^{{\tau}_1}_{{\tau}_0}d\tau'
e^{-i\omega(\tau-\tau')}\langle{0}|\phi^n(\tau)\phi^n(\tau')|0\rangle
 \ ,
\end{equation}
and the transition rate 
is given by
\begin{equation}
\label{eq:transitionnonlin}
\dot{F}_\tau(\omega)=
2\int^{\infty}_{0}ds\,
Re
\left(
e^{-i\omega{s}}\langle{0}|\phi^n(\tau)\phi^n(\tau-s)|0\rangle
\right) \ ,
\end{equation}
where we have used the property
$\langle{0}|\phi^n(\tau')\phi^n(\tau)|0\rangle=\langle{0}|\phi^n(\tau)\phi^n(\tau')|0\rangle^*$,
which follows from $\phi$ being a self adjoint operator, and have taken 
$\tau_0=-\infty$ and $\tau_1=\tau$. Finite time detectors are easily considered
by changing the upper limit in (\ref{eq:transitionnonlin}) to $\tau-\tau_0$.

For this non-linear coupling there are added issues not present in the linear 
case due to the interaction Lagrangian involving the product of the field 
taken at the same point. As a result the usual field theoretic divergence here
must be dealt with in a suitable manner. We explain here the
procedure on Minkowski space. The additional subtleties that arise for
automorphic fields will be addressed in section \ref{sec:nonlinauto} (see \cite{sz:suzuki} for a discussion on 
this point). 

Instead of the interaction Hamiltonian
(\ref{eq:interh2}), we are in fact led to consider a normal ordered version, $H_{\mathrm{int}}=cm(\tau):\phi^n(\tau):$. Then in 
(\ref{eq:transitionnonlin}) $\langle{0}|\phi^n(\tau)\phi^n(\tau')|0\rangle$
is replaced by $\langle{0}|:\phi^n(\tau)::\phi^n(\tau'):|0\rangle$ and by Wick's theorem we obtain
\begin{equation}
\label{eqn:nonlincorr}
\langle{0}|:\phi^n(\tau)::\phi^n(\tau'):|0\rangle
=
n!
{\langle{0}|\phi(\tau)\phi(\tau')|0\rangle}^n \ .
\end{equation}
Alternatively
using the original Hamiltonian (\ref{eq:interh2}) we are also led to (\ref{eqn:nonlincorr}) by expanding
the $2n$-point function in terms of $2$-point functions and replacing any divergent
quantities with their renormalized
expectation values, such as replacing $\langle{0}|\phi^2(x)|0\rangle$ with $\langle{0}|:\phi^2(x):|0\rangle$. 
These renormalized expectation values vanish on Minkowski space.

Putting (\ref{eqn:nonlincorr}) into (\ref{eq:transitionnonlin}), we are led
to a very similar expression for the transition rate as in the linear case.
Further it is clear from a direct calculation that if we consider the
uniformly accelerated worldline and the usual correlation function (\ref{eq:correusu})
in (\ref{eqn:nonlincorr}) we obtain a $\tau$-dependent expression. Further 
a numerical evaluation of the transition rate for $n=2$ shows that the transition
rate is also $\tau$-dependent even in the $\epsilon\rightarrow{0}$ limit. 
We see no reason why this would not also 
be the case for general $n$. This suggests
that as for the linear coupling we need to seek out a different regularization of the 
correlation function. We therefore consider a smeared field operator 
(\ref{eq:smear}) in the interaction Hamiltonian (\ref{eq:interh2}).
The transition rate is still given by (\ref{eq:transitionnonlin})
but now $\phi(\tau)$ is the smeared field operator, and 
we are led to an expression analogous to
(\ref{eqn:nonlincorr}) with no added complications due to the smearing.
The transition rate therefore depends on $\langle{0}|\phi(\tau)\phi(\tau')|0\rangle$
and so using the smeared field operator (\ref{eq:smear}) with the Lorentzian
profile function (\ref{eq:shape}) we are clearly led
to the regularization (\ref{eq:corre}) for this correlation function.

Consider again a uniformly accelerated detector. We have
already seen that Schlicht's correlation function on this worldline 
is given, for $d>2$, by (\ref{cronaccel}). Substituting (\ref{cronaccel}) into transition rate
(\ref{eq:transitionnonlin}) we obtain
\begin{equation}
\dot{F}_{\tau}(\omega)=
\left(
-
\frac{\Gamma[d/2-1]}{i^{d-2}(4\pi)^{d/2}}
\right)^n
n!
\lim_{\epsilon\rightarrow{0}}
\int^\infty_{-\infty}
ds\,
\frac{e^{-i\omega{s}}}{\left(\alpha\sinh\left(\frac{s}{2\alpha}\right)-i\epsilon\cosh\left(\frac{s}{2\alpha}\right)\right)^{n(d-2)}}.
\end{equation}
The integral may be performed by contour methods (see appendix B). The result is 
\begin{eqnarray}
\dot{F}_\tau(\omega)= 
&  &
\left(
\frac{\Gamma[d/2-1]}{4\pi^{d/2}}
\right)^n
n!
\frac{2\pi}{2^{d-2}\Gamma[m]\alpha^{m-1}}
\nonumber
\\
\label{thermalnonlin}
& \times &
\left\{
\begin{array}{ll}
\frac{\alpha\omega}{\left(e^{2\pi\omega\alpha}-1\right)}
\prod_{l=1}^{m/2-1}
\left(
\left(\frac{m}{2}-l\right)^2
+\alpha^2\omega^2
\right)
&
m\;\;\textrm{even}
\\
\frac{1}{\left(e^{2\pi\omega\alpha}+1\right)}
\prod_{l=1}^{(m-1)/2}
\left(
\left(\frac{m}{2}-l\right)^2
+\alpha^2\omega^2
\right)
&
m\;\;\textrm{odd}
\end{array}
\right. \ ,
\end{eqnarray}
where $m=n(d-2)$, and for $m=4$ the product over $l$ is set to $1$.
The response is independent of $\tau$ and satisfies the KMS condition
(\ref{eqn:kmscondition}) at the temperature $T=(2\pi\alpha)^{-1}$.
It also contains the expected statistics inversions for
different values of dimensions and couplings and agrees with the literature
(e.g \cite{sr:sriramk}).

\subsection{Automorphic fields}
\label{sec:nonlinauto}

For the non-linearly coupled detectors on quotients of Minkowski space,
we consider detectors coupled to automorphic fields in
Minkowski space as in section \ref{sec:automorphic}. We specialize in this
section to the case $n=2$ (though the case of higher order will 
follow in an analogous way).

The automorphic field is defined as previously by 
\begin{equation}
\label{eqn:autononlin}
\hat{\phi}(x)=
\frac{1}
{\left(\sum_{\gamma\in\Gamma}p(\gamma)^2\right)^{1/2}}
\sum_{\gamma\in\Gamma}p(\gamma)\phi(\gamma^{-1}x) \ .
\end{equation}
We consider a monopole detector coupled to the automorphic
field via the interaction Hamiltonian
\begin{equation}
H_{\mathrm{int}}
=
c
m(\tau)
\hat{\phi}^2(\tau) \ ,
\end{equation}
with
\begin{equation}
\label{eqn:autoschlicht}
\hat{\phi}(\tau)
=
\int
d^{d-1}\mathbf{\xi}\,
f(\mathbf{\xi})\hat{\phi}(x(\tau,\mathbf{\xi})) \ ,
\end{equation}
where $\mathbf{\xi}$ has the same meaning as in the previous sections.
Following the arguments of the previous sections, the transition rate
is found to be
\begin{equation}
\label{eq:transitionMboundnonlin}
\dot{F}_\tau(\omega)=
2\int^{\infty}_{0}ds\,
Re
\left(
e^{-i\omega{s}}\langle{0}|\hat{\phi}^2(\tau)\hat{\phi}^2(\tau-s)|0\rangle
\right) \ .
\end{equation}

We need to address the infinities in (\ref{eq:transitionMboundnonlin}).
We first note that
(\ref{eqn:autononlin}), (\ref{eqn:autoschlicht}) and the usual 
expansion of the Minkowski field operator $\phi$ in terms of annihilation
and creation operators leads to Wick's theorem for $\hat{\phi}$ in the form
\begin{eqnarray}
\langle{0}|\hat{\phi}(\tau_1)\hat{\phi}(\tau_2)\hat{\phi}(\tau_3)\hat{\phi}(\tau_4)|0\rangle & = & \langle{0}|\hat{\phi}(\tau_1)\hat{\phi}(\tau_4)|0\rangle
\langle{0}|\hat{\phi}(\tau_2)\hat{\phi}(\tau_3)|0\rangle
\nonumber
\\
&   & +\langle{0}|\hat{\phi}(\tau_1)\hat{\phi}(\tau_3)|0\rangle
\langle{0}|\hat{\phi}(\tau_2)\hat{\phi}(\tau_4)|0\rangle
\nonumber
\\
&   & +\langle{0}|\hat{\phi}(\tau_1)\hat{\phi}(\tau_2)|0\rangle
\langle{0}|\hat{\phi}(\tau_3)\hat{\phi}(\tau_4)|0\rangle \ .
\end{eqnarray}
Setting $\tau_1=\tau_2=\tau$ and $\tau_3=\tau_4=\tau'$ yields
\begin{equation}
\langle{0}|\hat{\phi}^2(\tau)\hat{\phi}^2(\tau')|0\rangle=2\langle{0}|\hat{\phi}(\tau)\hat{\phi}(\tau')|0\rangle^2
+\langle{0}|\hat{\phi}^2(\tau)|0\rangle\langle{0}|\hat{\phi}^2(\tau')|0\rangle \ .
\label{eqn:wicknonlin}
\end{equation}
The divergent term in (\ref{eqn:wicknonlin}) is the second term. 

We remove the divergence
by normal ordering the interaction Hamiltonian $H_{\mathrm{int}}=cm(\tau):\hat{\phi}^2(\tau):$
with respect to the
Minkowski vacuum. This renormalization procedure
is equivalent to subtracting off the divergent Minkowski terms.
In (\ref{eqn:wicknonlin})
$\langle{0}|\hat{\phi}^2(\tau)\hat{\phi}^2(\tau')|0\rangle$ 
is thus replaced by 
\begin{equation}
\label{eqn:ronormcorrelation}
\langle{0}|:\hat{\phi}^2(\tau)::\hat{\phi}^2(\tau'):|0\rangle=2\langle{0}|\hat{\phi}(\tau)\hat{\phi}(\tau')|0\rangle^2
+\langle{0}|:\hat{\phi}^2(\tau):|0\rangle\langle{0}|:\hat{\phi}^2(\tau'):|0\rangle \ ,
\end{equation}
where 
\begin{eqnarray}
\langle{0}|:\hat{\phi}^2(\tau):|0\rangle
& = &
\langle{0}|\hat{\phi}^2(\tau)|0\rangle-\langle{0}|\phi^2(\tau)|0\rangle \ ,
\nonumber
\\
& = &
\lim_{\tau'\rightarrow{\tau}}[\langle{0}|\hat{\phi}(\tau)\hat{\phi}(\tau')|0\rangle-\langle{0}|\phi(\tau)\phi(\tau')|0\rangle] \ ,
\nonumber
\\
\label{eqn:renorm}
& = &
\sum_{\gamma\in\Gamma\setminus\{I_d\}}
p(\gamma)
\langle{0}|\phi(\tau)\phi(\gamma^{-1}\tau)|0\rangle \ .
\end{eqnarray}
In the last equality in (\ref{eqn:renorm}) we have used the method
of images expression (\ref{eqn:automorcorre}) for the correlation function of the automorphic field.

Consider now, as an example,
a uniformly accelerated detector, following the trajectory (\ref{eq:accel}), 
on $4$-dimensional Minkowski space
with boundary at $x=0$. 
We have already seen in section \ref{sec:minkbound} that the two point function in this case
takes the form
\begin{equation}
\label{eqn:accboundpart}
\langle{0}|\hat{\phi}(\tau)\hat{\phi}(\tau')|0\rangle=
-\frac{1}{16\pi^2}
\frac{1}{\left(\alpha\sinh(\frac{\tau-\tau'}{2\alpha})-i\epsilon\cosh(\frac{\tau-\tau'}{2\alpha})\right)^2}
+
\frac{\beta}{16\pi^2\alpha^2}
\frac{1}{\cosh^2(\frac{\tau+\tau'}{2\alpha})}
 \ .
\end{equation}
Further from (\ref{eqn:renorm}) we have 
\begin{equation}
\label{eqn:renormpart}
\langle{0}|:\hat{\phi}^2(\tau):|0\rangle
=
\frac{\beta}{16\pi^2\alpha^2}
\frac{1}{\cosh^2(\frac{\tau}{\alpha})} \ ,
\end{equation}
where the $\epsilon\rightarrow{0}$ limit has been taken before
doing the transition rate integral as can be argued by the dominated
convergence theorem.
Substituting (\ref{eqn:accboundpart}) and (\ref{eqn:renormpart}) into
(\ref{eqn:ronormcorrelation}) and then (\ref{eq:transitionMboundnonlin})
we have a transition rate with 4 terms.
The first term comes from the Minkowski parts 
of the image sum 
and so leads to the usual thermal spectrum.
The other three terms however break the KMS
condition leading to a non-thermal response,
due to the presence of the boundary.
We shall not discuss the calculation further here.
For more details of the response function for
finite time observations in this case see \cite{sz:suzuki}.

\section{Causal detector for the Dirac field}
\label{sec:dirac}

\subsection{Minkowski space}
\label{sec:diracmink}

In this section we extend the causal detector to the massless Dirac field in $4$-dimensional
Minkowski space. The detector is
still a many-level quantum mechanical system with free Hamiltonian
$H_D$. However now the detector moves through a massless Dirac field
$\psi$ (with free Hamiltonian $H_\psi$) in Minkowski space to which it 
is coupled via the interaction Hamiltonian
\begin{equation}
\label{eqn:dirachint}
H_{\mathrm{int}}=
c
m(\tau)
\bar{\psi}(\tau)
\psi(\tau) \ ,
\end{equation}
where $\bar{\psi}=\psi^\dagger\gamma^0$, and $\psi(\tau)=\psi(x(\tau))$.
The equation of motion for the free field $\psi$ is the massless free
Dirac equation $i\gamma^\mu\partial_\mu\psi=0$. We choose a basis of solutions and expand the
field in terms of this basis. We work throughout with the standard representation
of $\gamma$ matrices,
\begin{eqnarray} \gamma^0=\left(\begin{array}{cc} 1 & 0 \\
		 0 & -1 \end{array}\right) \ , & \gamma^i=\left(\begin{array}{cc} 0 & \sigma_i \\
		-\sigma_i & 0 \end{array}\right) \ ,
\end{eqnarray}
where $\sigma_i$ are the Pauli matrices, $\sigma_1 =
\bigl(
\begin{smallmatrix} 0 & 1 \\ 1 & 0
\end{smallmatrix} 
\bigr)$,  
$\sigma_2 = 
\bigl(
\begin{smallmatrix} 
0 & -i \\ i & 0 
\end{smallmatrix}
\bigr)$
and 
$\sigma_3 = 
\bigl(
\begin{smallmatrix} 
1 & 0 \\ 0 & -1
\end{smallmatrix}
\bigr)$. Then
\begin{equation}
\label{eqn:diracexpanmode}
\psi(t,\mathbf{x})
=
\sum_{s=1,2}
\int
\frac{d^3k}{(2(2\pi)^3)^{\frac{1}{2}}}
\left[
b_s(\mathbf{k})u(\mathbf{k},s)e^{-i\omega{t}+i\mathbf{k}\mathbf{x}}+
d^\dagger_s(\mathbf{k})v(\mathbf{k},s)e^{i\omega{t}-i\mathbf{k}\mathbf{x}}
\right] \ ,
\end{equation}
where
\begin{eqnarray}
\label{eqn:uspinors}
u(\mathbf{k},1)=\left(\begin{array}{ccccccccccccccccccccc}1 \\ 0 \\ \frac{k_z}{\omega} \\ \frac{k_+}{\omega} \end{array} \right) \ ,
 &       \hspace{.25in}           u(\mathbf{k},2)=\left(\begin{array}{ccccccccccccccccccccc}0 \\ 1 \\ \frac{k_-}{\omega}\\ \frac{-k_z}{\omega} \end{array} \right) \ ,
\end{eqnarray}	
and
\begin{eqnarray}
v(\mathbf{k},1)=\left(\begin{array}{ccccccccccccccccccccc}\frac{k_z}{\omega} \\ \frac{k_+}{\omega} \\ 1 \\ 0 \end{array} \right) \ ,
 &       \hspace{.25in}
v(\mathbf{k},2)=\left(\begin{array}{ccccccccccccccccccccc}\frac{k_-}{\omega} \\ \frac{-k_z}{\omega} \\ 0 \\ 1 \end{array} \right) \ ,
\end{eqnarray}	
with $k_{\pm}=k_x\pm{ik_y}$. The modes in the expansion (\ref{eqn:diracexpanmode}) are expressed in
terms of a standard Minkowski vierbein, aligned along Minkowski coordinate axes,
and they are suitably 
orthonormal with respect to the usual Dirac inner product in Minkowski 
space,
\begin{equation}
\langle{\psi_1,\psi_2}\rangle=\int{d^3x}\,\psi_1^\dagger{\psi_2} \ .
\end{equation}
The free field is then quantized in the usual manner, imposing the
usual anticommutation relations on the annihilation/creation
operators,
\begin{equation}
\label{eqn:diracanti}
\left\{b_s(\mathbf{k}),b_{s'}^{\dagger}(\mathbf{k})\right\}
=
\delta_{ss'}
\delta(\mathbf{k}-\mathbf{k'}) \ ,
\end{equation}
and similarly for the $d$ operators, with all mixed anticommutators
vanishing.

We assume again that at time $\tau_0$ the full interacting field
is in the product state $|0,E_0\rangle=|0\rangle|E_0\rangle$. 
Working in the interaction picture we find, to first order in perturbation theory,
that the probability that at a later time $\tau_1>\tau_0$ the detector is found 
in state $|E_1\rangle$ is
\begin{eqnarray}
\sum_{\Psi}
|\langle\Psi,E_1|0,E_0\rangle|^2
& = & 
c^2
|\langle{E_1}|m(0)|E_0\rangle|^2
\int^{\tau_1}_{\tau_0}d\tau
\int^{{\tau}_1}_{{\tau}_0}d\tau'
e^{-i\omega(\tau-\tau')}
\nonumber
\\
&   &
\times
\langle{0}|\bar{\psi}(\tau)\psi(\tau)\bar{\psi}(\tau')\psi(\tau')|0\rangle
 \ ,
\end{eqnarray}
with $\omega=E_1-E_0$. Once again we shall concentrate on the response function part.
With the same change of coordinates as in section \ref{sec-linmink}, this part 
may be written as
\begin{equation}
F_{\tau_1,\tau_0}(\omega)=
2\int^{\tau_1}_{\tau_0}du
\int^{u-\tau_0}_{0}ds\,
Re(
e^{-i\omega{s}}\langle{0}\bar{\psi}(u)\psi(u)\bar{\psi}(u-s)\psi(u-s)|0\rangle
) \ .
\end{equation}
We have used the relation $\langle{0}|\bar{\psi}(\tau)\psi(\tau)\bar{\psi}(\tau')\psi(\tau')|0\rangle
=\langle{0}|\bar{\psi}(\tau')\psi(\tau')\bar{\psi}(\tau)\psi(\tau)|0\rangle^*$, 
which follows from the axioms of inner products and
$(\bar{\psi}(\tau')\psi(\tau'))^\dagger=\bar{\psi}(\tau')\psi(\tau')$.
Finally, as in the scalar case, we let the detector be 
switched on in the infinite past and differentiate with
respect to $\tau_1=\tau$ to get the transition rate, with the result
\begin{equation}
\label{eq:transitiondir}
\dot{F}_\tau(\omega)=
2\int^{\infty}_{0}ds\,
Re
\left(
e^{-i\omega{s}}\langle{0}|\bar{\psi}(\tau)\psi(\tau)\bar{\psi}(\tau-s)\psi(\tau-s)|0\rangle
\right) \ .
\end{equation}
Further here we find
\begin{equation}
\label{eq:diraccorr}
\langle{0}|\bar{\psi}(\tau)\psi(\tau)\bar{\psi}(\tau')\psi(\tau')|0\rangle
=
\mathrm{Tr}((S^+_M(\tau,\tau'))^2) \ ,
\end{equation}
where $\mathrm{Tr}$ is the trace and $S^+_M(\tau,\tau')=\langle{0}|\psi(\tau)\bar{\psi}(\tau')|0\rangle$ 
is the positive frequency Wightman function which is related to the
positive frequency scalar Wightman function by (see e.g \cite{bd:book})
\begin{equation}
S^+_M(\tau,\tau')=
i\gamma^{\mu}\partial_{\mu}G^+_M(\tau,\tau') \ .
\end{equation}
To prove expression (\ref{eq:diraccorr}) a straigtforward, but rather messy, calculation was
used where the expansion of the field (\ref{eqn:diracexpanmode}) was substituted directly into 
the correlation function $\langle{0}|\bar{\psi}(\tau_1)\psi(\tau_2)\bar{\psi}(\tau_3)\psi(\tau_4)|0\rangle$ and use was made
of the anticommutation relations (\ref{eqn:diracanti}), setting $\tau_1=\tau_2=\tau$, $\tau_3=\tau_4=\tau'$
at the end.
We note here that all expressions for the response are independent of the vierbein used
to express the $\psi$ field. This is due to the form of $H_{\mathrm{int}}$ (\ref{eqn:dirachint}) 
which is a Lorentz scalar.
In the case of a massive Dirac field in Minkowski space (\ref{eq:diraccorr}) contains a second
term proportional to $\mathrm{Tr}(S^+_M(\tau,\tau'))\mathrm{Tr}(S^-_M(\tau',\tau))$. 
Here in the massless case this term does not enter as $\mathrm{Tr}(S^+_M(\tau,\tau'))=0$
for any worldline.

Consider the uniformly accelerated worldline
(\ref{eq:accel}).
Again a numerical calculation shows that if we use the usual 
correlation function, i.e. with the $i\epsilon$ regularization, for the scalar field 
(\ref{eq:correusu}) in the expressions above we will 
get a $\tau$-dependent result for the transition rate (\ref{eq:transitiondir}) even in the $\epsilon\rightarrow{0}$ limit.
We are thus led once again to consider an alternative setup 
where we use a smeared form for the field operator in the interaction
Hamiltonian. That is we consider
\begin{equation}
\label{eq:smeardir}
\psi(\tau)=
\int
d^3\xi\,
W_\epsilon(\mathbf{\xi})
S(\tau,\mathbf{\xi})
\psi(x(\tau,\mathbf{\xi})) \ ,
\end{equation}
with the same definitions for $W_\epsilon(\mathbf{\xi})$ and $\mathbf{\xi}$
as in section \ref{sec-linmink}. In contrast to the scalar case, here
we include $S(\tau,\mathbf{\xi})$, which is the spinor transformation associated with the transformation 
from the Minkowski vierbein
to one adapted to the Fermi-Walker coordinates. However we may now argue that in this case
$S(\tau,\mathbf{\xi})$ may be dropped. Firstly we note that the metric written in 
Fermi Walker coordinates is \cite{sc:schlicht}
\begin{equation}
ds^2=
\left(1+2\left(\dot{t}\ddot{x}-\dot{x}\ddot{t}\right)\xi+\left(\ddot{x}^2-\ddot{t}^2\right)\xi^2\right)
d\tau^2
-
d\mathbf{\xi}^2 \ .
\end{equation}
Constant $\tau$ spatial sections are therefore flat. It then follows
that the transformation from Minkowski vierbein to that adapted to
these Fermi coordinates will be independent of $\mathbf{\xi}$, as Fermi Walker
transport along these spatial sections in a non rotating vierbein will
be trivial. It therefore follows that $S(\tau,\mathbf{\xi})$ may be taken outside
the integral in (\ref{eq:smeardir}). Further it follows from the form of $H_{\mathrm{int}}$
(\ref{eqn:dirachint}) that as $S$ may be taken outside the integral it may be dropped completely.
Therefore on $M$ we may work throughout with expressions written with respect
to the standard Minkowski vierbein and with 
\begin{equation}
\label{eqn:smeardiracfi}
\psi(\tau)=
\int
d^3\xi\,
W_\epsilon(\mathbf{\xi})
\psi(x(\tau,\mathbf{\xi})) \ ,
\end{equation}
as the expression for a smeared field operator.
By arguments similar to those that lead to (\ref{eq:transitiondir}) and (\ref{eq:diraccorr}) 
the transition rate is again
given by 
\begin{equation}
\label{eqn:smeartransdir}
\dot{F}_\tau(\omega)=
2\int^{\infty}_{0}ds\,
Re
\left(
e^{-i\omega{s}}\langle{0}|\bar{\psi}(\tau)\psi(\tau)\bar{\psi}(\tau-s)\psi(\tau-s)|0\rangle
\right) \ ,
\end{equation}
and we find
\begin{equation}
\label{eqn:smearcorrdirac}
\langle{0}|\bar{\psi}(\tau)\psi(\tau)\bar{\psi}(\tau')\psi(\tau')|0\rangle
=
\mathrm{Tr}(\langle{0}|\psi(\tau)\bar{\psi}(\tau')|0\rangle^2) \ ,
\end{equation}
where $\psi(\tau)$ is now the smeared field (\ref{eqn:smeardiracfi}).

Now suppose we consider again the Lorentzian profile function
\begin{equation}
W_{\epsilon}(\mathbf{\xi})
=
\frac{1}{\pi^{2}}
\frac{\epsilon}{(\mathbf{\xi}^2+\epsilon^2)^{2}} \ .
\end{equation}
The spinor correlation function is then given by
\begin{eqnarray}
\langle{0}|\psi(\tau)\bar{\psi}(\tau')|0\rangle & = & 
\frac{1}{2(2\pi)^3}
\sum_{s=1,2}
\int
d^3k\,
u(\mathbf{k},s)
u^\dagger(\mathbf{k},s)
\gamma^0
\nonumber
\\
&  & \times
\int
d^3\xi\,
W_\epsilon(\mathbf{\xi})
e^{-ik\cdot{x}(\tau,\mathbf{\xi})}
\int
d^3\xi'\,
W_\epsilon(\mathbf{\xi'})
e^{ik\cdot{x}(\tau,\mathbf{\xi'})} \ .
\label{eqn:diraccorrint}
\end{eqnarray}
The integrals over $\mathbf{\xi}$ and $\mathbf{\xi}'$ in (\ref{eqn:diraccorrint})
are the same as those in (\ref{eq:expan}). Proceeding as with (\ref{eq:expan}), we thus have
\begin{equation}
\label{eqn:dircorrinteq}
\langle{0}|\psi(\tau)\bar{\psi}(\tau')|0\rangle=
\frac{1}{2(2\pi)^3}
\sum_{s=1,2}
\int
d^3k\,
u(\mathbf{k},s)
u^\dagger(\mathbf{k},s)
\gamma^0
e^{-i\omega(t-t'-i\epsilon(\dot{t}+\dot{t}'))+i\mathbf{k}(\mathbf{x}-\mathbf{x'}-i\epsilon(\dot{\mathbf{x}}+\dot{\mathbf{x}}'))}
 \ .
\end{equation}
Further from (\ref{eqn:uspinors}) we have 
\begin{equation}
\sum_{s=1,2}
u(\mathbf{k},s)
u^\dagger(\mathbf{k},s)
\gamma^0 
= 
\left(
\begin{array}{cccc} 
1 & 0 & -\frac{k_z}{\omega} & \frac{-k_x+ik_y}{\omega}
\\ 
0 & 1 & \frac{-k_x-ik_y}{\omega} & \frac{k_z}{\omega}
\\
\frac{k_z}{\omega} & \frac{k_x-ik_y}{\omega} & -1 & 0 
\\
\frac{k_x+ik_y}{\omega} & -\frac{k_z}{\omega} & 0 & -1
\end{array} 
\right) \ .
\end{equation}
If we now compare (\ref{eqn:dircorrinteq}) with the scalar field expression
(\ref{eq:correlation}), with $d=4$, we find we have
\begin{equation}
\label{eq:SeqG}
S^+_M(\tau,\tau')=\langle{0}|\psi(\tau)\bar{\psi}(\tau')|0\rangle
=
i\gamma^\mu\partial_\mu{\langle{0}|\phi(\tau)\bar{\phi}(\tau')|0\rangle} \ ,
\end{equation}
where $\langle{0}|\phi(\tau)\bar{\phi}(\tau')|0\rangle$ is
the scalar field correlation function (\ref{eq:correlation})
and the partial derivative acts on the $(t,\mathbf{x})$
but NOT on the $(\dot{t},\dot{\mathbf{x}})$. Therefore
we find, from
(\ref{eq:corre}), that the spinor correlation function is given by
\begin{eqnarray}
\label{eqn:dirccorrfnsmeared}
S^+_M(\tau,\tau') & = & \frac{i}{4{\pi}^2}\frac{1}{[\tilde{t}^2-\tilde{x}^2-\tilde{y}^2-\tilde{z}^2]^2}
\nonumber
\\
&  &
\times
\left[
\begin{array}{cccc} 
2\tilde{t} & 0 & -2\tilde{z} & 2(i\tilde{y}-\tilde{x}) 
\\ 
0 & 2\tilde{t} & -2(i\tilde{y}+\tilde{x}) & 2\tilde{z} 
\\
2\tilde{z} & -2(i\tilde{y}-\tilde{x}) & 2\tilde{t} & 0 
\\
2(i\tilde{y}+\tilde{x}) & -2\tilde{z} & 0 & 2\tilde{t}
\end{array} 
\right] \ ,
\end{eqnarray}
where $\tilde{a}=(a(\tau)-a(\tau')-i\epsilon(\dot{a}(\tau)+\dot{a}(\tau')))$.
From this it is easy to show that
\begin{equation}
\label{eqn:diraccorrefunction}
\mathrm{Tr}(S^+_M(\tau,\tau')^2)
=
-\frac{1}{\pi^4}
\frac{1}{((t-t'-i\epsilon(\dot{t}+\dot{t}'))^2-(\mathbf{x}-\mathbf{x'}-i\epsilon(\dot{\mathbf{x}}+\dot{\mathbf{x}}'))^2)^3}
 \ .
\end{equation}

\subsubsection{Inertial detector}

First we consider the response of a Dirac field detector following the
inertial worldline (\ref{eqn:inertraj}) in Minkowski space. From
(\ref{eqn:diraccorrefunction}), (\ref{eqn:smeartransdir}) and (\ref{eqn:smearcorrdirac})
the transition rate is found to be
\begin{equation}
\label{eqn:diracinertmink}
\dot{F}_\tau(\omega)=
-\frac{1}{\pi^4}
\lim_{\epsilon\rightarrow{0}}
\int^{\infty}_{-\infty}ds\,
\frac{e^{-i\omega{s}}}{(s-2i\epsilon)^6} \ .
\end{equation}
The integral can be done by residues, with the result
\begin{equation}
\label{eqn:diracinertitrans}
\dot{F}_\tau(\omega)=
-\frac{\omega^5}{60\pi^3}\Theta(-\omega) \ .
\end{equation}

Consider also the ``power spectrum'' of the Dirac noise
as defined by Takagi in \cite{tk:takagi}. The noise $g(\tau,\tau')$
is defined by
\begin{equation}
g(\tau,\tau')
=
S(\tau)
S^+_M(\tau,\tau')
S({\tau'})^{-1} \ ,
\end{equation}
where $S(\tau)=S(\tau,\mathbf{\xi})$ as given in (\ref{eq:smeardir}). $S(\tau)$
is the spinor transformation which takes care of the Fermi-Walker transport
so that $S(\tau)\psi(\tau)$ does not rotate with respect to the detector's
proper reference frame.
The definition for the power spectrum on a stationary worldline,
where $S^+_M(\tau,\tau')$ depends on $\tau$ and $\tau'$ only through
$\tau-\tau'$, is
\begin{equation}
\label{eq:powerspec}
P(\omega)=\frac{1}{4}
\mathrm{Tr}\,\gamma^0
\int^{\infty}_{-\infty}
d\tau\,
{e^{-i\omega\tau}}g(\tau) \ .
\end{equation}
On the inertial trajectory (\ref{eqn:inertraj}), the transformation to the
Fermi frame is trivial and we find
\begin{eqnarray}
P(\omega) & = & \frac{i}{2\pi^2}
\int^{\infty}_{-\infty}
ds\,
\frac{e^{-i\omega{s}}}{(s-2i\epsilon)^3}
\nonumber
\\
& = & \frac{\omega^2}{2\pi}\Theta(-\omega) \ .
\end{eqnarray}
We note that the power spectrum is $-\omega$ times the transition
rate for the linearly coupled scalar field detector 
following the same trajectory.

\subsubsection{Uniformly accelerated detector}

Considering once again a detector following the uniformly accelerated worldline (\ref{eq:accel}).
We find, as expected, that the correlation function is invariant under translations in $\tau$ and the 
transition rate (\ref{eqn:smeartransdir}) is given by
\begin{equation}
\dot{F}_\tau(\omega)=
-\frac{1}{64\pi^4}
\lim_{\epsilon\rightarrow{0}}
\int^{\infty}_{-\infty}ds\,
\frac{e^{-i\omega{s}}}{\left(\alpha\sinh\left(\frac{s}{2\alpha}\right)-i\epsilon\cosh\left(\frac{s}{2\alpha}\right)\right)^6} \ .
\end{equation}
The integral may be done by contour integration (appendix \ref{ch:appendixB}). The result is
\begin{equation}
\label{eq:diractranm}
\dot{F}_\tau(\omega)=
\frac{1}{60\pi^3\alpha^4}
\frac{\omega}{\left(e^{2\pi\alpha\omega}-1\right)}
(4+5(\alpha\omega)^2+(\alpha\omega)^4) \ .
\end{equation}
The response is thermal in the sense that it satisfies the KMS
condition at the temperature $T=(2\pi\alpha)^{-1}$. It is interesting to note that there is no fermionic 
factor in the response, instead we have the usual Planckian factor 
found in the scalar case. 

We can see the fermionic factor appearing
however if we consider the power spectrum (\ref{eq:powerspec}) of the Dirac noise.
For the uniformly accelerated worldline we have
\begin{equation}
S(\tau)
=
\cosh\left(\frac{\tau}{2\alpha}\right)-\gamma^0\gamma^1\sinh\left(\frac{\tau}{2\alpha}\right) \ ,
\end{equation}
and the power spectrum (\ref{eq:powerspec}) is given by
\begin{equation}
P(\omega)=\frac{i}{16\pi^2}
\int^{\infty}_{-\infty}
d\tau\,
\frac{e^{-i\omega\tau}}{\left(\alpha\sinh\left(\frac{\tau}{2\alpha}\right)-i\epsilon\cosh\left(\frac{\tau}{2\alpha}\right)\right)^3} \ .
\end{equation}
Again the integral may be done by contour integration to give
\begin{equation}
\label{eqn:minkdiracpspec}
P(\omega)=
\frac{1}{2\pi}
\frac{\left(\omega^2+\frac{1}{4\alpha^2}\right)}{\left(1+e^{2\pi\alpha\omega}\right)} \ .
\end{equation}
Our expression here agrees with that found by Takagi \cite{tk:takagi}.
The power spectrum therefore does include the expected fermionic 
factor. 

\subsection{Dirac detector for automorphic fields}
\label{sec:diracauto}

Next we wish to consider this fermionic detector on $M_0$ and
$M_-$ and in particular address the issues concerning spin structure
on $M_-$. 
We consider an automorphic Dirac field on Minkowski space. The main
difference from the scalar case is that we must take care of what vierbeins
our expressions are written with respect to. In particular our vierbein
might not be invariant under the quotient group $\Gamma$. 

We begin with a massless 
Dirac field $\psi$
on $M$, expressed with respect to a vierbein
that is invariant under $\Gamma$. The automorphic field is then defined by
\begin{equation}
\label{eqn:diracautofield}
\hat{\psi}(x)=
\frac{1}
{\left(\sum_{\gamma\in\Gamma}p(\gamma)^2\right)^{1/2}}
\sum_{\gamma\in\Gamma}
p(\gamma)
\psi
(\gamma^{-1}x) \ ,
\end{equation}
where the normalization is such that, at equal times,
\begin{equation}
\left\{\hat{\psi}_\alpha(x),\hat{\psi}^\dagger_\beta(x')\right\}=\delta^{(d-1)}(x-x')\delta_{\alpha\beta} + \textrm{image terms} \ .
\end{equation}
The two-point function for the automorphic field is then given by the method of images,
\begin{equation}
\label{eq:diracautocorr}
S^+_{M/\Gamma}(x,x')=
\langle{0}|\hat{\psi}(x)\bar{\hat{\psi}}(x')|0\rangle
=
\sum_{\gamma\in\Gamma}
p(\gamma)
\langle{0}|\psi(x)\bar{\psi}(\gamma^{-1}x')|0\rangle \ .
\end{equation}

We consider a detector coupled to the automorphic field
via the interaction Hamiltonian
\begin{equation}
H_{\mathrm{int}}
=
cm(\tau)
\bar{\hat{\psi}}(\tau)\hat{\psi}(\tau) \ ,
\end{equation}
where
\begin{equation}
\label{eqn:smaereddirac}
\hat{\psi}(\tau)=
\int
d^3\xi
\,
W_\epsilon(\mathbf{\xi})
S(\tau,\mathbf{\xi})
\hat{\psi}(x(\tau,\mathbf{\xi})) \ .
\end{equation}
The transition rate is given by
\begin{equation}
\dot{F}_\tau(\omega)=
2\int^{\infty}_{0}ds\,
Re
\left(
e^{-i\omega{s}}\langle{0}|\bar{\hat{\psi}}(\tau)\hat{\psi}(\tau)\bar{\hat{\psi}}(\tau-s)\hat{\psi}(\tau-s)|0\rangle
\right) \ .
\end{equation}
Further we may show, with a calculation similar to that leading to (\ref{eq:diraccorr}) and (\ref{eqn:smearcorrdirac}), that
\begin{equation}
\langle{0}|\bar{\hat{\psi}}(\tau)\hat{\psi}(\tau)\bar{\hat{\psi}}(\tau')\hat{\psi}(\tau')|0\rangle
=
\mathrm{Tr}
\left(\langle{0}|\hat{\psi}(\tau)\bar{\hat{\psi}}(\tau')|0\rangle^2\right) \ ,
\end{equation}
with
\begin{equation}
\label{eqn:autodircorrfn}
\langle{0}|\hat{\psi}(\tau)\bar{\hat{\psi}}(\tau')|0\rangle
=
\sum_{\gamma\in\Gamma}
p(\gamma)
\langle{0}|\psi(\tau)\bar{\psi}(\gamma^{-1}\tau')|0\rangle \ .
\end{equation}
Therefore the method of images may be directly applied to our 
Minkowski space correlation functions here. 

It is important to note that the above mode sum expressions are changed when considering
a vierbein not invariant under the action of $\Gamma$. Suppose we
consider two vierbeins, one invariant under $\Gamma$ (labelled by
an $I$) and another not invariant (labelled by $N$). In the
vierbein $I$ the automorphic field is given by the mode sum expression
(\ref{eqn:diracautofield}). The transformation from $I$ to $N$ will
transform the spinors as $\hat{\psi}_I(x)\rightarrow\hat{\psi}_N(x)=S(x)\hat{\psi}_I(x)$.
Then from (\ref{eqn:diracautofield})
\begin{equation}
\hat{\psi}_N(x)=
\frac{1}
{\left(\sum_{\gamma\in\Gamma}p(\gamma)^2\right)^{1/2}}
\sum_{\gamma\in\Gamma}
p(\gamma)
S(x)
\psi_I
(\gamma^{-1}x) \ ,
\end{equation}
and hence the mode sum expression for the automorphic field in terms
of the non-invariant vierbein is
\begin{equation}
\hat{\psi}_N(x)=
\frac{1}
{\left(\sum_{\gamma\in\Gamma}p(\gamma)^2\right)^{1/2}}
\sum_{\gamma\in\Gamma}
p(\gamma)
S(x)S^{-1}(\gamma^{-1}x)
\psi_N
(\gamma^{-1}x) \ .
\end{equation} 

Similarly the two-point function tranforms as 
\begin{equation}
S^+_{IM/\Gamma}(x,x')
\rightarrow{S^+_{NM/\Gamma}(x,x')}=S(x)S^+_{IM/\Gamma}(x,x')S^{-1}(x') \ .
\end{equation}
From (\ref{eq:diracautocorr}), the mode sum expression for the two point function
in terms of the non-invariant vierbein is hence
\begin{eqnarray}
S^+_{NM/\Gamma}(x,x')
& = & 
\sum_{\gamma\in\Gamma}
p(\gamma)
S(x)\langle{0}|\psi_I(x)\bar{\psi}_I(\gamma^{-1}x')|0\rangle{S^{-1}(x')} \ ,
\nonumber
\\
& = & 
\sum_{\gamma\in\Gamma}
p(\gamma)
\langle{0}|\psi_N(x)\bar{\psi}_N(\gamma^{-1}x')|0\rangle{S^{-1}(\gamma^{-1}x')}S^{-1}(x') \ .
\end{eqnarray}
We have seen
explicit examples of this in sections \ref{sec-masslessspinm-} and
\ref{sec:massivediracemm-}, and on the $\rp$ geon in section
\ref{sec:geonbog}.

In sections \ref{sec:diracm0} and \ref{sec-diracM-} we
shall work throughout in vierbeins invariant under
$J_0$ and $J_-$ respectively.

\subsection{Dirac detector on $M_0$}
\label{sec:diracm0}

Consider now Dirac field theory on $M_0$ as an automorphic field theory
on $M$ where expressions are written with respect to the standard Minkowski 
vierbein. From (\ref{eqn:autodircorrfn}) the $M_0$ correlation function is 
given by
\begin{equation}
\label{eq:diraccorM0}
S^+_{M_0}(\tau,\tau')=
\sum_{n\in\Z}
\eta^n
S^+_{M}(\tau,J_0(\tau')) \ ,
\end{equation}
where $\eta=1,(-1)$ labels spinors with periodic (antiperiodic)
boundary conditions. We therefore find an explicit expression
for $S^+_{M_0}(\tau,\tau')$ from (\ref{eq:diraccorM0}) and (\ref{eqn:dirccorrfnsmeared}).
As we work throughout in the standard Minkowski vierbein here,
writing the smeared field operator as in (\ref{eqn:smaereddirac}),
we may again argue in an analogous way to in the
previous section that the spinor transformation
$S(\tau,\mathbf{\xi})$ can be dropped here.

\subsubsection{Inertial detector}

First consider a Dirac detector following the inertial worldline
(\ref{eqn:inertraj}) on $M_0$.
The power spectrum (\ref{eq:powerspec}) for the noise is found to be
\begin{equation}
P(\omega)=\left(\frac{\omega^2}{2\pi}
+2
\sum_{n=1}^{\infty}
\frac{\eta^n\omega}
{4\pi{n}a}
\sin(2n\omega{a})\right)
\Theta(-\omega) \ .
\end{equation}
As on Minkowski space the power spectrum is $-\omega$ 
times the transition rate of the linearly coupled scalar field detector
following the same trajectory (\ref{eqn:transm0scalarinert}).
The summation thus may be performed to give $-\omega$ times
(\ref{eqn:inerm0}) and (\ref{eqn:inerm2}). 

For the transition rate we find
\begin{equation}
\dot{F}_\tau(\omega)
=
-\frac{1}{\pi^4}
\sum_{n\in\Z}
\sum_{m\in\Z}
\int_{-\infty}^{\infty}
ds\,
\frac{\eta^n\eta^m((s-2i\epsilon)^2-4nma^2)}
{[(s-2i\epsilon)^2-(2na)^2]^2[(s-2i\epsilon)^2-(2ma)^2]^2} \ .
\end{equation}
The $n=0,m=0$ term gives the transition rate on Minkowski space
(\ref{eqn:diracinertitrans}). The integral for other terms may 
be done by residues, with the result
\begin{eqnarray}
\dot{F}_\tau(\omega) & = & 
\Bigg(
-\frac{\omega^5}{60\pi^3}
+
\frac{1}{32\pi}
\sum_{\substack{n.m=-\infty \\ n,m\neq{0}}}^{\infty}
\frac{\eta^n\eta^m}
{(m-n)(m+n)^3a^5}
\left[
\left(\left(\frac{2m\omega{a}}{n}+2\omega{a}\right)\cos(2\omega{n}a)
\right.
\right.
\nonumber
\\
& &\left.-\frac{(m+3n)}{n^2}\sin(2\omega{n}{a})\right)
+
\left(\left(\frac{2\omega{a}n}{m}+2\omega{a}\right)\cos(2\omega{m}a)
\right.
\nonumber
\\
& &
\left.
\left.
-\frac{(n+3m)}{m^2}\sin(2\omega{m}{a})\right)\right]\Bigg)
\Theta(-\omega) \ ,
\end{eqnarray}
where the $n=m$ and $n=-m$ terms are understood in the limiting sense 
and can be verified to be finite.
As expected the response does not depend on the velocity.

\subsubsection{Uniformly accelerated detector}

Next consider the power spectrum for the Rindler noise, that is 
we consider $g(\tau,\tau')$ on the uniformly accelerated worldline.
From (\ref{eq:powerspec}) we find 
\begin{equation}
P(\omega)=\frac{i}{16\pi^2}
\int^{\infty}_{-\infty}
d\tau
\sum_{n\in\Z}
\frac{\eta^n{e^{-i\omega\tau}}\left(\alpha\sinh\left(\frac{\tau}{2\alpha}\right)-i\epsilon\cosh\left(\frac{\tau}{2\alpha}\right)\right)}
{\left(\left(\alpha\sinh\left(\frac{\tau}{2\alpha}\right)-i\epsilon\cosh\left(\frac{\tau}{2\alpha}\right)\right)^2-(na)^2\right)^2} \ .
\end{equation}
The contributions to the integral from each term in the sum
may be calculated separately by contour integration. The result is
\begin{eqnarray}
P(\omega) & = & 
\frac{1}{2\pi}
\frac{(\omega^2+\frac{1}{4\alpha^2})}{(1+e^{2\pi\alpha\omega})}
\nonumber
\\
&  &
+2
\sum_{n=1}^\infty
\frac{\eta^n}{(1+e^{2\pi\alpha\omega})}
\left[
\frac{\alpha^2n^2a^2\cos\left(2\omega\alpha\mathrm{arctanh}\left(\frac{(\alpha^2n^2a^2+n^4a^4)^{1/2}}{\alpha^2+n^2a^2}\right)\right)}
{4\pi\left(\frac{\alpha^2}{\alpha^2+n^2a^2}\right)^{1/2}(2n^6a^6+4n^4a^4\alpha^2+2n^2a^2\alpha^4)}
\nonumber \right.\\
&  & 
\label{eqn:accepowerm0}
\left.+\frac{\alpha^3(\alpha^2n^2a^2+n^4a^4)^{1/2}\omega\sin\left(2\omega\alpha\mathrm{arctanh}\left(\frac{(\alpha^2n^2a^2+n^4a^4)^{1/2}}{\alpha^2+n^2a^2}\right)\right)}
{2\pi\left(\frac{\alpha^2}{\alpha^2+n^2a^2}\right)^{1/2}(2n^6a^6+4n^4a^4\alpha^2+2n^2a^2\alpha^4)}
\right] \ ,
\end{eqnarray}
where again $\eta$ labels the spin structure. We see that the power spectrum
is dependent on the spin structure. The $n=0$ term in (\ref{eqn:accepowerm0}) agrees 
with the Minkowski space power spectrum (\ref{eqn:minkdiracpspec}) as expected, and both the
$n=0$ and $n>0$ terms in (\ref{eqn:accepowerm0}) contain the fermionic factor.
Note that no simple relation holds between the power spectrum (\ref{eqn:accepowerm0})
and the transition rate of the linearly coupled scalar field detector (\ref{M0response}),
in contrast to the relation we observed on the inertial worldline.

For the transition rate of a fermionic detector on $M_0$ we have
\begin{equation}
\dot{F}_\tau(\omega)=
2\int^{\infty}_{0}ds\,
Re\left(e^{-i\omega{s}}\mathrm{Tr}\left(S^+_{M_0}(\tau,\tau-s)^2\right)\right) \ .
\label{eqn:transm0dircexp}
\end{equation}
We may evaluate (\ref{eqn:transm0dircexp}) on the uniformly accelerated worldline
by substituting the worldline into (\ref{eq:diraccorM0}) and (\ref{eq:SeqG}).
It is easy to show that the $n=0$ term leads to the transition rate
found on Minkowski space (\ref{eq:diractranm}) as expected.
The evaluation of the other terms is not so straightforward as the
residues are not so easy to calculate. We shall not present the
result here.

\subsection{Dirac detector on $M_-$}
\label{sec-diracM-}

On $M_-$ we can again build expressions from those on 
$M$ (or $M_0$) via the method of images.
The transition rate is given by
\begin{equation}
\label{eq:transM-}
\dot{F}_\tau(\omega)=
2\int^{\infty}_{0}ds\,
Re\left(e^{-i\omega{s}}\mathrm{Tr}\left(S^+_{M_-}(\tau,\tau-s)^2\right)\right) \ ,
\end{equation}
and
\begin{eqnarray}
\label{eq:m-intermsm0}
S^+_{M_-}(\tau,\tau') & = & S^+_{M_0}(\tau,\tau')+\rho{S^+_{M_0}(\tau,J_-(\tau'))} \ ,
\\
                      & = & \sum_{n\in\Z}\rho^n{S^+_{M}(\tau,J_-^n(\tau'))} \ ,
\end{eqnarray}
where $S^+_{M_0}(\tau,\tau')$ and $S^+_{M}(\tau,\tau')$ are written in terms
of a vierbein which rotates by $2\pi$ in the $(x,y)$ plane as $z\mapsto{z+2a}$
(i.e. the one spin structure on $M_0$ compatible with the two on $M_-$). 
$\rho=1(-1)$ labels spinors with periodic (antiperiodic) boundary conditions
on $M_-$ with respect to this vierbein. That is $\rho$ labels the two possible 
spin structures on $M_-$.

Now on $M_-$ our main question of interest is whether or not
our detector can distinguish the 
two possible spin structures.
As we saw in section \ref{sec-masslessspinm-} the stress tensor for the massless
spinor field in $M_-$ (\ref{eqn:tmunu}) has non-zero shear components, 
$\langle{0_-}|{T_{xz}}|0_-\rangle$ and $\langle{0_-}|{T_{yz}}|0_-\rangle$,
which change sign under a change of spin structure.
It is therefore conceivable that a detector with a non-zero $z$-component 
of angular momentum could detect the spin structure. However as the
relation between $\langle{0_-}|{T_{\mu\nu}}|0_-\rangle$ and the detector
response is not clear it is not possible to tell in advance whether or not our
detector model will be sensitive to the spin structure. 

We consider therefore a detector following the trajectory 
\begin{eqnarray}
t=t(\tau) \ ,\quad   x=x(\tau) \ ,\quad   y=y_0 \ ,\quad   z=z_0 \ ,
\label{eqn:m-trajdir}
\end{eqnarray}
where $y_0\neq{0}$ and $z_0$ are constants.
First we note that there is no direct 
analogue of the Rindler noise power spectrum here as the power spectrum is defined in 
\cite{tk:takagi} only for stationary trajectories. We therefore look
directly at the transition rate. 
From (\ref{eq:m-intermsm0}) and (\ref{eq:transM-})
the transition rate will contain four terms. 
The first term, coming from $\mathrm{Tr}\left(S^+_{M_0}(\tau,\tau')\right)^2$,
will give us the same response as on $M_0$ (for $\eta=-1$ in (\ref{eq:diraccorM0}),
as there expressions are written with respect to the standard Minkowski vierbein).
This part is independent of spin structure ($\rho$) on 
$M_-$. The fourth term, coming from $\mathrm{Tr}\left(\left(\rho{S^+_{M_0}}(\tau,J_-(\tau'))\right)^2\right)$,
will also be independent of spin structure, as it contains only
$\rho^2=1$ in both cases. 
Thus the only way in which the transition rate may be 
sensitive to the spin structure on $M_-$ is through the cross terms,
$\mathrm{Tr}(\rho{S^+_{M_0}(\tau,J_-(\tau'))}S^+_{M_0}(\tau,\tau'))$ and
$\mathrm{Tr}(\rho{S^+_{M_0}(\tau,\tau')}{S^+_{M_0}(\tau,J_-(\tau'))})$.
However it is a reasonably straightforward matter to show that these traces
are both $0$ on the trajectory (\ref{eqn:m-trajdir}), due to simple cancellations in the products of
the Wightman functions. Thus we see, even without an explicit calculation
on a specific trajectory, that the transition rate cannot depend on 
$\rho$, and so the detector is not sensitive to the spin structure,
for \emph{any} motion at constant $y$ and $z$.

Unfortunately an explict evaluation of the transition rate on the inertial
or uniformly accelerated worldlines, as on $M_0$, is difficult to obtain
and we shall not discuss it further here.

\section{Static detectors on the $\rp$ geon}
\label{sec:Geon}

In the recent literature Deser and 
Levin \cite{dl:deser3,dl:deser2,dl:deser}
have presented kinematical arguments for the calculation
of the Hawking-Unruh effects in a large class of black hole
and cosmological spaces by mapping the trajectories of 
detectors in these spacetimes to Rindler trajectories 
in higher dimensional embedding spaces (known as
GEMS, or global embedding Minkowski spacetimes) in which these
spacetimes have global embeddings. In \cite{dl:deser3} uniformly
accelerated observers in de Sitter and Anti de Sitter space are considered. It is
seen that in de Sitter space their experience is thermal with temperature $T=a_5/(2\pi)$ 
where $a_5$ is their associated acceleration in the $5$-dimensional embedding space.
In Anti de Sitter space their experience is thermal provided the acceleration
is above a certain threshold. In \cite{dl:deser2} static observers in
Schwarzschild space are considered via a $6$-dimensional flat embedding space 
and the expected temperature and entropy are recovered. In
\cite{dl:deser} this GEMS approach for the derivations of temperature
and entropy is extended to 
Schwarzschild-(anti) de Sitter and Reissner-Nordstr\"om spaces in $4$ dimensions and rotating BTZ spaces
in $3$ dimensions, and the methods of \cite{dl:deser} can
be readily adapted to other cases. We note that indeed any
Einstein geometry has a GEMS \cite{go:gems}.

\cite{sl:gems2} considers  
GEMS calculations on a large class of higher dimensional black
holes, generalising the $4$-dimensional results of Deser and Levin
(and the results for the $4$-dimensional AdS hole and others
in  \cite{hkp:sohkimpark}).
In particular, $d$-dimensional Schwarzschild and Reissner-Nordstrom
in asymptotically flat, de Sitter and Anti de Sitter spaces are discussed.
The case of $4$-dimensional asymptotically locally anti-de Sitter is
particularly interesting as solutions with planar, cylindrical,
toroidal and hyperbolic horizon topology exist. The higher dimensional
versions of these non-spherical AdS black holes are also considered.
Their global embeddings in higher dimensional Minkowski spaces are found
and the associated temperatures and entropies obtained.
Other references on GEMS come from the group 
of Hong, Park, Kim, Soh and Oh 
\cite{hkp:sohkimpark,hkp:sohkimpark1,hkp:sohkimpark1.5,hkp:sohkimpark2,hkp:sohkimpark3}. These include
the $4$-dimensional AdS hole as mentioned above, static rotating and charged BTZ holes,
$(2+1)$ de Sitter holes, scalar tensor theories, charged dilatonic
black holes in $1+1$ dimensions, charged and uncharged black
strings in (2+1) dimensions, and a few other cases.
A recent paper by Chen and Tian \cite{ct:nonstatgems} argues that
the GEMS approach holds for general stationary motions in curved spacetimes.
However these authors argue further that the approach in general fails for 
non-stationary motions. The example they use is that of a freely falling
observer in the Schwarzschild geometry. We note here that although their
argument does prove that the GEMS argument is not valid for some
non-stationary trajectories by use of an example, it does not prove that
the GEMS approach is useless for all such trajectories.

Within the kinematical arguments employed in all the work reviewed above
the great simplification in working with these GEMS is that
we are mapping situations in curved spacetimes to corresponding
ones in a flat spacetime, where calculations are always simpler,
both conceptually and technically.
It seems reasonable following the success of the GEMS programme
that the responses of particle detectors in black hole
and cosmological backgrounds
could also be related in some way to responses of 
corresponding detectors in their GEMS. We note 
immediately that such a mapping of detector responses is
clearly not trivial as we would expect different responses
to occur due to the different dimensions which the spacetimes and 
their GEMS have, however some relation is still expected.
In this section then we present an argument which should
be relevant to the response of a static detector in the 
single exterior of the $\rp$ geon black hole (and the Kruskal
spacetime) via an embedding of the Kruskal manifold into
a $7$-dimensional Minkowski space. This embedding space is 
different to the $6$-dimensional embedding of Kruskal
so far used in the GEMS literature \cite{fr:Fronsdal}, but we use
it as it is more easily adapted to the $\rp$ geon.

We begin by first presenting the embedding (see \cite{gb:embed}).
The complexified Kruskal manifold $M_C$ here is considered to be an algebraic variety 
in $\C^7$. With coordinates $(z_1,\ldots,z_7)$ and metric
\begin{equation}
ds^2
=
-(dz_1)^2-(dz_2)^2
-
\ldots
-(dz_6)^2
+(dz_7)^2 \ ,
\end{equation}
$z_7$ being the timelike coordinate, $M_C$ is determined by
\begin{eqnarray}
(z_6)^2-(z_7)^2+4/3(z_5)^2 & = & 16M^2 \ ,
\nonumber
\\
\left((z_1)^2+(z_2)^2+(z_3)^2\right)(z_5)^4 & = & 576M^6 \ ,
\nonumber
\\
\sqrt{3}z_4z_5+(z_5)^2 & = & 24M^2 \ .
\end{eqnarray}
The Lorentzian section of $M_C$, denoted by $\hat{M}_L$, is the subset
stabilised by $J_L:(z_1,\ldots,z_7)\mapsto(z_1^*,\ldots,z_7^*)$.
$\hat{M}_L$ consists of two connected components, one with $z_5>0$
and one $z_5<0$, both of which are isometric to the Kruskal manifold,
which we denote by $M_L$.
An explicit embedding of $M_L$ into $\hat{M}_L$ with $z_5>0$ is
given by
\begin{eqnarray}
z_1 & = & r\sin\theta\cos\phi \ ,
\nonumber
\\
z_2 & = & r\sin\theta\sin\phi \ ,
\nonumber
\\
z_3 & = & r\cos\theta \ ,
\nonumber
\\
z_4 & = & 4M\left(\frac{r}{2M}\right)^{1/2}-2M\left(\frac{2M}{r}\right)^{1/2} \ ,
\nonumber
\\
z_5 & = & 2M\left(\frac{6M}{r}\right)^{1/2} \ ,
\nonumber
\\
z_6 & = & 4M\left(\frac{2M}{r}\right)^{1/2}\mathrm{exp}\left(-\frac{r}{4M}\right)X \ ,
\nonumber
\\ 
z_7 & = & 4M\left(\frac{2M}{r}\right)^{1/2}\mathrm{exp}\left(-\frac{r}{4M}\right)T \ ,
\label{eqn:z1toz7}
\end{eqnarray}
with $X^2-T^2>-1$ and $r=r(T,X)$ defined as the unique solution to
\begin{equation}
\left(\frac{r}{2M}-1\right)\mathrm{exp}\left(\frac{r}{2M}\right)=X^2-T^2 \ .
\end{equation}
Here $(T,X,\theta,\phi)$ are a set of usual Kruskal 
coordinates, giving the usual Kruskal metric on 
$M_L$.
In each of the four regions of $M_L$, $|X|\neq|T|$, one
can introduce as usual local Schwarzschild coordinates
$(t,r,\theta,\phi)$. For $X>|T|$, the transformation reads
\begin{eqnarray}
T & = & \left(\frac{r}{2M}-1\right)^{1/2}\mathrm{exp}\left(\frac{r}{4M}\right)\sinh\left(\frac{t}{4M}\right) \ ,
\nonumber
\\
X & = & \left(\frac{r}{2M}-1\right)^{1/2}\mathrm{exp}\left(\frac{r}{4M}\right)\cosh\left(\frac{t}{4M}\right) \ ,
\end{eqnarray}
where $r>2M$, and the expressions for $z_1,z_2,\ldots,z_5$ are as in (\ref{eqn:z1toz7}) while those for $z_6$ and $z_7$ become
\begin{eqnarray}
z_6 & = & 4M\left(1-\frac{2M}{r}\right)^{1/2}\cosh\left(\frac{t}{4M}\right) \ ,
\nonumber
\\
z_7 & = & 4M\left(1-\frac{2M}{r}\right)^{1/2}\sinh\left(\frac{t}{4M}\right) \ .
\end{eqnarray}
Recalling that $z_7$ is the timelike coordinate in the embedding space, we
see immediately that an observer static in the exterior region $X>|T|$
at $r=\textrm{const}, \theta=\textrm{const}, \phi=\textrm{const}$ is a Rindler
observer in the $7$-dimensional embedding space with $(z_1,\ldots,z_5)$ constant
and acceleration in the $z_6$-direction of magnitude 
\begin{equation}
\label{eqn:accelembed}
a=1/\alpha=\frac{1}{4M\left(1-\frac{2M}{r}\right)^{1/2}} \ .
\end{equation}
As we have seen, the response of such a Rindler
detector in the embedding space is thermal with the
associated temperature
\begin{equation}
\label{eqn:localgemshawk}
T=\frac{a}{2\pi}=\frac{1}{2\pi\alpha}=\frac{1}{8\pi{M}\left(1-\frac{2M}{r}\right)^{1/2}} \ .
\end{equation}
This gives the Hawking temperature as seen by the static observer in the
black hole spacetime.
The associated black hole temperature, i.e. the temperature as seen at infinity,
in the Kruskal spacetime is
given by the Tolman relation
\begin{equation}
\label{eqn:infgems hawk}
T_0=g_{00}^{1/2}T=\frac{1}{8\pi{M}} \ ,
\end{equation}
(\ref{eqn:localgemshawk}) and (\ref{eqn:infgems hawk}) are the expected expressions
on Kruskal space \cite{bd:book}. Thus the black hole temperature as seen by a
static observer has been derived from the Unruh temperature seen by the
associated Rindler observer in the global embedding Minkowski spacetime.

Next we consider the $\rp$ geon. This is built as a quotient
of the Kruskal manifold under the involutive isometry
$J_G:(T,X,\theta,\phi)\mapsto(T,-X,\pi-\theta,\phi+\pi)$.
We now extend the action of the group generated by $J_G$ 
to the $7$-dimensional embedding space $M_C$ in order to
obtain a suitable embedding space for the geon.
A suitable extension of $J_G$ is
$\bar{J}_G:(z_1,z_2,z_3,z_4,z_5,z_6,z_7)\mapsto(-z_1,-z_2,-z_3,z_4,z_5,-z_6,z_7)$,
which is an involution on $M_C$.
Again the worldline of a static detector in the $\rp$ geon exterior $X>|T|$ is
mapped to the worldline of a Rindler observer with acceleration in the
$z_6$-direction with magnitude (\ref{eqn:accelembed}) in this embedding space.
We suggest therefore that the calculations of the time dependent responses
of an accelerated observer in the $d$-dimensional quotients of Minkowski space, 
done in section \ref{sec:quotient}, should have relevance to the response of a static detector in the
exterior of the $\rp$ geon (although the exact nature of the relationship is 
not clear).
In particular if we specialise the results of section \ref{sec:quotient} to 
a detector with uniform acceleration (\ref{eqn:accelembed}) in the 
quotient of a $7$-dimensional Minkowski space under involution $\bar{J}_G$ we
see that the response has two parts. The thermal time-independent part
is given by (\ref{thermal}), which in the present case reads
\begin{equation}
\dot{F}_{M\tau}(\omega)
=
\frac{a^4}{64\pi^2(e^{\frac{2\pi\omega}{a}}+1)}
(1/4+\omega^2/a^2)(9/4+\omega^2/a^2) \ .
\end{equation}
Again this is a thermal response associated with a temperature
$T=\frac{a}{2\pi}$. Clearly the response is different to that of the
static detector on 
Kruskal due to the higher number of dimensions though the two should be related.
The image part of the response for this Rindler detector is, 
in the case of a detector switched on in the infinite past,
\begin{equation}
\dot{F}_{I\tau}(\omega)
=
\frac
{3}
{8\pi^{3}}
\int_0^\infty
ds
\;
\frac{\cos(\omega{s})}
{\left(\frac{4}{a^2}\cosh^2\left(\frac{a(2\tau-s)}{2}\right)+4r^2\right)^{5/2}} \ .
\end{equation}

\begin{figure}[htbp]
\includegraphics[angle=0, width={4in}]{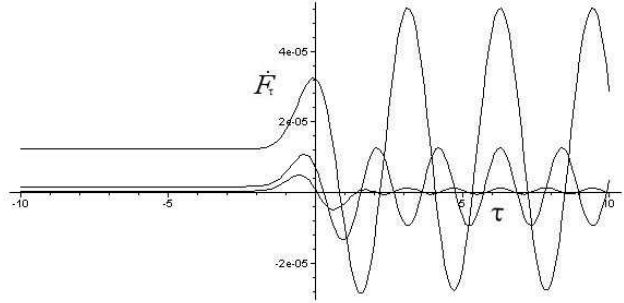}
\centering
\caption[Transition rate associated with the response of a static detector on the $\rp$ geon.]{Transition rate 
for a detector uniformly accelerated in the $z_1$ direction on the
quotient of $7$-dimensional Minkowski space under the involution $\bar{J}_G$.
The parameters are $\alpha=1$, $C=(2z_1)^2+(2z_2)^2+(2z_3)^2=4r^2=64/9$, and $\omega=1$ (upper curve), 
$\omega=1.5$ and $\omega=2$ (lower curve).}
\label{fig:GEMSGEON}
\end{figure}

The total response for certain values of the parameters is shown in figure \ref{fig:GEMSGEON}.
The comments of section \ref{sec:automorphic} then follow. The image part
consists of a term periodic in $\tau$ with period $\pi/\omega$ plus a term
bounded by a function which dies off exponentially for large $\tau$.
The numerical evidence exhibits behaviour
qualitatively very similar to that of figure \ref{fig:accel}.
The comments made in the section \ref{sec:minkbound} about finite time 
detections also follow here. In particular the oscillatory behaviour of the 
boundary part as $\tau\rightarrow\infty$ is a property only of the case of infinite time detection.
For a detector switched on at $-\infty<\tau_0<0$ the transition rate oscillates for some
time period with $\tau>0$, but eventually it will fall to the thermal response and so
at late times the difference between the response on the Minkowski space
and that on the quotient space vanishes. This implies for finite time static detectors 
on the $\rp$ geon the difference between the response there and that on Kruskal spacetime
falls off also to $0$ at late times, which is in agreement with the comments made in \cite{lm:geon}.
Note however the different behaviour in the case of the infinite time detection.

\section{Detectors in conformally flat spacetimes}
\label{sec:conformal}

In this section we consider monopole detectors in conformally flat spacetimes coupled
to a massless conformally coupled scalar field. Conformally flat spacetimes are of particular interest
as they include all two-dimensional spacetimes and all spatially flat Friedmann-Robertson-Walker models,
which include a particular coordinatization of half of the de Sitter manifold as a special case. We will
be considering detectors in de Sitter space, and $\rp$ de Sitter space, in section \ref{sec-desitrp3}.
The main aim of this section is to argue for a regularization of the Wightman function
which is similar to that considered for the scalar field in Minkowski space in the earlier sections. Transition 
rates of monopole detectors can then be calculated for detectors travelling along arbitrary 
trajectories. The discussion at the start of this section follows that of~\cite{bd:book}.

The metric on a conformally flat spacetime satisfies 
\begin{equation}
\label{eqn:conffieldmet}
g_{\mu\nu}
=
\Omega^2(x)
\eta_{\mu\nu} \ ,
\end{equation}
where $\eta_{\mu\nu}$ is the usual Minkowski metric and $\Omega(x)$
is a continuous, non-vanishing, finite, real function. We use associated Minkowski
coordinates $(\eta,\mathbf{x})$ in which (\ref{eqn:conffieldmet}) reads
\begin{equation}
\label{eqn:conffieldmet2}
ds^2
=
\Omega^2
(d\eta^2-d\mathbf{x}\cdot{d}\mathbf{x}) \ .
\end{equation}

We consider here a 
conformally coupled massless scalar field. The wave equation in
$d$ dimensions is
\begin{equation}
\label{eqn:conform}
\left[\Box+\frac{(d-2)R}{4(d-1)}\right]\phi=0 \ .
\end{equation}
From (\ref{eqn:conffieldmet}) it follows that (\ref{eqn:conform})
takes the form
\begin{equation}
\label{eqn:conmink}
\eta^{\mu\nu}\partial_\mu\partial_\nu\bar{\phi}=0 \ ,
\end{equation}
where $\bar{\phi}=\Omega^{(d-2)/2}\phi$.
The exponent $(d-2)/2$ is called the conformal weight of the scalar field.
Equation (\ref{eqn:conmink}) has the familiar Minkowski form and admits the usual plane wave 
mode solutions, 
\begin{equation}
\label{eqn:minkspacemodesol}
\bar{\phi}_\mathbf{k}(\eta,\mathbf{x})
=
\frac{1}{(2\omega(2\pi)^{d-1})^{1/2}}
e^{-i\omega\eta+i\mathbf{k}\cdot\mathbf{x}} \ ,
\end{equation}
where $\omega=|\mathbf{k}|$. These modes are positive frequency with
respect to the timelike conformal Killing vector $\partial_\eta$.
A set of mode solutions to (\ref{eqn:conform}) are then clearly 
given by
\begin{equation}
\label{eqn:conformspacemodesol}
\phi_\mathbf{k}(\eta,\mathbf{x})
=
\frac{1}{(2\omega(2\pi)^{d-1}\Omega^{d-2}(\eta,\mathbf{x}))^{1/2}}
e^{-i\omega\eta+i\mathbf{k}\cdot\mathbf{x}} \ .
\end{equation}
The field is then expanded in these mode solutions
\begin{equation}
\label{eqn:confexp}
\phi(x)=
\int
d^{d-1}k
\left[a_\mathbf{k}\phi_\mathbf{k}(x)+a^\dagger_\mathbf{k}\phi^*_\mathbf{k}(x)\right] \ ,
\end{equation}
and canonically quantized in the usual way, by imposing the usual commutation
relations on annihilation and creation operators. The vacuum state annihilated 
by all the annihilation operators above $|0\rangle$ is known as the
conformal vacuum state.

Now we consider a model monopole particle detector coupled
to this conformally invariant scalar field. The model of the
unregularized detector is similar to that in the previous sections.
The transition rate for a detector turned on
at time $\tau_0$, when the field was in the conformal vacuum,
and off at $\tau>\tau_0$, is hence 
\begin{equation}
\dot{F}_{\tau}(\omega)=
2\int^{\tau-\tau_0}_{0}ds\,
Re
\left(
e^{-i\omega{s}}\langle{0}|\phi(\tau)\phi(\tau-s)|0\rangle
\right) \ .
\label{eqn:conftransrate}
\end{equation}
From (\ref{eqn:confexp}) it follows that the Wightman function
in (\ref{eqn:conftransrate}) is given by
\begin{eqnarray}
\langle{0}|\phi(x)\phi(x')|0\rangle
\label{eqn:correcon}
& = &
\int
d^{d-1}k\,
\phi_\mathbf{k}(x)
\phi_\mathbf{k}^*(x')
\\
& = &
\Omega^{(2-d)/2}(x)
\int
d^{d-1}k\,
\bar{\phi}_\mathbf{k}(x)
\bar{\phi}_\mathbf{k}^*(x')
\Omega^{(2-d)/2}(x') \ .
\end{eqnarray}
The curved spacetime Wightman function is hence
related to the Minkowski Wightman function by
\begin{equation}
\label{eqn:conrelat}
\langle{0}|\phi(x)\phi(x')|0\rangle
=
\Omega^{(2-d)/2}(x)
\langle{0}|\bar{\phi}(x)\bar{\phi}(x')|0\rangle
\Omega^{(2-d)/2}(x') \ .
\end{equation}

We now ask: What is a suitable regularization for
(\ref{eqn:correcon}) on a given trajectory $x=x(\tau)$?
That is, we wish to regularize the integral
\begin{eqnarray}
\langle{0}|\phi(x(\tau))\phi(x(\tau'))|0\rangle
& = & 
\frac{\Omega^{(2-d)/2}(x(\tau))
\Omega^{(2-d)/2}(x(\tau'))}
{(2\pi)^{d-1}}
\nonumber
\\
\label{eqn:correconf}
&    &
\times
\int
\frac{d^{d-1}k}{2|\mathbf{k}|}\,
e^{-i|\mathbf{k}|(\eta(\tau)-\eta(\tau'))+i\mathbf{k}\cdot(\mathbf{x}(\tau)-\mathbf{x}(\tau'))} \ .
\end{eqnarray}
We argue, as with the case discussed in Minkowski space, that 
a naive $i\epsilon$-prescription where we just introduce an 
ultraviolet cut-off factor $e^{-\epsilon|\mathbf{k}|}$ (as seems to have
been considered by Birrell and Davies in the case of de Sitter space in 
the spatially flat coordinates~\cite{bd:book}) is not suitable here. 
Indeed we have shown via a numerical analysis that in the case of a 
comoving observer in de Sitter space 
in the spatially flat coordinates the transition rate for a detector 
switched on in the infinite past is time dependent. We interpret this, as 
Schlicht does for the uniformly accelerated detector in Minkowski 
space, as an unphysical result.

We introduce instead a regularization based on a frequency cut-off in a frame 
adapted to the detector's motion.
Our proposal is to take the mode solutions in the curved space (\ref{eqn:conformspacemodesol})
and multiply them by the factor $\Omega^{(d-2)/2}(\eta,\mathbf{x})$ to give the
modes (\ref{eqn:minkspacemodesol}). For a given mode $\phi$ (\ref{eqn:minkspacemodesol}) 
we then define the frequency as seen by the 
moving detector operationally by
\begin{equation}
i\frac{d\phi}{d\tau}=\omega'\phi \ ,
\end{equation}
where $\tau$ is the detector's proper time in the curved space. We thus
find from (\ref{eqn:minkspacemodesol}) that 
$\omega'=\omega\dot{\eta}(\tau)-\mathbf{k}\cdot\dot{\mathbf{x}}(\tau)$ and 
we regularize the integral (\ref{eqn:correconf}) by 
introducing a cut-off of the form
$e^{-\epsilon(\omega(\dot{\eta}(\tau)+\dot{\eta}(\tau'))-\mathbf{k}\cdot(\dot{\mathbf{x}}(\tau)+\dot{\mathbf{x}}(\tau')))}$.
The integral is then exactly as in (\ref{eq:correlation}) and the result is just
the correlation function with conformal relation (\ref{eqn:conrelat}) to
(\ref{eq:corre}) and (\ref{eqn:2dimcorrfun}). That is,
\begin{eqnarray}
& \langle{0}|\phi(\tau)\phi(\tau')|0\rangle=
\frac{\Gamma[d/2-1]}{4\pi^{d/2}}
\frac{\Omega^{(2-d)/2}(x(\tau))
\Omega^{(2-d)/2}(x(\tau'))}{A^{d/2-1}} \ , \nonumber
\\
& A=\left[i^2(\eta(\tau)-\eta(\tau')-i\epsilon(\dot{\eta}(\tau)+\dot{\eta}(\tau')))^2
+(\mathbf{x}(\tau)-\mathbf{x}(\tau')-i\epsilon(\dot{\mathbf{x}}(\tau)+\dot{\mathbf{x}}(\tau')))^2\right] \ ,
\nonumber
\\
\label{eqn:conf2233}
\end{eqnarray}
for $d>2$, while for $d=2$ the expression is just (\ref{eqn:2dimcorrfun}) with $t$ replaced by $\eta$.

If we were to define the frequency as seen by the detector by
\begin{equation}
\label{eqn:freqdef}
i\frac{d\bar{\phi}}{d\tau}=\omega'\bar{\phi} \ ,
\end{equation}
that is, using (\ref{eqn:conformspacemodesol}) we would find 
$\omega'=\omega\dot{\eta}(\tau)-\mathbf{k}\cdot\dot{\mathbf{x}}(\tau)-i\frac{(d-2)}{2}\frac{\dot{\Omega}(\tau)}{\Omega(\tau)}$.
This frequency would have an additional imaginary part. Introducing a cut-off $e^{-\epsilon(\omega'(\tau)+\omega'(\tau'))}$
into (\ref{eqn:correconf}) would give a correlation function without the relation (\ref{eqn:conrelat}) to the
Minkowski one. Our prescription leading to (\ref{eqn:conf2233}) is then equivalent to introducing the cut-off
$e^{-\epsilon\mathrm{Re}(\omega'(\tau)+\omega'(\tau'))}$, where $\mathrm{Re}$ denotes the real part.
It is interesting to note that in the cases considered in section \ref{sec-desitrp3}, that is, the 
comoving and uniformly accelerated observers on de Sitter space, the imaginary part of $\omega'$
is found to be a constant, which then does not contribute to the transition rate integrals,
and thus makes no difference to the results in the $\epsilon\rightarrow{0}$ limit.
On a more general trajectory however the imaginary part of $\omega'$, defined by (\ref{eqn:freqdef}), could depend on
$\tau$ and so could conceivably affect the transition rate.

We shall examine the consequences of the proposal (\ref{eqn:conf2233})
in section \ref{sec-desitrp3}.
In particular we shall recover the expected thermal responses
for comoving and uniformly accelerated detectors in de Sitter space.
Further we will show that for the comoving observer, 
in the coordinates that admit flat
spatial sections, this regularization may also arise from 
considering a spatially averaged field operator
and so from a model of an extended detector.

\section{De Sitter and $\rp$ de Sitter spaces}
\label{sec-desitrp3}

In this section we begin by considering a model detector in de Sitter space. 
Throughout we consider a conformally coupled massless scalar field moving 
through the Euclidean vacuum \cite{ba:desitstates}.
We see that here a similar situation is
encountered to that found by Schlicht for the uniformly accelerated
detector in Minkowski space.
It is seen that in the case of a comoving detector switched on in
the infinite past and off at finite $\tau$, if the
correlation function is regularized by a naive $i\epsilon$ prescription, as for
example is done in Birrell and Davies~\cite{bd:book}, we are led to an unphysical,
$\tau$-dependent response. 
We therefore use the regularization introduced in section \ref{sec:conformal}. 
Further, for a comoving detector,
we show that such a regularization can arise also by considering a model 
detector with spatial extent, that is by considering
a smeared field operator in the interaction Hamiltonian.
We recover the usual time independent thermal response for comoving 
and uniformly accelerated detectors.

Subsection \ref{sec:rp3} then considers comoving observers in $\rp$ de Sitter space,
such that the motion is orthogonal to the distinguished foliation \cite{jk:desit}.
In addition to the thermal part seen in de Sitter space, the transition rate contains
an image part, related to that found in section \ref{sec:automorphic} for a uniformly accelerated 
detector on a $4$-dimensional Minkowski space with a planar boundary.
We also address a comoving detector in de Sitter and $\rp$ de Sitter space in a GEMS approach, by considering
the response of the associated uniformly accelerated detectors in higher dimensional
Minkowski (with boundaries in the case of $\rp$ de Sitter) embedding spaces.
As we are able to do the calculations both in the original curved spaces
and in the global embedding spaces, the results help to clarify the
relation and validity of relating detector responses to those in embedding spaces.

\subsection{Detectors in de Sitter space}
\label{sec-desit}

We represent $d$-dimensional de Sitter space as the hyperboloid 
\begin{equation}
\label{eqn:desithyper}
z_0^2-z_1^2-\cdots-z_{d}^2=-\alpha^2 \ ,
\end{equation}
embedded in the ${d+1}$-dimensional Minkowski space,
\begin{equation}
ds^2=
dz_0^2-dz_1^2-\cdots-dz_{d}^2 \ ,
\end{equation}
with $z_i$ real-valued coordinates.
Let us consider the coordinates $(t,\mathbf{x})$ defined by
\begin{eqnarray}
z_0 & = & \alpha\sinh(t/\alpha)+\frac{e^{t/\alpha}}{2\alpha}|\mathbf{x}|^2 \ ,
\nonumber
\\
z_{d} & = & \alpha\cosh(t/\alpha)-\frac{e^{t/\alpha}}{2\alpha}|\mathbf{x}|^2 \ ,
\nonumber
\\
z_i & = & e^{t/\alpha}x_i \ .
\end{eqnarray}
These coordinates cover the half of the de Sitter hyperboloid
given by $z_0+z_{d}>0$.
The line element is that of a $d$-dimensional
Friedman-Robertson-Walker spacetime with exponentially expanding flat
spatial sections,
\begin{equation}
ds^2
=
dt^2
-
e^{2t/\alpha}
\left(dx_1^2+\cdots+d_{d-1}^2\right) \ .
\end{equation}
Introducing the conformal time $\eta=-\alpha{e^{-t/\alpha}}$, the line
element becomes conformal to Minkowski,
\begin{equation}
\label{eqn:conflineel}
ds^2
=
\frac{\alpha^2}{\eta^2}\left[d\eta^2-\sum_{i=1}^{d-1}(dx_i)^2\right] \ ,
\end{equation}
where $-\infty<\eta<0$.

Consider a conformally coupled scalar field in the
line element (\ref{eqn:conflineel}). As we have a conformally coupled field in a conformally flat spacetime
the situation is conformally trivial, and we shall apply the discussion 
in section \ref{sec:conformal}, with $\Omega^2(x)=\alpha^2/\eta^2$.
The vacuum state associated with a complete set of modes positive
frequency with respect to conformal Killing time $\eta$, that is the conformal
vacuum, coincides with the state known as the Euclidean vacuum~\cite{bd:book}.
The Euclidean vacuum $|0_E\rangle$ is uniquely characterised
as the state whose correlation function $\langle{0_E}|\phi(x)\phi(x')|0_E\rangle$
is invariant under the connected component of the de Sitter group,
and the only singularities of the correlation function are when $x'$ is
on the lightcone of $x$~\cite{ba:desitstates}.
Even though we have here defined the Euclidean vacuum in
coordinates that only cover half of the de Sitter hyperboloid, it 
is worth mentioning that the state is well defined on the whole hyperboloid~\cite{bd:book}.

We shall now specialize to $4$-dimensional de Sitter space (although the
extension to higher dimensions is straightforward) and 
consider a uniformly accelerated detector following the worldline
\begin{eqnarray}
z_0 & = & \alpha\sinh(t/\alpha)+\frac{e^{-t/\alpha}}{2\alpha}r^2 \ ,
\nonumber
\\
z_4 & = & \alpha\cosh(t/\alpha)-\frac{e^{-t/\alpha}}{2\alpha}r^2 \ ,
\nonumber
\\
z_1 & = & z_2  = 0 \ ,
\nonumber
\\
\label{eqn:accelobs}
z_3 & = & r,
\end{eqnarray}
with $r=\textrm{constant}$. The worldline of such an observer in the 
embedding space is a hyperbola $(z_4)^2-(z_0)^2=\alpha^2-r^2$, $z_1=z_2=0$, $z_3=r$.
In the de Sitter space the observer has constant proper
acceleration $a$, where $a^2=-g_{\mu\nu}\dot{u}^\mu\dot{u}^\nu$,
$\dot{u}^\mu={u}^\nu(\nabla_\nu{u}^\mu)$ and $u^\mu$ is the tangent vector of the
trajectory, of magnitude
\begin{equation}
a=\frac{r}{\alpha(\alpha^2-r^2)^{1/2}} \ .
\end{equation}
The proper time for the accelerated observer is
$\tau=(\alpha^2-r^2)^{1/2}t/\alpha$.
The response function for the detector is
\begin{equation}
\label{eqn:desittrans2}
\dot{F}_{\tau}(\omega)=
2\int^{\tau-\tau_0}_{0}ds\,
Re
\left(
e^{-i\omega{s}}\langle{0}_E|\phi(\tau)\phi(\tau-s)|0_E\rangle
\right) \ ,
\end{equation}
and
\begin{eqnarray}
& \langle{0}_E|\phi(\tau)\phi(\tau')|0_E\rangle=
-\frac{1}{4\pi^2\alpha^2}
\frac{\eta(\tau)\eta'(\tau)}{A} \nonumber
\\
& A=[(\eta(\tau)-\eta(\tau')-i\epsilon(\dot{\eta}(\tau)+\dot{\eta}(\tau'))^2
-(\mathbf{x}(\tau)-\mathbf{x}(\tau')-i\epsilon(\dot{\mathbf{x}}(\tau)+\dot{\mathbf{x}}(\tau')))^2] \ .
\nonumber
\\
\label{eq:euclidcorre}
\end{eqnarray}
Substituting (\ref{eqn:accelobs}) into (\ref{eq:euclidcorre}) we find 
\begin{equation}
\langle{0}_E|\phi(\tau)\phi(\tau')|0_E\rangle=
-\frac{1}{16\pi^2\left((\alpha^2-r^2)^{1/2}\sinh\left(\frac{\tau-\tau'}{2(\alpha^2-r^2)^{1/2}}\right)
-i\epsilon\cosh\left(\frac{\tau-\tau'}{2(\alpha^2-r^2)^{1/2}}\right)\right)^2} \ ,
\end{equation}
and the transition rate
of the detector switched on in the infinite past and off at time $\tau$ is 
independent of $\tau$ and is given by
\begin{equation}
\dot{F}_{\tau}(\omega)=
\frac{\omega}
{2\pi(e^{2\pi\omega(\alpha^2-r^2)^{1/2}}-1)} \ .
\end{equation}
The accelerated detector thus experiences a
thermal response at temperature
\begin{equation}
T=\frac{1}{2\pi}
\left(\frac{1}{\alpha^2}+a^2\right)^{1/2} \ .
\end{equation}
The response of a comoving detector is obtained by setting $a=0$. The transition rate
is still thermal at temperature
$T=1/(2\pi\alpha)$.
These results agree with the previous literature (e.g \cite{bd:book,dl:deser3}).
What is new is that we have obtained these results in a causal way for a 
detector switched on in the infinite past and read at a finite time, as opposed to the
case usually considered of a detection over the entire worldline.

We end this section by showing that the regularization in (\ref{eq:euclidcorre}), in
the case of a comoving observer, may be obtained by considering the monopole detector as the
limit of an extended 
detector in de Sitter space.
The reason why this is simple in the case of a comoving observer but not for
other trajectories is that spatial hypersurfaces of constant $t$ in the
coordinates $(t,\mathbf{x})$ are flat Euclidean spaces, which allow us to 
introduce a detector with infinite spatial extent along these slices. 
Care must be taken when defining the shape function for the detector
however, because the hypersurfaces are expanding with increasing $t$,
with a shape which is rigid in the proper distance the regularization follows.
The averaging over spatial hypersurfaces in effect introduces a short distance,
high frequency cut-off in the modes. 

The detector model is that of section \ref{sec-linmink}. It is a multi-level
quantum mechanical system coupled to a massless conformally coupled
scalar field via the interaction Hamiltonian
\begin{equation}
\label{eqn:comovhint}
H_{\mathrm{int}}=cm(\tau)\phi(\tau) \ .
\end{equation}
We consider a detector following the trajectory
$t=\tau$ that is $\eta=-\alpha{e^{-\tau/\alpha}}$
and $\mathbf{x}=0$. In (\ref{eqn:comovhint})
we consider the field smeared with a detector profile
function over constant $\tau$ hypersurfaces, that is
\begin{equation}
\label{eqn:comovshape}
\phi(\tau)=\int
d^3x\,
W_\epsilon(\mathbf{x})
\phi(\tau,\mathbf{x}) \ .
\end{equation}
For the profile function we choose
\begin{equation}
\label{eq:shapedesitt}
W_{\epsilon}(\mathbf{x})
=
\frac{1}{\pi^2}
\frac{\epsilon{e^{-\tau/\alpha}}}{(\mathbf{x}^2+\epsilon^2{e^{-2\tau/\alpha}})^{2}} \ .
\end{equation}
The detector shape (\ref{eq:shapedesitt}) is now time dependent! The reason for this
is that we want a detector which is rigid in its rest frame.
That is, we want a detector which is rigid with
respect to proper distance and not comoving distance.
The two distances are related by
$L_{\textrm{prop}}=e^{\tau/\alpha}L_{\textrm{comov}}$.
In (\ref{eqn:comovshape}) the integration is done over
$\mathbf{x}$, which is a comoving coordinate, and using 
a time independent shape function there would mean that
the detector is rigid with respect to comoving distance.
It is a simple matter to show that a shape function which selects
a distance scale $L'$ may be obtained from one which selects
a distance scale $L$ by
\begin{equation}
W_{L'}(\mathbf{x})=\frac{L^3}{{L'}^3}
W_L\left(\frac{L}{L'}\mathbf{x}\right) \ .
\end{equation}
If we write (\ref{eq:shapedesitt}) now in terms
of proper distance we find
\begin{equation}
W_{\epsilon_{\textrm{prop}}}(\mathbf{\xi})
=
\frac{1}{\pi^2}
\frac{\epsilon_{\textrm{prop}}}{\left(\mathbf{x}^2+\epsilon^2_{\textrm{prop}}\right)^{2}} \ ,
\end{equation}
and so in terms of proper distance (\ref{eq:shapedesitt}) is
a rigid shape in the sense that it is time independent.
Using this shape function we find, using the mode expansion
of $\phi$,
\begin{eqnarray}
\langle{0}_E|\phi(\tau)\phi(\tau')|0_E\rangle
& = &
\frac{\eta\eta'}{(2\pi)^{3}\alpha^2}
\int
\frac{d^{3}k}{2\omega}
\int
d^{3}x\,
W_\epsilon(\mathbf{x})
e^{-i(\omega{\eta(\tau)}-\mathbf{k}\cdot\mathbf{x})}
\nonumber
\\
& & \times\int
d^{3}x'\,
W_\epsilon(\mathbf{x'})
e^{i(\omega{\eta(\tau')}-\mathbf{k}\cdot\mathbf{x})} \ .
\end{eqnarray}
Further we find
\begin{eqnarray}
g_\epsilon(\mathbf{k};\tau)
& = &
\int
d^{3}x\,
W_\epsilon(\mathbf{x})
e^{-i(\omega{\eta(\tau)}-\mathbf{k}\cdot\mathbf{x})} \ ,
\nonumber
\\
& = & e^{-i|\mathbf{k}|\eta(\tau)}e^{-\epsilon|\mathbf{k}|e^{-\tau/\alpha}}
\nonumber
\\
& = & e^{-i|\mathbf{k}|\eta(\tau)}e^{-\epsilon|\mathbf{k}|\dot{\eta}(\tau)}
 \ ,
\end{eqnarray}
where the integration is done by transforming to spherical coordinates 
(appendix \ref{ch:appendixA}).
Hence
\begin{equation}
\label{eqn:corrdesitexten}
\langle{0}_E|\phi(\tau)\phi(\tau')|0_E\rangle
 = 
\frac{\eta\eta'}{(2\pi)^{3}\alpha^2}
\int
\frac{d^{3}k}{2\omega}\,
e^{-i\omega(\eta-\eta'-i\epsilon(\dot{\eta}+\dot{\eta}'))} \ .
\end{equation}
The expression (\ref{eqn:corrdesitexten}) agrees with that found above from
the ultraviolet cut-off regularization.

\subsection{$\rp$ de Sitter space}
\label{sec:rp3}

In this section we consider an inertial detector
that is linearly coupled to a conformally coupled massless scalar field in
$\rp$ de Sitter spacetime~\cite{jk:desit}.~\footnote{See also \cite{bm:mcinnes}
for a nice discussion on de Sitter space vs $\rp$ de Sitter.}

$\rp$ de Sitter space is built as a quotient of de Sitter space
under the group generated by the discrete isometry
\begin{equation}
\label{eqn:rpgroup}
J:(z_0,z_1,z_2,z_3,z_4)\mapsto(z_0,-z_1,-z_2,-z_3,-z_4) \ ,
\end{equation}
which induces a map $\tilde{J}$ on the hyperboloid (\ref{eqn:desithyper}).
Although $J$ has fixed points on $M$, $\tilde{J}$ acts freely on the
hyperboloid.
The isometry group of $4$-dimensional de Sitter space is $O(1,4)$, being the largest subgroup 
of the isometry group of the $5$-dimensional Minkowski embedding space which preserves (\ref{eqn:desithyper}).
The isometry group of $\rp$ de Sitter space is then the largest subgroup of $O(1,4)$ which
commutes with $J$. That is, $\Z_2\times{O(4)}/\Z_2$ where the non-trivial element
of the first $\Z_2$ factor sends $z_0$ to $-z_0$ while the non-trivial element
of the $\Z_2$ in the second factor is given by $J$ which clearly acts trivially on
$\rp$ de Sitter space. The connected component of the isometry group is 
$SO(4)$. The foliation given by $z_0=\mathrm{constant}$ hypersurfaces
is a geometrically distinguished one as it is the only foliation whose
spacelike hypersurfaces are orbits of the connected component of
the isometry group. This is made clearer by introducing the
globally defined coordinates $(t,\chi,\theta,\phi)$ 
\begin{eqnarray}
z_0 & = & \alpha\sinh(t/\alpha) \ ,
\nonumber
\\
z_{4} & = & \alpha\cosh(t/\alpha)\cos\chi \ ,
\nonumber
\\
z_1 & = & \alpha\cosh(t/\alpha)\sin\chi\cos\theta \ ,
\nonumber
\\
z_2 & = & \alpha\cosh(t/\alpha)\sin\chi\sin\theta\cos\phi \ ,
\nonumber
\\
z_3 & = & \alpha\cosh(t/\alpha)\sin\chi\sin\theta\sin\phi \ ,
\end{eqnarray}
in which the metric reads
\begin{equation}
ds^2
=
dt^2
-
\alpha^2\cosh^2(t/\alpha)
[d\chi^2+\sin^2\chi(d\theta^2+\sin^2\theta\,{d\phi^2})] \ ,
\end{equation} 
where $(\chi,\theta,\phi)$ on de Sitter ($\rp$ de Sitter) space are
a set of hyperspherical coordinates on $S^3$ ($\rp$) respectively.
$(t,\chi,\theta,\phi)$ make manifest the $O(4)$ isometry subgroup.

We denote by $|0_G\rangle$ the vacuum state induced by the
Euclidean vacuum $|0_E\rangle$ on de Sitter space (see \cite{jk:desit} for more details).
We consider a particle detector that is linearly coupled to a massless
conformally coupled scalar field. 
The detector and field are assumed to be in the states $|E_0\rangle$ and 
$|0_G\rangle$ respectively at time $\tau_0=-\infty$, and we seek the 
probability that at time $\tau>\tau_0$ the detector is found in the
state $|E_1\rangle$. Through arguments analogous to those in section \ref{sec-linmink},
we find that the transition rate is 
\begin{equation}
\label{eqn:desittrans}
\dot{F}_{\tau}(\omega)=
2\int^{\infty}_{0}ds\,
Re
\left(
e^{-i\omega{s}}\langle{0_G}|\phi(\tau)\phi(\tau-s)|0_G\rangle
\right) \ .
\end{equation}
By the method of images we have
\begin{equation}
\label{eqn:methoddesit}
\langle{0_G}|\phi(x)\phi(x')|0_G\rangle
=
\langle{0_E}|\phi(x)\phi(x')|0_E\rangle
+
\langle{0_E}|\phi(x)\phi(Jx')|0_E\rangle \ ,
\end{equation}
where on the RHS expressions live in de Sitter space, and the correlation function
$\langle{0_E}|\phi(x)\phi(x')|0_E\rangle$ is given by  (\ref{eq:euclidcorre}).

Consider now a detector that follows the geodesic worldline 
\begin{eqnarray}
z_0 & = & \alpha\sinh(\tau/\alpha) \ ,\nonumber \\
z_4 & = & \alpha\cosh(\tau/\alpha) \ , \nonumber \\
\label{eqn:geodtraj}z_1 & = z_2 = z_3 =0 \ .
\end{eqnarray}
On $\rp$ de Sitter space this represents the motion of any geodesic observer
whose motion is orthogonal to the distinguished foliation.
The transition rate (\ref{eqn:desittrans}) is in two parts, a de Sitter part
and an image part. We have calculated already the de Sitter part, coming from
the first term in (\ref{eqn:methoddesit}), in 
section \ref{sec-desit}. The result was the usual thermal, Planckian, response at 
temperature $T=1/(2\pi\alpha)$.
We need the image term. In order to find $\langle{0_E}|\phi(x)\phi(Jx')|0_E\rangle$
on this worldline we first write $\langle{0_E}|\phi(x)\phi(x')|0_E\rangle$ in terms of the
coordinates $(z_0,z_1,z_2,z_3,z_4)$ of the embedding space and then act on $x'$ with $J$, finding
\begin{equation}
\label{eqn:desitcorr}
\langle{0_E}|\phi(x)\phi(Jx')|0_E\rangle
=\frac{1}{8\pi^2}
\frac{1}{\left(
+\mathbf{z}\cdot\mathbf{z'}
+z_0z'_0+\alpha^2\right)} \ ,
\end{equation}
where $\mathbf{z}=(z_1,z_2,z_3,z_4)$. The regularization has been omitted
as in the image part of the correlation function 
no regularization is required: The geometric reason for this, as with the uniformly accelerated
detector on $M$ with boundary, is that the wordline and its image under $J$ are completely 
spacelike separated.
The image term gives to the transition rate the contribution
\begin{equation}
\label{eqn:imagerpdesit}
\dot{F}_{I\tau}(\omega)=
2\int^{\infty}_{0}ds\,
\frac{\cos(\omega{s})}
{16\pi^2\alpha^2\cosh^2\left(\frac{2\tau-s}{2\alpha}\right)} \ .
\end{equation}

We see that the image term contribution (\ref{eqn:imagerpdesit}) is
exactly the same as the image term contribution in the response of a uniformly 
accelerated detector on a $4$-dimensional Minkowski space with boundary 
at $x=0$ ((\ref{eqn:boundaryaccel})
with $d=4$).
Therefore the total response of our inertial detector 
in $\rp$ de Sitter space (with Ricci scalar $R=12/\alpha^2$) is identical to the response
of a uniformly accelerated detector travelling in $4$-dimensional Minkowski space with
a boundary at $x=0$ with the acceleration $1/\alpha$ perpendicular
to the boundary.
It follows that our numerical results in figure \ref{fig:accel} 
also give the response on $\rp$ de Sitter space, with the 
appropriate interpretation for $\alpha$. In particular, 
the image term breaks the KMS condition and the 
response is non-thermal and non-Planckian. When the detector is switched
on in the infinite past, the response
at large $\tau$ is oscillatory
in $\tau$ with period $\pi/\omega$. 
When the detector is switched on at a finite time $\tau_0$, 
$\tau_0\ll{-1}$, numerical evidence indicates that
the response is approximately periodic in the region
$0<\tau<-\tau_0$, with period $\pi/\omega$, but it falls 
to the thermal response as $\tau\rightarrow\infty$,
as discussed further in section \ref{sec:minkbound}
and illustrared in figures \ref{fig:finitetime}, \ref{fig:finitetimewindow} and \ref{fig:finitetimewindow2}. 
This clarifies
and adds to the discussion given in~\cite{jk:desit}.

\begin{figure}[htbp]
\includegraphics[angle=0, width={4in}]{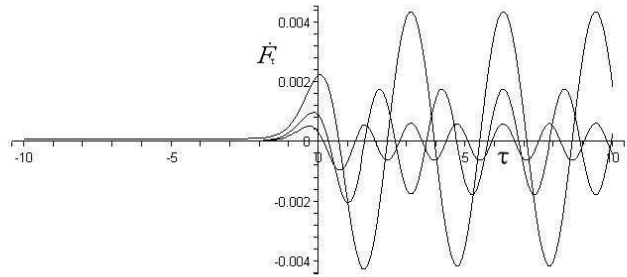}
\centering
\caption[Transition rate associated with the response of a comoving detector on $\rp$ de Sitter space.]{Transition rate 
for a detector uniformly accelerated in the $z_1$ direction
on the quotient space of $5$-dimensional Minkowski space under the involution $J$.
The parameters are $\alpha=1$, $C=(2z_1)^2+(2z_2)^2+(2z_3)^2=0$ and $\omega=1$ (upper curve), 
$\omega=1.5$ and $\omega=2$ (lower curve).}
\label{fig:GEMSDESITTER}
\end{figure}

We wish to compare these particle detector results in 
$\rp$ de Sitter space to the asociated GEMS particle detector.
We see from (\ref{eqn:geodtraj}) that the GEMS worldline of interest 
is a Rindler trajectory with acceleration $a=1/\alpha$ in the $5$-dimensional
embedding space. Therefore in the $5$-dimensional Minkowski embedding space of de Sitter space
we see that the response of
a detector following this worldline is a thermal one 
with associated temperature $T=1/(2\pi\alpha)$ and so we expect, as indeed we saw in
section \ref{sec-desit},
the response of the detector in de Sitter space to also be a thermal one with
temperature $T=1/(2\pi\alpha)$.
Again the actual responses of detectors in the two situations are not
identical (as seen in sections \ref{sec-linmink} and \ref{sec-desit}) due 
to the different dimensions of the spaces, the most obvious difference being the 
presence of the Planckian factor in the de Sitter response and the
Fermi factor in Rindler response on the odd dimensional embedding space. 
As the $\rp$ de Sitter spacetime is built as a quotient of
de Sitter space under the map $J:(z_0,z_1,z_2,z_3,z_4)\mapsto(z_0,-z_1,-z_2,-z_3,-z_4)$
we have immediately the action of this map on the embedding space.
The geodesic worldline of interest maps to a Rindler worldline in this embedding space
with acceleration $a=1/\alpha$,
so in the GEMS approach we consider a Rindler particle detector with this acceleration
in this $5$-dimensional embedding space. The transition rate 
was found in section \ref{sec:quotient}. The thermal part of the transition rate 
is constant in time and is given by
\begin{equation}
\dot{F}_{M\tau}(\omega)
=
\frac{a^2}{8\pi(e^{\frac{2\pi\omega}{a}}+1)}
(1/4+\omega^2/a^2) \ ,
\end{equation}
corresponding to the temperature
$T=\frac{a}{2\pi}$.
The image part of the transition rate depends on the proper time and is given by
\begin{equation}
\dot{F}_{I\tau}(\omega)
=
\frac
{1}
{4\pi^{2}}
\int_0^\infty
ds
\,
\frac{\cos(\omega{s})}
{\left(4/a^2\cosh^2\left(\frac{a(2\tau-s)}{2}\right)\right)^{3/2}} \ .
\end{equation}
The total response is shown for various values of the parameters
in figure \ref{fig:GEMSDESITTER}.
The qualitative similarities to the $\rp$ de Sitter transition rate are apparent.
They provide evidence that, at least in some cases, the GEMS procedure may be applied
to quotient spaces such as $\rp$ de Sitter space and the $\rp$ geon where the 
embedding spaces are Minkowski spaces with suitable identifications.
\chapter{Conclusions}
\label{ch:discussion}

This thesis has been concerned with quantum field theory on topologically
non-trivial manifolds and with particle detector models. 
Chapters \ref{ch:unruhbog} and \ref{ch:hawkingbog} discussed thermal effects 
for the free Dirac field on
the $\rp$ geon and on a topologically analogous flat
spacetime $M_-$ via a Bogolubov transformation analysis. Compared with
the scalar field~\cite{lm:geon}, the main new issue with fermions is
that the spacetimes admit two inequivalent spin structures, and there
are hence two inequivalent Hartle-Hawking like vacua on the geon and
two inequivalent Minkowski-like vacua on~$M_-$. We showed that an
observer in the exterior region of the geon can detect both the
nonthermality of the Hartle-Hawking like state and the spin structure
of this state by suitable interference measurements, and similar
results hold for a Rindler observer on~$M_-$. When probed with
suitably restricted operators, such as operators at asymptotically
late Schwarzschild (respectively Rindler) times, these states
nevertheless appear thermal in the usual Hawking (Unruh) temperature,
for the same geometric reasons as in the scalar
case~\cite{lm:geon}. As a by-product of the analysis, we presented 
the Bogolubov transformation for the Unruh effect for the massive 
Dirac field in $(3+1)$-dimensional Minkowski space, complementing 
and correcting the previous literature. 

As a late time observer in the geon exterior sees a thermal state in
the usual Hawking temperature, the classical laws of black hole
mechanics lead the observer to assign to the geon the same entropy as
to a conventional Schwarzschild hole with the same mass. It was found
in \cite{lm:geon} that an attempt to evaluate the geon entropy by
path-integral methods leads to half of the Bekenstein-Hawking entropy
of a Schwarzschild hole of the same mass, and it was suggested
that state-counting computations of the geon entropy could shed light
on this discrepancy. Our work says little of what the full framework
of such a computation would be, but our work would presumably provide
part of the fermionic machinery in the computation. In particular, the
issue of the spin structure would need to be faced seriously: 
Does an entropy computation by state-counting need to count the two
spin structures as independent degrees of freedom? 

In chapter \ref{ch:emchapter} we computed the stress-energy expectation
value on $M_0$ and $M_-$ in Minkowski-like vacuum states, for both
massive scalar and spinor fields. 
For the scalar field our results on $M_0$ agree with those in
\cite{th:scalar}. For the spinor field on $M_0$ they agree with
\cite{an:spinor}. On $M_-$ the results for the
massive fields are new. 
Further in the massless limit our expectation values agree
with the previous literature \cite{lm:geon,bd:em,dw:dewitthi}. 
We presented also the calculation for $2$-component massless
spinors. It was shown on both $M_0$ and $M_-$ that the stress
tensor is independent of the handedness of the spinor.
In all cases the values fall
off exponentially in the large $m$ limit, and the leading order 
correction for small mass is $O(m^2)$.
Further it is noted that for the scalar field in the large $a$ limit on $M_-$
there is an exponential decay in the massive case while for the massless field
the behaviour is $a^{-4}$. For the massive field the difference between $\langle{0}|T_{\mu\nu}|0\rangle$
on $M_-$ and the corresponding values on $M_0$ falls off exponentially in the limit of large $r^2:=x^2+y^2$, while
for the massless field it behaves as $O(r^{-3})$.
It is seen that for spinors the two spin
structures are distinguished by the sign of a non-vanishing shear
component.

Our underlying interest in the expectation values arises from the role of 
$M_0$ and $M_-$ in modelling, in the context of accelerated observers on 
flat spacetimes, the Hawking(-Unruh) effect on respectively the Kruskal 
manifold and the $\rp$ geon. As we have seen in 
chapters \ref{ch:unruhbog}, \ref{ch:hawkingbog} and \ref{ch:pdchapt} certain aspects of the thermal 
and non-thermal effects for scalar and spinor fields on $M_0$ and $M_-$ 
are at present understood from the viewpoint of Bogolubov transformations 
and particle detector analyses (see also \cite{lm:geon}), but the connections 
between (non-)thermality and stress-energy remain less clear. We view our 
results in chapter \ref{ch:emchapter}, in conjunction with those in chapters
\ref{ch:unruhbog}, \ref{ch:hawkingbog}, \ref{ch:pdchapt} and in \cite{lm:geon}, 
as data points to which we anticipate 
future work on this question to provide a deeper understanding.

In chapter \ref{ch:pdchapt} we considered particle detector models in the
context of quantum field theory in curved spacetime. In particular we 
investigated the model of Schlicht \cite{sc:schlicht,sc:schlicht2}. The 
model is that of a monopole detector linearly coupled to a massless scalar
field which is smeared with a window function in order to regularize the
correlation function in the transition rate.
We extended the regularization of the correlation function for the massless
linearly coupled scalar field to $d$-dimensional Minkowski space, and we showed
that it leads to the expected responses for inertial and uniformly accelerated
detectors switched on in the infinite past and off at $\tau<\infty$. Further we
extended the regularization of Schlicht to the massive scalar field in 
Minkowski space and have shown that it reduces to that of \cite{sc:schlicht,sc:schlicht2}
in the massless limit. Next we introduced a model of a linearly coupled
massless scalar field detector on
spacetimes built as quotients of Minkowski space under certain discrete
isometries. In a number of cases the model, when switched on at $\tau_0=-\infty$
and read at $\tau<\infty$, was shown to reproduce the known asymptotic
responses. These cases include the uniformly accelerated detector on $M_0$ 
\cite{lm:geon,dlo:detecbound} as
well as the inertial and uniformly accelerated detectors on Minkowski space
with boundary when the motion is parallel to the boundary \cite{dlo:detecbound}.
These results suggest that our model is reasonable.
Further we presented a number of new responses, the most interesting of
which are the time dependent responses on Minkowski space with boundary and
on $M_-$. An inertial detector approaching the boundary on Minkowski space with
boundary was considered and seen to react in a qualitatively similar way to one
travelling parallel to the boundary but taking progressively smaller distances
(ie comparing figures \ref{fig:inertialparal} and \ref{fig:inertialwl0}).
The main difference is that in the detector approaching the boundary a divergence
in the transition rate occurs as the boundary is reached.
A detector with uniform acceleration perpendicular to the boundary was
also considered and an interesting observation made. For
a detector which is switched on in the infinite past the transition rate 
is found to oscillate in $\tau$ at late times with period $\pi/\omega$, 
never tending to the Minkowski
thermal response, no matter how far from the boundary the detector
gets in the future. However for a detector switched on at a finite time
(that is $\tau_0>-\infty$), the response will at late times tend to the
thermal Minkowski response. For instantaneous, exponential and Gaussian 
switching functions the conclusion is the same.
Responses were also considered on the quotient spaces of Minkowski
space under the involution 
$J_{c_k}:(t,x_1,x_2,\ldots\,x_{d-1})\mapsto(t,-x_1,-x_2,-\ldots\,-x_k,x_{k+1},\ldots,x_{d-1})$
where $1<k<d$ and certain relations to the responses on Minkowski space with 
boundary noted.
The responses of uniformly accelerated detectors, where the motion is
in the $x_1$ direction, are relevant for the discussion of static detectors
in the $\rp$ geon exterior as well as comoving detectors in $\rp$ de Sitter
via their global embedding Minkowski spacetimes (GEMS).
Next we extended the detector model and regularization to the 
nonlinearly coupled scalar field and to the massless Dirac field.
With a few minor technicalities the extension is quite 
straightforward.
In the case of the Dirac field again we obtained the transition
rate and the power spectrum of the Dirac noise for a detector switched 
on in the infinite past on inertial and a uniformly accelerated trajectories.  
The power spectrum for the accelerated detector
agrees with the previous literature (see e.g \cite{tk:takagi}) 
and so suggests our model is reasonable. Further we briefly considered
the response of the Dirac detector on $M_0$ and $M_-$.
One aim was to see whether a uniformly accelerated Dirac detector
on $M_-$ could distinguish the two spin structures there. Unfortunately
we have shown that this is not the case for our particular model.

In section \ref{sec:Geon} we considered the response of a static detector
in the exterior region of the $\rp$ geon via a global embedding Minkowski
space. Although the GEMS program has so far only been applied in
a kinematical setting, our aim was to examine the possibility that the response of the detector
in the embedding space is related to that in the underlying curved space.
We found that the response is related to that of uniformly accelerated detectors given 
in section \ref{sec:quotient}. In particular it is shown in the 
embedding space, and expected on the geon, that the response is not thermal, 
in the sense that it does not satisfy the KMS condition, for most times.
Further it is seen that for a detector switched on in the infinite past 
the response is approximately thermal at early times but does not return to
the thermal response at late times in contrast to expectations
(see e.g \cite{lm:geon} and sections \ref{ch:unruhbog} and \ref{ch:hawkingbog}
here). If the detector however is turned on at some finite time in the
distant past then the response is approximately thermal
when turned on and returns to being approximately thermal in the distant future.

Lastly we extended our model to conformally flat spacetimes and
considered some responses on de Sitter space and $\rp$ de Sitter
space.
The regularization of \cite{sc:schlicht} is not easily adaptable to
general motions in conformally flat spacetimes, due to the possibility
of spatially closed hypersurfaces in the detector's rest frame.
We argued however for a similar regularization by reinterpreting the
regularization
as an ultraviolet cut-off in the high ``frequency'' modes.
On de Sitter space the transition rate for a detector switched
on in the infinite past is found for a uniformly accelerated 
detector, and it is found to agree with previous literature \cite{dl:deser3}.
This result suggests our regularization is reasonable. In the
case of an inertial detector in de Sitter space the regularization we introduced is shown
to also come from the consideration of a detector with spatial 
extension where the detector is rigid in its rest frame.
On $\rp$ de Sitter space the response of a detector following the
comoving worldline was considered in two ways.
Firstly a direct calculation showed that the response is
exactly that of a uniformly accelerated detector approaching the
boundary on Minkowski space with boundary (identifying $1/\alpha$ with
the acceleration). Again therefore the response of a detector 
switched on in the infinite past has an oscillatory behaviour in the
distant future and does not tend to the expected thermal result, the image
term breaking the KMS condition, in contrast to what was expected \cite{jk:desit}.
The response of a detector switched on at a finite time does
however tend to the expected thermal response at late times.
Secondly we considered the same calculation from the GEMS perspective.
Although the response is clearly different to that in the original space, 
due to the different dimensions of $\rp$ de Sitter space and the embedding
space, it is seen that the response is qualitatively very similar.
The calculation therefore provides a good example for investigating
detector responses in curved spacetimes and those in their GEMS, and
suggests also that the response found in the GEMS of the $\rp$ geon in section
\ref{sec:Geon} is indeed closely related to the response of the static detector in the
geon itself. 

\subsubsection{Future Work}

There are many possible directions for future work on the topics
discussed in this thesis.
On the $\rp$ geon for instance one interesting subject which we have
not really discussed here is its entropy. In \cite{lm:geon} it is argued
from the observed temperature and the classical laws of black hole thermodynamics
that an observer at late Schwarzschild time in the exterior will assign an entropy
to the hole of $1/4$ its late time horizon area, which is the entropy of a Kruskal hole with the same mass.
However it was also shown that from a path integral approach the entropy assigned
to the hole is $1/8$ the late time horizon area. The reason for the difference is not clear.
It would be very interesting therefore to consider a state counting argument for the
entropy of the geon in the context of a quantum gravity model such as loop quantum
gravity, as was done for large non-rotating black holes in \cite{ak:lqgentropy}.

Further work for the particle detector models discussed could include a 
thorough investigation of the dependency of Schlicht's model on the
window function (detector shape). 
Also it would be interesting to investigate in depth the most general case
of an arbitrary trajectory in a general curved spacetime.
The GEMS calculations discussed here are also worthy of further investigation.
In particular, to what extent do the responses in the GEMS relate to the
responses in the original curved spaces? We know that the GEMS procedure
may not be applied reliably to the response of an arbitrary non-stationary
trajectory in an arbitrary spacetime \cite{ct:nonstatgems} but we expect for 
some classes of trajectories and spacetimes the GEMS approach is useful.

Finally, as mentioned when discussing our stress energy results, the
relation between the (non)-thermal behaviour seen in various black hole
(and flat) spacetimes and the stress energy expectation values is not clear 
and a deeper understanding is required.

\appendix

\chapter{Analytic continuation of Rindler modes across the horizons}
\label{ch:appendix1}

In this appendix we provide more details on the analytic continuation 
of the Rindler modes (\ref{eqn:rinmodes}) on $M_0$ into the future and
left hand wedges.

(\ref{eqn:rinmodes}) are Rindler modes in $R_0$ expressed with respect to the
globally defined Minkowski vierbein. We build from them a complete set of
positive frequency Minkowski modes (i.e. positive frequency with respect to
$\partial_t$) by continuing them (with $\Omega\in{\R}$) across the horizons into $F$ and $L$ in the
lower half complex $t$ plane.
(\ref{eqn:rinmodes}) are functions of $t-x$ and $t+x$ where in $R$
\begin{eqnarray}
\xi & = & (x^2-t^2)^{1/2} = ((x-t)(x+t))^{1/2} \ , 
\\
\tau & = & \mathrm{arctanh}\left(\frac{t}{x}\right) = \frac{1}{2}\ln\left(\frac{x+t}{x-t}\right) \ , 
\end{eqnarray}
Continuing these across the future horizon $t=x$, from $(t-x)<0$ to $(t-x)>0$ 
we note that $t=x$ is 
a branch point.
In order to analytically continue into the lower half complex $t$ plane
we must make the substitutions $\sqrt{x-t}\rightarrow{i}\sqrt{t-x}$
and $\ln(x-t)\rightarrow\ln(t-x)+i\pi$. This gives
\begin{eqnarray}
\xi & = & (x^2-t^2)^{1/2} \rightarrow i((t-x)(x+t))^{1/2} = i\xi \ ,
\\
\tau & = & \mathrm{arctanh}\left(\frac{t}{x}\right) \rightarrow \frac{1}{2}\ln\left(\frac{x+t}{t-x}\right)-\frac{i\pi}{2} = \tau-\frac{i\pi}{2} \ , 
\end{eqnarray}
where on the left hand side of these expressions $(\tau,\xi)$ are in $R$
and on the right $(\tau,\xi)$ are in $F$ as defined by (\ref{eqn:milne}).
Under this substitution it is easy to check that the Rindler
modes in $R$ (\ref{eqn:rinmodes}) become (\ref{eqn:fmode}) and
(\ref{eqn:fmode2}) in $F$.

The continuation further to the $L$ wedge is similar. We must cross the
horizon $t=-x$ which again is a branch point in the expressions
for $(\tau,\xi)$ in $F$. Crossing from $(x+t)>0$ to $(x+t)<0$ in the lower
half $t$-plane we make the substitutions
$\sqrt{x+t}\rightarrow{-i}\sqrt{-(x+t)}$ and
$\ln(x+t)\rightarrow\ln(-(x+t))-i\pi$. Then
\begin{eqnarray}
\xi & = & (t^2-x^2)^{1/2} \rightarrow -i((x-t)(x+t))^{1/2} = -i\xi \ , 
\\
\tau & = & \mathrm{arctanh}\left(\frac{x}{t}\right) \rightarrow \frac{1}{2}\ln\left(\frac{x+t}{x-t}\right)-\frac{i\pi}{2} = \tau-\frac{i\pi}{2} \ ,
\end{eqnarray}
where the expressions on the left are in $F$ and those
on the right are in $L$ with $(\tau,\xi)$ defined by (\ref{eqn:rinl}).
Under this continuation the modes (\ref{eqn:fmode}) and
(\ref{eqn:fmode2}) in $F$ become (\ref{eqn:modefrigl}).

A complete set of negative frequency Minkowski modes can be obtained
in a similar manner by continuing the Rindler modes in $R$ (\ref{eqn:rinmodes}), 
with $\Omega\in{\R}$, across the horizons in the upper half complex $t$-plane.
Alternatively we may just take the charge conjugates of our positive frequency modes.

\chapter{Evaluating integrals using hyperspherical coordinates}
\label{ch:appendixA}

In this appendix we show how to evaluate integrals of the form
\begin{equation}
\label{eqn:integral}
\int
d^{d-1}k
\,
F(|\mathbf{k}|)
e^{i\mathbf{k}\cdot\mathbf{R}}e^{\epsilon\mathbf{k}\cdot\mathbf{\tilde{R}}} \ ,
\end{equation}
where $d>2$, the integration is over $\mathbf{k}\in\R^{d-1}$,
$\mathbf{R}$ and $\mathbf{\tilde{R}}$ are vectors in $\R^{d-1}$ and 
$F(|\mathbf{k}|)$ is an arbitrary function.

First we transform to hyperspherical coordinates (polar and spherical coordinates for 
$d=3$ and $d=4$ respectively) in $\mathbf{k}$ space
$(|\mathbf{k}|,\theta_1,\ldots,\theta_{d-2})$, where
\begin{eqnarray}
k_1& = &|\mathbf{k}|\sin\theta_1\ldots\sin\theta_{d-3}\sin\theta_{d-2} \ , \nonumber
\\
k_2& = &|\mathbf{k}|\sin\theta_1\ldots\sin\theta_{d-3}\cos\theta_{d-2} \ , \nonumber
\\
k_3& = &|\mathbf{k}|\sin\theta_1\ldots\sin\theta_{d-4}\cos\theta_{d-3} \ , \nonumber
\\
   &\vdots& \nonumber
\\
k_{d-1}& = &|\mathbf{k}|\cos\theta_1 \ ,
\end{eqnarray}
with $0<|\mathbf{k}|<\infty$, $0<\theta_i<\pi$ for $i=1,\ldots,d-3$ and $0<\theta_{d-2}<2\pi$. Then
\begin{equation}
\label{eqn:intangmodk}
\int
d^{d-1}k
\,
F(|\mathbf{k}|)
e^{i\mathbf{k}\cdot\mathbf{R}}e^{\epsilon\mathbf{k}\cdot\mathbf{\tilde{R}}}
=
\int_0^\infty
d|\mathbf{k}|
\,
|\mathbf{k}|^{d-2}F(|\mathbf{k}|)
\int
d^{d-2}\Omega
\,
e^{i\mathbf{k}\cdot\mathbf{R}}e^{\epsilon\mathbf{k}\cdot\mathbf{\tilde{R}}} \ .
\end{equation}
The angular integral may now be performed by choosing a suitable set of axes in $\mathbf{k}$
space. In particular we choose the $k_{d-1}$ axis in the direction of $\mathbf{R}$
and the $k_{d-2}$ axis such that $\mathbf{\tilde{R}}$ lies in the 
$(k_{d-2},k_{d-1})$-plane. Then $\mathbf{k}\cdot\mathbf{R}=|\mathbf{k}||\mathbf{R}|\cos\theta_1$
and $\mathbf{k}\cdot\mathbf{\tilde{R}}=|\mathbf{k}||\mathbf{\tilde{R}}|(\cos\theta_1\cos\psi
+\sin\theta_1\cos\theta_2\sin\psi)$, where $\psi$ is the angle between $\mathbf{R}$ and 
$\mathbf{\tilde{R}}$, i.e. $|\mathbf{R}||\mathbf{\tilde{R}}|\cos\psi=\mathbf{R}\cdot\mathbf{\tilde{R}}$.
Considering first $d>4$ we have\footnote{For $d=5$ the expression is (\ref{eqn:b4}) without the product term.}
\begin{eqnarray}
\int
d^{d-2}\Omega
\,
e^{i\mathbf{k}\cdot\mathbf{R}}e^{\epsilon\mathbf{k}\cdot\mathbf{\tilde{R}}}
& = &
\int_0^{2\pi}
d\theta_{d-2}
\prod_{i=3}^{d-3}
\int_0^\pi
d\theta_i
\,
(\sin\theta_i)^{d-2-i}
\int_0^\pi
d\theta_1
\,
(\sin\theta_1)^{d-3}
e^{i|\mathbf{k}||\mathbf{R}|\cos\theta_1}
\nonumber
\\
&   & \times
\int_0^\pi
d\theta_2
\,
(\sin\theta_2)^{d-4}
e^{\epsilon|\mathbf{k}||\mathbf{\tilde{R}}|(\cos\theta_1\cos\psi
+\sin\theta_1\cos\theta_2\sin\psi)}
 \ .
\label{eqn:b4}
\end{eqnarray}
From \cite{gr:inttables} we have
\begin{equation}
\label{eqn:ibessexp}
\int_0^\pi
d\theta
\,
(\sin\theta)^{2\nu}
e^{{\pm}a\cos\theta}
=
\pi^{1/2}
\left(
\frac{2}
{a}
\right)^{\nu}
\Gamma[\nu+1/2]
I_{\nu}(a) \ ,
\end{equation}
for $Re(\nu)>-1/2$, where $I_{\nu}(a)$ is a Bessel function, and so
\begin{eqnarray}
\int_0^\pi
d\theta_2
\,
(\sin\theta_2)^{d-4}
e^{\epsilon|\mathbf{k}||\mathbf{\tilde{R}}|\sin\theta_1\cos\theta_2\sin\psi}
& = &
\pi^{1/2}
\left(
\frac{2}
{\epsilon|\mathbf{k}||\mathbf{\tilde{R}}|\sin\theta_1\sin\psi}
\right)^{\frac{d-4}{2}}
\Gamma[(d-3)/2]
\nonumber
\\
&   & \times
I_{\frac{d-4}{2}}(\epsilon|\mathbf{k}||\mathbf{\tilde{R}}|\sin\theta_1\sin\psi) \ .
\end{eqnarray}
The integral over $\theta_3\ldots\theta_{d-2}$ in (\ref{eqn:b4}) may be performed by noting that it is the surface area of 
a unit $(d-4)$-dimensional sphere (in topologists notation), which is given by
\begin{equation}
S_{d-4}
=
2
\frac{\pi^{(d-3)/2}}
{\Gamma[(d-3)/2]} \ ,
\end{equation}
see e.g \cite{zw:zwillinger}.
Thus we have
\begin{eqnarray}
\int
d^{d-2}\Omega
\,
e^{i\mathbf{k}\cdot\mathbf{R}}e^{\epsilon\mathbf{k}\cdot\mathbf{\tilde{R}}}
& = &
\frac{(2\pi)^{d/2-1}}
{(\epsilon|\mathbf{k}||\mathbf{\tilde{R}}|\sin\psi)^{(d-4)/2}}
\int^{\pi}_{0}
d\theta_1
\,
(\sin\theta_1)^{d/2-1}
e^{(i|\mathbf{k}||\mathbf{R}|+\epsilon|\mathbf{k}||\mathbf{\tilde{R}}|\cos\psi)\cos\theta_1}
\nonumber
\\
&  & \times
I_{\frac{d-4}{2}}(\epsilon|\mathbf{k}||\mathbf{\tilde{R}}|\sin\theta_1\sin\psi) \ .
\label{eqn:theta1intexp}
\end{eqnarray}
The integral over $\theta_1$ may be done by expanding the Bessel function
as a series and doing the
integral over $\theta_1$ for each term in the sum. We have
\begin{eqnarray}
\int_0^{\pi}
d\theta
\,
(\sin\theta)^{\nu+1}
e^{a\cos\theta}
I_{\nu}(b\sin\theta)
& = &
\!\sum_{k=0}^\infty
\left(\frac{b}{2}\right)^{\nu+2k}
\!\!\!\!\!\!\!\!\frac{1}{k!\Gamma[\nu+k+1]}
\int_0^{\pi}
\!\!d\theta
\,
(\sin\theta)^{2\nu+2k+1}
e^{a\cos\theta} \ ,
\nonumber
\\
& = & 
(2\pi)^{1/2}
\sum_{k=0}^\infty
\frac{b^{\nu+2k}}{k!2^k}
a^{-(\nu+k+1/2)}
I_{\nu+k+1/2}(a) \ ,
\nonumber
\\
& = & 
(2\pi)^{1/2}b^\nu
\sum_{k=0}^\infty
\frac{b^{2k}}{k!}
\left(
\frac{d}{dq}
\right)^k
\left(q^{-\frac{1}{2}(\nu+\frac{1}{2})}I_{\nu+1/2}(\sqrt{q})\right)
 \ ,
\nonumber
\\
\label{eqn:tayl}
\end{eqnarray}
where $q=a^2$, and the second line follows from (\ref{eqn:ibessexp}).
Recognizing the sum in the last expression of (\ref{eqn:tayl}) as
a Taylor series we thus have
\begin{equation}
\label{eqn:intermI}
\int_0^{\pi}
d\theta
\,
(\sin\theta)^{\nu+1}
e^{a\cos\theta}
I_{\nu}(b\sin\theta)
=
(2\pi)^{1/2}
\frac{b^{\nu}}{(\sqrt{a^2+b^2})^{\nu+1/2}}
I_{\nu+1/2}(\sqrt{a^2+b^2}) \ .
\end{equation}
Further we find 
\begin{equation}
-(i|\mathbf{R}|+\epsilon|\mathbf{\tilde{R}}|\cos\psi)^2-(\epsilon
|\mathbf{\tilde{R}}|\sin\psi)^2=(\mathbf{R}-i\epsilon\mathbf{\tilde{R}})\cdot(\mathbf{R}-i\epsilon\mathbf{\tilde{R}})
=(\mathbf{R}-i\epsilon\mathbf{\tilde{R}})^2 \ .
\end{equation}

Collecting these results together we have
\begin{equation}
\label{eqn:fint}
\int
d^{d-1}k
\,
F(|\mathbf{k}|)
e^{i\mathbf{k}\cdot\mathbf{R}}e^{\epsilon\mathbf{k}\cdot\mathbf{\tilde{R}}}
=
(2\pi)^{(d-1)/2}
\int_0^\infty
d|\mathbf{k}|
\,
|\mathbf{k}|^{d-2}
F(|\mathbf{k}|)
\frac{J_{{(d-3)}/{2}}\left(|\mathbf{k}|\sqrt{(\mathbf{R}-i\epsilon\mathbf{\tilde{R}})^2}\right)}
{\left(|\mathbf{k}|\sqrt{(\mathbf{R}-i\epsilon\mathbf{\tilde{R}})^2}\right)^{(d-3)/2}} \ ,
\end{equation}
where we have replaced the $I$ Bessel in (\ref{eqn:intermI}) by a
$J$ Bessel using $I_\nu(z)=e^{-\frac{\nu\pi{i}}{2}}J_\nu(ze^\frac{\pi{i}}{2})$.

For $d=4$ the expression analogous to (\ref{eqn:b4}) for the angular integral is
\begin{eqnarray}
\int
d^{2}\Omega
\,
e^{i\mathbf{k}\cdot\mathbf{R}}e^{\epsilon\mathbf{k}\cdot\mathbf{\tilde{R}}}
& = &
\int_0^\pi
d\theta_1
\,
\sin\theta_1
e^{i|\mathbf{k}||\mathbf{R}|\cos\theta_1}
\nonumber
\\
&   & \times
\int_0^{2\pi}
d\theta_2
\,
e^{\epsilon|\mathbf{k}||\mathbf{\tilde{R}}|(\cos\theta_1\cos\psi
+\sin\theta_1\cos\theta_2\sin\psi)}
\nonumber
\\
& = & 
2\pi
\int_0^\pi
d\theta_1
\,
\sin\theta_1
e^{i|\mathbf{k}||\mathbf{R}|\cos\theta_1+\epsilon|\mathbf{k}||\mathbf{\tilde{R}}|\cos\theta_1\cos\psi}
I_0(\epsilon|\mathbf{k}||\mathbf{\tilde{R}}|\sin\theta_1\sin\psi)
 \ ,
\nonumber
\\
\label{eqn:b4d4}
\end{eqnarray}
where we have used (\ref{eqn:ibessexp}) for the integral over $\theta_2$.
Expression (\ref{eqn:b4d4}) agrees with (\ref{eqn:theta1intexp}) when $d=4$
and thus we are led again to the result (\ref{eqn:fint}).

For $d=3$ we take the $k_2$ axis in the direction of $\mathbf{R}$
in (\ref{eqn:intangmodk}) then $\mathbf{k}\cdot\mathbf{R}=|\mathbf{k}|
|\mathbf{R}|\cos\theta_1$, $\mathbf{k}\cdot\mathbf{\tilde{R}}=|\mathbf{k}|
|\mathbf{\tilde{R}}|(\cos\theta_1\cos\psi+\sin\theta_1\sin\psi)$
and 
\begin{equation}
\int
d\Omega
\,
e^{i\mathbf{k}\cdot\mathbf{R}}e^{\epsilon\mathbf{k}\cdot\mathbf{\tilde{R}}}
=
\int_0^{2\pi}
d\theta_1
e^{i|\mathbf{k}|
|\mathbf{R}|\cos\theta_1}
e^{\epsilon|\mathbf{k}|
|\mathbf{\tilde{R}}|(\cos\theta_1\cos\psi+\sin\theta_1\sin\psi)} \ . 
\end{equation}
Hence, \cite{gr:inttables},
\begin{equation}
\int
d^2k
\,
F(|\mathbf{k}|)
e^{i\mathbf{k}\cdot\mathbf{R}}e^{\epsilon\mathbf{k}\cdot\mathbf{\tilde{R}}}
=
2\pi
\int_0^\infty
d|\mathbf{k}|
\,
|\mathbf{k}|
F(|\mathbf{k}|)
{J_{0}\left(|\mathbf{k}|\sqrt{(\mathbf{R}-i\epsilon\mathbf{\tilde{R}})^2}\right)}
\ ,
\end{equation}
which also agrees with (\ref{eqn:fint}). Thus (\ref{eqn:fint})
is the expression we require for all $d>2$.

In the case of equation (\ref{eqn:gequation}) we want
\begin{eqnarray}
g_\epsilon(\mathbf{k};\tau)
& = &
\frac{\epsilon\Gamma(d/2)}
{\pi^{d/2}}
e^{-i(\mathit{k}\cdot\mathit{x}(\tau))}
\int
d^{d-1}\xi\,
\frac{1}
{(\mathbf{\xi}^2+\epsilon^2)^{d/2}}
e^{-i\bar{\mathbf{k}}\cdot\mathbf{\xi}} \ ,
\nonumber
\\
& = &
\frac{\epsilon\Gamma(d/2)}
{\pi^{d/2}}
e^{-i(\mathit{k}\cdot\mathit{x}(\tau))}
\frac{(2\pi)^{(d-1)/2}}
{|\bar{\mathbf{k}}|^{{(d-3)}/{2}}}
\int_0^\infty
d|\mathbf{\xi}|
\,
\frac{|\mathbf{\xi}||\mathbf{\xi}|^{{(d-3)}/{2}}}
{(\mathbf{\xi}^2+\epsilon^2)^{d/2}}
{J_{{(d-3)}/{2}}(|\mathbf{\xi}||\bar{\mathbf{k}}|)}
 \ ,
\nonumber
\\
{}
\end{eqnarray}
where the second line follows from (\ref{eqn:fint}).
Again the integral may be found in integral tables (see \cite{as:abromowitz} p488 no.11.4.44)
\begin{equation}
\int_0^\infty
d|\mathbf{\xi}|
\,
\frac{|\mathbf{\xi}||\mathbf{\xi}|^{{(d-3)}/{2}}}
{(\mathbf{\xi}^2+\epsilon^2)^{d/2}}
{J_{{(d-3)}/{2}}(|\mathbf{\xi}||\bar{\mathbf{k}}|)}
=
\frac{|\bar{\mathbf{k}}|^{(d/2-1)}\epsilon^{-1/2}}
{2^{(d/2-1)}\Gamma(d/2)}
K_{1/2}(|\bar{\mathbf{k}}|\epsilon) \ ,
\end{equation}
$K$ is a modified Bessel function and $K_{1/2}(z)=(\pi/(2|\bar{\mathbf{k}}|\epsilon))^{1/2}e^{-|\bar{\mathbf{k}}|\epsilon}$
so finally we find
\begin{equation}
g_\epsilon(\mathbf{k};\tau)
=
e^{-i(\mathit{k}\cdot\mathit{x}(\tau))}
e^{-\epsilon|\bar{\mathbf{k}}|} \ .
\end{equation}

For equation (\ref{eq:correlation}) we find
\begin{eqnarray}
\langle{0}|\phi(\tau){\phi}(\tau')|0\rangle
& = & 
\frac{1}{(2\pi)^{d-1}}
\int
\frac{d^{d-1}k}{2|\mathbf{k}|}\,
e^{-i|\mathbf{k}|(t-t'-i\epsilon(\dot{t}+\dot{t}'))
+i\mathbf{k}\cdot(\mathbf{x}-\mathbf{x'}-i\epsilon(\dot{\mathbf{x}}+\dot{\mathbf{x}}'))} \ ,
\nonumber
\\
& = &
\frac{1}{2(2\pi)^{(d-1)/2}}
\int_0^\infty
d|\mathbf{k}|
\,
|\mathbf{k}|^{d-3}
{e^{-i|\mathbf{k}|(t-t'-i\epsilon(\dot{t}+\dot{t}'))}}
\nonumber
\\
\label{eqn:integral2}
&   & \times\frac{J_{{(d-3)}/{2}}\left(|\mathbf{k}|\sqrt{(\mathbf{x}-\mathbf{x'}-i\epsilon(\dot{\mathbf{x}}+\dot{\mathbf{x}}'))^2}\right)}
{\left(|\mathbf{k}|\sqrt{(\mathbf{x}-\mathbf{x'}-i\epsilon(\dot{\mathbf{x}}+\dot{\mathbf{x}}'))^2}\right)^{(d-3)/2}}
 \ .
\end{eqnarray}
The $|\mathbf{k}|$ integral may be found in \cite{gr:inttables}, using
\begin{equation}
\int_0^\infty
dx\,
e^{-\alpha{x}}
x^\nu
J_\nu(\beta{x})
=
\frac{(2\beta)^\nu\Gamma(\nu+1/2)}
{\pi^{1/2}(\alpha^2+\beta^2)^{\nu+1/2}} \ ,
\end{equation}
for $Re(\alpha)>0$, the result is
(\ref{eq:corre}).
The same integrations are required in finding the usual correlation
function (\ref{eq:correusu}) where $\epsilon$ is taken to $0$ in (\ref{eqn:integral2})
and then a cut-off $e^{-\epsilon|\mathbf{k}|}$ is introduced.

For equation (\ref{eqn:ggaussian}) we want
\begin{eqnarray}
g_\epsilon(\mathbf{k};\tau)
& = &
\frac{1}{\epsilon^{d-1}(2\pi)^{(d-1)/2}}
e^{-i(\mathit{k}\cdot\mathit{x}(\tau))}
\int
d^{d-1}\xi\,
e^{-\frac{\mathbf{\xi}^2}{2\epsilon^2}}
e^{-i\bar{\mathbf{k}}\cdot\mathbf{\xi}} \ ,
\nonumber
\\
& = & 
\frac{1}{\epsilon^{d-1}|\bar{\mathbf{k}}|^{(d-3)/2}}
e^{-i(\mathit{k}\cdot\mathit{x}(\tau))}
\int_0^\infty
d|\mathbf{\xi}|
\;
|\mathbf{\xi}||\mathbf{\xi}|^{(d-3)/2}
e^{-\frac{|\mathbf{\xi}|^2}{2\epsilon^2}}
J_{{(d-3)}/{2}}(|\bar{\mathbf{k}}||\mathbf{\xi}|)
 \ .
\nonumber
\\
{}
\end{eqnarray}
Again from \cite{as:abromowitz} (p486 no.11.4.29) we have
\begin{equation} 
\int_0^\infty
dt
\,
e^{-\alpha^2t^2}
t^{\nu+1}
J_\nu(bt)
=
\frac{b^\nu}{(2\alpha^2)^{\nu+1}}e^{-\frac{b^2}{4\alpha^2}} \ ,
\end{equation}
and so we find
\begin{equation}
g_\epsilon(\mathbf{k};\tau)
=
e^{-i(\mathit{k}\cdot\mathit{x}(\tau))}
e^{-\frac{\epsilon^2|\bar{\mathbf{k}}|^2}{2}} \ .
\end{equation} 

\chapter{Expressions in $2$ dimensions}
\label{ch:appendixA2}

In this appendix we present the calculation of some of the expressions
appearing in chapter \ref{ch:pdchapt} in $2$ dimensions.

\section{The correlation function}

In $2$-dimensional Minkowski space the solutions to the 
massless Klein Gordon equation may be expressed as the
mode expansion
\begin{equation}
\label{eqn:2dimexpan}
\phi(t,x)=
\int^{\infty}_{-\infty}
\frac{dk}{(4|k|\pi)^{1/2}}
\left(
a(k)e^{-i(|k|{t}-kx)}
+a^{\dagger}(k)e^{i(t-kx)}
\right) \ .
\end{equation}
The Wightman function is then
\begin{equation}
\label{eqn:2dimwhi}
\langle{0}|\phi(x)\phi(x')|0\rangle
=
\frac{1}{4\pi}
\int^{\infty}_{-\infty}\,
\frac{dk}{|k|}
e^{-i(|k|{(t-t')}-k(x-x'))} \ .
\end{equation}
We see immediately from (\ref{eqn:2dimexpan}) and (\ref{eqn:2dimwhi})
that there is an infrared divergence due to the $k=0$ mode (as well 
as the usual ultraviolet divergences found in all dimensions).
In calculating the Wightman function (\ref{eqn:2dimwhi}) we could
introduce an infrared cut-off in order to regularize this divergence,
by for example considering
\begin{equation}
\langle{0}|\phi(x)\phi(x')|0\rangle
=
\frac{1}{4\pi}
\int^{\infty}_{-\infty}
\frac{dk}{|k|}
\left(e^{-i(|k|{(t-t')}-k(x-x'))}-\Theta(C-|k|)\right) \ ,
\end{equation}
where $\Theta(x)$ is a step function and $C$ a
positive constant (see e.g. \cite{na:nakanishi} for more details).
However as we shall see later the divergence does not
contribute to the detector response provided we include a temporal window
function and consider an asymptotic (infinite time)
detection. Without the window function the contribution from these terms
may be easily found and so we do not regularize it
here but simply drop the divergent term in any calculations done. The ultraviolet
divergence in (\ref{eqn:2dimwhi}) still requires regularization.
In the usual  $i\epsilon$-prescription
\begin{eqnarray}
\langle{0}|\phi(x)\phi(x')|0\rangle
& = &
\frac{1}{4\pi}
\int^{\infty}_{-\infty}
\frac{dk}{|k|}\,
e^{-i(|k|{(t-t'-i\epsilon)}-k(x-x'))} \ ,
\nonumber
\\
& = &
\frac{1}{4\pi}
\left(
\int^{\infty}_{0}
\frac{dk}{k}\,
e^{-ik((t-t'-i\epsilon)-(x-x'))}
+
\int^{\infty}_{0}
\frac{dk}{k}\,
e^{-ik((t-t'-i\epsilon)+(x-x'))}
\right) \ .
\nonumber
\\
{} 
\end{eqnarray}
The integrals may be done by parts
\begin{eqnarray}
\int^{\infty}_{0}
\frac{dk}{k}\,
e^{-i\mu{k}} 
& = & 
\left[
\ln(k)e^{-i\mu{k}}
\right]^\infty_0
+i\mu
\int^{\infty}_{0}
dk
\ln(k)
e^{-i\mu{k}} \ ,
\nonumber
\\
& = &
-\ln(0)-(\mathbf{C}+\ln(i\mu)) \ ,
\end{eqnarray}
when $Im(\mu)<0$ \cite{gr:inttables}, where $\mathbf{C}$ is Euler's constant. 
So
\begin{eqnarray}
\langle{0}|\phi(x)\phi(x')|0\rangle
& = &
\frac{1}{4\pi}(-2\ln(0)-2\mathbf{C}-\ln(i[(t-t'-i\epsilon)-(x-x')])
\nonumber
\\
&   & -\ln(i[(t-t'-i\epsilon)+(x-x')]))
\nonumber
\\
\label{eqn:2dimcorrefunc}
& = &
-\frac{1}{4\pi}\ln((t-t'-i\epsilon)^2-(x-x')^2)+C \ ,
\end{eqnarray}
where $C$ is an ill-defined constant. The usual approach is to say that
derivatives of the correlation function are well defined and 
are given by the derivatives of (\ref{eqn:2dimcorrefunc}). We will
show later however that in our particle detector model we may work
with (\ref{eqn:2dimcorrefunc}) in cases where the constant term does
not contribute to the response.

Next consider the correlation function for the smeared field in two dimensions. 
The discussion follows as in sections \ref{sec-linmink}.
For a Lorentzian shape function 
\begin{equation}
g_\epsilon(\mathbf{k};\tau)
=
\frac{\epsilon}
{\pi}
e^{-i(\mathit{k}\cdot\mathit{x}(\tau))}
\int^\infty_{-\infty}
d\xi\,
\frac{e^{-i\bar{k}\xi}}
{(\xi^2+\epsilon^2)}
 \ ,
\end{equation}
where $\bar{k}:=\mathit{k}\cdot{e_\xi(\tau)}$.
Now
\begin{equation}
\int^\infty_{-\infty}
dx\,
\frac{e^{-ipx}}
{(x^2+a^2)}
=\frac{\pi}{a}e^{-|ap|}
 \ ,
\end{equation}
for $a\neq{0}$ and $p$ real \cite{gr:inttables}, so
\begin{equation}
g_\epsilon(\mathbf{k};\tau)
=
e^{-i(\mathit{k}\cdot\mathit{x}(\tau))}
e^{-\epsilon|\mathit{k}\cdot{e_\xi(\tau)}|}
=
e^{-i(|k|t(\tau)-kx(\tau))}
e^{-\epsilon(|k|\dot{t}-k\dot{x})}
 \ .
\end{equation}
Then
\begin{equation}
\langle{0}|\phi(\tau)\phi(\tau')|0\rangle
=
\frac{1}{4\pi}
\int^{\infty}_{-\infty}
\frac{dk}{|k|}\,
e^{-i(|k|{(t-t'-i\epsilon(\dot{t}+\dot{t}'))}-k(x-x'-i\epsilon(\dot{x}+\dot{x}')))} \ .
\end{equation}
An analogous calculation to that leading to (\ref{eqn:2dimcorrefunc}), noting that
on a timelike worldline $Im[k(t-t'-i\epsilon(\dot{t}+\dot{t}'))
\pm{k}(x-x'-i\epsilon(\dot{x}+\dot{x}'))]<0$,
gives
\begin{equation}
\label{eqn:corrfun2dimsch}
\langle{0}|\phi(\tau)\phi(\tau')|0\rangle
= 
-\frac{1}{4\pi}\ln((t-t'-i\epsilon(\dot{t}+\dot{t}'))^2
-(x-x'-i\epsilon(\dot{x}+\dot{x}'))^2)+C \ ,
\end{equation}
where $C$ is an ill-defined constant.

\section{Inertial detector}

Next we consider an inertial detector in $2$ dimensions, firstly
for a finite time detector switched on instantaneously.
From (\ref{eq:transitiontindep}) and (\ref{eqn:corrfun2dimsch})
the transition rate is
\begin{equation}
\dot{F}_{\Delta\tau}(\omega)
=  
-\frac{1}{4\pi}
\int^{\Delta\tau}_{-\Delta\tau}ds\,
e^{-i\omega{s}}
\left(
\ln((s-2i\epsilon)^2)
+
C
\right) \ ,
\end{equation}
where $\Delta\tau=\tau-\tau_0$.
The contribution of the constant is
\begin{eqnarray}
\dot{F}_{\Delta\tau}(\omega)
& = & 
C
\int^{\Delta\tau}_{-\Delta\tau}ds\,
e^{-i\omega{s}} \ ,
\nonumber
\\
& = & 
\frac{2C}{\omega}\sin(\omega(\Delta\tau)) \ .
\end{eqnarray}
Thus for instant switching constant terms in the Wightman 
function do contribute to the transition rate, even in the
limit as $(\tau-\tau_0)\rightarrow\infty$.

Consider then the transition rate for a detector switched 
smoothly with an exponential switching function.
\begin{eqnarray}
\dot{F}_{\Delta\tau}(\omega)
& = & 
C
\int^{\infty}_{-\infty}ds\,
e^{-i\omega{s}-\frac{|s|}{\Delta\tau}} \ ,
\nonumber
\\
& = & 
C
\left(
\int^{\infty}_{0}ds\,
e^{-i\omega{s}-\frac{s}{\Delta\tau}}
+
\int^{\infty}_{0}ds\,
e^{i\omega{s}-\frac{s}{\Delta\tau}}
\right) \ ,
\nonumber
\\
& = &
\frac{2C\Delta\tau}{(1+\Delta\tau^2\omega^2)} \ .
\end{eqnarray}
Again for finite time detections the constant terms do contribute but in the 
limit as the time of detection $\tau-\tau_0$ goes to infinity these 
contributions vanish. Further as the constant terms are infinite it seems
sensible only to consider the rate of a smoothly switched detector in
this asymptotic limit.
The remaining transition rate is then
\begin{eqnarray}
\dot{F}_{\Delta\tau}(\omega)
& = & 
-\frac{1}{4\pi}
\int^{\infty}_{-\infty}ds\,
e^{-i\omega{s}-\frac{|s|}{\Delta\tau}}
\ln((s-2i\epsilon)^2) \ ,
\nonumber
\\
& = & 
-\frac{1}{4\pi}
\left(
\int^{\infty}_{0}ds\,
e^{-i\omega{s}-\frac{s}{\Delta\tau}}
\ln((s-2i\epsilon)^2)
+
\int^{\infty}_{0}ds\,
e^{i\omega{s}-\frac{s}{\Delta\tau}}
\ln((s+2i\epsilon)^2)
\right) \ ,
\nonumber
\\
& = &
-\frac{1}{4\pi}
\left[
\frac{2\ln(2i\epsilon)}{i\omega+\frac{1}{\Delta\tau}}-
\frac{2\ln(2i\epsilon)}{i\omega-\frac{1}{\Delta\tau}}
+
\frac{2}{i\omega+\frac{1}{\Delta\tau}}
\int_0^\infty
ds\,
\frac{e^{-i\omega{s}-\frac{s}{\Delta\tau}}}
{(s-2i\epsilon)}
\right.
\nonumber
\\
&  & 
\left.
-
\frac{2}{i\omega-\frac{1}{\Delta\tau}}
\int_0^\infty
ds\,
\frac{e^{i\omega{s}-\frac{s}{\Delta\tau}}}
{(s+2i\epsilon)}
\right] \ ,
\end{eqnarray}
where the last line follows by parts.
As we are only interested in the asymptotic response
we now let $\Delta\tau\rightarrow\infty$.
Then 
\begin{equation}
\dot{F}_\tau(\omega)
=
-\frac{1}{2\pi{i}\omega}
\int^{\infty}_{-\infty}ds\,
\frac{e^{-i\omega{s}}}{(s-2i\epsilon)} \ .
\end{equation}
The integral may be done simply by residues closing the
contour in the upper half complex plane for $\omega<0$ and
the lower half plane for $\omega>0$. The result is
\begin{equation}
\dot{F}_\tau(\omega)
=
-\frac{1}{\omega}
\Theta(-\omega) \ .
\end{equation}

Finally we consider the relation with the local density
of states (\ref{eqn:localdensity}) which in two dimensions
is
\begin{eqnarray}
\rho_M(\omega,x)
& = & 
\int^\infty_{-\infty}
\frac
{dk}
{2\pi}\,
\delta(|k|-\omega)
=
\int^\infty_0
\frac
{dk}
{2\pi}\,
\delta(k-\omega)
+
\int^0_{-\infty}
\frac
{dk}
{2\pi}\,
\delta(-k-\omega) \ ,
\nonumber
\\
& = & \frac{1}{\pi}\Theta(\omega) \ .
\end{eqnarray}
Thus
\begin{equation}
\dot{F}_\tau(\omega)=
-\frac{\pi}{\omega}
\Theta(-\omega)
\rho_M(|\omega|,x) \ .
\end{equation}

\section{Uniformly accelerated detector}

Now let us consider the response of a uniformly 
accelerated detector in $2$ dimensions. 
Again due to the infinite contributions from
the constant terms in the correlation function we
consider only the asymptotic limit of a smoothly switched 
detector. The transition rate, dropping the constant terms
which do not contribute in the asymptotic limit, is
\begin{eqnarray}
\dot{F}_{\Delta\tau}(\omega)
& =  & 
-\frac{1}{4\pi}
\int^{\infty}_{-\infty}ds\,
e^{-i\omega{s}-\frac{|s|}{\Delta\tau}}
\ln\left(4\left(\alpha\sinh\left(\frac{s}{2\alpha}\right)-i\epsilon\cosh\left(\frac{s}{2\alpha}\right)\right)^2\right) \ ,
\nonumber
\\
& =  & 
-\frac{1}{4\pi}
\left[
\int^{\infty}_{0}ds\,
e^{-i\omega{s}-\frac{s}{\Delta\tau}}
\ln\left(4\left(\alpha\sinh\left(\frac{s}{2\alpha}\right)-i\epsilon\cosh\left(\frac{s}{2\alpha}\right)\right)^2\right)
\right.
\nonumber
\\
&   &
\left.
+
\int^{\infty}_{0}ds\,
e^{i\omega{s}-\frac{s}{\Delta\tau}}
\ln\left(4\left(\alpha\sinh\left(\frac{-s}{2\alpha}\right)-i\epsilon\cosh\left(\frac{s}{2\alpha}\right)\right)^2\right)
\right]
 \ .
\end{eqnarray}
Integrating by parts twice and then taking the $\Delta\tau\rightarrow\infty$ limit
we find 
\begin{equation}
\label{acceltran2dim}
\dot{F}_{\tau}(\omega)
=
-\frac{(\alpha^2+\epsilon^2)}{8\pi\omega^2\alpha^2}
\int^{\infty}_{-\infty}
ds\,
\frac{e^{-i\omega{s}}}{\left(\alpha\sinh\left(\frac{s}{2\alpha}\right)
-i\epsilon\cosh\left(\frac{s}{2\alpha}\right)\right)^2} \ .
\end{equation}
Comparing (\ref{acceltran2dim}) with (\ref{transaccel})
when $\Delta\tau\rightarrow\infty$ we see that
the response is proportional to the response of a 
uniformly accelerated detector in $4$-dimensional 
Minkowski space. In $4$ dimensions we have already recovered the 
usual thermal response (\ref{thermal}) thus in $2$ dimensions 
the response is
\begin{equation}
\dot{F}_{\tau}(\omega)
=
-\frac{1}{\omega(e^{2\pi\alpha\omega}-1)} \ .
\end{equation}

\chapter{Integrals on uniformly accelerated worldline}
\label{ch:appendixB}

\begin{figure}[htbp]      
\includegraphics{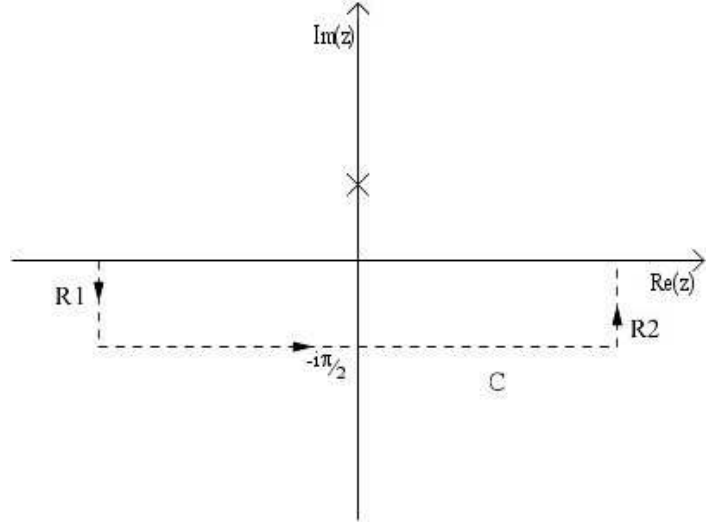}
\centering
\caption[Contour used in the integration of (\ref{eqn:contiq}).]{Contour used in the integration of $I(q)$.}
\label{fig:contour}
\end{figure}

In this appendix we evaluate the integral
\begin{equation}
I(\omega)=
\lim_{\epsilon\rightarrow{0^+}}
\int_{-\infty}^{\infty}
ds
\,
\frac{e^{-i\omega{s}}}
{(\alpha\sinh(\beta{s})-i\epsilon\cosh(\beta{s}))^k} \ ,
\end{equation}
for arbitrary $\alpha>0$, $\beta>0$, $\omega\in{\R}$ and $k\in{\Z}^+$.
Writing $s=z/\beta$, $\epsilon=\alpha\eta$, $q=\omega/\beta$, we 
may instead consider
\begin{equation}
\label{eqn:contiq}
I(q)=
\lim_{\eta\rightarrow{0^+}}
\frac{1}{\beta\alpha^k}
\int_{-\infty}^{\infty}
dz
\,
\frac{e^{-iqz}}
{(\sinh(z)-i\eta\cosh(z))^k} \ .
\end{equation}
The integrand has a pole at $z=\mathrm{arctanh}(i\eta)$ of order $k$.
We consider the integral as a contour integral and deform the contour to that
shown in figure \ref{fig:contour}.
Now we are free to take $\lim_{\eta\rightarrow{0^+}}$. Further we may easily show that in the
limit as the contour goes off to $\pm\infty$ the contributions from $R1$ and $R2$ vanish and so
\begin{equation}
I(q)=
\frac{1}{\beta\alpha^k}
\int_C
dz
\,
\frac{e^{-iqz}}
{(\sinh(z))^k} \ ,
\end{equation}
where $C$ is the line $z=-i\pi/2$ in the complex $z$-plane.
Changing the integration variable by $z=W-i\pi/2$ gives
\begin{eqnarray}
I(q) & = &
\frac{e^{-\pi{q}/2}}{\beta\alpha^k(-i)^k}
\int_{-\infty}^{\infty}
dW
\,
\frac{e^{-iqW}}
{(\cosh(W))^k} \ ,
\nonumber
\\
& = &
\frac{e^{-\pi{q}/2}}{\beta\alpha^k(-i)^k}
\left(
\int_{-\infty}^{\infty}
dW
\,
\frac{\cos(qW)}
{(\cosh(W))^k}
-i
\int_{-\infty}^{\infty}
dW
\,
\frac{\sin(qW)}
{(\cosh(W))^k}
\right) \ .
\end{eqnarray}
In the second integral the integrand is odd in $W$ and so the integral
is $0$. The first integral may be found in \cite{gr:inttables}, with the
result
\begin{equation}
I(q)=
\frac{2^{k-1}e^{-\pi{q}/2}}{\Gamma(k)\beta\alpha^k(-i)^k}
\Gamma((k+iq)/2)
\Gamma((k-iq)/2) \ .
\end{equation}
We may further use the properties of $\Gamma$ functions \cite{gr:inttables}, 
\begin{eqnarray}
\Gamma(z+1) & = & z\Gamma(z) \ ,
\nonumber
\\
\Gamma(1-z)\Gamma(z) & = & \frac{\pi}{\sin{\pi{z}}} \ ,
\nonumber
\\
\Gamma(1/2+z)\Gamma(1/2-z) & = & \frac{\pi}{\cos{\pi{z}}} \ ,
\end{eqnarray}
to write the $\Gamma$ functions that enter the 
transition rate (\ref{transaccel}) in terms of products of polynomials
to obtain the expression (\ref{thermal}).

\chapter{Details for automorphic field detector}
\label{ch:appendixd}

In this appendix we give some more detail on some statements made in 
section \ref{sec:autodetector} regarding automorphic fields.

The automorphic field on Minkowski space, which represents fields
on $M/\Gamma$ is defined by
\begin{equation}
\hat{\phi}(x)
:=
\frac{1}{\left(\sum_{\gamma\in\Gamma}p(\gamma)^2\right)^{1/2}}
\sum_{\gamma\in\Gamma}p(\gamma)\phi(\gamma^{-1}x) \ ,
\end{equation}
where $p(\gamma)$ is a representation of $\Gamma$ in $SL(\R)$.
Then
\begin{equation}
\left[\hat{\phi}(x),\dot{\hat{\phi}}(x')\right]
=
\frac{1}{\sum_{g\in\Gamma}p(g)^2}
\sum_{g\in\Gamma}
\sum_{h\in\Gamma}
p(g)p(h)
\left[\phi(g^{-1}x),\dot{\phi}(h^{-1}x')\right] \ .
\end{equation}
Now writing $h\rightarrow{hg}$ we have, at equal times
\begin{eqnarray}
\label{eqn:commuauto}
\left[\hat{\phi}(x),\dot{\hat{\phi}}(x')\right]
& = & 
\frac{1}{\sum_{g\in\Gamma}p(g)^2}
\sum_{g\in\Gamma}
\sum_{h\in\Gamma}
p(g)p(hg)
[\phi(g^{-1}x),\dot{\phi}(g^{-1}h^{-1}x')] \ ,
\nonumber
\\
& = & 
\frac{1}{\sum_{g\in\Gamma}p(g)^2}
\sum_{g\in\Gamma}
p(g)^2
\sum_{h\in\Gamma}
p(h)
[\phi(x),\dot{\phi}(h^{-1}x')] \ ,
\nonumber
\\
& = &
\sum_{h\in\Gamma}p(h)
[\phi(x),\dot{\phi}(h^{-1}x')]=i\delta^{(d-1)}(x-x')+\mathrm{image\;terms} \ .
\end{eqnarray}

Next consider the two-point function
$\langle{0}|\hat{\phi}(\tau)\hat{\phi}(\tau')|0\rangle$,
where 
\begin{equation}
\label{eqn:smearedautofield}
\hat{\phi}(\tau)
=
\int
d^{d-1}\xi
\,
W_\epsilon(\mathbf{\xi})
\hat{\phi}(x(\tau,\mathbf{\xi})) \ .
\end{equation}
With a similar calculation we find
\begin{equation}
\langle{0}|\hat{\phi}(\tau)\hat{\phi}(\tau')|0\rangle
=
\frac{1}{\sum_{g\in\Gamma}p(g)^2}
\sum_{g\in\Gamma}
\sum_{h\in\Gamma}
p(g)p(h)
\langle{0}|\phi(g^{-1}\tau)\phi(h^{-1}\tau')|0\rangle \ .
\end{equation}
Writing $h\rightarrow{hg}$ and following through the
steps which lead to (\ref{eqn:commuauto}) we have
\begin{equation}
\langle{0}|\hat{\phi}(\tau)\hat{\phi}(\tau')|0\rangle
=
\sum_{h\in\Gamma}p(h)
\langle{0}|\phi(\tau)\phi(h^{-1}\tau')|0\rangle
 \ .
\end{equation}
That is, the two-point function for the smeared 
automorphic field (\ref{eqn:smearedautofield})
is given by the method of images applied to the
two point function for the smeared Minkowski field.

\chapter{Evaluating (\ref{eqn:boundpartmboundinet}) for inertial detector approaching boundary}
\label{ch:appendixC}

\begin{figure}[htbp]      
\includegraphics{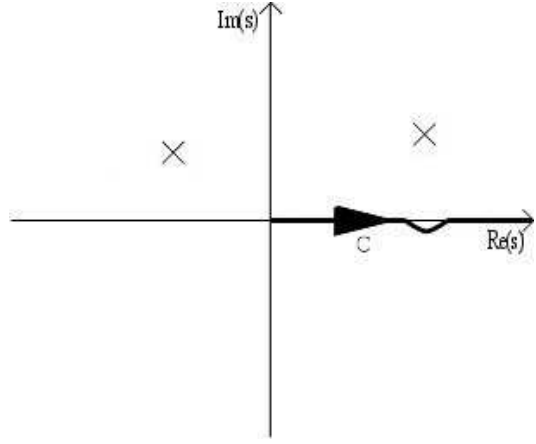}
\centering
\caption[Contour used in the integration of (\ref{eqn:contofint}).]{Contour used in the integration. The right
 hand pole is at $s=(2i\epsilon+2v\tau)/(1+{v})$, the left
at $s=(2i\epsilon-2v\tau)/(1-{v})$. Recall that $-\infty<\tau<0$ and $-1<v<0$.}
\label{fig:contour3}
\end{figure}

In this appendix we evaluate the integral 
\begin{equation}
\label{eqn:contofint}
\dot{F}_\tau(\omega)=
-\frac{\beta}{2\pi^2}
(1-v^2)
\lim_{\epsilon\rightarrow{0^+}}
\int^{\infty}_{0}ds\,
Re
\left(
\frac
{e^{-i\omega{s}}}
{(s-2i\epsilon)^2-v^2(s-2\tau)^2}
\right) \ ,
\end{equation}
with $-1<v<0$ and $-\infty<\tau<0$.
The integrand has two poles, at $s=(2i\epsilon\pm2v\tau)/(1\pm{v})$. When the limit is taken 
only the pole with $(+)$ lies in the integration region. We consider the integration
before the limit is taken as a contour integral and we deform the contour as shown in figure \ref{fig:contour3}.
Then we are free to take the limit and obtain
\begin{equation}
\label{eqn:F2}
\dot{F}_\tau(\omega)=
-\frac{\beta}{2\pi^2}
\int_{C}ds\,
Re
\left(
\frac
{e^{-i\omega{s}}}
{(s+c)(s-b)}
\right) \ ,
\end{equation}
where $c=2v\tau/(1-v)$ and $b=2v\tau/(1+v)$. 

We compute the integral in (\ref{eqn:F2}) by rotating the 
contour onto the imaginary $s$ axis. For $\omega>0$ we rotate to the negative 
imaginary axis, $s\rightarrow{-it}$.
We find after separating the integrand by partial fractions
\begin{eqnarray}
\dot{F}_\tau(\omega) & = & 
-\frac{\beta}{2\pi^2(c+b)}
\int_{0}^\infty
dt
\,
Re
\left(
\frac
{e^{-\omega{t}}}
{(t-ib)}-
\frac
{e^{-\omega{t}}}
{(t+ic)}
\right) \ ,
\nonumber
\\
& = & 
-\frac{\beta}{2\pi^2(c+b)}
\int_{0}^\infty
dt
\,
\frac
{e^{-\omega{t}}t}
{(t^2+b^2)}-
\frac
{e^{-\omega{t}}t}
{(t^2+c^2)} \ .
\end{eqnarray}
The integrals may be computed in terms of sine and cosine integrals
\cite{gr:inttables},
giving the result (\ref{eqn:inertonbound}). For $\omega<0$ we rotate to the positive
imaginary axis $s\rightarrow{it}$. The method is similar to that above
except now a contribution is picked up from a pole on the positive real axis.
This contribution is given by $2\pi{i}\textrm{Res}(e^{-i\omega{s}}/((s+c)(s-b)),s=b)$.
The residue is easily found as $s=b$ is a simple pole. 
The result is as shown in (\ref{eqn:inertonbound}).

\chapter{Asymptotic behaviour of (\ref{eqn:boundaryaccel}) and (\ref{eqn:m-accelbound})}
\label{ch:appendixe}

In this appendix we show that (\ref{eqn:boundaryaccel}) of section \ref{sec:minkbound}
consists of a function periodic in $\tau$ with period $\pi/\omega$ plus
a function $B_\tau(\omega)$ bounded in absolute value by 
$\frac{\beta\Gamma(d/2-1)}{(d-2)\pi^{d/2}\alpha^{d-3}}e^{-\frac{(d-2)\tau}{\alpha}}$.
In particular we show when $d=4$ (\ref{eqn:boundaryaccel}) is given by
\begin{equation}
\dot{F}_{B\tau}(\omega)=
\frac{\beta\omega\cos(2\tau\omega)}{2\pi\sinh(\omega\pi\alpha)}+B_\tau(\omega) \ .
\end{equation}
Further we present an analytic proof that when the detector is
switched on at a finite $\tau_0$ (\ref{eqn:boundaryaccel}) and 
(\ref{eqn:m-accelbound}) (with the
upper limit replaced by $\tau-\tau_0$) vanish as $\tau\rightarrow\infty$.

In this appendix we also discuss the behaviour of (\ref{eqn:m-accelbound})
as $\tau\rightarrow\infty$. We show that (\ref{eqn:m-accelbound}) also consists
of a function periodic in $\tau$ with period $\pi/\omega$ plus a function
which decays exponentially as $\tau\rightarrow\infty$.

Consider (\ref{eqn:boundaryaccel}),
\begin{equation}
\dot{F}_{B\tau}(\omega)=
\frac{\beta\Gamma(d/2-1)}
{2\pi^{d/2}}
\int^{\infty}_{0}ds\,
\frac{\cos(\omega{s})}
{(2\alpha\cosh(\frac{2\tau-s}{2\alpha}))^{d-2}} \ .
\end{equation}
Changing the integration variables by $s=2\alpha{x}+2\tau$ gives
\begin{eqnarray}
\dot{F}_{B\tau}(\omega) & = & 
\frac{\beta\Gamma(d/2-1)}
{2\pi^{d/2}(2\alpha)^{d-3}}
\int^{\infty}_{-\tau/\alpha}dx\,
\frac{\cos(\omega(2\alpha{x}+2\tau))}
{\cosh^{d-2}x} \ ,
\nonumber
\\
& = & 
\frac{\beta\Gamma(d/2-1)}
{2\pi^{d/2}(2\alpha)^{d-3}}
\left(
\int^{\infty}_{-\infty}dx\,
\frac{\cos(\omega(2\alpha{x}+2\tau))}
{\cosh^{d-2}x}-
\int^{-\tau/\alpha}_{-\infty}dx\,
\frac{\cos(\omega(2\alpha{x}+2\tau))}
{\cosh^{d-2}x}
\right) \ ,
\nonumber
\\
& = & 
\frac{\beta\Gamma(d/2-1)}
{2\pi^{d/2}(2\alpha)^{d-3}}
\Bigg(
\cos(2\tau\omega)\int^{\infty}_{-\infty}dx\,
\frac{\cos(2\alpha\omega{x})}
{\cosh^{d-2}x}
-
\sin(2\tau\omega)\int^{\infty}_{-\infty}dx\,
\frac{\sin(2\alpha\omega{x})}
{\cosh^{d-2}x}
\nonumber
\\
&   &
-
\int^{-\tau/\alpha}_{-\infty}dx\,
\frac{\cos(\omega(2\alpha{x}+2\tau))}
{\cosh^{d-2}x}
\Bigg) \ .
\label{eqn:G3}
\end{eqnarray}
In the last expression the integrand in the second term is odd in $x$ and hence integrates
to $0$. The first term is periodic in $\tau$ with period $\pi/\omega$, further the integral
may be performed in terms of Gamma functions \cite{gr:inttables}.
When $d=4$ the result is
\begin{equation}
\label{eqn:G4}
\dot{F}_{B\tau}(\omega)=
\frac{\beta\omega\cos(2\tau\omega)}{2\pi\sinh(\omega\pi\alpha)}
 -
\frac{\beta}
{4\pi^2\alpha}
\int^{-\tau/\alpha}_{-\infty}dx\,
\frac{\cos(\omega(2\alpha{x}+2\tau))}
{\cosh^2x} \ .
\end{equation}
Writing $y=-x$, we find that the third term in (\ref{eqn:G3}) is bounded in absolute 
value by
\begin{eqnarray}
\frac{\beta\Gamma(d/2-1)}
{2\pi^{d/2}(2\alpha)^{d-3}}
\int^{\infty}_{\tau/\alpha}dy\,
\frac{1}
{\cosh^{d-2}y}
& \leq & 
\frac{\beta\Gamma(d/2-1)}
{\pi^{d/2}\alpha^{d-3}}
\int^{\infty}_{\tau/\alpha}dy\,
\frac{1}
{e^{(d-2)y}} \ ,
\nonumber
\\
& = & 
\frac{\beta\Gamma(d/2-1)}
{(d-2)\pi^{d/2}\alpha^{d-3}}
{e^{-\frac{(d-2)\tau}{\alpha}}} \ .
\label{eqn:boundsecon}
\end{eqnarray}

Next consider (\ref{eqn:boundaryaccel}) for the detector switched
on instantaneously at finite $\tau_0$,
\begin{equation}
\dot{F}_{B\tau}(\omega)=
\frac{\beta\Gamma(d/2-1)}
{2\pi^{d/2}}
\int^{\tau-\tau_0}_{0}ds\,
\frac{\cos(\omega{s})}
{(2\alpha\cosh(\frac{2\tau-s}{2\alpha}))^{d-2}} \ .
\end{equation}
We consider the behaviour as $\tau\rightarrow{\infty}$ for fixed $\tau_0$.
\begin{eqnarray}
\dot{F}_{B\tau}(\omega)=
\frac{\beta\Gamma(d/2-1)}
{2\pi^{d/2}}
\left(
\int^{\tau-\tau_0}_{-\infty}ds\,
\frac{\cos(\omega{s})}
{(2\alpha\cosh(\frac{2\tau-s}{2\alpha}))^{d-2}}
-
\int^{0}_{-\infty}ds\,
\frac{\cos(\omega{s})}
{(2\alpha\cosh(\frac{2\tau-s}{2\alpha}))^{d-2}}
\right)
\ .
\nonumber
\\
\label{eqn:boundvanish}
\end{eqnarray}
Changing variables by $s=-2\alpha{y}+2\tau$
we have seen in (\ref{eqn:boundsecon}) that 
the second term is bounded in absolute value 
by $\frac{\beta\Gamma(d/2-1)}{(d-2)\pi^{d/2}\alpha^{d-3}}e^{-\frac{(d-2)\tau}{\alpha}}$.
In the first term in (\ref{eqn:boundvanish}) we change
variables by $s=\tau-\tau_0-v$. The result is
\begin{eqnarray}
\frac{\beta\Gamma(d/2-1)}
{2\pi^{d/2}}
\int^{\infty}_{0}dv\,
\frac{\cos(\omega{(\tau-\tau_0-v)})}
{(2\alpha\cosh(\frac{\tau+\tau_0+v}{2\alpha}))^{d-2}} 
& \leq & 
\frac{\beta\Gamma(d/2-1)}
{2\pi^{d/2}}
\int^{\infty}_{0}dv\,
\frac{1}
{(\alpha{e}^{\frac{\tau+\tau_0+v}{2\alpha}})^{d-2}} \ ,
\nonumber
\\
& = & 
\frac{\beta\Gamma(d/2-1)}
{(d-2)\pi^{d/2}\alpha^{d-3}}
{e^{-\frac{(d-2)(\tau+\tau_0)}{2\alpha}}} \ .
\end{eqnarray}
This term is therefore also bounded by a function
which exponentially decays as $\tau\rightarrow\infty$.
Hence, for the detector switched
on instantaneously at finite $\tau_0$,
$\dot{F}_{B\tau}(\omega)\rightarrow{0}$
as $\tau\rightarrow\infty$.

Consider now (\ref{eqn:m-accelbound}), that is
\begin{eqnarray}
\dot{F}_{I\tau}(\omega)
& = & 
\frac{1}{2\pi^2}
\sum_{n=-\infty}^\infty
\int^{\infty}_{0}ds\,
\frac{\cos(\omega{s})}
{4\alpha^2\cosh^2\left({\frac{2\tau-s}{2\alpha}}\right)+4y_0^2+a^2(2n-1)^2} \ ,
\nonumber
\\
& = & 
\frac{1}{2\pi^2}
\sum_{n=-\infty}^\infty
\left(
\int^{\infty}_{-\infty}ds\,
-
\int^{0}_{-\infty}ds\,
\right)
\frac{\cos(\omega{s})}
{4\alpha^2\cosh^2\left({\frac{2\tau-s}{2\alpha}}\right)+4y_0^2+a^2(2n-1)^2} \ .
\nonumber
\\
\label{eqn:sumint}
\end{eqnarray}
Changing integration variables by $s=2\alpha{x}+2\tau$ we find
the first term in (\ref{eqn:sumint}) becomes
\begin{equation}
\frac{2\alpha}
{2\pi^2}
\cos(2\tau\omega)
\sum_{n=-\infty}^{\infty}
\int_{-\infty}^{\infty}
dx\,
\frac{\cos(2\alpha\omega{x})}{4\alpha^2\cosh^2x+y_0^2+a^2(2n-1)^2} \ ,
\end{equation}
which is periodic in $\tau$ with period $\pi/\omega$.
Changing coordinates by $s=-2\alpha{z}$ the second term
in (\ref{eqn:sumint}) becomes
\begin{equation}
\label{eqn:bounded2}
I(\omega)
= 
-\frac{1}{2\alpha\pi^2}
\sum_{n=1}^\infty
\int_0^\infty
dz\,
\frac{\cos(2\alpha\omega{z})}
{\cosh^2\left(z+\frac{\tau}{\alpha}\right)
+\left(\frac{y_0}{\alpha}\right)^2
+\left(\frac{a}{2\alpha}\right)^2(2n-1)^2} \ .
\end{equation}
Thus
\begin{eqnarray}
|I(\omega)|
& \leq &
\frac{1}{2\alpha\pi^2}
\left(
\int_0^\infty
dz\,
\frac{1}
{\cosh^2\left(z+\frac{\tau}{\alpha}\right)
+\left(\frac{y_0}{\alpha}\right)^2
+\left(\frac{a}{2\alpha}\right)^2}
\right. 
\nonumber
\\
&     &+\left.
\sum_{n=2}^\infty
\int_0^\infty
dz\,
\frac{1}
{\cosh^2\left(z+\frac{\tau}{\alpha}\right)
+\left(\frac{y_0}{\alpha}\right)^2
+\left(\frac{a}{2\alpha}\right)^2(2n-1)^2}
\right) \ .
\label{eqn:G11}
\end{eqnarray}
It is simple to show that the first term is 
(\ref{eqn:G11}) is bounded in absolute value by
$\frac{1}{\alpha\pi^2}e^{-2\tau/\alpha}$.
The integrand in the second term of (\ref{eqn:G11}), call it 
$I_2(\omega)$, is a monotonically decreasing function of $n$. We can
therefore bound the sum by an integral
\begin{eqnarray}
I_2(\omega)
& \leq & 
\frac{1}{2\alpha\pi^2}
\int_0^\infty
dz\,
\int_1^\infty
dn\,
\frac{1}{\cosh^2\left(z+\frac{\tau}{\alpha}\right)
+\left(\frac{y_0}{\alpha}\right)^2
+\left(\frac{a}{2\alpha}\right)^2(2n-1)^2} \ ,
\nonumber
\\
& \leq &
\frac{1}{2\alpha\pi^2}
\int_0^\infty
dz\,
\int_{1/2}^\infty
dn\,
\frac{1}{\cosh^2\left(z+\frac{\tau}{\alpha}\right)
+\left(\frac{y_0}{\alpha}\right)^2
+\left(\frac{a}{2\alpha}\right)^2(2n-1)^2} \ .
\nonumber
\\
\end{eqnarray}
Changing variables by $n=\frac{1}{2}+\frac{\alpha}{a}v$
\begin{eqnarray}
I_2(\omega)
& \leq & 
\frac{1}{2a\pi^2}
\int_0^\infty
dz\,
\int_0^\infty
dv\,
\frac{1}{\cosh^2\left(z+\frac{\tau}{\alpha}\right)
+\left(\frac{y_0}{\alpha}\right)^2
+v^2} \ ,
\nonumber
\\
& = &
\frac{1}{4a\pi}
\int_0^\infty
dz\,
\frac{1}{\left(\cosh^2\left(z+\frac{\tau}{\alpha}\right)
+\left(\frac{y_0}{\alpha}\right)^2\right)^{1/2}} \ ,
\nonumber
\\
& \leq & 
\frac{1}{2a\pi}
\int_0^\infty
dz\,
e^{-(z+\tau/\alpha)} 
=
\frac{1}{2a\pi}
e^{-\tau/\alpha} \ . 
\label{eqn:boundsum}
\end{eqnarray}
Thus we see that $I(\omega)\leq\frac{1}{\alpha\pi^2}e^{-2\tau/\alpha}+\frac{1}{2\pi{a}}e^{-\tau/\alpha}$
and so $I(\omega)$ vanishes as $\tau\rightarrow\infty$ and (\ref{eqn:m-accelbound}) becomes
periodic in $\tau$ with period $\pi/\omega$ at late times.

Finally consider (\ref{eqn:m-accelbound}) when the detector is switched on at $\tau_0>-\infty$, 
that is
\begin{eqnarray}
\dot{F}_{I\tau}(\omega)
& = & 
\frac{1}{2\pi^2}
\sum_{n=-\infty}^\infty
\int^{\tau-\tau_0}_{0}ds\,
\frac{\cos(\omega{s})}
{4\alpha^2\cosh^2\left({\frac{2\tau-s}{2\alpha}}\right)+4y_0^2+a^2(2n-1)^2} \ ,
\nonumber
\\
& = & 
\frac{1}{2\pi^2}
\sum_{n=-\infty}^\infty
\left(
\int^{\tau-\tau_0}_{-\infty}ds\,
-
\int^{0}_{-\infty}ds\,
\right)
\frac{\cos(\omega{s})}
{4\alpha^2\cosh^2\left({\frac{2\tau-s}{2\alpha}}\right)+4y_0^2+a^2(2n-1)^2} \ .
\nonumber
\\
\label{eqn:sumintfixed0}
\end{eqnarray}
We consider the limit as $\tau\rightarrow\infty$ for fixed $\tau_0$.
We saw in the calculation (\ref{eqn:bounded2}) to (\ref{eqn:boundsum}) that 
the second term in (\ref{eqn:sumintfixed0}) is bounded in absolute value by 
$\frac{1}{\alpha\pi^2}e^{-2\tau/\alpha}+\frac{1}{2\pi{a}}e^{-\tau/\alpha}$. Consider
then the first term in (\ref{eqn:sumintfixed0}). With a change of variables
by $s=\tau-\tau_0-v$ this term becomes
\begin{eqnarray}
I(\omega)
& = & 
\frac{1}{2\pi^2}
\sum_{n=-\infty}^\infty
\int^{\infty}_{0}dv\,
\frac{\cos(\omega{(\tau-\tau_0-v)})}
{4\alpha^2\cosh^2\left({\frac{\tau+\tau_0+v}{2\alpha}}\right)+4y_0^2+a^2(2n-1)^2} \ ,
\nonumber
\\
& = & 
\frac{1}{4\pi^2\alpha^2}
\sum_{n=1}^\infty
\int^{\infty}_{0}dv\,
\frac{\cos(\omega{(\tau-\tau_0-v)})}
{\cosh^2\left({\frac{\tau+\tau_0+v}{2\alpha}}\right)+\left(\frac{y_0}{\alpha}\right)^2+\left(\frac{a}{2\alpha}\right)^2(2n-1)^2} \ ,
\nonumber
\\
& = & 
\frac{1}{4\pi^2\alpha^2}
\left[
\int^{\infty}_{0}dv\,
\frac{\cos(\omega{(\tau-\tau_0-v)})}
{\cosh^2\left({\frac{\tau+\tau_0+v}{2\alpha}}\right)+\left(\frac{y_0}{\alpha}\right)^2+\left(\frac{a}{2\alpha}\right)^2} \right.
\nonumber
\\
&   &
+
\left.
\sum_{n=2}^\infty
\int^{\infty}_{0}dv\,
\frac{\cos(\omega{(\tau-\tau_0-v)})}
{\cosh^2\left({\frac{\tau+\tau_0+v}{2\alpha}}\right)+\left(\frac{y_0}{\alpha}\right)^2+\left(\frac{a}{2\alpha}\right)^2(2n-1)^2} \right] \ .
\label{eqn:bounded333}
\end{eqnarray}
It is a simple matter to show the first term in (\ref{eqn:bounded333}) is bounded in
absolute value by $\frac{e^{-\frac{\tau+\tau_0}{\alpha}}}{\pi^2\alpha}$.
In the second term in (\ref{eqn:bounded333}), call it $I_2(\omega)$,
the integrand is a monotonically decreasing function of $n$. The sum
can be bounded by an integral and we find
\begin{equation}
|I_2(\omega)|\leq
\frac{1}{4\pi^2\alpha^2}
\int^{\infty}_{0}dv\,
\int^{\infty}_{1/2}dn\,
\frac{\cos(\omega{(\tau-\tau_0-v)})}
{\cosh^2\left({\frac{\tau+\tau_0+v}{2\alpha}}\right)+\left(\frac{y_0}{\alpha}\right)^2+\left(\frac{a}{2\alpha}\right)^2(2n-1)^2} \ .
\end{equation}
Changing variables by $n=\frac{1}{2}+\frac{\alpha}{a}z$
\begin{eqnarray}
|I_2(\omega)| & \leq & 
\frac{1}{4\pi^2\alpha{a}}
\int^{\infty}_{0}dv\,
\int^{\infty}_{0}dz\,
\frac{1}
{\cosh^2\left({\frac{\tau+\tau_0+v}{2\alpha}}\right)+\left(\frac{y_0}{\alpha}\right)^2+z^2} \ ,
\nonumber
\\
& = &
\frac{1}{8\pi\alpha{a}}
\int^{\infty}_{0}dv\,
\frac{1}
{\left(\cosh^2\left({\frac{\tau+\tau_0+v}{2\alpha}}\right)+\left(\frac{y_0}{\alpha}\right)^2\right)^{1/2}} \ ,
\nonumber
\\
& \leq & 
\frac{1}{4\pi\alpha{a}}
\int^{\infty}_{0}dv\,
e^{-\frac{\tau+\tau_0+v}{2\alpha}} \ ,
\nonumber
\\
& = & 
\frac{1}{2\pi{a}}
e^{-\frac{\tau+\tau_0}{2\alpha}} \ .
\end{eqnarray}
Thus finally we see that $|I_2(\omega)|\rightarrow{0}$
as $\tau\rightarrow\infty$, and so (\ref{eqn:m-accelbound})
also vanishes in this limit.

\newpage
\addcontentsline{toc}{chapter}
         {\protect\numberline{Bibliography\hspace{-96pt}}}

\end{document}